\documentclass[useAMS,usenatbib,usegraphicx]{mn2e}
\usepackage{epsfig}
\usepackage{times}

\DeclareRobustCommand{\ion}[2]{%
\relax\ifmmode
\ifx\testbx\f@series
{\mathbf{#1\,\mathsc{#2}}}\else
{\mathrm{#1\,\mathsc{#2}}}\fi
\else\textup{#1\,{\mdseries\textsc{#2}}}%
\fi}
\def\mteff{T_{\rm eff}}

\newcommand{\vsini}{$v \sin i$}
\newcommand{\dfrac}[2]{\frac{\displaystyle #1}{\displaystyle #2}}

\def\teff{\hbox{$\,T_{\rm eff}$}}
\def\kms{\hbox{$\,{\rm km}\,{\rm s}^{-1}$}}
\def\ms{\hbox{$\,{\rm m}\,{\rm s}^{-1}$}}

\def\degr{\hbox{$^\circ$}}
\def\halpha{\hbox{${\rm H}\alpha$}}

\def\bet{\hbox{$\beta$}}

\def\arcmin{\hbox{$^\prime$}}
\def\arcsec{\hbox{$^{\prime\prime}$}}
\def\sun{\hbox{$\odot$}}

\def\mteff{T_{\rm eff}}
\def\teff{\hbox{$\,T_{\rm eff}$}}

\def\logg{\hbox{$\,\log\,g$}}
\def\kms{\hbox{$\,{\rm km}\,{\rm s}^{-1}$}}

\def\ndiii{\hbox{\ion{Nd}{iii}}}
\def\prii{\hbox{\ion{Pr}{ii}}}

\def\euii{\hbox{\ion{Eu}{ii}}}
\def\priii{\hbox{\ion{Pr}{iii}}}

\title[A search for rapid pulsations among 9 luminous Ap stars]{A search
for rapid pulsations among 9 luminous Ap stars\thanks{Based on observations 
collected at the  European Southern Observatory, Paranal, Chile, as part of 
programmes 075.D-0145 (A), 078.D-0080(A), 072.D-0138(A) and 077.D-0150(A).}}
\author[L. M. Freyhammer et al.]
 {L. M. Freyhammer$^{1}$\thanks{E-mail: lmfreyhammer\,@\,uclan.ac.uk},
 D. W. Kurtz$^{1}$, M. S. Cunha$^{2}$, G. Mathys$^{3}$  
 \newauthor{V. G. Elkin$^{1}$ and J.D. Riley$^{1}$} \\
 $^{1}$Centre for Astrophysics, University of Central Lancashire, Preston
 PR1 2HE\\
 $^{2}$Centro de Astrof\'isica da Universidade do Porto, Rua das Estrelas,
 4150 Porto, Portugal\\
 $^{3}$European Southern Observatory, Casilla 19001, Santiago 19, Chile}

\begin{document}

\date{Draft \today ; Accepted . Received ; in original form }

\pagerange{\pageref{firstpage}--\pageref{lastpage}} \pubyear{2002}

\maketitle

\label{firstpage}

\begin{abstract}
  The rapidly oscillating Ap stars are of importance for studying the
  atmospheric structure of stars where the process of chemical element
  diffusion is significant. We have performed a survey for rapid
  oscillations in a sample of 9 luminous Ap stars, selected from their
  location in the colour-magnitude diagram as more evolved
  main-sequence Ap stars that are inside the instability strip for
  rapidly oscillating Ap (roAp) stars.  Until recently this region was
  devoid of stars with observed rapid pulsations.  We used the VLT
  UV-Visual Echelle Spectrograph (UVES) to obtain high time resolution
  spectroscopy to make the first systematic spectroscopic search for
  rapid oscillations in this region of the roAp instability strip.  We
  report 9 null-detections with upper limits for radial-velocity
  amplitudes of $20-65$\,\ms\ and precisions of $\sigma=7-20$\,\ms\
  for combinations of Nd and Pr lines.  Cross-correlations confirm
  these null-results.  At least six stars are magnetic and we provide
  magnetic field measurements for four of them, of which three are
  newly discovered magnetic stars. It is found that four stars have
  magnetic fields smaller than $\sim2$\,kG, which according to
  theoretical predictions might be insufficient for suppressing
  envelope convection around the magnetic poles for more evolved Ap
  stars. Suppression of convection is expected to be essential for the
  opacity mechanism acting in the hydrogen ionisation zone to drive
  the high-overtone roAp pulsations efficiently. Our null-results
  suggest that the more evolved roAp stars may require particularly
  strong magnetic fields to pulsate. Three of the studied stars do,
  however, have magnetic fields stronger than 5\,kG.
\end{abstract}


\section{Introduction}

Why do some stars oscillate while others do not? This question is
particularly relevant for the class of rapidly oscillating Ap (roAp)
stars for which the discovery of high-frequency oscillations came as a
surprise \citep{kurtz82}. The $\delta$\,Scuti pulsations of stars in
this area of the colour-magnitude diagram (CMD) were not theoretically
predicted to produce the observed high-frequency, high-overtone
modes. Subsequently, considerable observational and theoretical
efforts have been able to describe most pulsation properties of roAp
stars through i) the oblique pulsator model \citep{kurtz82}, ii) the
probable driving mechanism of the pulsations \citep{balmforthetal01}
and iii) their link with the magnetic fields present in Ap stars (see,
e.g., \citealt{cunha06}; \citealt{saio05}).

Ap stars are chemically peculiar stars that range from early-B to
early-F spectral types with the majority having detectable magnetic
fields with strengths of a few $10^2$ to a few $10^4$\,G
(\citealt{bychkovetal03}, according to whom $\sim55$ per cent have
mean longitudinal fields above 400\,G).  For the cool Ap stars (also
called CP2 stars), abundance anomalies are typically visible in their
spectra as abnormally strong absorption lines, particularly of
rare-earth elements (REEs). Candidate roAp stars can therefore be
selected photometrically due to the influence of these spectral
features on narrow-band photometric indices such as those of the
Str\"omgren filter system (\citealt{martinez93}, see also
Sect.\,\ref{sec:selection}).

The high-overtone roAp oscillations are thought to be driven by the
opacity-mechanism acting in the partial hydrogen ionisation zone
\citep{dziembowskietal96,balmforthetal01,cunha02,saio05}.  The
frequency range of the excited modes is related to the presence of
strong magnetic fields in roAp stars that i) directly stabilise the
low-frequency, low-order $\delta$\,Scuti oscillations, excited by the
$\kappa$-mechanism acting in the \ion{He}{ii} partial ionisation zone
\citep{saio05}, and ii) indirectly enhance the driving of high-order
oscillations by the $\kappa$-mechanism acting in the partial hydrogen
ionisation zone \citep{balmforthetal01}. In particular, the
stabilisation of the $\delta$\,Scuti modes can be explained by the
dissipation of slow Alfv\'en waves \citep{saio05} and by the
suppression of turbulent motion in the outer layers of the magnetic
pole regions of roAp stars, which speeds up gravitational settling of
helium and drains it from the \ion{He}{ii} ionisation zone
\citep{theado05}.  An implication of the high radial order of the
observed modes of roAp stars is that the observable pulsation
amplitudes become depth dependent.  This pulsational structure
provides a unique opportunity for studying the magneto-acoustic
structure of the pulsations in 3D because of the element
stratification in the atmospheres of these stars. Different elements,
such as Fe, Pr, Nd, Eu and H, act as tracers for the pulsation
structure due to their different properties regarding formation depth
and extent of formation regions. Examples of depth-dependent
measurements of pulsations are now numerous, such as
\citet{kurtzetal05, elkinetal2005b, ryabchikovaetal07}.  The roAp
stars are thus unique objects for studying the interactions among
stellar pulsation, magnetic fields and atomic diffusion. The latter is
important for, e.g., estimates of globular cluster ages where He
settling is significant, pulsation driving mechanisms in sdB and
$\beta$\,Cephei stars where radiative levitation of Fe is needed for
instability, and the Standard Solar Model where both He settling and
radiative levitation of some metals are included.

Only 37 roAp stars are known at present
(\citealt{kurtzetal06b,tiwarietal07}) despite several searches for
rapid pulsation in Ap stars, such as those by \citet{nelsonetal93,
  martinezetal94, handleretal99, ashokaetal00, weissetal00,
  dorokhovaetal05,joshietal06}.  The selection of roAp candidates is
mostly based on photometric indices pointing to chemical peculiarities
in the spectra.  There is at present no clear correlation between
radial velocity (RV) amplitude and photometric amplitude in roAp stars
(see Table~1 of \citealt{kurtzetal06b}).  For a (non-exhaustive) list
of spectroscopic studies of roAp stars, see \citet{kurtzetal06a}.
\citet{cunha02} calculated a theoretical instability strip for roAp
stars and compared it to locations of 16 known roAp stars.  The region
around the terminal age main-sequence is remarkably devoid of
pulsators which, as Cunha points out, could be an observational bias.
She predicted that more luminous and evolved Ap stars may pulsate with
lower frequencies ($0.67 - 0.83$\,mHz) which makes it harder to detect
them in the typically short high-speed observing runs.  Note that this
frequency range overlaps with the highest $\delta$\,Sct frequencies,
such as for HD\,34282 (0.92\,mHz, \citealt{amadoetal04}).
Alternatively, she proposed that this absence of known evolved roAp
stars could reflect a real deficiency in the fraction of pulsators in
that region of the HR diagram, resulting from the fact that the
magnetic field is less likely to suppress envelope convection in more
evolved stars.  The photometric survey by \citet{martinezetal94} was
indeed less sensitive to low-amplitude roAp periods longer than
15\,min and got a null-result for HD\,116114 for which a 21-min
oscillation was recently detected spectroscopically by
\citet{elkinetal05}.  A dedicated survey of luminous Ap stars
therefore seems pertinent.

Ap stars inside the roAp instability strip that do not exhibit any
detectable variability are called non-oscillating Ap (noAp) stars. As
seen in, e.g., the astrometric HR-diagram by \citeauthor{hubrigetal05}
(\citeyear{hubrigetal05}, their figure 2) for roAp and Ap stars, the
apparent noAp stars occupy essentially the same regions as the roAp
stars. However, the noAp stars appear to be systematically more
evolved than the roAp stars \citep{north97,handleretal99,hubrig00}.
To fully understand the mechanism responsible for driving oscillations
in roAp stars, it is essential to confirm and understand why so many
stars with similar characteristics are apparently stable against
pulsations. The apparent absence of oscillations in noAp stars does
not, however, necessarily mean that oscillations are suppressed.  For
instance, detection of oscillations may depend on lifetimes of excited
modes (may be days for some roAp stars, \citealt{handler04}), on
beating between multiple modes, on the geometry of the magnetic
field's orientation and the aspect in which the stellar surface is
observed together with the surface distribution of the chemical
elements used to detect the oscillations.  Further, the
signal-to-noise ratio ($S/N$) of the obtained photometry or
spectroscopy may simply be too low to detect the typically
low-amplitude roAp pulsations.
Finding high precision null-results for a statistically significant
sample of roAp candidates would be a safe basis for concluding whether
noAp stars exist or whether the current lack of known luminous roAp
stars is an observational bias. From a theoretical point of view,
stars located inside the theoretical instability strip may be noAp
stars if the magnetic field is too weak to effectively suppress
convection in the stellar envelope. A good test sample should,
therefore, also include stars with strong magnetic fields.

To try to answer the question of whether the absence of observed
variability among the luminous Ap stars is an observational selection
effect only, or in fact due to intrinsic properties, we have studied 9
such stars at high radial-velocity precision. Our immediate goal with
UVES was to search these luminous Ap stars for roAp oscillations at
high radial velocity precision. Table\,\ref{tab:obslog} lists the
known properties of the 9 targets and gives an observing log for the
collected spectra. We find no pulsation in any of the 9 stars;
Tables\,\ref{tab:cogpow} and \ref{tab:ccpow} below present the
null-results of the frequency analyses.

In the following sections, we describe selection and observation of
the targets along with the data reduction in Sect.\,2, then follows
the data analyses including estimation of physical parameters and the
radial-velocity analysis in Sect.\,3. We show that our analyses reach
the same precision as other radial-velocity studies of roAp stars in
the literature and we then give the results for each target on a
star-by-star basis. Finally we discuss the null-results in Sect.\,4.

\begin{table*}
  \centering
  \begin{minipage}{140mm}
    \caption{\label{tab:obslog}Observing log indicating number of
      observations and the mid-series heliocentric Julian Date for
      each series of spectra. Exposure times were 40\,s, except for
      the fainter stars HD\,110072 and HD\,170565 where 80\,s was
      used; readout and setup times were $24-27$\,s.  The last two
      columns indicate $S/N$ measured in the continuum based on the
      standard deviation in the residual flux from two consecutive
      spectra. {\it Lower} and {\it Upper} respectively refer to
      wavelengths below or above the gap between the two CCDs at
      6000\,\AA, while the ranges in parentheses correspond to the
      $S/N$ range for all spectra in that series. The $S/N$ estimates
      assume random noise only.}
    \begin{tabular}{@{}lrccccll@{}}
      \hline
      Star        &\#spec& $\alpha_{2000.0}$&$\delta_{2000.0}$ &HJD\,obs&Time&S/N 
      Lower& $S/N$ Upper \\ 
      &     & (h:m:s)    & (\degr:\,\arcmin\,:\,\arcsec\,) &(d)&(h)&  &   \\ \hline
      HD\,107107 & 111 & 12:19:04.8 & -40:09:47 & 2453509.504 &1.94 & 79\phantom{1} (77--82)   & 62\phantom{1}  (56--77) \\
      HD\,110072 &  69 & 12:39:50.2 & -34:22:30 & 2453509.591 &2.06 & 50\phantom{1} (43--55)   & 38\phantom{1}  (32--49) \\
      HD\,131750 & 111 & 14:56:20.7 & -30:52:36 & 2453509.681 &1.98 & 80\phantom{1} (74--87)   & 66\phantom{1}  (57--74) \\
      HD\,132322 & 111 & 15:01:36.0 & -63:55:35 & 2453510.519 &2.03 &151            (147--155) & 141            (132--152)\\
      HD\,151301 & 111 & 16:49:28.3 & -54:26:47 & 2453510.715 &1.95 & 82\phantom{1} (76--84)   & 67\phantom{1}  (60--79) \\
      HD\,170565 &  85 & 18:30:08.2 & -02:35:26 & 2453509.778 &2.50 & 89\phantom{1} (83--94)   & 77\phantom{1}  (69--86) \\
      HD\,197417 & 125 & 20:48:48.1 & -72:12:44 & 2453509.879 &2.20 &107            (94--113)  & 93\phantom{1}  (79--110)\\
      HD\,204367 & 111 & 21:28:41.2 & -25:38:39 & 2453510.798 &1.96 &113            (108--120) & 101            (93--115)\\
      HD\,208217 & 138 & 21:56:56.4 & -61:50:44 & 2453510.896 &2.52 &156            (142--165) & 148            (128--159)\\
      \hline
      \hline
    \end{tabular}
  \end{minipage}
\end{table*}

\section{Selection of the candidates, observations and data reduction}

\subsection{Selection}
\label{sec:selection}

\begin{figure*}
  \hspace{-13pt}
  \includegraphics[height=0.95\textwidth, angle=270]{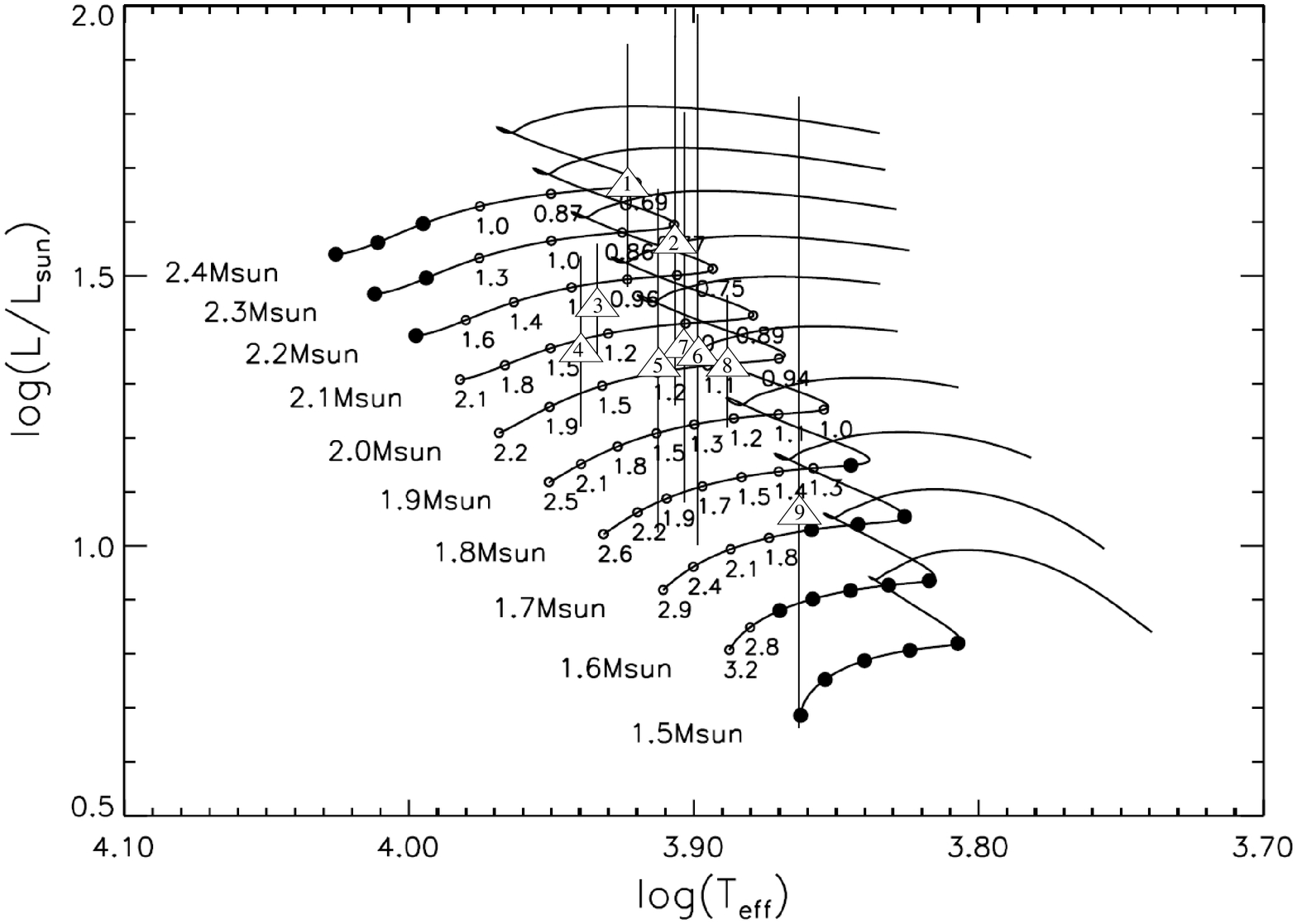}
  \caption{\label{fig:cmd}Colour magnitude diagram with the 9 UVES
    targets superposed and labelled as in Table\,\ref{tab:targets}:
    \#1 (HD\,197417), \#2 (HD\,131750), \#3 (HD\,132322), \#4
    (HD\,204367), \#5 (HD\,107107), \#6 (HD\,170565), \#7
    (HD\,151301), \#8 (HD\,208217) and \#9 (HD\,110072).  Stellar
    evolution tracks are from \citet{christensen-dalsgaard93} and
    circles indicate theoretically predicted locations for stars with
    pulsationally stable (filled) and unstable (open)
    modes. Luminosities are based on {\it Hipparcos} parallaxes and
    the associated uncertainties are indicated with error
    bars. Predicted pulsation frequencies are given in mHz.  }
\end{figure*}

We selected 9 stars based on Str\"omgren and $\beta$ indices from the
Cape catalogue of cool Ap stars by \citet{martinez93}, and
luminosities from {\it Hipparcos} parallaxes. For Str\"omgren
photometry, the $c_1$ index is normally an indicator of luminosity
while $m_1$ is an indicator of metallicity. The $\beta$ index is
sensitive to temperatures for stars in the range A3--F2. Also $b-y$
indicates temperature, but is influenced by reddening. The relative
indices $\delta c_1$ and $\delta m_1$ indicate how a given star
deviates in $c_1$ and $m_1$ from `normal' zero-age main sequence stars
\citep[see also][]{crawford79}. A more negative value of $\delta m_1$
indicates stronger metallicity. For normal stars $\delta c_1$ is a
luminosity indicator, while for Ap stars it is depressed by heavy line
blocking and hence is not a reliable indicator of luminosity.  A
negative $\delta c_1$ is, in fact, indicative of strong
peculiarity. However, because $c_1$ increases with luminosity,
luminous Ap stars may have positive $\delta c_1$ which, without
independent information on the luminosity, makes their chemical
composition appear `normal' in this index. A negative $\delta c_1$ has
frequently been used for target selection in previous searches for
roAp stars; as a consequence, luminous Ap stars have seldom been
tested for pulsation.  Our sample of 9 roAp candidates was therefore
hand-picked among Ap stars with {\it Hipparcos} parallaxes; their main
physical properties are given in Table\,\ref{tab:targets}.
Fig.\,\ref{fig:cmd} shows that the astrometric luminosities
(Table\,\ref{tab:targets}, column\,11) place the chosen sample among
the more evolved Ap stars (towards the end of their core-hydrogen
burning phase). The corresponding temperatures (column 9) are
described in Sect.\,\ref{sec:teffrot}.  The absolute magnitude in the
$V$-band and the stellar luminosity were derived using the standard
relations:
\begin{equation}
  M_V = m_V + 5 + 5\/\log\pi - A_V,
\end{equation}
where the trigonometric parallax $\pi$ is measured in arc seconds and
the interstellar extinction in the $V$-band is
$A_V=4.3\,E(b-y)=4.3\cdot0.74\,E(B-V)$, and
\begin{equation}
  \log\dfrac{L}{L_{\sun}} = - \dfrac{M_V + BC - M_{{\rm bol},\sun}}{2.5},
\end{equation}
where we used $M_{{\rm bol},\sun}=4.72$.  Column 8 in
Table\,\ref{tab:targets} gives reddening read from the maps by
\citet{bursteinetal82} and columns 11 and 12 give the luminosities
derived with and without reddening.  The reddening maps have errors in
$E(B-V)$ of 0.01 mag or 10 per cent (whichever is the greatest) and
resolution of 0.03 mag.  HD\,170565, HD\,132322 and HD\,151301 are
near the galactic plane and outside these maps so a lower limit on
reddening was used. These reddening estimates are in all cases
conservative values but contribute little to the luminosities where
errors are dominated by the errors from the parallaxes. The only
exception is HD\,132322 which has a precise distance. No Lutz-Kelker
correction \citep{lutzetal73} was applied to the luminosities.  The
values for projected rotational velocities and magnetic fields in
Table\,\ref{tab:targets} are described further in
Sects.\,\ref{sec:teffrot} and \ref{sec:mag}.

\begin{table*}
  \caption{\label{tab:targets}
    Main properties of the 9 luminous Ap stars. The columns give:
    ($2-6$) Str\"omgren indices from \citet{martinez93},
    ($7$) the parallax $\pi$ from {\it Hipparcos} -- 
    the values in columns $10-12$ are 
    based on this value,
    (8) reddening from \citet{bursteinetal82} 
    fixed to 0.21 for HD\,170565, HD\,132322 and HD\,151301,
    (9) temperatures derived from the $\beta$ index,
    ($10,~12$) absolute magnitude and luminosity assuming no
    reddening, while luminosity in Col.\,11 accounts for reddening.
    Bolometric correction is from interpolation in the tables of \citet{flower96}. 
    Projected rotation velocities \vsini\ (13) with estimated errors were
    obtained with the measurements of the mean quadratic magnetic field (14). 
    Magnetic field measurements from the literature are specified with superscripts.
    The numbers in the last column (15) associate the stars with labels in 
    Fig.\,\ref{fig:cmd}. 
  }
  
\begin{tabular}{@{}l@{~~}c@{~~}c@{~~}c@{~~}c@{~~}c@{~~}c@{~~}c@{~~}c@{~~}l@{~}c@{~
}c@{}l@{~~}c@{~}c@{}}
\hline
Star    &$V  $ &$b-y$ &$m_1$ &$c_1$ &\bet  &  $\pi$         &$E_{B-V}$ &\teff 
&$M_V$  &$\log L$/L$_\odot$&$\log L$/L$_\odot$& ~\vsini& $ \langle B_{\rm 
  q}\rangle$&Note\\
HD      & mag  & mag  & mag  & mag  & mag  & (mas)          &  mag     & K    
& mag   &\scriptsize E$_{B-V}$$\ne$0& & ~\kms  & kG & \\  \hline
107107 &8.734 &0.074 &0.234 &0.835 &  2.864& $3.42\pm1.11$ &0.10--0.14& 8300 & 1.404 & 1.33 & 1.68 &  $10.5\pm0.5$        & $5.2\pm0.4$ &\#5\\ 
110072 &10.104\phantom{0}&0.227&0.281&0.487&2.759&$2.45\pm1.44$&0.06--0.09&7300&2.050& 1.07 & 1.25 &$\phantom{2}3.3\pm0.5$& $1.5\pm0.6$ &\#9\\ 
131750 &8.549 &0.121 &0.260 &0.710 & 2.847&  $2.86\pm1.21$ &0.06--0.12& 8100 & 0.831 & 1.56 & 1.79 &  $25.3\pm1.8$        & $5.3\pm3.3$ &\#2\\
132322 &7.357 &0.129 &0.227 &0.943 & 2.894&  $5.78\pm0.73$ &$>$0.21   & 8600 & 1.167 & 1.44 & 2.03 &  $34.3\pm1.8$        & $ \le6.0^a$ &\#3\\ 
151301 &8.541 &0.222 &0.244 &0.666 & 2.827&  $3.56\pm1.40$ &$>$0.21   & 8000 & 1.298 & 1.37 & 1.90 &  $13.7\pm1.1$        & $ \le2.4  $ &\#7\\
170565 &9.130 &0.269 &0.253 &0.628 & 2.825&  $2.72\pm1.45$ &$>$0.21   & 7900 & 1.303 & 1.37 & 1.90 &  $18.0\pm3.0$        &     N/A$^b$ &\#6\\ 
197417 &8.009 &0.038 &0.233 &0.939 & 2.877&  $3.23\pm0.84$ &0.03--0.06& 8400 & 0.555 & 1.68 & 1.82 &  $25.5\pm0.4$        & $ \le2.2  $ &\#1\\ 
204367 &7.830 &0.039 &0.220 &0.996 & 2.899&  $5.09\pm0.92$ &0.03--0.06& 8700 & 1.364 & 1.36 & 1.49 &  $10.8\pm0.3$        & $ \le1.4  $ &\#4\\
208217 &7.199 &0.102 &0.274 &0.626 & 2.816&  $6.83\pm0.90$ &$<0.03$   & 7700 & 1.371 & 1.34 & 1.37 &  $10.8\pm0.7$        &$8.0\pm0.5^c$&\#8\\ 
\hline \hline
\multicolumn{15}{l}
{$^a \langle B_{\rm z}\rangle=0.36\pm0.05$\,kG  \citep{hubrigetal06}, 
  $^b \langle B_{\rm z}\rangle=1.76\pm0.17$\, kG \citep{kudryavtsevetal06}, }  
\\
\multicolumn{15}{l}{$^c \langle B\rangle=7.96\pm0.59$\,kG
  \citep{mathysetal97} }                                    \\ 
\end{tabular}
\end{table*}

Fig.\,\ref{fig:cmd} shows the locations of these stars in a
colour-magnitude diagram (CMD) superposed with evolutionary tracks by
\citet{christensen-dalsgaard93} and locations indicated for stars that
are predicted to be pulsationally stable or unstable
(\citealt{cunha02}). Luminosities are from distances based on {\it
  Hipparcos} measurements \citep{hip} with considerable errors from
the parallaxes indicated. Temperatures were estimated from the grids
by \citet{moonetal85} based on $\beta$ photometry \citep{martinez93}
alone.

\subsection{Observations and data reduction}

Our spectroscopic observations were obtained at the Very Large
Telescope (VLT) using the Ultraviolet and Visual Echelle Spectrograph
(UVES). The fast readout mode of 625 kpix\,s$^{-1}$ was used for
reading out the two CCD-mosaics on the red arm of UVES in the 4-port
low gain mode. We observed our 9 targets about 2\,h each (see the
observing log in Table\,1) on the two nights of $18 - 19$ May and $19
- 20$ May 2005 (JD2453509 and JD2453510), using exposure times of
40\,s with $24 - 27$\,s readout and overhead times, giving a mean time
resolution of 64\,s. However, for the faintest stars HD\,110072 and
HD\,170565, exposure times were doubled to 80\,s, which reduced the
time resolution to 107\,s. The total number of spectra per target
ranges from 69 to 138. To ensure high count rates, an image slicer was
used. The spectra were processed with the UVES pipeline and ESO MIDAS
package to extract 1D spectra using nightly flatfield spectra and
thorium-argon calibrations obtained at each telescope pointing. The
spectra were extracted using the `average extraction' option and the
flatfielded spectra from the two CCDs on the mosaic were rebinned to
same step size in wavelength and combined in one spectrum. The
rectification was performed in 3 steps. First, all spectra from the
same night were normalised with a spline-fit to a high-$S/N$ spectrum
of that night, only considering large-scale patterns, such as slopes.
Next an undulating continuum pattern that followed the spectral orders
was fitted with splines using averaged spectra of all spectra each
night and then applied to eliminate the undulations in individual
spectra.  Finally, all spectra of the same series (i.e. of the same
star) were rectified relative to a mean spectrum of that
series. Removal of effects from cosmic rays and CCD blemishes in the
spectra was made in two steps: cosmic rays were identified and removed
by comparing each 2D CCD image with the previous and subsequent one,
and the final, rectified 1D spectrum was corrected by a routine that
identified and removed sharp emission features of non-stellar
origin. For all spectra, the region $\lambda\lambda\,6515 -
6535$\,\AA\ suffers from spectrum-to-spectrum variability on the level
of 10 per cent.  A gap occurs in the region $\lambda\lambda\,5963 -
6032$\,\AA\ and is caused by the gap between the two CCD mosaic halves
(referred to as `lower' and `upper').

For unknown reasons, our UVES reductions resulted in an increased
noise in the spectra redwards of the gap, which varies from star to
star but is particularly pronounced for HD\,110072. This affects the
accuracy of \vsini, magnetic and radial-velocity measurements by
increasing the noise for this spectral region, in some cases by 33 per
cent.  The effect is most clear when comparing the noise of
cross-correlation results in the spectral regions bluer and redder
than the gap (Table\,\ref{tab:ccpow}), or when comparing noise in
co-added series of spectra of individual stars on both sides of the
gap. The effect varies for individual stars and we suspect the origin
of the problem to be found in the reduction of the spectra.  During
observations, it is more difficult to keep an object on the slicer
when observing very close to zenith. This also affects the $S/N$ and
was indeed the case for HD\,110072 and HD\,131750 that were both
observed at about 10 degree distance from zenith.  However, as the
performed radial velocity analyses utilise regions on both sides of
the gap, including the unaffected 1000\,\AA\ wide region below the
gap, and do reach precisions comparable to those published in the
literature for roAp stars, the current quality of the reductions
suffice for the aims of this investigation.

The zero point of the absolute wavelength calibration is of little
importance for the differential radial-velocity analysis, but was
found accurate to the level of $300\pm200$\,\ms\ based on the telluric
line list by \citet{griffinetal73}. This is as good as one may expect
from radial velocities of telluric lines that depend on conditions in
the Earth's upper atmosphere. Barycentric velocity corrections were
recorded in the FITS headers of the reduced spectra for the later
analysis of the velocity fields, and not included in the wavelength
calibration.
\begin{figure*}
  \vspace{-3pt} \hspace{-13pt}
  \includegraphics[height=0.98\textwidth, angle=270]{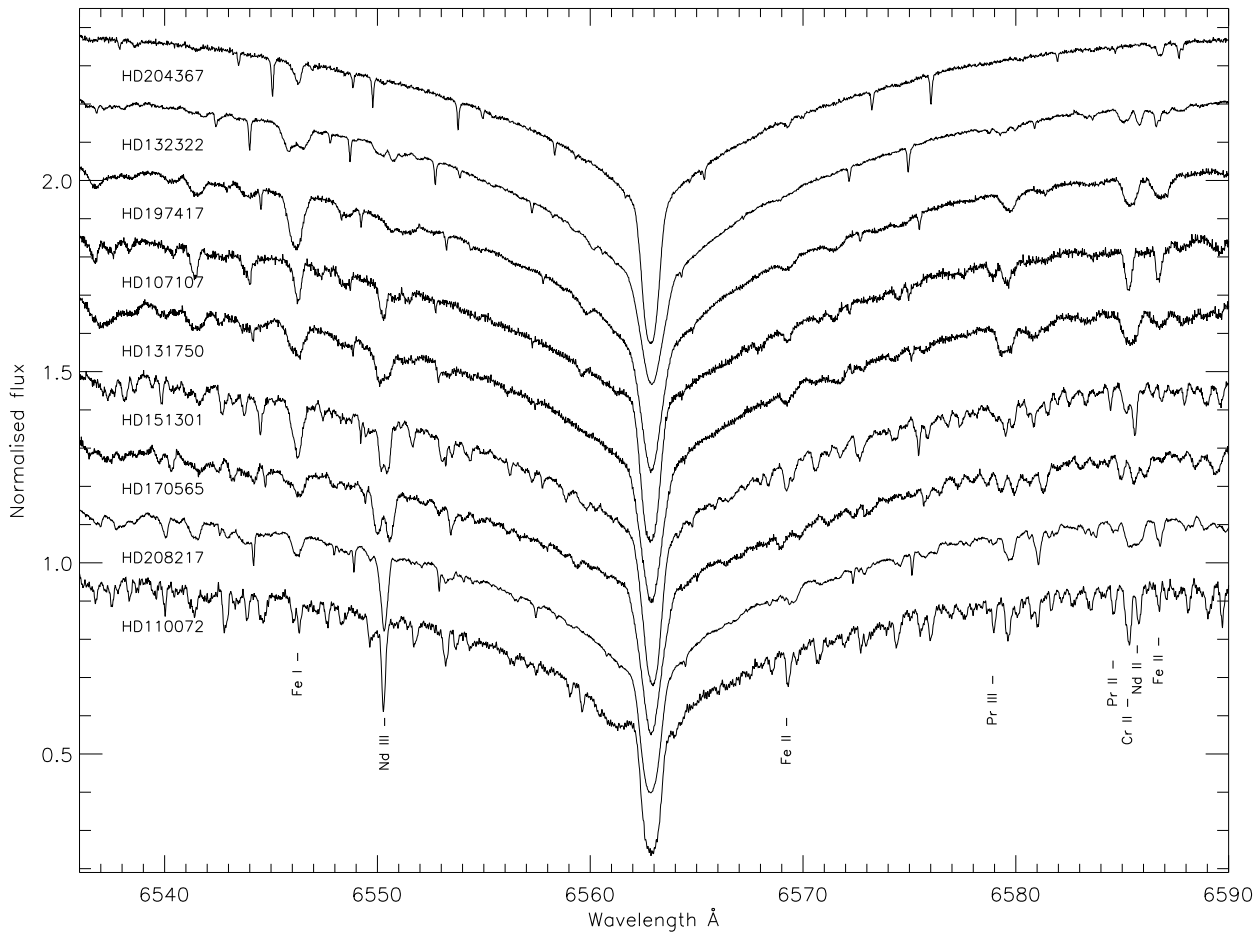}
  \caption{\label{fig:ha}Comparison of the \halpha\ profiles of all
    examined Ap stars. To increase the readability, the individual
    spectra have been offset in flux, and appear with 
    increasing temperature upwards.  Note the secondary spectrum superposed on
    that of HD\,110072, and the clearly double \ndiii\,6550.32\,\AA\
    lines for HD\,170565 and HD\,151301 (indicative of abundance
    spots). Each spectrum was made by co-adding all spectra in the
    series for that star. The spectrum of HD\,110072 was smoothed over
    two pixels.}
\end{figure*}

\section{Data analysis and results}

Our aim with this study was to search the spectra of nine roAp
candidates for rapid oscillations and through simple estimates of
physical properties to characterise the detected noAp or roAp stars.
For this purpose, we used the cross-correlation technique and
centre-of-gravity procedure to search for rapid pulsation using,
respectively, Doppler shifts of whole wavelength regions and of
individual and combined line profiles.  Stellar properties were
estimated with photometry and/or by comparison of observed spectra to
synthetic line profiles.

\begin{table}
  \centering
  \caption{\label{tab:linelist}List of laboratory wavelengths of the most
    important  lines used in the radial-velocity measurements based on
    the Centre-of-Gravity method. Note
    that due to broadening and abundance differences among the roAp candidates,
    not all lines could be used for all stars, and in some cases line blending
    made it necessary to include close lines in the measurements. Laboratory
    wavelengths have been taken from NIST, DREAM and VALD. For comparison with the
    stellar lines a few telluric lines were used ($\lambda$\, 5919.2, 5946.0, 
    5949.2, 6278.9 and 6889.9\,\AA). 
  }
  \begin{tabular}{@{}l@{\,\,}l@{}}
    \hline
    Line  &Wavelength (\AA)  \\ \hline
    \ion{H}{i}      &6562.852  \\
    \ion{Li}{i}     &6707.761    \\
    \ion{Na}{i}     &5895.924, 5889.951  \\
    \ion{Ca}{i}     &6122.217, 6162.173 \\
    \ion{Cr}{ii}    &5280.054, 5310.687 \\
    \ion{Fe}{i}     &5434.524, 6136.614, 6137.694 \\
    \ion{Fe}{ii}    &5061.718, 5284.109 \\
    \ion{Ba}{ii}    &5853.675, 6141.713   \\
    \ion{Ce}{ii}    &5077.854, 5117.946, 5147.565, 5459.193, 5468.371, \\
                    &5613.694, 5680.261, 5711.437, 5858.546  \\
    \ion{Pr}{ii}    &5220.108, 5605.642, 6114.381 \\
    \ion{Pr}{iii}   &5208.507, 5284.693, 5299.993, 5844.408, 5956.043, 6053.003,  \\
                    &6090.010, 6160.233, 6195.619, 6866.793  \\
    \ion{Nd}{ii}    &5319.811, 5356.967, 5361.467, 5620.594, 6425.779, 6550.178   \\
    \ion{Nd}{iii}   &5085.001, 5203.902, 5286.764, 5825.857, 5845.068, \\
                    &6145.072, 6327.244, 6550.326 \\
    \ion{Eu}{ii}    &6437.680, 6645.064  \\
    \ion{Gd}{ii}    &5860.727  \\
    \hline
    \hline
  \end{tabular}
\end{table}

For the purpose of spectral line identification, a synthetic
comparison spectrum was produced with {\small SYNTH}
\citep{piskunov92} using a Kurucz stellar atmosphere model for $T_{\rm
  eff}=8750\,$K, turbulent velocity 2\,km\,s$^{-1}$, $\log\,g = 4.0$
and slightly increased solar metal abundance $\log{N_{\rm{Z}}/N_{\rm
    Z,\sun}}=+0.5$. Atomic line data were taken from the Vienna Atomic
Line Database (VALD, \citealt{kupkaetal99}) for increased abundances
mainly for the elements Nd, Pr, Sr, Cr, Eu, as given in
Table\,\ref{tab:model}. Other sources used for line data were the
atomic database NIST\footnote{http://physics.nist.gov} and the
Database on Rare Earth Elements at Mons University
(DREAM\footnote{http://w3.umh.ac.be/$\sim$astro/dream.shtml}) through
its implementation in the VALD.  This model's temperature is less
optimal for the coolest of the studied stars, but still appropriate
for selecting lines for the radial-velocity analyses.

In the following section we describe the estimation of stellar
physical parameters for the examined sample, the radial-velocity
measurements and frequency analyses, and a performance test carried
out on UVES data for two known roAp stars.

\subsection{Analysis}
The present study is a survey for variability, and therefore an
analysis of abundances, temperatures, surface gravities and projected
rotational velocities is outside the scope of this paper. Such an
analysis must take the chemical element stratification, and possibly
also temperature gradients, into account. We do, however, in
Table\,\ref{tab:targets} give estimates of \vsini, effective
temperatures and magnetic fields of the studied sample.

\subsubsection{Temperatures and rotational velocities}
\label{sec:teffrot}
The spectra of Ap stars have flux distributions that are deformed by
strongly peculiar elemental abundances. Calibrations based on
Str\"omgren indices 
may therefore not provide reliable estimates of temperatures and
surface gravity, \logg. The \bet\ index, however, remains largely
unaffected by this.  Effective temperatures were thus estimated with
the $c_0,\beta$ grids by \citet{moonetal85} using \bet\ from
Table\,\ref{tab:targets} (from \citealt{martinez93}). We allowed the
value of $c_0$ to be free in these grids, fixed \bet\ and assumed
$3.5\le\log\,g\le4.0$.  The resulting photometric temperatures are
given in Table\,\ref{tab:targets}.  An additional check on \teff\ was
made by comparing synthetic line profiles to the observed spectra.
This method is sensitive to normalisation errors in the \halpha\
region (the above-mentioned undulations), and the core-wing anomaly in
this line \citep{cowleyetal01} that prohibits fitting the atomic line
profile with normal models. However, the average deviation of the
photometric and spectroscopic temperatures is only 5 per cent, with
the exception of HD\,131750 which has a 17 per cent higher
spectroscopic temperature. Moon \& Dworetsky's grids are furthermore
ambiguous for $\mteff > 8500$\,K and in those cases the spectroscopic
temperatures of HD\,132322 and HD\,204367 were $8250-9250$\,K and
$8000-8250$\,K, respectively.  The synthetic line profiles were based
on Kurucz atmospheres \citep{castellietal03} and atomic line data from
the VALD database using increased REE abundances (identical for all
models). The grid of models had \teff\,$=7000 - 10500$\,K in steps of
500\,K with $\log\,g = 3.5$, 4.0 and, in a few cases, 4.5. Best fits
were obtained for $\log\,g = 3.5 - 4.0$, mainly to the lower side of
this range.  Better estimates of $\mteff$ require a detailed abundance
analysis that takes stratification and magnetic fields into account,
which will not be done here.

\begin{table*}
  \centering
  \caption{\label{tab:cogpow}
    Frequency analyses from combination of individual line shifts determined
    for multiple lines of different species with the Centre-of-Gravity
    method. The highest amplitude in each stacked amplitude spectrum is given as
    $A_{\rm max}$. `n' indicates number of measurements,
    typically number of lines or components in case of double structures. Linear
    shifts have been fitted and removed separately from each time series. See
    Table\,\ref{tab:linelist} for a listing of the lines typically used. Low-frequency
    peaks (0.4\,mHz or below) were excluded from the noise estimates and
    not considered for significance. `Nd,Pr' indicates combined radial velocity 
    series for all possible Nd and Pr lines.}
  
\begin{tabular}{@{\,}ll@{\,}r@{\,\,}rl@{\,}r@{\,\,}rl@{\,}r@{\,\,}rl@{\,}r@{\,\,}r
l@{\,}r@{\,\,}rl@{\,}r@{\,\,}rl@{\,}r@{\,\,}rl@{\,}r@{\,\,}rl@{}r@{}}
\hline
Star
    
&\multicolumn{3}{c}{\ion{Nd}{ii}}&\multicolumn{3}{c}{\ion{Nd}{iii}}&\multicolumn{3
}{c}{\ion{Ce}{ii}}
&\multicolumn{3}{c}{\ion{Pr}{ii}}
    
&\multicolumn{3}{c}{\ion{Pr}{iii}}&\multicolumn{3}{c}{Nd,Pr}&\multicolumn{3}{c}{\ion{Eu}{ii}}&\multicolumn{3}{c}{Telluric}&\multicolumn{2}{c}{H core}\\
& $\sigma$  &${A}_{\rm max}$ & n &
  $\sigma$  &${A}_{\rm max}$ & n &
  $\sigma$  &${A}_{\rm max}$ & n &
  $\sigma$  &${A}_{\rm max}$ & n &
  $\sigma$  &${A}_{\rm max}$ & n &
  $\sigma$  &${A}_{\rm max}$ & n &
  $\sigma$  &${A}_{\rm max}$ & n &
  $\sigma$  &${A}_{\rm max}$ & n &
  $\sigma$  &${A}_{\rm max}$ \\ 
HD     & \multicolumn{3}{l}{(\ms)}  &
\multicolumn{3}{l}{(\ms)}  &
\multicolumn{3}{l}{(\ms)}  &
\multicolumn{3}{l}{(\ms)}  &
\multicolumn{3}{l}{(\ms)}  &
\multicolumn{3}{l}{(\ms)}  &
\multicolumn{3}{l}{(\ms)}  &
\multicolumn{3}{l}{(\ms)}  &
\multicolumn{2}{l}{(\ms)}\\ 
\hline
107107 & 28 &  82   & 4 & 28 & 108 & 6  & 32      & 118   & 6  & 44   & 133   & 2  & 16 & 59  & 10  & 12 & 39  & 25 & 23 & 88 & 2 & 20 & 62 & 2  & 20& 65 \\ 
110072 & 27 &  92   & 5 & 16 &  52 & 8  & 30      & 100   & 5  & 31   & 110   & 5  & 15 & 55  & 10  & 10 & 42  & 28 & 31 & 95 & 2 & 20 & 57 & 2  & 52&135  \\  
131750 & 38 &  132  & 4 & 30 & 102 & 5  & 40      & 150   & 6  & 59   & 177   & 2  & 29 & 112 & 12  & 18 & 58  & 23 & 48 &161 & 2 & 20 & 57 & 3  & 21& 72 \\  
132322 & 32 &  90   & 5 & 20 &  69 & 11 & {\scriptsize N/A}    & {\scriptsize N/A} & 0  & 58  & 196 & 1  & 22  & 70 & 8  & 13 & 37& 26 & 22 & 72 & 4 & 10 & 30  & 3  & 19 & 59 \\  
151301 & 24 &  84   & 3 & 11 &  35 &  9 & 62      & 237   & 1  & 41   & 136   & 2  & 12 & 39  &  9  & \phantom{0}8  & 24 & 25 & 20& 86 & 4  & 13 & 42& 4  & 19  & 54 \\  
170565 & 24 &  79   & 7 & 15 &  52 & 13 & 23      & 64    & 12 & 53   & 163   & 4  & 16 & 44  & 13  & 10 & 32  & 41 & 37 &111 & 4 & \phantom{0}9 & 30& 2  & 24  & 75 \\  
197417 & 49 &  181  & 3 & 41 & 114 &  6 & {\scriptsize N/A}    & {\scriptsize N/A} & 0  & 67  & 228 & 1  & 25  & 96 & 9  & 20 & 65& 19 & 54 & 185& 2 & 17 & 59  & 4  & 17 & 52 \\  
204367 & 53 &  181  & 4 & 33 &  86 &  6 & {\scriptsize N/A}    & {\scriptsize N/A} & 0  & 80  & 249 & 1  & 51  & 143 & 5 & 25 & 72& 16 & 76 & 236& 3 & \phantom{0}7  & 18 & 2  & 14& 46 \\  
208217 & 10 &  34   & 7 & \phantom{0}5  &  17 & 12 & {\scriptsize N/A}  & {\scriptsize N/A}  & 0  & 20   & 69  & 4  & \phantom{0}7& 29 & 13 & \phantom{0}4 & 15 & 37 & 14 & 70 & 6 & 10   & 33  & 6  & 13   & 40 \\  
\hline
\hline
\end{tabular}
\end{table*}
\begin{table*}
  \centering
  \caption{\label{tab:ccpow}Frequency analyses from cross
    correlations. Low-frequency peaks (0.4\,mHz or below) were excluded from
    the noise estimates and not considered for significance. These 6 regions
    essentially cover: the spectrum bluer than the 6000\,\AA\ gap, two regions 
    redder than the gap and avoiding
    the weak telluric line region at $\lambda\lambda\,6275-6320$\,\AA, the 
    \halpha-region, the
    region with the \ion{Na}{d} doublet, and a comparison region dominated
    by telluric lines. }

  \begin{tabular}{@{}lllllllllllll@{}}
    \hline
    Star
    &\multicolumn{2}{c}{5150--5800\,\AA}&\multicolumn{2}{c}{6035--
      6273\,\AA}&\multicolumn{2}{c}{6350--6700\,\AA}&\multicolumn{2}{c}{6510--
      6570\,\AA}&\multicolumn{2}{c}{5888--5898\,\AA}&\multicolumn{2}{c}{6873--6899\,\AA}\\
    &$\sigma$ &$A_{\rm max}$       &$\sigma$ &$A_{\rm max}$ &$\sigma$ &$A_{\rm max}$              
    &$\sigma$ &$A_{\rm max}$       &$\sigma$ &$A_{\rm max}$ &$\sigma$ &$A_{\rm max}$   \\ 
    HD     & \multicolumn{2}{c}{(\ms)} &
    \multicolumn{2}{c}{(\ms)} &
    \multicolumn{2}{c}{(\ms)} &
    \multicolumn{2}{c}{(\ms)} &
    \multicolumn{2}{c}{(\ms)} &
    \multicolumn{2}{c}{(\ms)}\\ \hline
    107107 & 3.2     &  8.9        & 4.4   & 12.5    & 4.1   &  13.9     & 4.2     
    & 13.2     & 5.0  & 13.4        & 5.0  & 13.4  \\  
    110072 & 3.2     &  12.3$^a$   & 4.7   & 12.7    & 5.0   &  12.8     & 5.4     
    & 14.2     & 5.9  & 17.3        & 5.2  & 16.7  \\  
    131750 & 3.0     &  9.1        & 4.9   & 14.9    & 5.3   &  16.1     & 4.8     
    & 15.0     & 3.1  & 12.9$^b$    & 3.6  & 10.9  \\  
    132322 & 2.0     &   6.3       & 6.2   & 20.9    & 5.4   &  13.5     & 6.7     
    & 16.4     & 1.4  & 5.4         & 1.2  & 3.4   \\  
    151301 & 1.4     &   4.6       & 8.6   &  31.7   &10.0   &  38.3     & 8.4     
    & 30.9     & 2.1  &  6.5        & 1.4  & 5.9   \\  
    170565 & 4.7     &  9.7        & 6.9   & 26.1    &6.6    &   27.7    & 5.4     
    & 19.8     & 2.2  &  7.9        & 1.8  & 7.7   \\  
    197417 & 1.7     &   5.3       & 3.5   & 9.8     &9.4    &  10.1     & 3.7     
    & 10.4     & 2.2  &  7.1        & 2.2  & 8.9   \\  
    204367 & 2.0     &   5.3       & 17.5  &  53.0   &16.2   &   48.9    & 9.3     
    & 27.1     & 3.4  &  11.0       & 2.4  & 6.3   \\  
    208217 & 1.4     &   3.0       & 2.8   &  7.8    &2.8    &   7.5     & 2.5     
    & 7.4      & 9.0  &  38.0$^c$   & 1.3  & 3.4   \\  
    \hline
    \hline
    \multicolumn{13}{l}{$^a$ at 0.45\,mHz, next-highest peak has 
      $A_{\rm max}=7.7$\,\ms} \\
    \multicolumn{13}{l}{$^b$ at 0.45\,mHz, next-highest peak has 
      $A_{\rm max}=8.6$\,\ms} \\
    \multicolumn{13}{l}{$^c$ at 0.48\,mHz, next-highest peak has 
      $A_{\rm max}=31.3$\,\ms} \\
  \end{tabular}
\end{table*}

Projected rotation velocity estimates were made by measuring the
\ion{Fe}{i} lines $\lambda\,$5434.52 and 5576.08\,\AA\ that are rather
insensitive to magnetic broadening. Synthetic models for $\log\,g =
3.5$ were compared to the spectra for a range of iron abundances and
rotational broadening, using models corresponding in temperature to
the individual stars. The macroturbulence was varied in the range
$1-4$\,\kms. When strong line blending hampered the measurements,
additional iron lines were used to constrain the estimates. The
derived velocities were then refined with synthetic models that took
magnetic broadening into account ({\small SYNTHMAG})
\citep{piskunov99}. The analyses of several, typically 20--30, iron
lines that determine the mean quadratic magnetic fields also give
precise rotation velocities \citep{mathysetal06}. These values are
formally more precise than, but consistent with, the above estimates
and are 
therefore given in Table\,\ref{tab:targets}, except for the case of
HD\,170565 where only the first estimate is given. It should be noted,
though, that the values of 
the rotation velocities that are obtained as part of the analysis
performed to determine the mean
quadratic magnetic field are upper limits on the \vsini, since they
may actually include contributions from other line broadening effects
that are proportional to wavelength, such as micro- or
macroturbulence. (Instrumental and thermal broadening are however duly
isolated -- see \citet{mathysetal06} for details.) However these
contributions are mostly negligible, except in the slowest-rotating
stars. 

\subsubsection{Magnetic fields}
\label{sec:mag}
Known roAp stars have strong magnetic fields \citep{kurtzetal06a}, and
we therefore searched for magnetically resolved or broadened lines in
the spectra of the roAp candidates.  Many Ap stars have fields that
are measurable only with polarimetry. It takes a strong field,
typically exceeding $\sim 1.5$\,kG, combined with a slow projected
rotation rate (smaller than \vsini\ $\sim\,10$\,\kms) to produce
magnetically resolved lines by the Zeeman effect \citep{mathysetal97}.

In the simplest cases of spectral lines corresponding to doublet or
triplet Zeeman patterns, simple formulae can be applied to determine
in a virtually approximation-free manner the mean magnetic field
modulus $\langle B\rangle$ from measurement of the wavelength
separation of the resolved Zeeman components
\citep{mathys89}. $\left<B\right>$ is the average of the modulus of
the magnetic vector, over the visible stellar hemisphere, weighted by
the local line intensity. For a triplet pattern, its value (in G) is
obtained from the wavelength separation between the central $\pi$
component and either of the $\sigma$ components, $\Delta\lambda$, by
application of the formula:
\begin{equation}
  \centering
  \left<B\right>=\Delta\lambda/(4.67\cdot10^{-13}\;\lambda_{c}^{2}\;g_{\rm eff}),
  \label{eq:zeeman1}
\end{equation}
where $\lambda_{\rm c}$ is the central wavelength of the line and
$g_{\rm eff}$ is the effective Land\'e factor of the transition.  Both
$\Delta\lambda$ and $\lambda_{\rm c}$ are expressed in \AA.  For a
doublet pattern, the relation between the field modulus and the
wavelength separation of the split components (each of which is the
superposition of a $\pi$ and a $\sigma$ component) is:
\begin{equation}
  \centering
  \left<B\right>=\Delta\lambda/(9.34\cdot10^{-13}\;\lambda_{c}^{2}\;g_{\rm eff}).
  \label{eq:zeeman}
\end{equation}
Note that Eq.\,\ref{eq:zeeman} also describes the relation between
$\langle B\rangle$ and the wavelength separation of the red and blue
$\sigma$ components of a triplet.

Mainly due to smearing by rotational broadening, only one of the
studied stars (HD\,208217) has clearly resolved lines from magnetic
splitting that can be used with Eqs.\,\ref{eq:zeeman1} or
\ref{eq:zeeman} to estimate the field strength directly. In one other
star, HD\,107107, magnetic splitting and rotational broadening are
comparable for the lines with the highest magnetic sensitivity, so
that it is possible to obtain an estimate of the surface magnetic
field by fitting synthetic profiles to these lines. This estimate,
which we shall denote by $\langle B_{\rm synth}\rangle$, should be of
the same order of magnitude as the mean magnetic field modulus, but it
is not entirely equivalent to the latter. (In particular, $\langle
B_{\rm synth}\rangle$ is model-dependent, while $\langle B\rangle$ is
not.)

For the other stars we estimated the surface magnetic field from
consideration of the magnetic broadening and intensification of
magnetically sensitive lines. By assuming that a line's full-width at
half maximum (FWHM) increases linearly with the separation of its
(unresolved) Zeeman-split components, we used Eq.\,\ref{eq:zeeman} to
obtain an estimate $\langle B_{\rm FWHM}\rangle$ of the magnetic field
by comparing lines with different effective Land\'e factors \citep[see
also][]{Preston71}.  The assumption is justified as long as the
magnetic splitting of the analysed lines is small compared to their
overall width (in particular, due to rotational broadening); then,
$\langle B_{\rm FWHM}\rangle$ should typically be comparable to
$\langle B\rangle$ or $\langle B_{\rm q}\rangle$ (see below).

Other estimates of the magnetic fields
were obtained by comparing
the observed spectra to magnetically resolved or broadened line
profiles to determine $\langle B_{\rm synth}\rangle$. These models were synthesised with {\small SYNTHMAG} for a
temperature grid of \teff\,$=7500$, 8000 and 8500\,K.  For this
comparison, Cr and Fe lines with low Land\'e factors were used to fix
abundances and the rotation rate of the models prior to applying them
to magnetically resolved or broadened lines. Instrumental broadening
of 0.05\,\AA\ was adopted while the macroturbulence was fixed for each
star within the range $1-4$\,\kms. This approach worked well for
magnetically broadened lines because both the equivalent width and the
FWHM are affected by magnetic broadening; when absorption in a line is
intrinsically spread over a larger wavelength range, the absorption
becomes more effective, increasing the equivalent width (magnetic
intensification). Again, the quantity that is derived should be
comparable to the mean magnetic field modulus, but in general is not
exactly equal to it.

Finally, the values of the mean quadratic magnetic field, $\langle
B_{\rm q}\rangle 
\equiv (\langle B^2\rangle + \langle B^2_{\rm z}\rangle)^{1/2}$, were
derived through application of the method described in
\citet{mathysetal06}. Here $\langle B^2\rangle$ is the mean square
magnetic field modulus (the average over the stellar disk of the
square of the modulus of the field vector, weighted by the local
emergent line intensity), while $\langle B^2_{\rm z}\rangle$ is the
mean square longitudinal field (the average over the stellar disk of
the square of the line-of-sight component of the magnetic vector,
weighted by the local emergent line intensity). The analysis was based
on consideration of samples of reasonably unblended lines of
\ion{Fe}{i} and \ion{Fe}{ii}; the number of analysed lines varies from
star to star and ranges from 14 to 33.  This approach could not be
successfully applied to HD\,170565, for which the number of usable
diagnostic lines (9) proved insufficient to untangle the contributions
of the magnetic field (probably fairly strong), non-negligible
rotation, and a possibly inhomogeneous distribution of iron on the
stellar surface. The mean quadratic magnetic field is typically a few
percent greater than the mean magnetic field modulus.

For 3 stars, we also indicate in Table~\ref{tab:targets} values of the
mean longitudinal magnetic field $\langle B_{\rm z}\rangle$ and the
magnetic field modulus $\langle B\rangle$ from the
literature. $\langle B_{\rm z}\rangle$ is the average over the stellar
disk of the component of the magnetic vector along the line of sight,
weighted by the local emergent line intensity. This field moment
strongly depends on the geometry of the observation, and contrary to,
e.g., $\langle B\rangle$ or $\langle B_{\rm q}\rangle$, it typically
varies considerably during a stellar rotation cycle.  Accordingly, it
is not well suited to characterise the strength of the magnetic field
in a star, other than to give (through its absolute value) a generally
very conservative lower limit of the latter. But significant
measurements of $\langle B_{\rm z}\rangle$ at least provide definitive
evidence that a star is magnetic.

\subsubsection{Radial-velocity shifts}
\label{sect-rvshift}
Precise radial velocity shifts were measured for several spectral
lines with the centre-of-gravity method and by fitting with Gaussian
profiles.  We used the local continuum in the selected sub-regions of
measured lines, which gave consistent amplitudes for both methods in
tests on UVES spectra for known roAp stars (see below).  The
centre-of-gravity method was preferred due to its stability and better
ability to deal with blended line profiles. For strong and isolated
lines, the two methods were comparable.  Table\,\ref{tab:linelist}
gives the most important spectral lines used, but the actual selection
of lines depended on line blending and composition for each stellar
case.

Additionally, line shifts were determined by cross-correlating the
spectra with averaged spectra of each series. The maxima of the
correlation functions were determined with a spline fit, which was
found more reliable and stable than with Gaussian or 4th-order
polynomial functions.  The cross-correlation regions were chosen to be
free of static features: $\lambda\lambda\,5150-5800$, $6035-6273$,
$6350-6700$ and $6510-6570$\,\AA.  However, to check for non-stellar
periodicities, we also used two regions dominated by telluric or the
interstellar lines of \ion{Na}{d}: $\lambda\lambda\,5888-5898$ and
$6873-6899$\,\AA. Because the UVES pipeline produces merged spectra in
the linear wavelength scale only, we rebinned the selected regions of
the spectra to $\log$ wavelength which is more appropriate
\citep{tonryetal79} when determining velocity shifts using large
wavelength regions.  For each region, the size of the bins was
optimised according to pixel size and local spectral resolution.

\begin{figure}
  \vspace{-3pt}
  \includegraphics[width=5.5cm,height=7.9cm,
  angle=270]{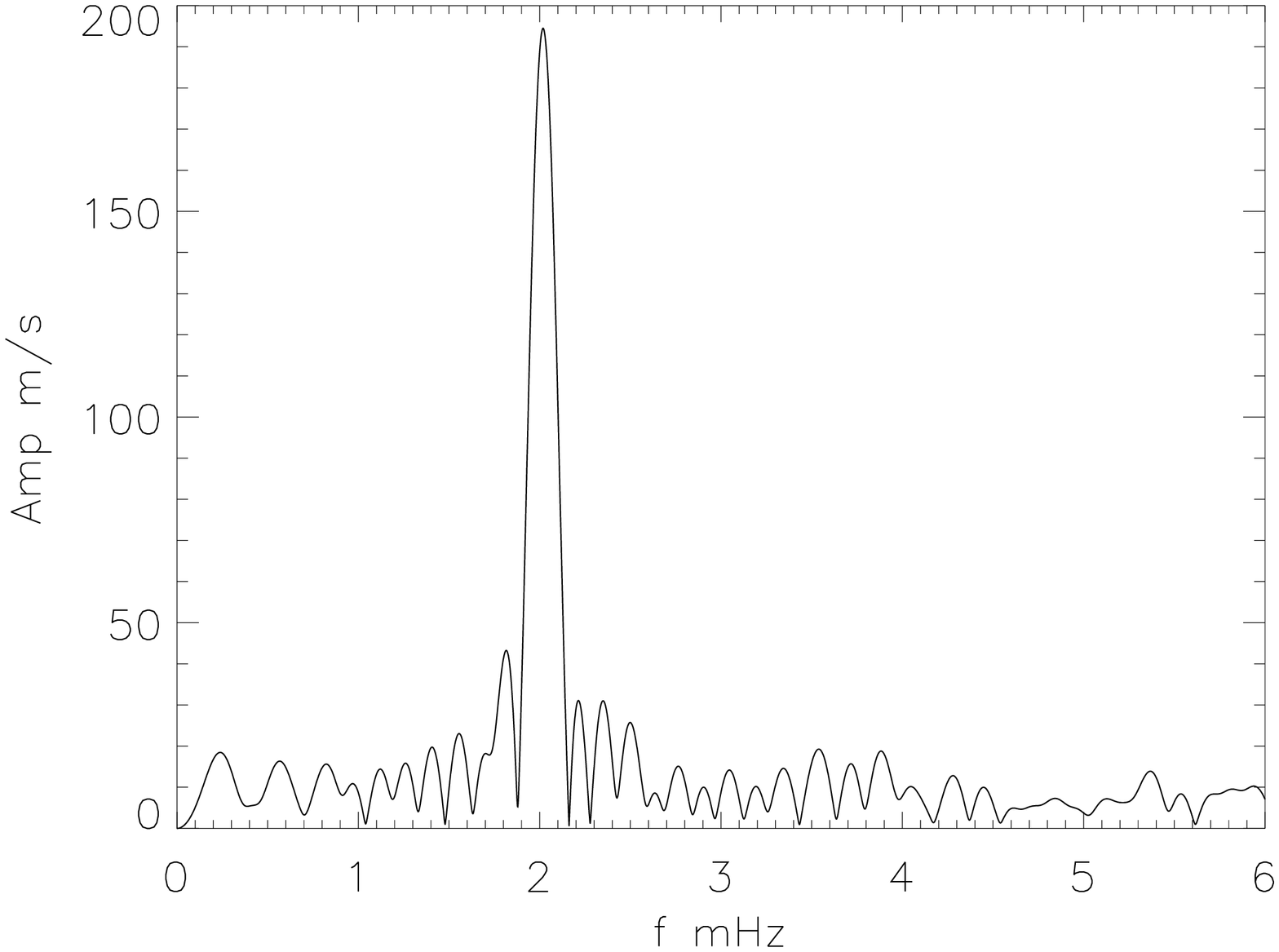}
  \includegraphics[width=5.5cm,height=7.9cm,
  angle=270]{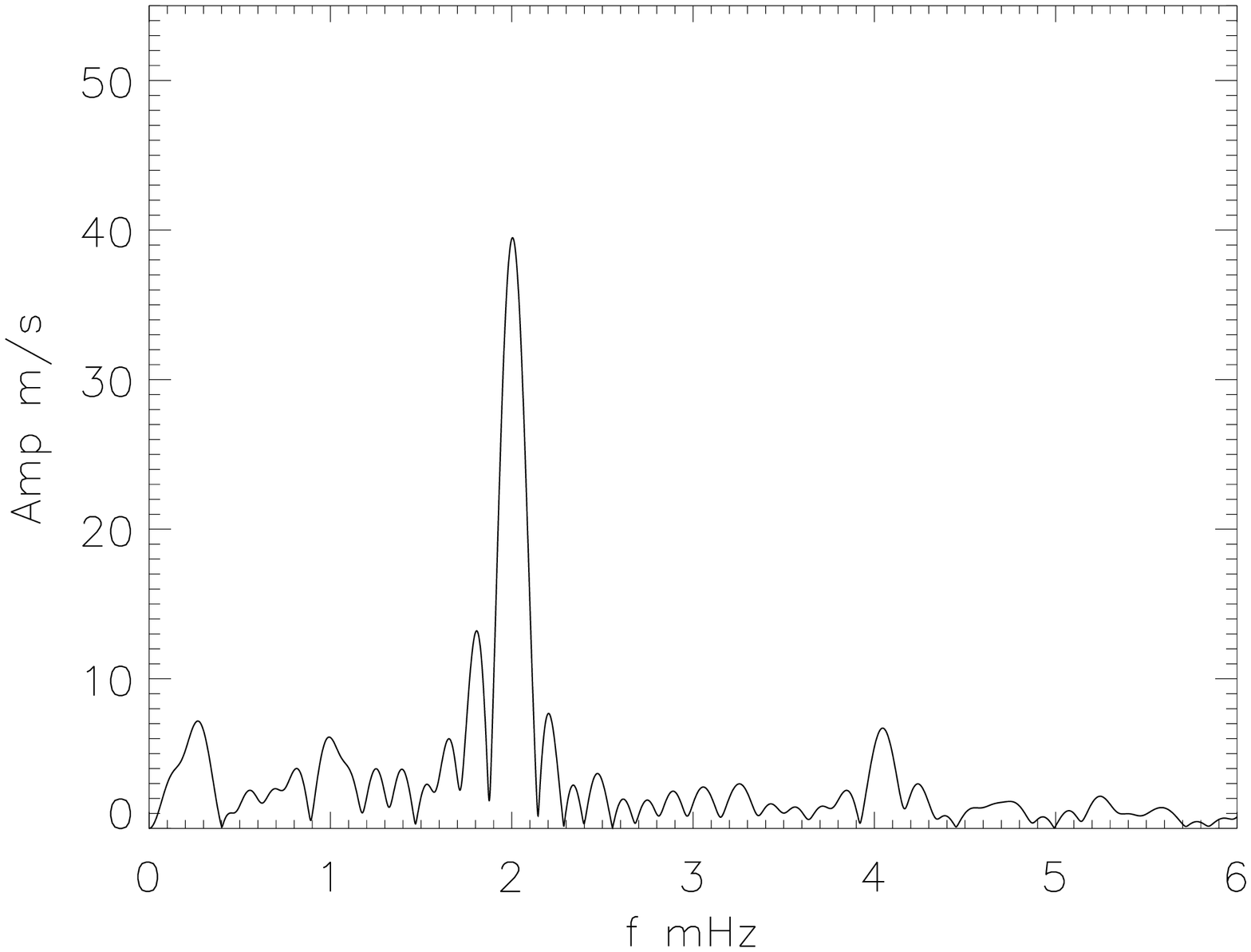}
  \caption{\label{fig:33lib}Amplitude spectra for the known roAp star
    33\,Lib. Top: the 2.015\,mHz frequency for the
    \euii\,6645.06\,\AA\ line. Bottom: Cross correlation (see text)
    for the region $\lambda\lambda\,5150-5800$\,\AA. The frequencies
    2.015\,mHz, its harmonic 4.030\,mHz, and 1.769\,mHz are
    recovered. Note different ordinate scales. }
\end{figure}

Because of the stratification of Ap star atmospheres, high-order mode
pulsation amplitude and phase may change from element to
element. Therefore the amplitudes of cross-correlation velocities
cannot be directly compared to measurements of individual lines.
Nevertheless, cross-correlation is efficient for detecting pulsations
by using long wavelength regions with numerous lines to obtain high
$S/N$.

\begin{figure}
  \vspace{-3pt}
  \includegraphics[width=5.5cm,height=7.9cm,
  angle=270]{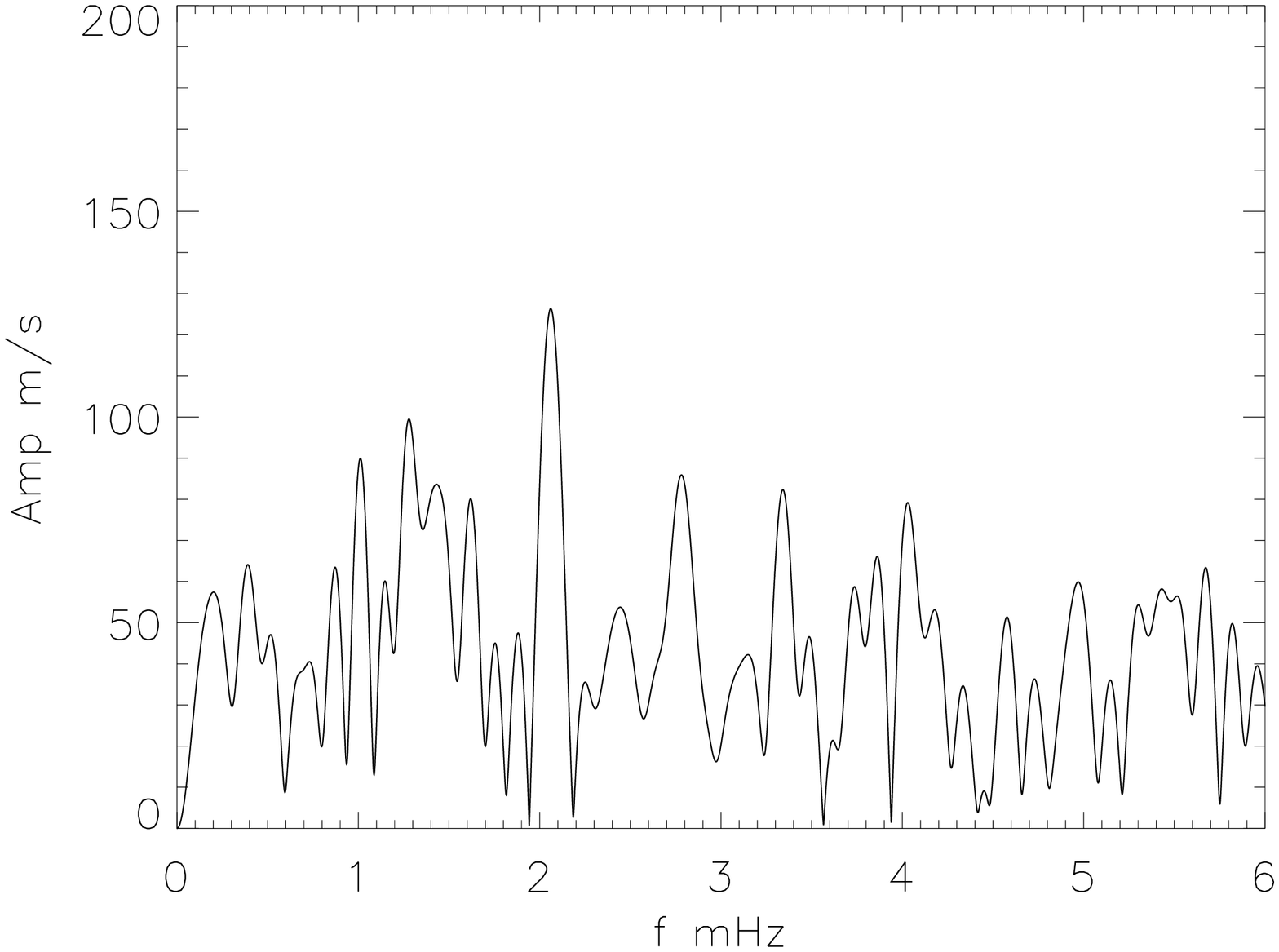}
  \includegraphics[width=5.5cm,height=7.9cm,
  angle=270]{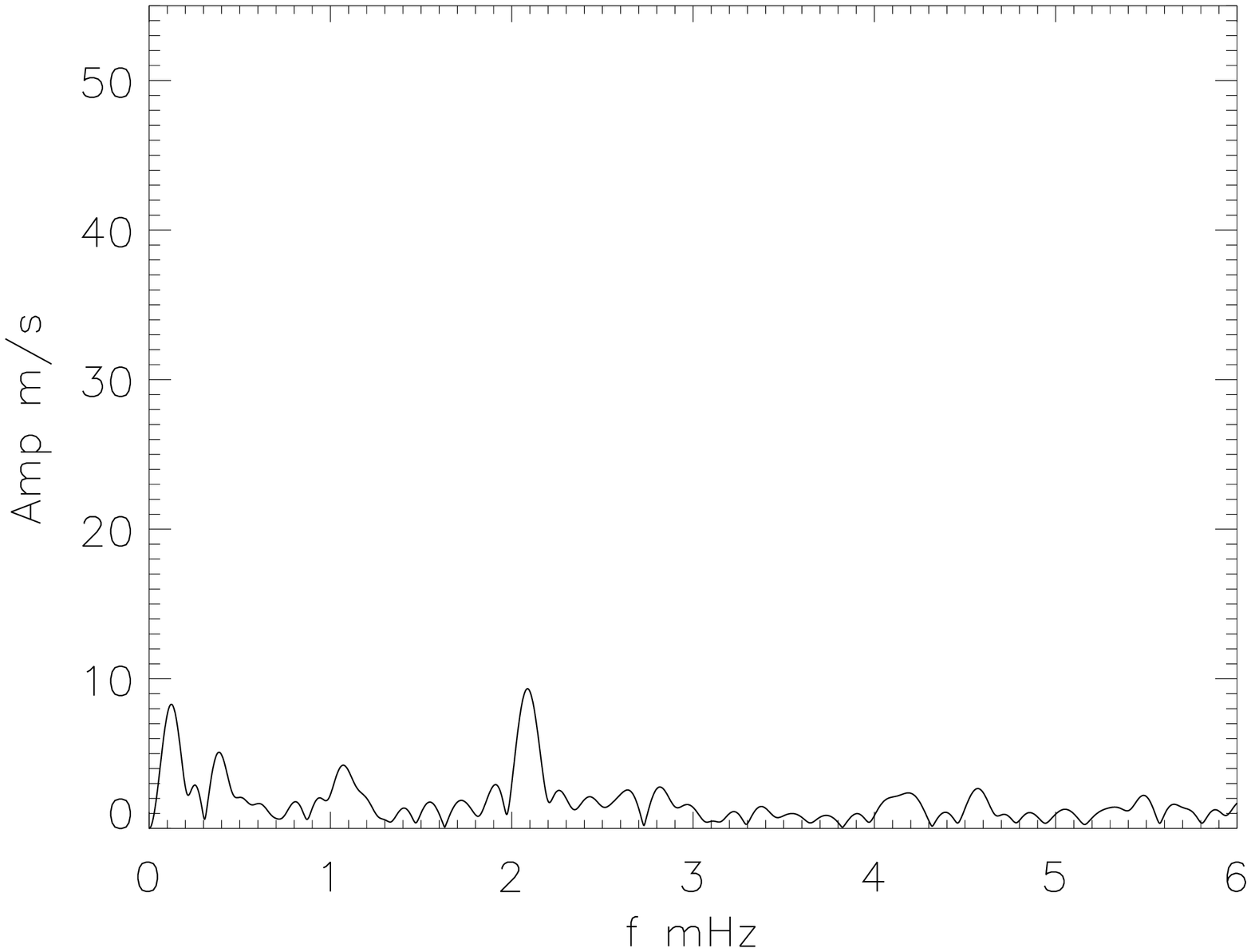}
  \caption{\label{fig:hd154708} Same as Fig.\,\ref{fig:33lib}, but for
    the known roAp star HD\,154708. Top: significant amplitude is seen
    at the known frequency 2.088\,mHz for the
    \priii\,5299.99\,\AA-line. Bottom: cross-correlation
    ($\lambda\lambda\,5150-5800$\,\AA) unambiguously detects the
    2.088\,mHz frequency. }
\end{figure}

{\small IDL} tools for line and cross-correlation
measurements and analyses were developed and tested on UVES spectra of 
the known roAp
stars 33\,Lib and HD\,154708, published by \citet{kurtzetal05} and
\citet{kurtzetal06b}, respectively.  For 33\,Lib, we confirm two
frequencies, 2.015\,mHz and 1.769\,mHz (Fig.\,\ref{fig:33lib}), that
are in excellent agreement with the published frequencies, amplitudes
and noise levels for lines of \euii\ and H$\alpha$. In particular, we
also confirm the low-amplitude oscillation for \ion{Fe}{i}\,5434.52
($31\pm4$\,\ms). Similarly, we find amplitudes of $32\pm5$\,\ms\ for
\ion{Ca}{i} and \ion{Ti}{ii} lines and $130\pm13$\,\ms\ for
\ion{Ba}{ii}. A telluric line (6888.96\,\AA) shows only noise with a
highest amplitude of $9\pm3$\,\ms. We also made cross-correlation
measurements of the above-mentioned wavelength regions and recovered
the 2.015 frequency with amplitudes in the range $20\pm2$ to $50\pm5$\,\ms.
All cross-correlation and line measurements gave similar significance
$S/N=10$, although at very different amplitudes (see also
Fig.\,\ref{fig:33lib}).

The second test, the new roAp star HD\,154708, was even stronger as
this star pulsates with amplitudes that are among the smallest known
for roAp stars.  \citet{kurtzetal06b} needed to combine RV
measurements for several lines of this star in order to detect its
rapid oscillation.  We confirm that no individual line (including
\halpha) shows signal on or above the 4\,$\sigma$ level. Yet, as seen
in Fig.\,\ref{fig:hd154708}, we directly recover the known 2.088\,mHz
mode with cross-correlations in the lower region of the spectra,
$\lambda\lambda$$5150-5800$\,\AA, with $S/N = 6.2$, and a marginal
detection ($110\pm32$\,\ms) using the single line
\priii\,5299.99\,\AA.

\subsubsection{Frequency analysis}
\label{sec:drift}
Frequency analyses were performed using a Discrete Fourier Transform
programme \citep{kurtz85} and the {\mbox{\small PERIOD04}}
\citep{lenzetal05} programme.  Linear trends in the individual
$\sim2$\,hr radial-velocity series were fitted and removed with linear
least-squares fitting.  The noise $\sigma$ in the amplitude spectra (see
Tables\,\ref{tab:cogpow} and \ref{tab:ccpow}) is determined from
least-squares fitting of harmonics to the data following
\citet{deeming75}.  Because the barycentric velocity correction is
approximately linear for each series of spectra, and rather small (the
correction varies $44-150$\,\ms\ per hr for the 9 stars), it was
eliminated with other drifts by a linear fit before the frequency
analysis.  With the 0.3 arcsec slit and seeing conditions of $0.9-1.4$
arcsec, the centring error for UVES is $50-100$\,\ms\ according to
\citet{bouchyetal2004}.  Furthermore, a 1-mbar change in pressure may
induce drifts of 90\,\ms.  During each of our observing nights, the
pressure changed $1.5-2.0$ mbar.  We therefore expect instrumental
drifts of up to 280\,\ms\ per night, depending on seeing and pressure,
and less during a 2-hr series on a star.  The drift during a series of
spectra may be non-linear (which is what we actually see in some
cases). We noticed that comparison lines or regions with non-stellar
constant lines occasionally exhibit drifts that are not seen for other
regions of the same spectra (such as in Fig.\,\ref{fig:204367cog},
panel `Tell').  This difference may be because the strong and sharp
telluric lines result in higher sensitivity, and they are influenced
by fast wind speeds in the high layers of the Earth's atmosphere where
telluric lines are formed.

The null-results are presented on a star-by-star basis in Sections
\ref{sec:107107}--\ref{sec:208217}.  We emphasise the statistical fact
that when calculating about 50 amplitude spectra for each of nine
candidate roAp stars, the chance for a spurious peak to reach the
4\,$\sigma$ level increases. Furthermore, the analyses show several
combined amplitude spectra with single, prominent peaks reaching
$3-4\,\sigma$. However, the reality of such peaks is that they often
originate from shallow and blended lines.  Our criteria for detecting
rapid oscillations are therefore: {\em i}) a peak of 4 times the noise (see
\citealt{bregeretal93, kuschnigetal97}) in an amplitude spectrum, {\em ii})
confirmation in an amplitude spectrum for either a line, or
combination of lines, of a different ionisation or element, or
cross-correlation region, and {\em iii}) only frequencies above 0.4\,mHz
(42\,min) are considered. The latter is because we have less control
over drifts on these time scales in the wavelength calibration
(i.e. we did not observe simultaneous reference spectra).  Lower
frequencies are not typical for known roAp stars and may be due to,
e.g., stellar rotation and surface spots. As an upper limit of the
studied frequency range we use 6\,mHz. The sampling frequency is
either 9.5\,mHz or 15.6\,mHz, and the Nyquist frequency is half of
that but still above the frequencies in known roAp stars.

In the following sections, we comment case-by-case on the individual
stars in our sample.

\subsection{HD\,107107}
\label{sec:107107}

\begin{figure}
  \vspace{3pt}
  \includegraphics[width=60mm,height=82mm,
  angle=270]{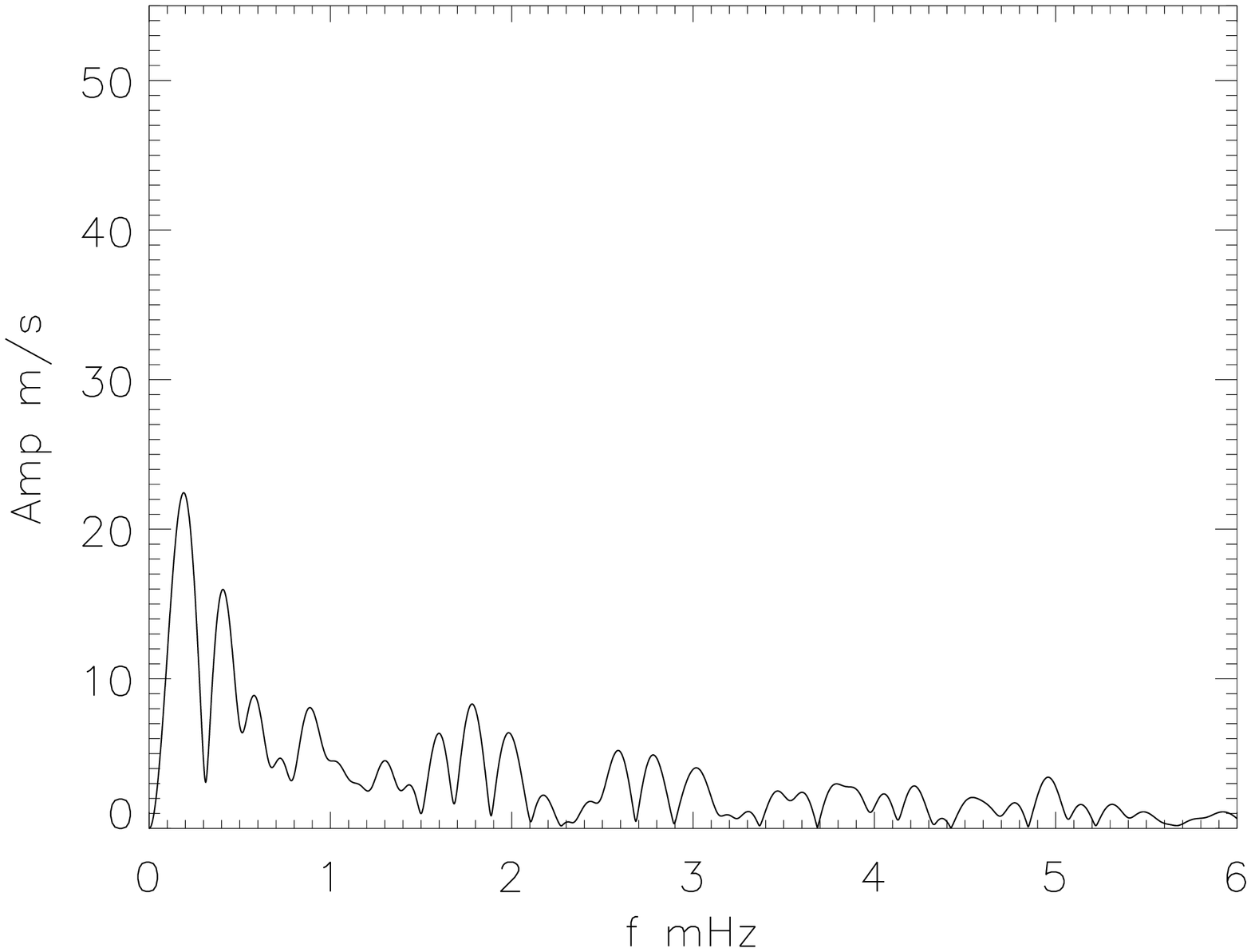}
  \includegraphics[width=60mm,height=82mm,
  angle=270]{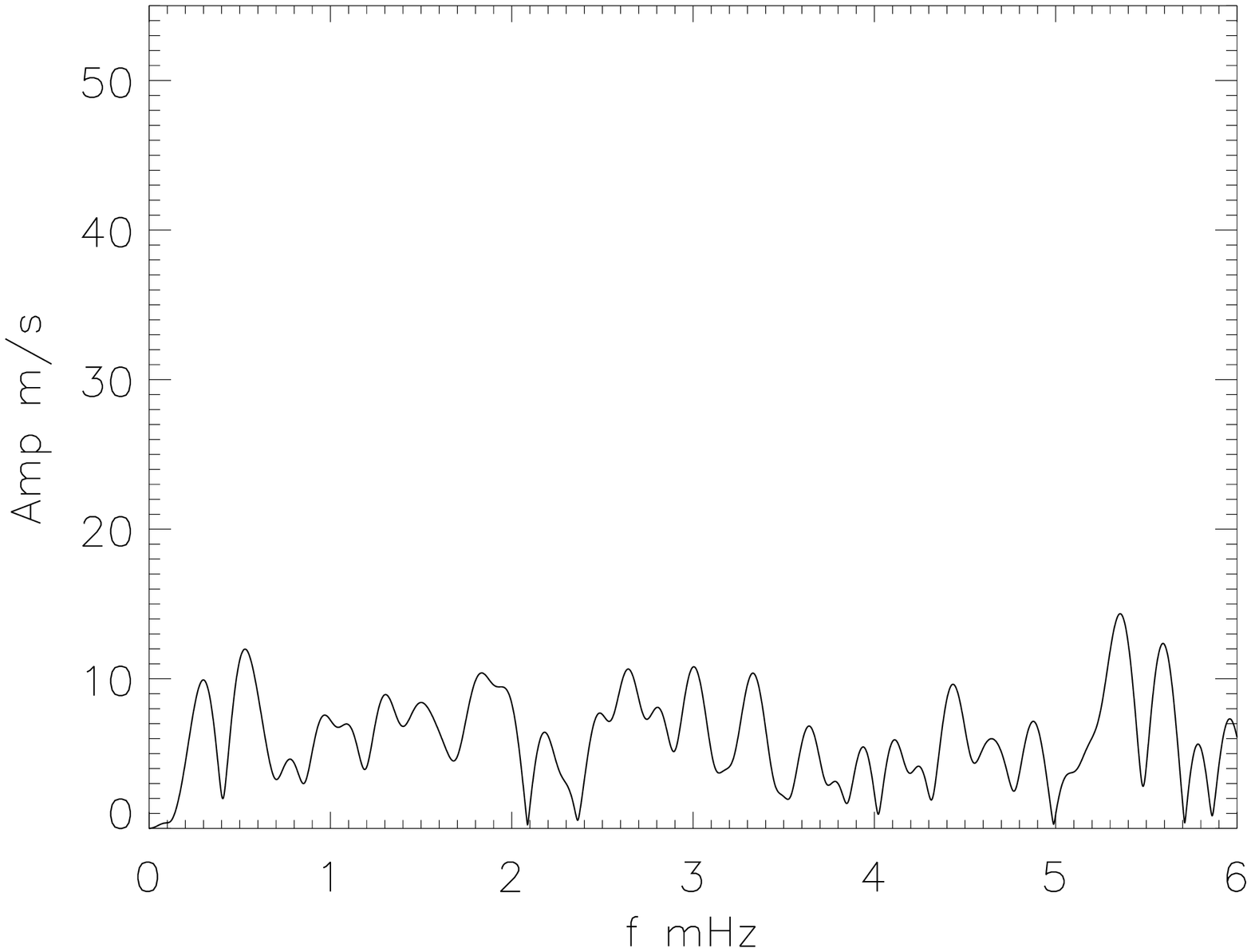}
  \caption{\label{fig:107107cc}Amplitude spectra from
    cross-correlation for the HD\,107107 spectra. Top: for the
    wavelength region $\lambda\lambda\,5150-5800$\,\AA; Bottom: for
    the wavelength region $\lambda\lambda\,6350-6700$\,\AA.}
\end{figure}
With a magnitude of $V=8.734$, this star is one of the faintest in our
sample. \citet{martinez93} obtained 1.89\,h photometry during a single
night and excluded periodic variability above 0.3 and 0.7\,mmag for
frequencies higher than 1.0 and 0.4\,mHz respectively. Based on his
$\beta$ photometry (Table\,\ref{tab:targets}) the corresponding
temperature from the grids by \citet{moonetal85} is 8300\,K.  The {\it
  Hipparcos} mission obtained 100 useful measurements; an amplitude
spectrum of those has a noise level of 3.1\,mmag and excludes peaks
above 9\,mmag. The distribution of {\it Hipparcos} data is, however,
not well suited for detecting rapid oscillations, but may instead show
rotation periods of spotted stars.  The Michigan Spectral Catalogue
\citep{houketal75} classifies the star as Ap CrEuSr.  No spectroscopy
has been previously published for HD\,107107.

Our observations of this star comprise 111 UVES spectra obtained over
a time span of 1.94\,hr with 63-s time resolution.  The spectra are
rotationally broadened to \vsini\,=\,10.5\,\kms\
(Table\,\ref{tab:targets}) and have single and strong REE lines.  The
star is very peculiar with strong \ndiii\ and \priii\ lines, probably
with a large ratio between abundances of singly- and doubly-ionised
REEs. The latter could indicate ionisation disequilibria of REEs, a
common feature among known roAp stars \citep{ryabchikovaetal04}.  The
\ion{Na}{d} $\lambda\lambda$\,5889.95 and 5895.92\,\AA\ lines each
have a stellar and two interstellar components that
are sharper, stronger and red-shifted with respect to the stellar one.

Radial velocity shifts of 47 stellar lines were measured
(Table\,\ref{tab:cogpow}) using the full line profiles where line
blending permitted it.  The core of \halpha\ is constant to 65\,\ms\
($\sigma=20$\,\ms) and other lines also exhibit no detectable
variability in the considered frequency range ($0.4 - 6.0$\,mHz).
Table\,\ref{tab:cogpow} and Figure\,\ref{fig:107107cog} show selected
results of the atomic lines measured in the frequency analysis. As
indicated in the table, radial velocity series for different lines of
same species were combined following \citet{kurtzetal06a} to reduce
the noise. Combining all line measurements, including different
species, we reach $\sigma=8$\,\ms\ and a maximum amplitude of 34\,\ms.
One peak at 1.8\,mHz (panel `HD\,107107 all') is just above the
4\,$\sigma$ detection limit, but originates from weak and blended
\ndiii\ lines. Yet, this ion has some of the highest amplitudes in
roAp stars and is excellent for detecting rapid pulsations. However,
this peak cannot be confirmed by lines of other elements, so following
the criteria in Sect.\,\ref{sec:drift} we rather label it a possible
detection that needs confirmation.  Other lines (such as \euii) show
peaks near the detection limit, but again the frequencies are
unconfirmed elsewhere. This demonstrates the difficulty in reliable
detection of signal at this noise level.  Selected results of the
cross-correlation analysis, presented in Table\,\ref{tab:ccpow} and
Fig.\,\ref{fig:107107cc}, strengthen this null-result. For the 5
stellar wavelength regions defined in Sect.\,\ref{sect-rvshift}, the
noise in the cross-correlation radial velocity shifts is around
4\,\ms\ without significant peaks above 15\,\ms. This is comparable to
the result for the stable telluric line region 
($\lambda\lambda 6873 - 6899$\,\AA).

\begin{figure}
  \vspace{3pt}
  \includegraphics[width=65mm, angle=270]{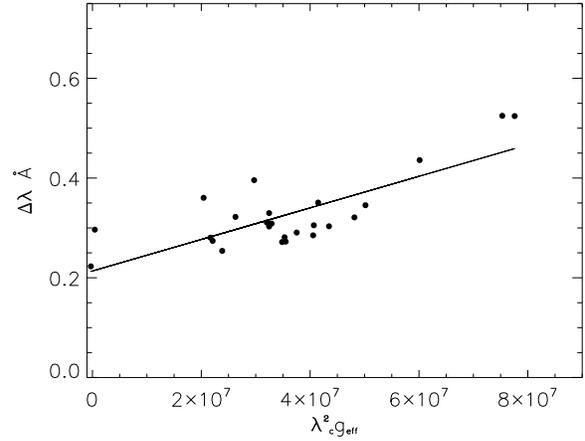}
  \caption{\label{fig:107107mag2}Magnetic broadening measurements of
    25 Fe lines of HD\,107107.  Measured Gaussian FWHM
    ($\Delta\lambda$) vs product of laboratory wavelength squared and
    Land\'e factor ($\lambda^2_c\,g_{\rm eff}$).  A significant
    relation ($r=0.77$) fitted with a least-squares linear fit, is
    indicated with a line.}
\end{figure}

Some \ion{Cr}{ii} and \ion{Fe}{ii} lines are magnetically broadened or
partially resolved and the mean quadratic field determined for 27 iron
lines is $\left<B_{\rm q}\right>=5.2\pm0.4$\,kG.  Using measured FWHM of 25
Fe lines, we derive (Fig.\,\ref{fig:107107mag2}) a field strength of
$3.4\pm0.6$\,kG (1\,$\sigma$ error).  Due to the combination of
marginally resolved lines, rotational broadening and line blending,
the use of apparently double lines gave less consistent results that,
however, support existence of a strong magnetic field.  {\small
  SYNTHMAG} synthetic profiles for different magnetic field strengths
were computed and compared to \ion{Cr}{ii}\,5116.049,
\ion{Cr}{ii}\,5318.382, \ion{Cr}{i}\,5247.566 and
\ion{Fe}{ii}\,6149.25\,\AA.  The best fit was
obtained for a magnetic field strength of 
$\langle B_{\rm synth}\rangle=4.5\pm0.5$\,kG (estimated error).  
The magnetic field modulus was further estimated directly
to be $\left<B\right>=5.6\pm2.3$\,kG from separations of the partially
resolved components of 5 \ion{Cr}{i}, \ion{Cr}{ii} and \ion{Fe}{i}
double lines using Gaussian fitting.

HD\,107107 is, therefore, a chemically peculiar A star that is
pulsationally stable above 39\,\ms\ ($\sigma=12$\,\ms) for all Nd and
Pr lines combined, and 9\,\ms\ ($\sigma=3$\,\ms) for
cross-correlations.  The star is a new magnetic star, with
marginally resolved Zeeman-split lines and a field of
$\left<B_{\rm q}\right>=5.2\pm0.4$\,kG.

\subsection{HD\,110072}
\label{sec:110072}

This is the faintest star in our sample. \citet{martinez93} excluded
photometric variability above about 1.4 and 0.8\,mmag for frequencies
above 0.4 and 0.9\,mHz respectively, based on 67\,min of photometry on
a single night. The {\it Hipparcos} data show no significant peaks
above 22\,mmag at shorter frequencies. \citet{houketal75} correctly
note that HD\,110072 is type Ap Sr(Cr) rather than K0
(\citealt{sao66}, classification source: M. W. Mayall).

We collected 69 spectra with UVES in 2.06\,hr with a time resolution
of 107\,s.  The spectra are sharp lined: the estimated \vsini\ is only
$3.3\pm0.5$ \kms\ and many strong REE lines are visible, e.g., Pr, Nd,
Y, \euii\ and \ion{Ce}{ii}.  Also Cr, \ion{Fe}{ii}, \ion{Ni}{ii},
\ion{Co}{i} and \ion{Al}{ii} are strong. In addition to a stellar
component, the \ion{Na}{d} lines each have 3 sharper and bluer
(with respect to the stellar component) interstellar components of 
comparable strengths.  The average spectrum
is very similar to those of two known roAp stars 33\,Lib and
HD\,176232 (Fig.\,\ref{fig:110072cmproap}). These have temperatures of
\teff\,$=7550\pm150$ and $7550\pm100$\,K respectively
\citep{ryabchikovaetal04}, which is supported by the indistinguishable
shapes of the \halpha\ wings of 33\,Lib and HD110072
(\teff\,$=7300$\,K). All three stars have similar peculiarities for
REEs, and \ion{Ba}{ii}, \ion{Si}{i} and \ion{Ca}{i} are considerably
weaker in HD\,110072.

The \halpha\ profile is strongly asymmetric (see Fig.\,\ref{fig:ha})
with a dip $\sim70$\,\kms\ blueward of the \halpha\ core. This dip is
only $3-4$ per cent below the \halpha\ wing, but about 40 per cent
broader (FWHM) than the \halpha\ core.  We re-observed the star 2 yr
later with FEROS at the ESO 2.2-m telescope and found this feature to have
disappeared. This indicates that HD\,110072 is a binary with a
secondary star that may be less luminous than the
primary. 
The broadness of the core 
of the secondary's \halpha\ line can either be due to faster rotation
($\sim 50$\,\kms) or a later spectral type (\halpha\ weakens toward
G0). At other wavelengths, the UVES and FEROS spectra are largely
identical and no secondary spectrum is seen (see also
Fig.\,\ref{fig:110072cmproap}). However, a few lines appear only in
the recent FEROS spectrum (resolution $R=48000$) such as at locations
of \ion{Sc}{i}\,6151.20\,\AA, \ion{O}{i}\,6156.77\,\AA,
\ion{Sm}{ii}\,6157.53\,\AA, \ion{Nd}{ii}\,6549.52\,\AA\ and
\ion{Fe}{ii}\,6552.33\,\AA.  These lines have a broadening similar to
the primary's spectrum.  We suspect they originate from this star and
their appearance is a result of viewing different aspects of a
magnetic field and/or spotted chemical surface distribution combined
with line blending.

\begin{figure}
  \vspace{3pt}
  \includegraphics[width=60mm,height=82mm,
  angle=270]{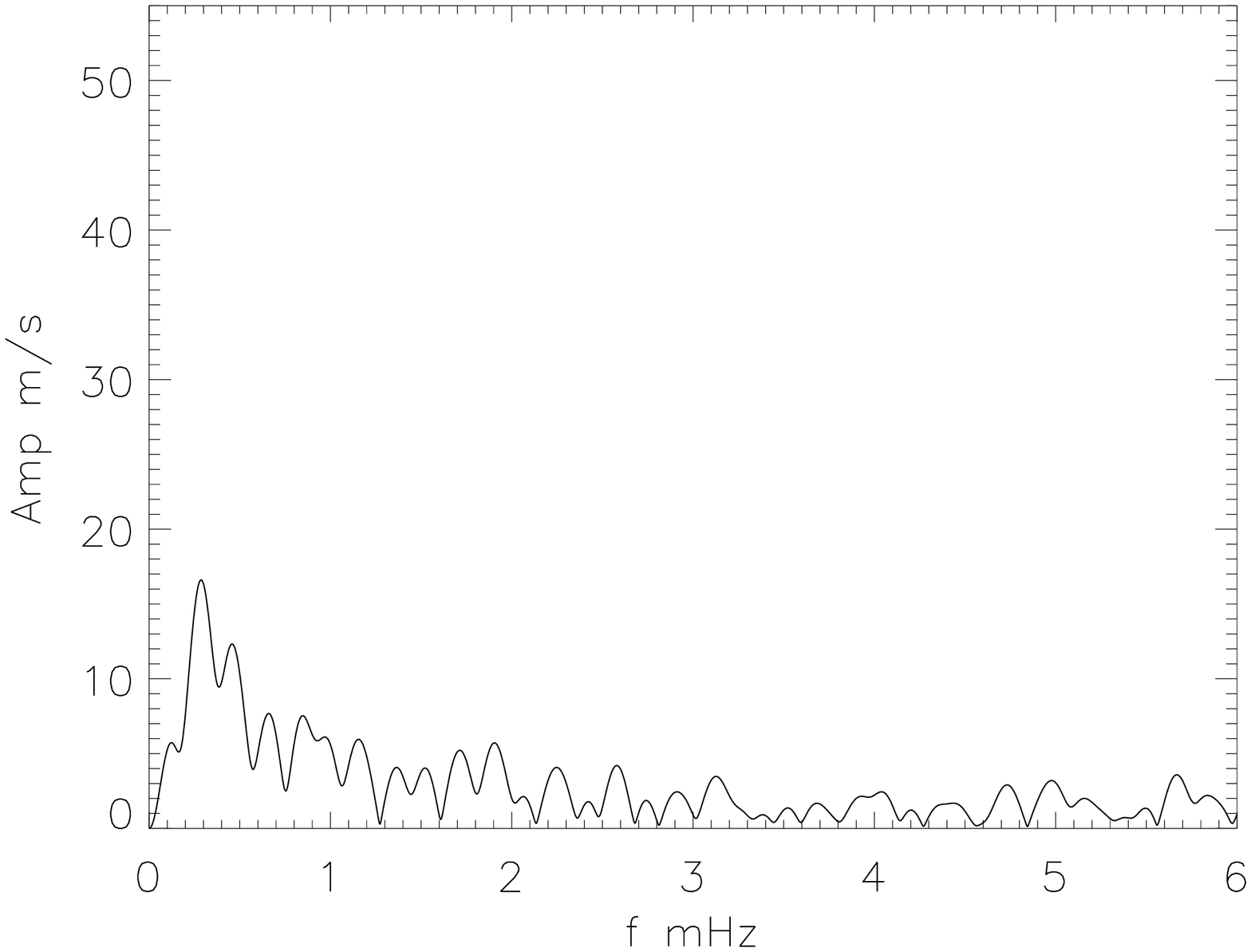}
  \includegraphics[width=60mm,height=82mm,
  angle=270]{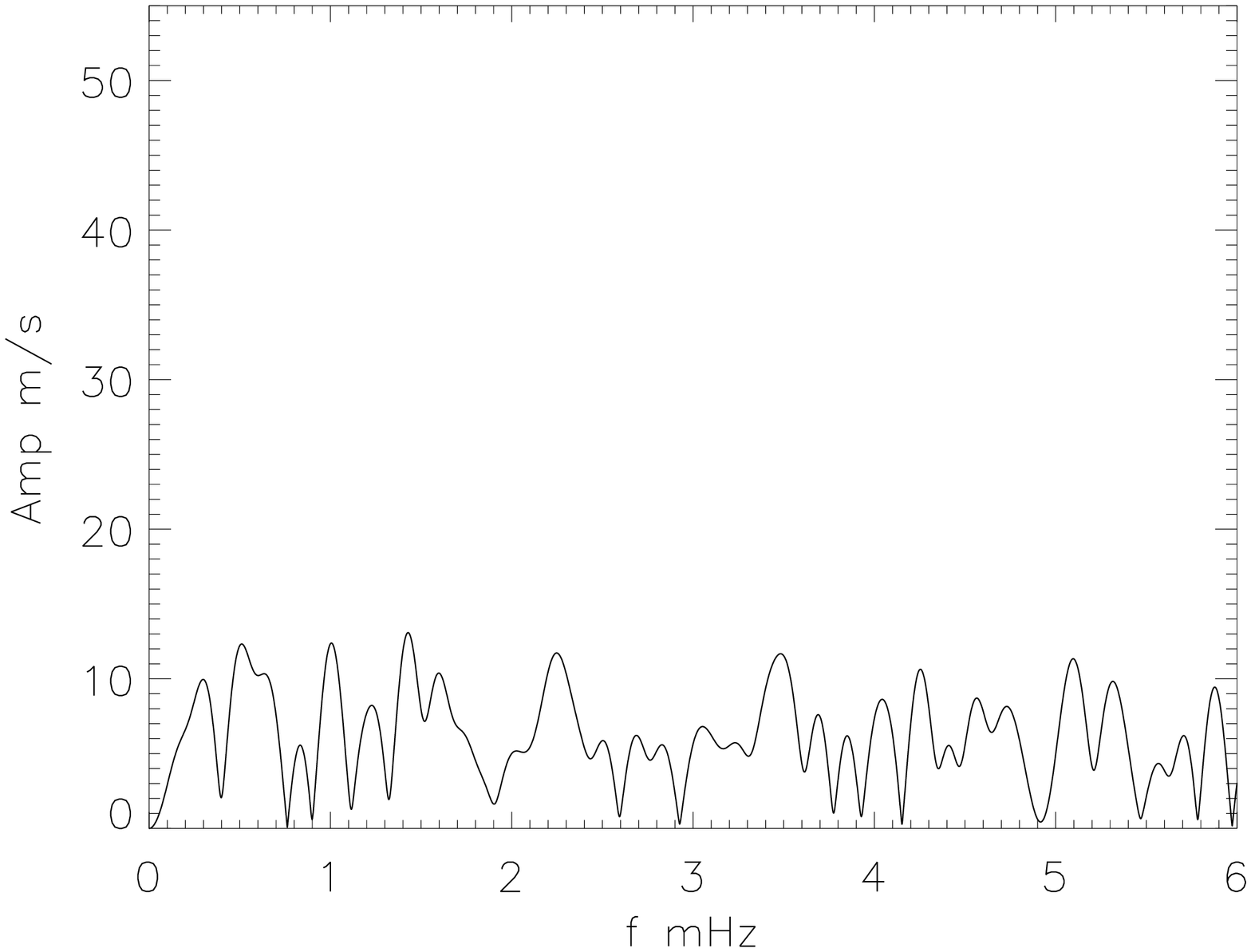}
  \caption{\label{fig:110072cc}Amplitude spectra from
    cross-correlation for the HD\,110072 spectra. Top: for the
    wavelength region $\lambda\lambda\,5150-5800$\,\AA; Bottom: for
    the wavelength region $\lambda\lambda\,6350-6700$\,\AA.}
\end{figure}

Radial velocity shifts were measured for 49 stellar lines
(Table\,\ref{tab:cogpow}) and show no detectable signal.  The blended
H$\alpha$ core is stable to 135\,\ms\ but the noise of
$\sigma=52$\,\ms\ in this radial velocity series is considerable,
caused by fewer spectra and lower $S/N$ than obtained for the other
roAp candidates.  Highest significance ($3.9\,\sigma$) is seen for the
combined \ion{Fe}{ii} lines at 1.86\,mHz, but iron is known in other
roAp stars to have low amplitudes, and there is no support for this
frequency from other lines. Combining all 49 lines
(Fig.\,\ref{fig:110072cog}) reduces the noise to 8\,m\,s$^{-1}$,
excluding peaks above 25\,\ms.  The cross-correlations in
Table\,\ref{tab:ccpow}, with examples in Fig.\,\ref{fig:110072cc},
result in flat amplitude spectra above 0.45\,mHz. A 4\,$\sigma$ peak
at 0.45\,mHz is caused by non-linear drifts. The telluric line region
is stable to 17\,\ms.

\begin{figure}
  \vspace{3pt}
  \includegraphics[width=65mm, angle=270]{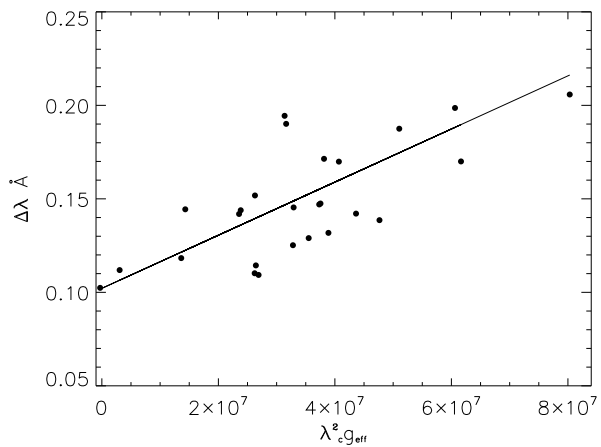}
  \caption{\label{fig:110072mag1} Magnetic broadening measurements of
    27 Fe lines in HD\,110072 plotting the measured Gaussian FWHM
    ($\Delta\lambda$) vs the product of laboratory wavelength squared
    and Land\'e factor ($\lambda^2_c\,g_{\rm eff}$).  A significant
    relation ($r=0.70$), fitted with a least-squares linear fit, is
    indicated with a line.  }
\end{figure}

The mean quadratic field determined for 33 iron lines is
$\left<B_{\rm q}\right>=1.5\pm0.6$\,kG.  Comparison of 14 Cr and Fe lines
with {\small SYNTHMAG} models indicates broadening by a magnetic field
of $1-3$\,kG, in particular for lines such as \ion{Cr}{ii}\,5246,
\ion{Cr}{i}\,5247, \ion{Fe}{i}\,5324 and \ion{Fe}{ii}\,6149.25\,\AA.
Measurements of FWHM of 27 iron lines with {\small IRAF}'s {\small
  onedspec.splot} task give (Fig.\,\ref{fig:110072mag1}) a mean
magnetic field modulus of $\left<B_{\rm FWHM}\right>=1.5\pm0.3$\,kG
(1\,$\sigma$ error).

HD\,110072, the coolest star in our sample, with
$T_{\rm{eff}}=7300$\,K, is thus a sharp-lined, double-lined binary
with strongly peculiar lines of, e.g., Nd, Pr and Eu.  HD\,110072 is a
new magnetic star with a field of $\left<B_{\rm q}\right>=1.5\pm0.6$\,kG and
has spectral features very similar to the roAp stars 33\,Lib and
HD\,176232.  It is therefore intriguing that the star is pulsationally
stable to 42\,\ms\ ($\sigma=10$\,\ms) for all Nd and Pr lines
combined, and 12\,\ms\ ($\sigma=3$\,\ms) for cross-correlations.  The
slow rotation of HD\,110072, its peculiar abundances and the rare
combination of a magnetic field and its binary status makes it an
important case for studying stellar evolution and diffusion processes.

\subsection{HD\,131750}
\label{sec:131750}

For this star, \citet{houketal75}'s classification is Ap CrEuSr.
\citet{martinez93} observed it $1-2$\,h on each of three nights. The
first night showed a clear peak around 8.5\,mHz (which is outside the
range we consider), but the other nights showed no variability above
0.6\,mmag for $f>0.6$\,mHz. \citet{strohmeyeretal66} listed an
unconfirmed 0.35\,mag variability from photographic plates (no period
given). {\it Hipparcos} data do not support this.

Our 111 spectra, obtained in 1.98\,hr with a time resolution 64\,s,
show HD\,131750 to be rotating with \vsini\,=\,25.3\,\kms\ with strong
lines of Nd, Eu and Pr.  The \ion{Na}{d} doublet is strong with
multiple sharp interstellar components.
\begin{figure}
  \vspace{3pt}
  \includegraphics[width=60mm,height=82mm,
  angle=270]{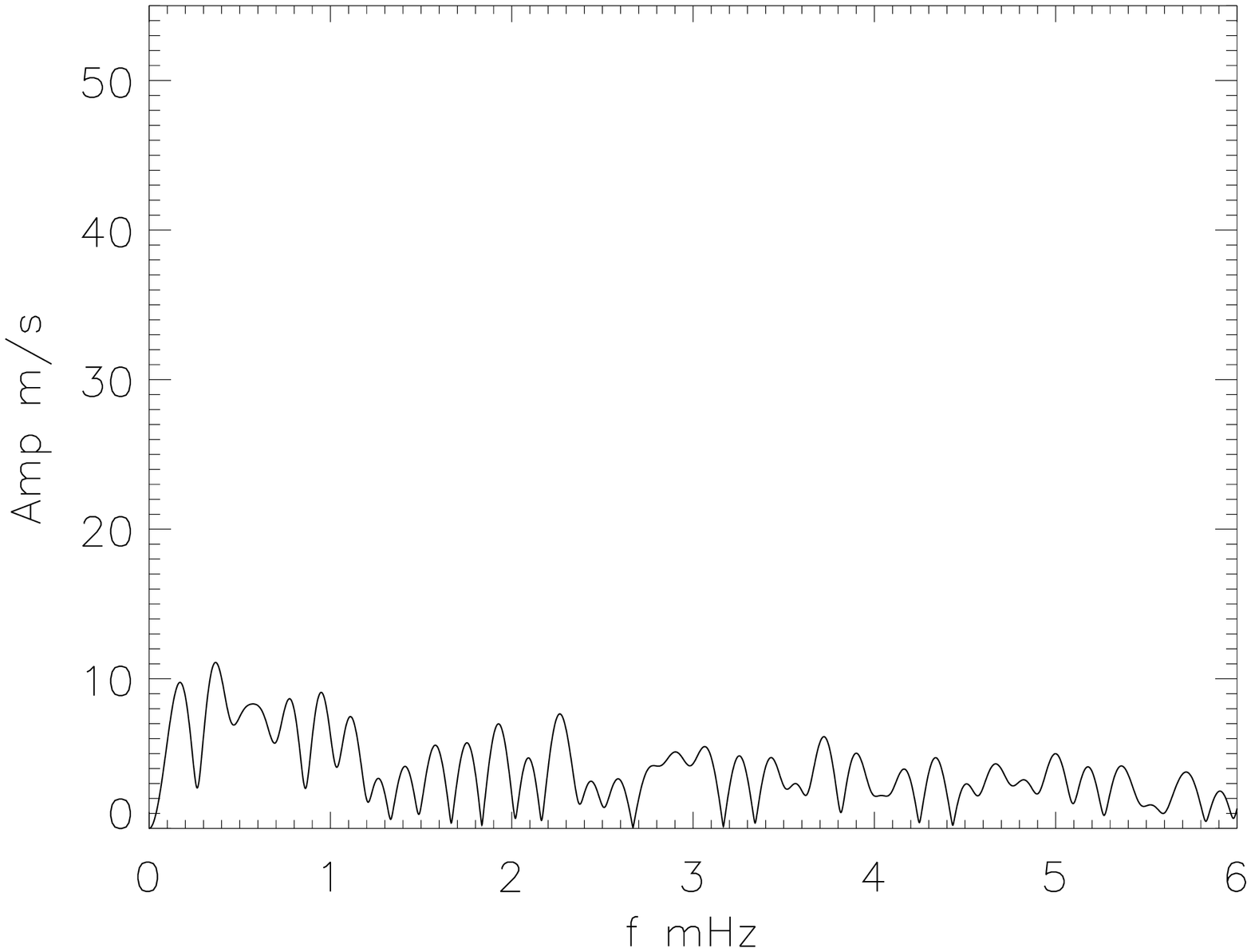}
  \includegraphics[width=60mm,height=82mm,
  angle=270]{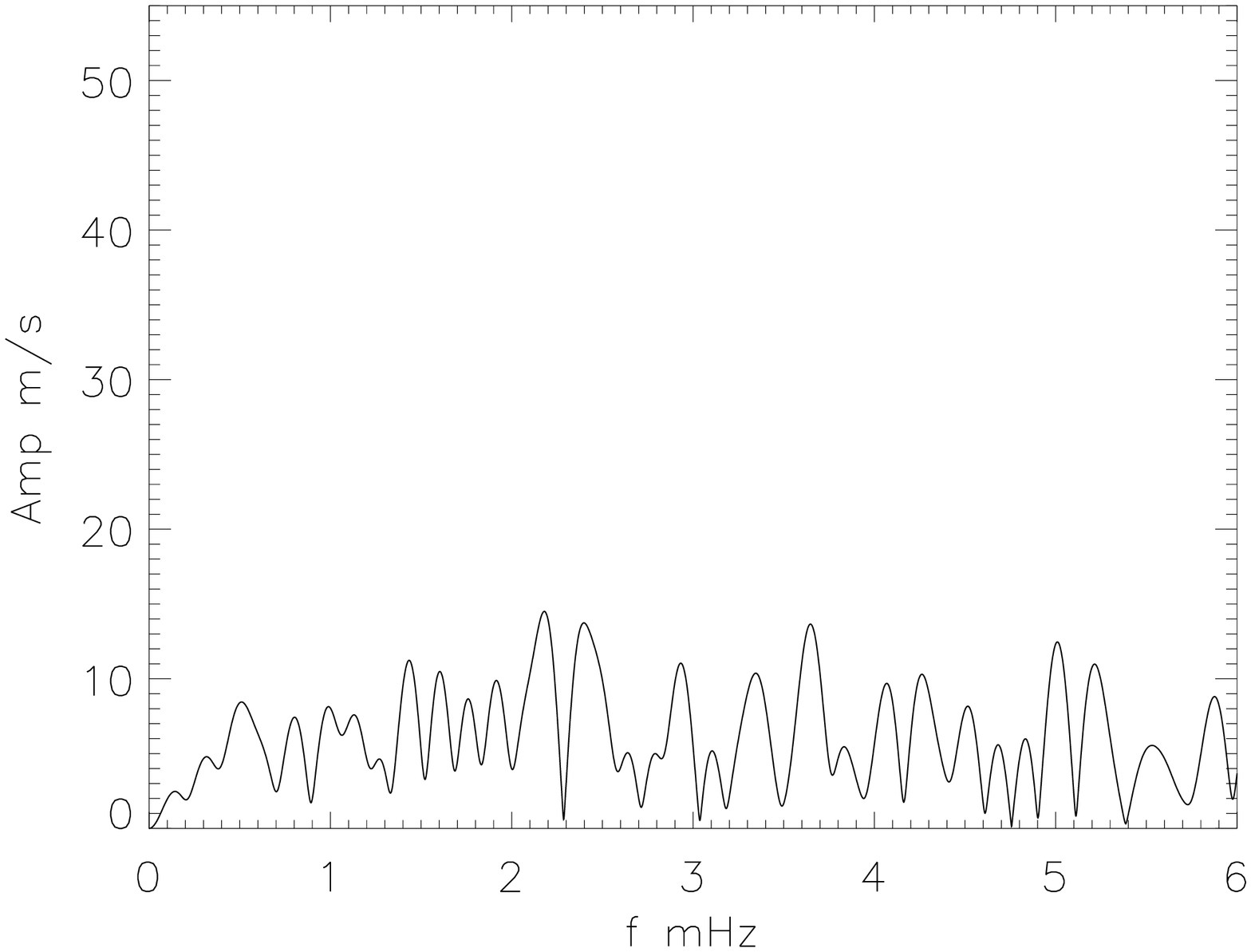}
  \caption{\label{fig:131750cc}Amplitude spectra from
    cross-correlation for the HD\,131750 spectra. Top: for the
    wavelength region $\lambda\lambda\,5150-5800$\,\AA; Bottom: for
    the wavelength region $\lambda\lambda\,6350-6700$\,\AA.}
\end{figure}
Radial velocity shifts of 40 stellar lines were measured
(Table\,\ref{tab:cogpow}). The noise in the radial-velocity
measurements is relatively high due to the rotation rate.  The core of
H$\alpha$ is stable to 72\,\ms\ ($\sigma=21$\,\ms), while all measured
lines combined (Fig.\,\ref{fig:131750cog}) reveal no rapid pulsation
above 39\,\ms\ ($\sigma=8$\,\ms).  The cross-correlations provide an
upper limit of $9-15$\,\ms\ to the averaged radial velocity shifts of
the measured regions. Flat amplitude spectra of two of these regions
are given in Fig.\,\ref{fig:131750cc}.
The \ion{Na}{d} region has a significant peak at 0.45\,mHz caused by
non-linear instrumental drifts in the data.  Elsewhere, the amplitude
spectra are flat, devoid of significant peaks.

Some of the stellar line profiles have flat `squared' cores, such as
\ion{Cr}{ii}\,5613.18 and \ion{Fe}{ii}\,5854.19, and others have
apparently split cores such as \ion{Fe}{ii}\,5457.73,
\ion{Cr}{ii}\,5472.60 and \ion{Fe}{i}\,5862.35.  Magnetically resolved
lines require a Zeeman splitting at least comparable to rotational
broadening, in this case around $\left<B\right>=8$ kG.  The mean
quadratic field was derived to $\left<B_{\rm q}\right>=5.3\pm3.3$\,kG by
using 9 iron lines.  It is our impression that the upper limit on any
magnetic field modulus (at this rotation phase) is 8\,kG.

HD\,131750 is thus an Ap star with strong REEs, no rapid oscillations
above 58\,\ms\ ($\sigma=18$\,\ms) amplitude for all Nd and Pr lines
combined, and 9\,\ms\ ($\sigma=3$\,\ms) for cross-correlations.  The
rotation rate \vsini\,=\,25.3\,\kms\ results in considerable line
blending and the star may have a magnetic field of a few kG.

\subsection{HD\,132322}
\label{sec:132322}
\citet{houketal75} classify this star as `Ap SrCrEu,\,A1' and note
that Sr is extremely strong.  Photometric Str\"omgren indices also
indicate strong peculiarity.  \citet{martinez93} examined the star
during a single night and excluded rapid photometric variability above
0.4\,mmag for the frequency range $0.4 -
10$\,mHz. \citet{levatoetal96} obtained 2 spectra of the star and
found \vsini\,=\,85\,\kms, a somewhat high value for an Ap
star. Additionally, \citet{hubrigetal06} discovered a mean
longitudinal magnetic field of $357\pm51$\,G.

\begin{figure}
  \vspace{3pt}
  \includegraphics[width=65mm,height=88mm,
  angle=270]{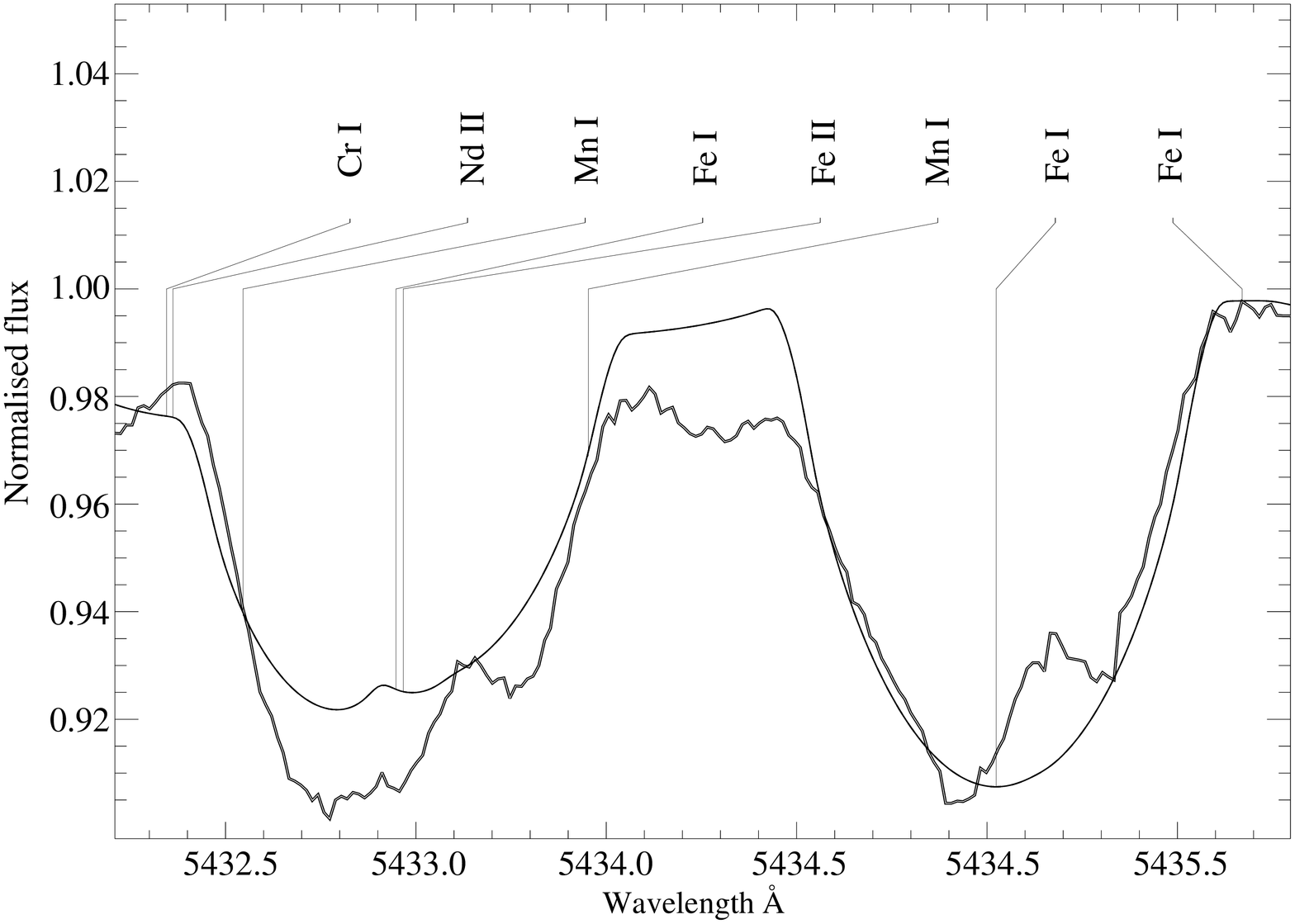}
  \caption{\label{fig:132322_5434}Double line structures in the
    magnetically insensitive line \ion{Fe}{i}\,5434.52 in the averaged
    spectrum of HD\,132322 (thick line). A synthetic model for
    \vsini\,=\,30\,\kms\ and $\left<B\right>=0$\,G is superposed (thin
    line) with its dominant atomic lines indicated.  }
\end{figure}

HD\,132322, $V=7.357$\,mag, is the second brightest star in our sample
and the individual spectra have $S/N$ well above 100. A total of 111
spectra were obtained in 2.03\,hr, providing a 66-s time resolution.
The spectra appear to show splitting of all lines, such as Ba, Fe and
Cr lines, and also of all REEs, such as Nd and \euii. Surprisingly,
even lines with small Land\'e factors are double
(Fig.\,\ref{fig:132322_5434}) which would indicate a spotted surface
distribution rather than splitting of lines by a magnetic field.  It
is, however, improbable that all elements, in particular Fe, are split
due to a spotted surface.  In the course of the 2 hr of observations,
we do not find any systematic change in the radial velocity difference
of the components, nor in their centre of gravity.  The estimated
rotational broadening is \vsini\,=\,25--35\,\kms, based on the full
double profiles, and \vsini\,$=34.3\pm1.8$\,\kms\ from the quadratic
magnetic field analysis, which together with the peculiarity results
in considerable blending.  Lines of Pr are not very strong but may
also be double.  The otherwise useful lines of \ion{Ce}{ii} are
absent. For some roAp stars this ion has the highest pulsation
amplitude, such as for $\beta$\,CrB (Kurtz, Elkin \& Mathys 2007).
The \ion{Na}{d} doublet is strong with a broad and a sharp
(interstellar) component.

\begin{figure}
  \vspace{-1pt}
  \includegraphics[width=60mm,height=82mm,
  angle=270]{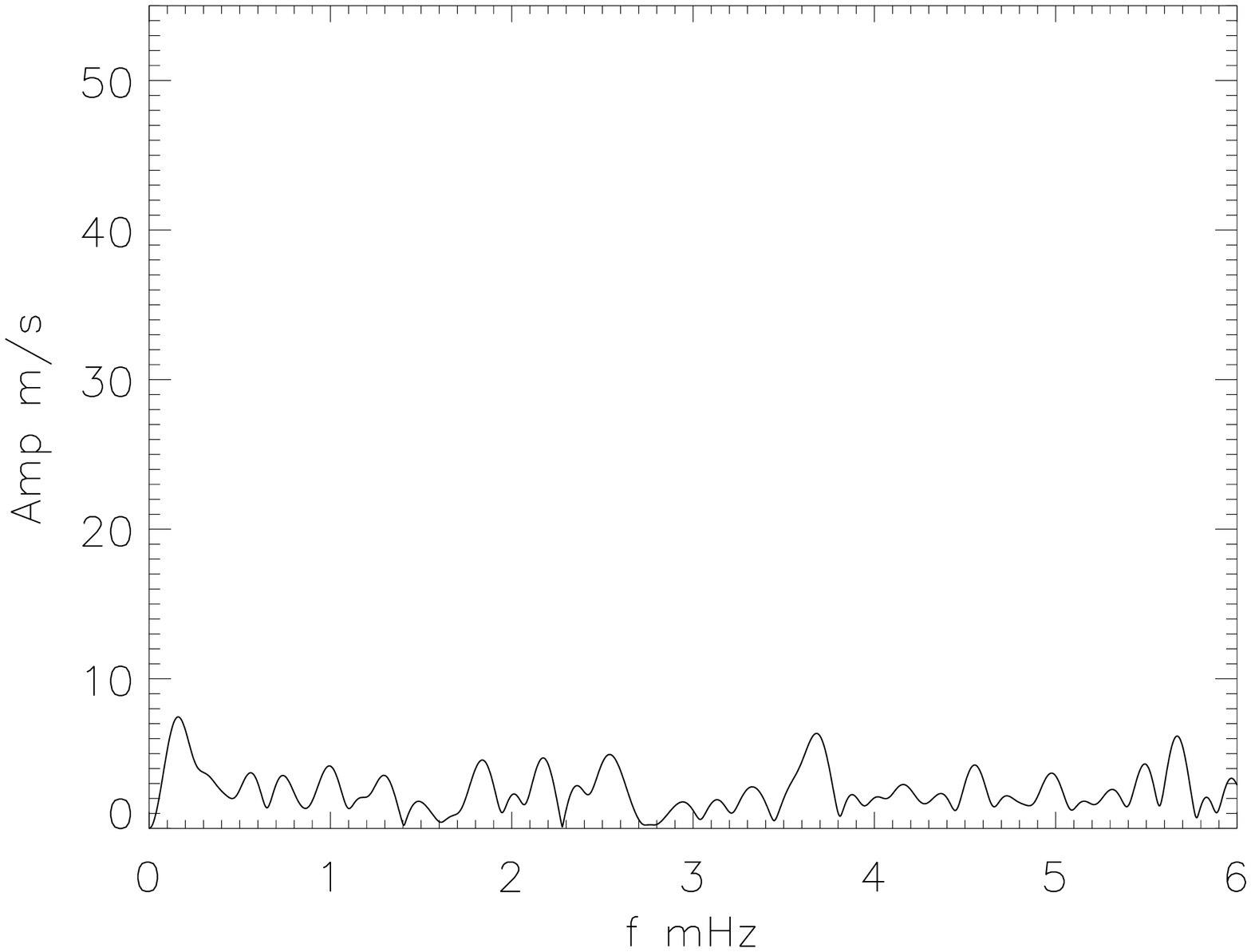}
  \includegraphics[width=60mm,height=82mm,
  angle=270]{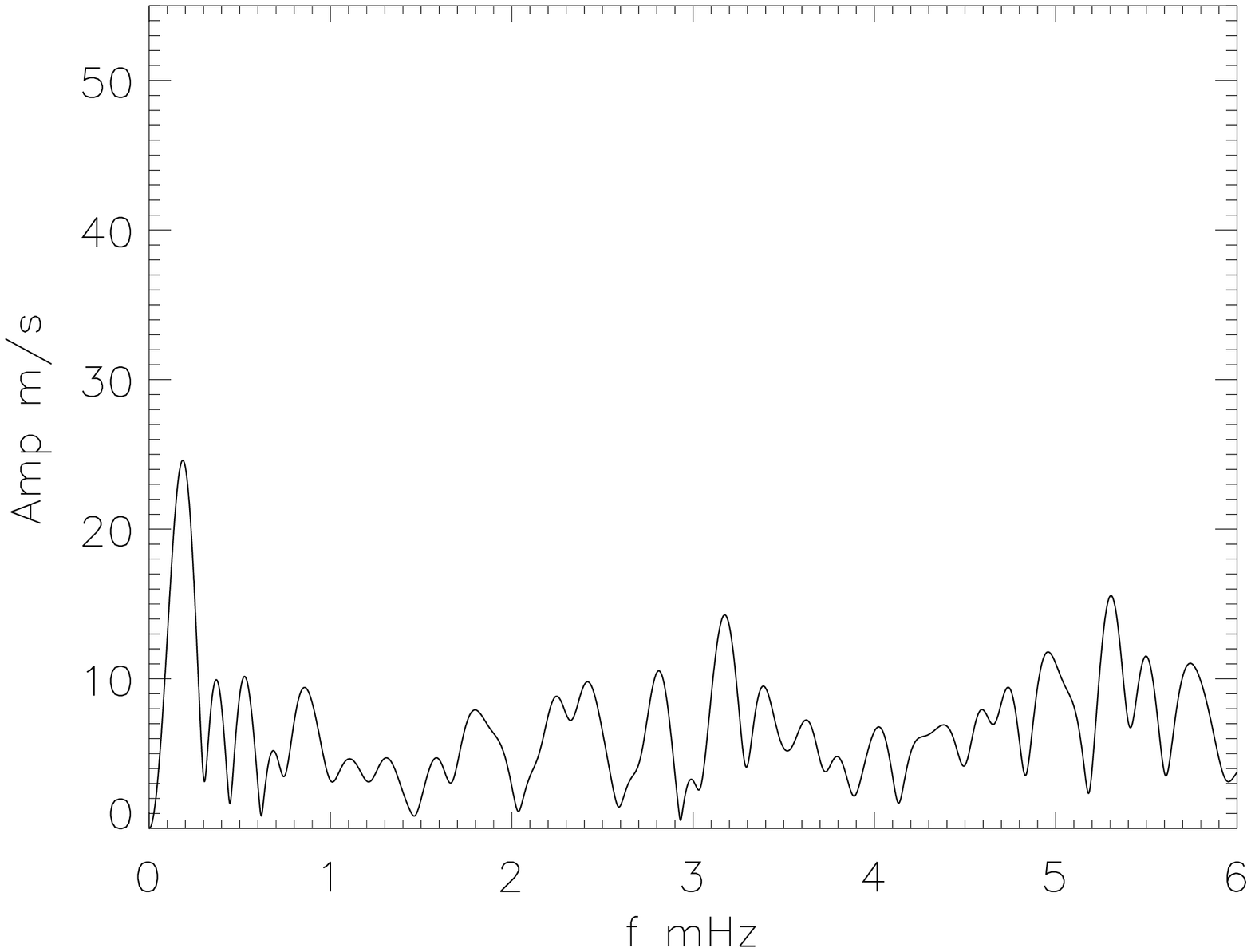}
  \caption{\label{fig:132322cc}Amplitude spectra from
    cross-correlation for the HD\,132322 spectra. Top: for the
    wavelength region $\lambda\lambda\,5150-5800$\,\AA; Bottom: for
    the wavelength region $\lambda\lambda\,6350-6700$\,\AA.}
\end{figure}

Radial velocity shifts measured for 34 stellar lines put an upper
limit of 24\,\ms\ ($\sigma=8$\,\ms) to rapid pulsation when combining
all lines (Fig.\,\ref{fig:132322cog}).  The core of H$\alpha$ is
stable to 59\,\ms\ ($\sigma=19$\,\ms, Table\,\ref{tab:cogpow}).  All
amplitude spectra are flat, except for that of combined \ion{Cr}{ii}
which has a non-significant peak at 1.31\,mHz.  The cross-correlation
analysis (see, e.g., Fig.\,\ref{fig:132322cc}) shows flat amplitude
spectra down to amplitudes of $6-15$\,\ms\ without any significant
peaks.

Using 14 iron lines, the mean quadratic field could only be constrained
to a maximum intensity of $\left<B_{\rm q}\right>\le6.0$\,kG. This weak
constraint results from the fact that, because of the considerable
broadening and distortion of the spectral lines, only a small number
of them could be identified as sufficiently free from blends to be
used in the analysis. Indeed many of the lines, including those that
are magnetically insensitive, show 
double structures, which typically consist of a component
separated $30\pm3$\,\kms\ from a redder and $45\pm15$ per cent weaker
component (in equivalent width).  The FWHM of the weaker component is
$31\pm18$ per cent less than for the other.  This pattern is similar
for lines of REEs, Cr and Fe (cf. Figs.\,\ref{fig:spot} and
Fig.\,\ref{fig:132322_5434}).  A comparison of the \halpha\ profile to
a synthetic spectrum, shifted in wavelength corresponding to the
separation in the double structures, firmly excludes a secondary
spectrum of a star of comparable brightness as it would have
introduced a strong asymmetry.  New high-resolution spectra are
required to understand these double structures.

HD\,132322 is a magnetic Ap star with projected rotation velocity
\vsini\,=\,34\,\kms. All lines, even those of REEs, are double except
for \halpha.  This can partly, but not fully, be explained by
abundance spots or a secondary spectrum. Pulsations are excluded down
to 37\,\ms\ ($\sigma=13$\,\ms) for all Nd and Pr lines combined, and
6\,\ms\ ($\sigma=2$\,\ms) for cross-correlations.


\subsection{HD\,151301}
\label{sec:151301}
The star is classified Ap SrCrEu by \citet{houketal75}.
\citet{martinez93} observed it on 6 nights for $0.9-1.4$\,hr each, and
excluded pulsations down to $0.5-0.8$ mmag for frequencies above
0.5\,mHz.

We obtained 111 UVES spectra in 1.95\,hr at a time resolution of 63\,s.
The lines of REEs are strong (Nd, Pr, Eu) and many are double (see
also Fig.\,\ref{fig:ha}) indicating abundance spots.  The \ion{Na}{d}
lines have a stellar component and a sharp, strong, interstellar line
at longer wavelength.  The photometric temperature is $T_{\rm eff} =
8000$\,K and the spectroscopy provides an upper limit $T_{\rm
  eff}=9000$\,K for $\log\,g = 3.5$.  With the astrometric luminosity
of the star, this agrees with HD\,151301 being more than halfway
through its main sequence lifetime.

\begin{figure}
  \vspace{3pt}
  \includegraphics[width=60mm,height=82mm,
  angle=270]{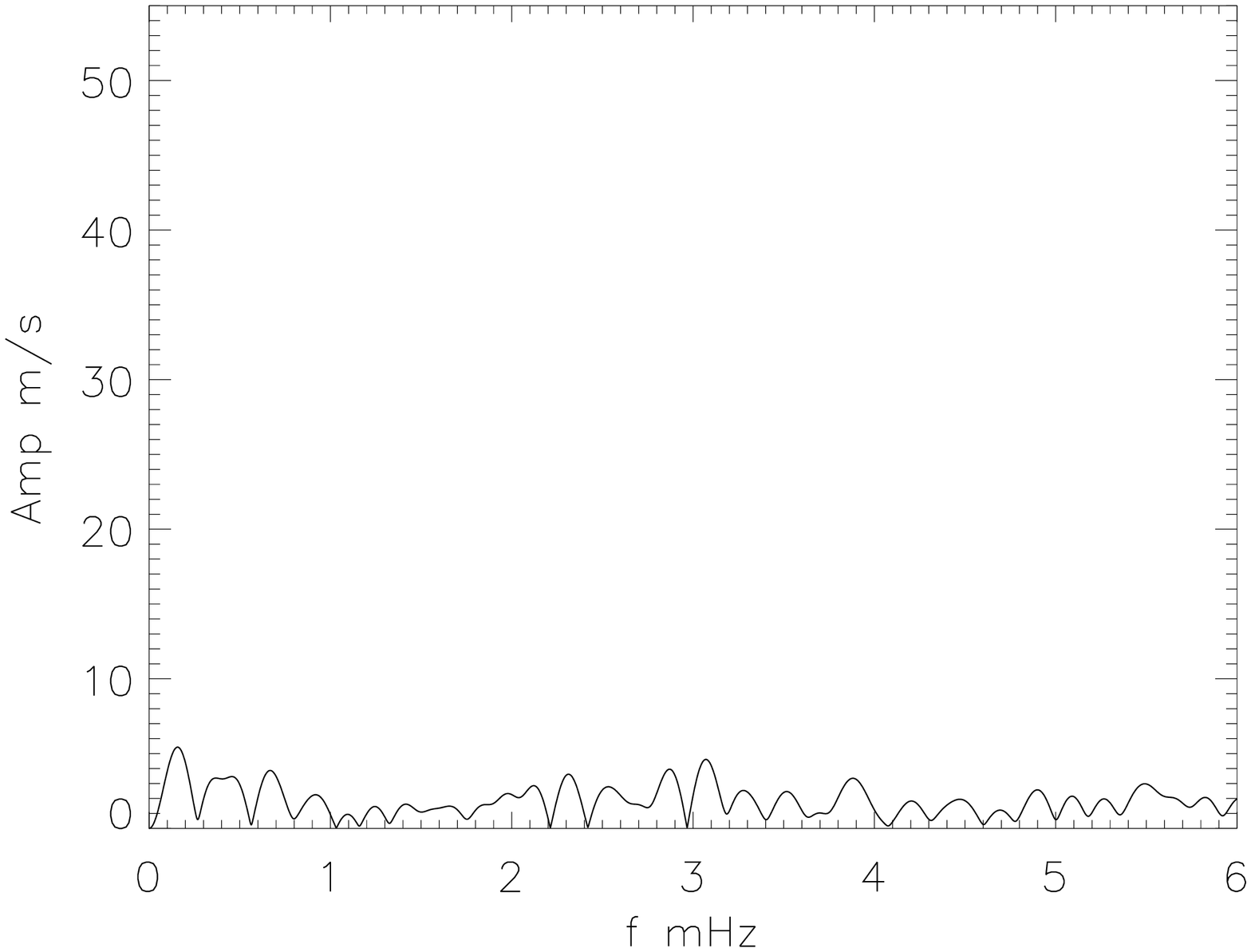}
  \includegraphics[width=60mm,height=82mm,
  angle=270]{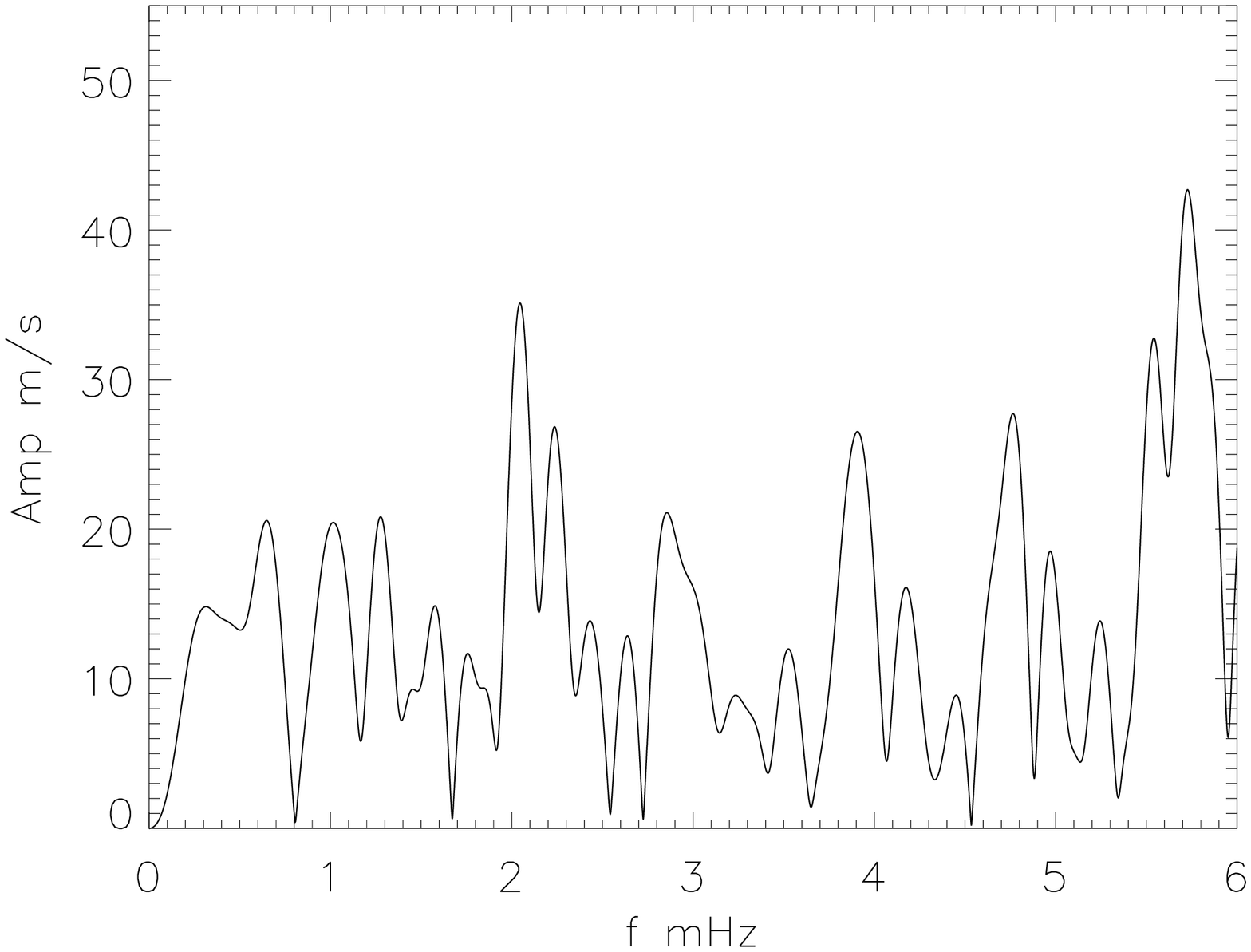}
  \caption{\label{fig:151301cc}Amplitude spectra from
    cross-correlation for the HD\,151301 spectra. Top: for the
    wavelength region $\lambda\lambda\,5150-5800$\,\AA; Bottom: for
    the wavelength region $\lambda\lambda\,6350-6700$\,\AA.}
\end{figure}

Radial velocity shifts of 39 stellar lines were measured
(Table\,\ref{tab:cogpow}). Lines of \prii, \ion{Ce}{ii} and
\ion{Nd}{ii} are weak with considerable scatter in their
radial-velocity series. When combining all 39 lines, we find an upper
limit on rapid pulsation of 18\,\ms\ ($\sigma=6$\,\ms), while the core
of \halpha\ is stable to 54\,\ms\ ($\sigma=19$\,\ms). Each double
component of the two strongest \euii\ lines was measured and the
combined amplitude spectrum (Panel `EuII', Fig.\,\ref{fig:151301cog})
shows a significant ($4.2\,\sigma$) peak at 2.22\,mHz.  It is equally
significant for the individual \euii\ lines and when combining three
available \euii\ lines.  No other line or combination of lines confirm
this peak, including a few weak lines of \ion{La}{ii} and \ion{Ce}{ii}
added to the analysis. Even cross-correlation of the whole \euii\
profiles did not recover the 2.22\,mHz frequency and it is therefore
considered a probable, but unconfirmed detection.  The spectrum of
HD\,151301 is rich in lines, and the cross-correlations reach a low
noise level (Table \ref{tab:cogpow}), in particular in the bluer
spectrum below the 6000\,\AA\ gap (1.4\,\ms,
Fig.\,\ref{fig:151301cc}). The 
spectrum region above the 6000\,\AA\ gap results in considerably larger 
noise ($8-10$\,\ms) but also excludes significant rapid pulsations.
The rotational broadening is \vsini\,$=13.7\pm1.1$\,\kms.  Some
\ion{Fe}{i} and \ion{Fe}{ii} lines show additional broadening or
asymmetric profiles.  An upper limit of the mean quadratic field is
found to $\left<B_{\rm q}\right>\le2.4$\,kG, using 18 Fe lines.  
{\small SYNTHMAG} fitting to \ion{Cr}{ii}\,5313.56, 
\ion{Fe}{i}\,6137.69 and \ion{Fe}{i}\,5266.55
rejects magnetic fields stronger than 2 kG.  However, from 
line-width measurements of magnetic broadening of 13 Fe lines,
a weak, possibly insignificant relation
($r=0.50$) indicates $\left<B_{\rm FWHM}\right>=1.2\pm0.60$\,kG.

HD\,151301 is a strongly chemically peculiar star possibly having a
magnetic field of up to 2\,kG. The surface distribution of REEs is
spotted. No pulsations are seen down to amplitudes of 24\,\ms\
($\sigma=8$\,\ms) for all Nd and Pr lines combined, and 5.4\,\ms\
($\sigma=1.4$\,\ms) for cross-correlations.

\subsection{HD\,170565}
\label{sec:170565}

\citet{martinez93} observed this star on 5 nights. In general, no
pulsations are seen down to 0.7\,mmag for frequencies above
0.5\,mHz. However, the last two nights show two peaks around 1.7 and
2.6\,mHz at $3 - 4$ times the noise. The first of these is also
present on the first and most intensively observed
night. \citet{kudryavtsevetal06} detected a magnetic field in this
star with a mean longitudinal field of $1.76\pm0.17$\,kG.

\begin{figure}
  \vspace{3pt}
  \includegraphics[width=60mm,height=82mm,
  angle=270]{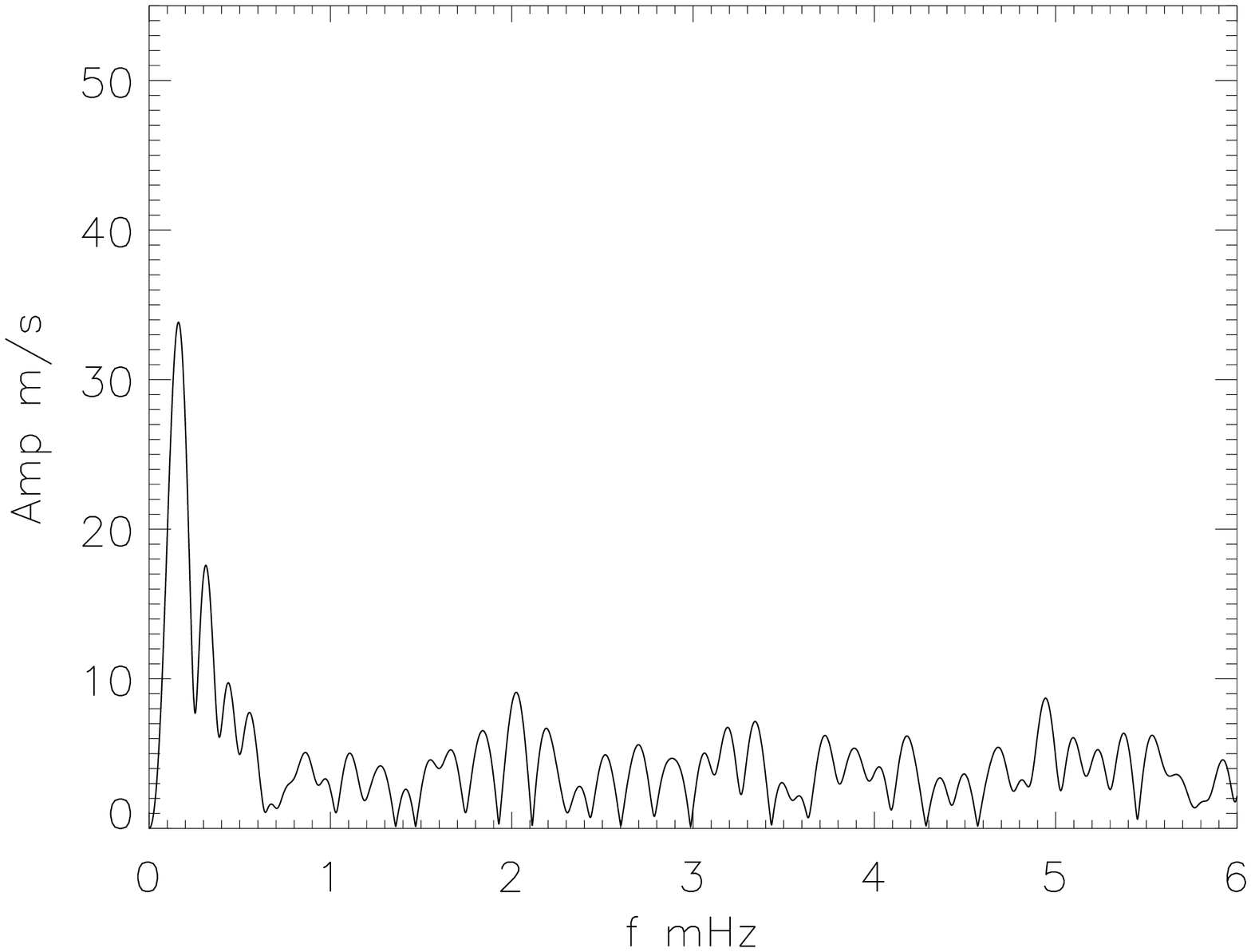}
  \includegraphics[width=60mm,height=82mm,
  angle=270]{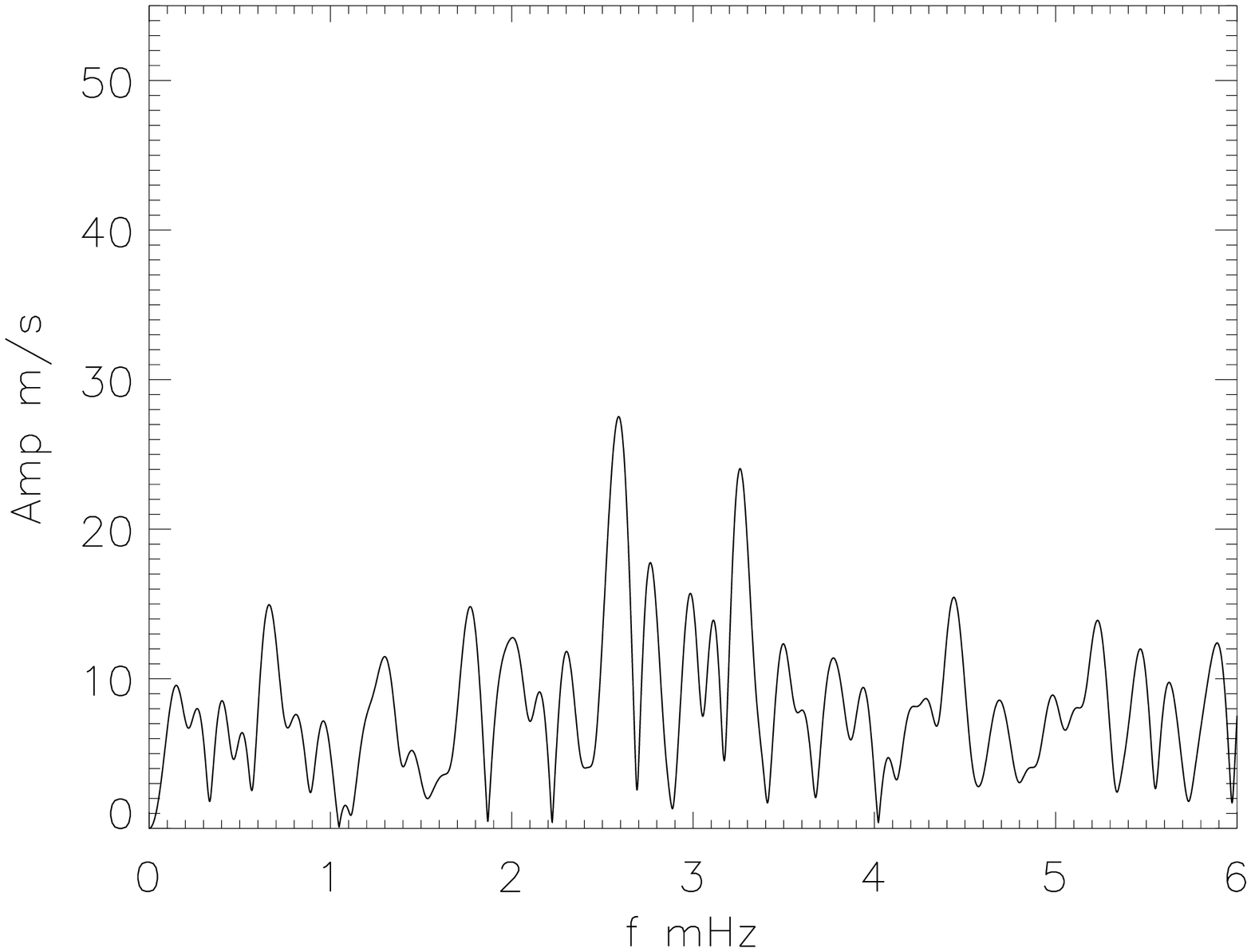}
  \caption{\label{fig:170565cc}Amplitude spectra from
    cross-correlation for the HD\,170565 spectra. Top: for the
    wavelength region $\lambda\lambda\,5150-5800$\,\AA; Bottom: for
    the wavelength region $\lambda\lambda\,6350-6700$\,\AA.}
\end{figure}

We obtained 85 spectra of the star in 2.50\,h with a time resolution
of 106\,s. The star is highly peculiar and has
\vsini\,=\,18\,\ms. There are many strong REE lines, such as those of
\euii, \ion{Ce}{ii} and Nd that are all double, indicating abundance
spots. \prii\ is less strong, but also double. As a result
line-blending is considerable.  Radial velocity shifts of 41 stellar
lines were measured and reject pulsations down to 32\,\ms\
($\sigma=8$\,\ms) when combining all lines (Fig.\,\ref{fig:170565cog}
and Table\,\ref{tab:cogpow}).  The core of \halpha\ shows stability to
75\,\ms\ ($\sigma=24$\,\ms).  There are no confirmed significant peaks
in periodograms for any combination of radial velocity series.
Cross-correlations show stability down to $10-30$\,\ms\
(Table\,\ref{tab:ccpow} and Fig.\,\ref{fig:170565cc}). Only the
$\lambda\lambda 6350-6700$\,\AA\ region shows a significant
(4.2\,$\sigma$) peak at 2.59\,mHz (the same as the second `transient'
photometric period).  No other line measurements or cross-correlations
support this unconfirmed detection.
Cr is double or broadened while Fe is broadened in several cases.
Magnetic measurements were tried with three methods: mean quadratic
field measurements, {\small SYNTHMAG} fitting to 4 Cr and Fe lines,
and from widths of 22 Fe lines. However, due to the rotation, line
blending, and a possible inhomogeneous stellar surface distribution of
iron, it is not possible to constrain the known magnetic field with
these data.

This star is a known magnetic star which is consistent with the
present data. It is a chemically peculiar Ap star with a spotted
surface distribution of Cr and REEs, and is pulsationally stable down
to 32\,\ms\ ($\sigma=10$\,\ms) for all Nd and Pr lines combined, and
10\,\ms\ ($\sigma=5$\,\ms) for cross-correlations.

\subsection{HD\,197417}
\label{sec:197417}

The classification for this object is Ap CrEu(Sr) \citep{houketal75},
and the star has been investigated earlier due to its chemical
peculiarities and photometric variability associated with its rotation
period.  \citet{martinez93} observed the star on two nights. The 
corresponding
amplitude spectra are flat and place an upper limit of 0.6\,mmag on
photometric variability.  \citet{floquetetal84} studied this star and
found a photometric rotation period of $4.551\pm0.002$\,d. With
spectra obtained at 12\,\AA\ mm$^{-1}$ dispersion, they used Balmer
and \ion{Ca}{ii} lines to determine $T_{\rm eff} = 9500$\,K and $\log
g =4.0$, and from \ion{Mg}{ii}\,4481\,\AA\ they measured
\vsini\,=\,23\kms.  A strong variability was noticed in the intensity
and profile of \ion{Ca}{ii}\,3922.6\,\AA\ and by assuming an oblique
rotator geometry and a spotted surface distribution of elements, they
proposed an inclination of $i=64$\degr (angle of the stellar rotation
axis to the line of sight) and that \ion{Ca}{ii}, \ion{Eu}{ii} and
\ion{Sr}{ii} in particular seemed located in a common spot.  From 156
{\it Hipparcos} measurements we do not find this period significant,
but cannot reject it based on the noisier {\it Hipparcos} data alone.
Based on 4 spectra, \citet{levatoetal96} listed the star as probably
single (25 per cent chance for random velocity distribution).
\citet{paunzenetal05} list the star as a confirmed chemically peculiar
star with $\Delta a=0.054$ and $(b-y)_0=0.032$.

\begin{figure}
  \vspace{3pt}
  \includegraphics[width=60mm,height=82mm,
  angle=270]{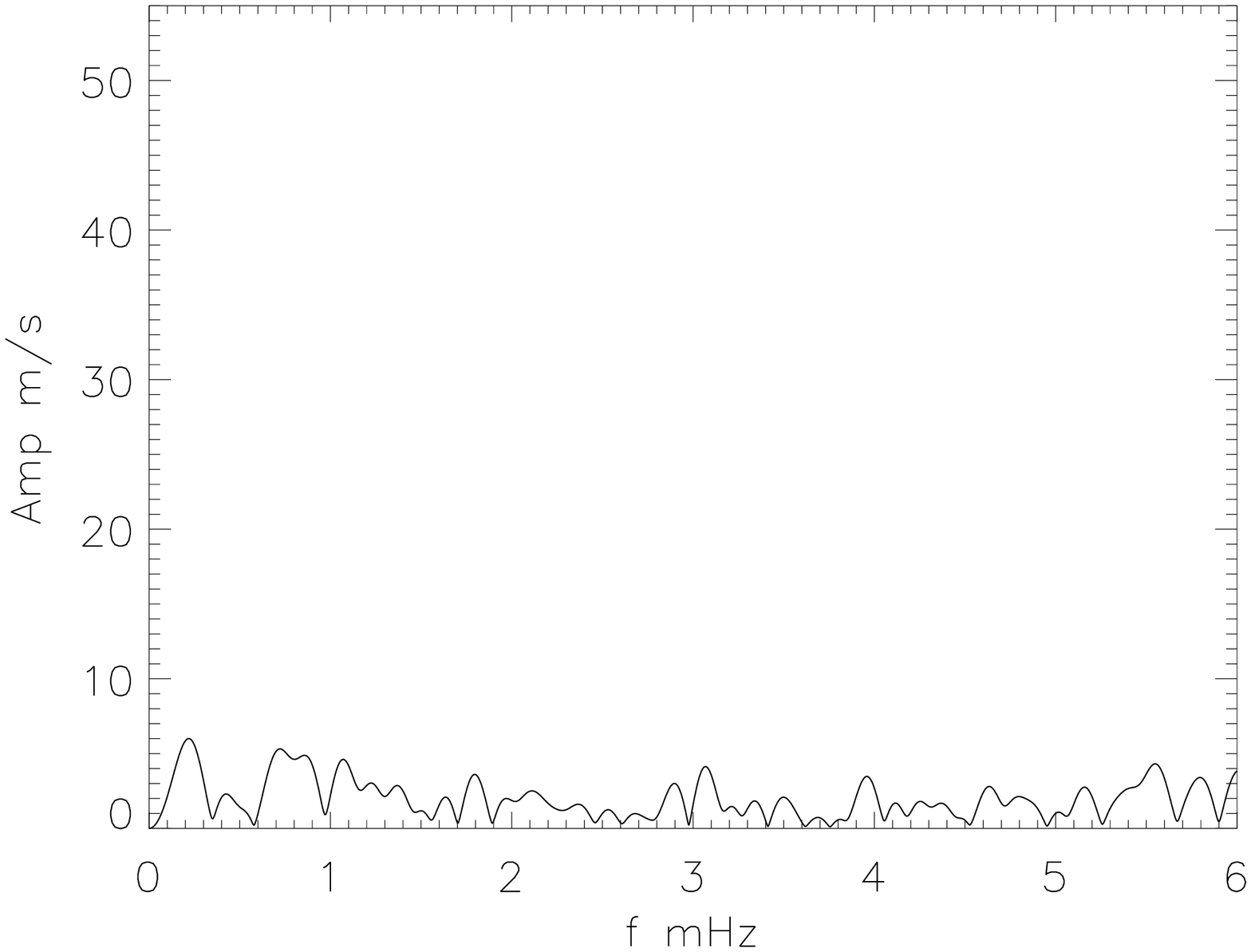}
  \includegraphics[width=60mm,height=82mm,
  angle=270]{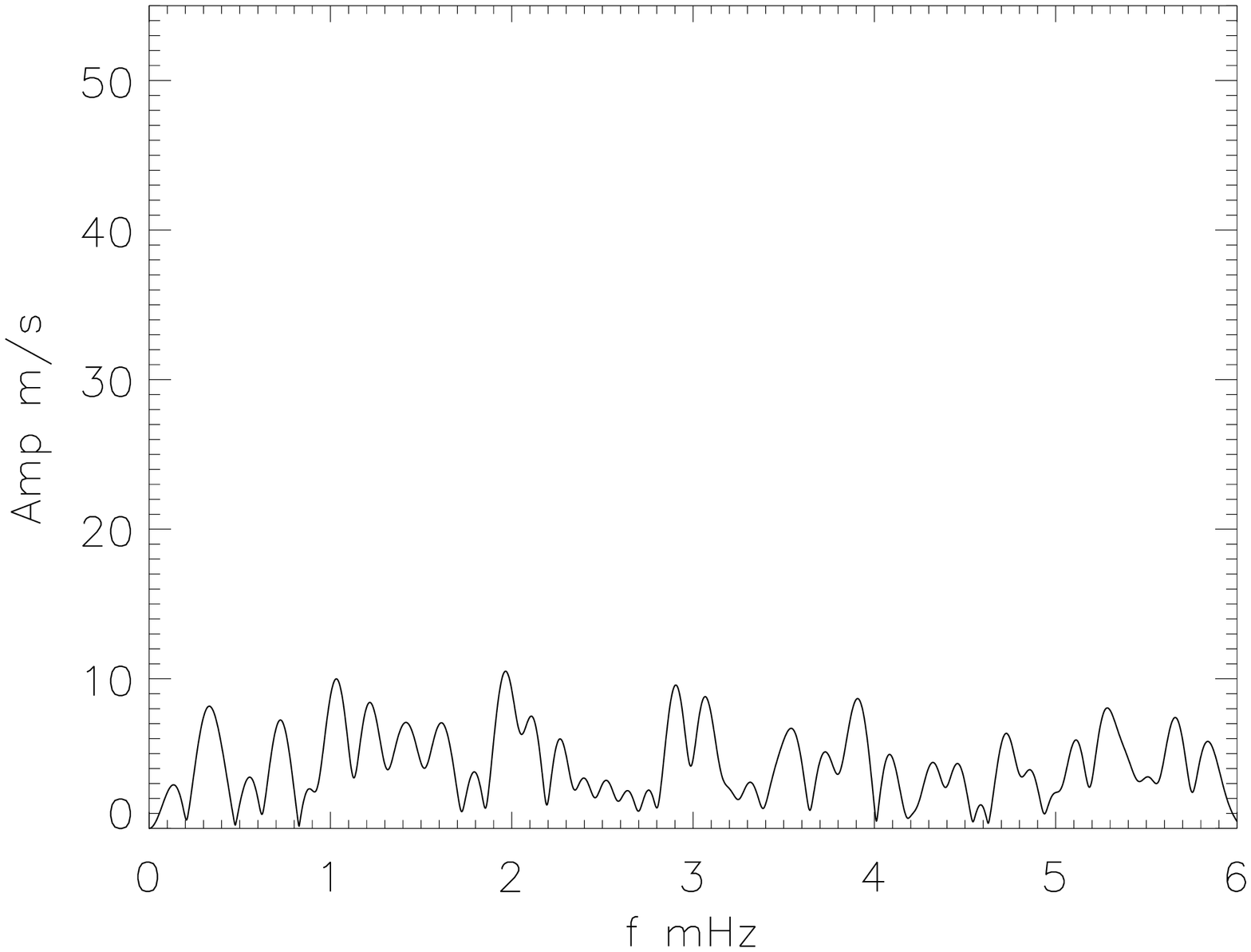}
  \caption{\label{fig:197417cc}Amplitude spectra from
    cross-correlation for the HD\,197417 spectra. Top: for the
    wavelength region $\lambda\lambda\,5150-5800$\,\AA; Bottom: for
    the wavelength region $\lambda\lambda\,6350-6700$\,\AA.}
\end{figure}

HD\,197417 has the second-highest $\delta c_1$ (0.015) of our sample,
and the 125 UVES spectra, obtained in 2.20\,h at 63-s time resolution,
show weak lines of REEs; \prii, \priii\ and, e.g., \ndiii\,6550.32,
6145.07 and 6327.24\,\AA\ are almost absent ($2 - 4$ per cent below
the continuum).  Both \euii\,6437.68 and 6645.06\,\AA\ lines are weak
(less than 5 per cent below the continuum). Some REE lines are partially
split, e.g. \priii\,6866.80, \euii\,6437.64 and \euii\,6645.06\,\AA,
which may indicate that the stellar surface is spotted. This may in
combination with the rotation \vsini\,=\,25.5\,\ms\ partly explain the
weak REE lines of this Ap star. Lines of \ion{Co}{i}, \ion{Fe}{ii} and
\ion{La}{ii} are strong, while a weak \ion{Ca}{i} supports the high
photometric temperature, $T_{\rm eff} = 8400$\,K.  \ion{Ce}{ii} is
absent.  The \ion{Na}{d} lines have a stellar component and a sharper
interstellar component at longer wavelengths.

Radial velocity shifts of 33 stellar lines were measured
(Table\,\ref{tab:cogpow} and Fig.\,\ref{fig:197417cog}). The noise is
considerable due to the low peculiarities and line blending due to
rotational broadening, but the radial-velocity series for the core of
\halpha\ alone, or all 33 lines combined, exclude rapid pulsation to
48\,\ms\ ($\sigma=13$).
Cross-correlations result in flat amplitude spectra down to
$5-10$\,\ms\ (see, e.g., Fig.\,\ref{fig:197417cc}).
Magnetically sensitive lines, such as \ion{Cr}{ii}\,5116.04,
\ion{Fe}{ii}\,6149.25, and \ion{Fe}{i}\,6232.64\,\AA, are not
magnetically resolved. Comparison of 8 Fe and Cr lines with
magnetically broadened {\small SYNTHMAG} profiles, and also an 
attempt to measure the mean quadratic field using 20 lines, exclude
magnetic fields above $\sim2$\,kG. Due to considerable line blending,
smaller fields cannot be quantified without polarimetric measurements.

HD\,197417 does not appear to have strong REE abundances in the
wavelength regions we cover. This may, however, be an effect from a
spotted surface distribution of REEs and rotational broadening.  The
existence of an inhomogeneous surface distribution of REEs is
supported by the star's known (rotational) photometric period and the
many double REE lines.  A spotted surface on Ap stars can often be
related to presence of a magnetic field, for which we in this case put
an upper limit at 2\,kG.  The star is stable to 65\,\ms\
($\sigma=20$\,\ms) for all Nd and Pr lines combined, and 5\,\ms\
($\sigma=2$\,\ms) for cross-correlations.  The significant rotation
\vsini\,=\,25.5\,\kms\ does, combined with blending from the many
double lines, limit the lines available for analysis.

\subsection{HD\,204367}
\label{sec:204367}

The Michigan Spectral Catalogue lists the star as
A(p SrEuCr) and notes that it is either a weak Ap star, or a normal
star of spectral type A0IV/Vs. Supposedly due to this, \citet{martinez93} did not
make any time series photometry of this object. The star has the
highest $\delta c_1$ (0.020\,mag) and $c_0$ index in our sample.  Because
HD\,204367 is one of the least evolved of the studied stars
(Fig.\,\ref{fig:cmd}), the high indices rather indicate a less
peculiar spectrum rather than a higher
luminosity. \citet{manfroidetal98} obtained 21 measurements (one per
night) in the Geneva photometric system of this Ap candidate and found
no variability (mmag level) over a 23-night run. With 90 {\it
  Hipparcos} measurements, we find the star stable to 6.5\,mmag for
periods longer than a day.

We obtained 111 spectra of this moderate rotator
(\vsini\,=\,10.8\,\kms) in 1.96\,hr at a time resolution of 64\,s.
The spectra revealed the least chemically peculiar star in this study,
with considerably fewer lines and longer continuum windows than any of
the other stars. Lines of Nd, Pr, Ce and Cr are either absent or much
weaker than in the other studied stars. The absorption of \ion{Eu}{ii}
is in all cases less than 2 per cent below the continuum and only a few
REE lines could be used in the velocity analysis. Lines of Ni, Si, S,
Na, Ba and Zn are strong.

\begin{figure}
  \vspace{3pt}
  \includegraphics[width=60mm,height=82mm,
  angle=270]{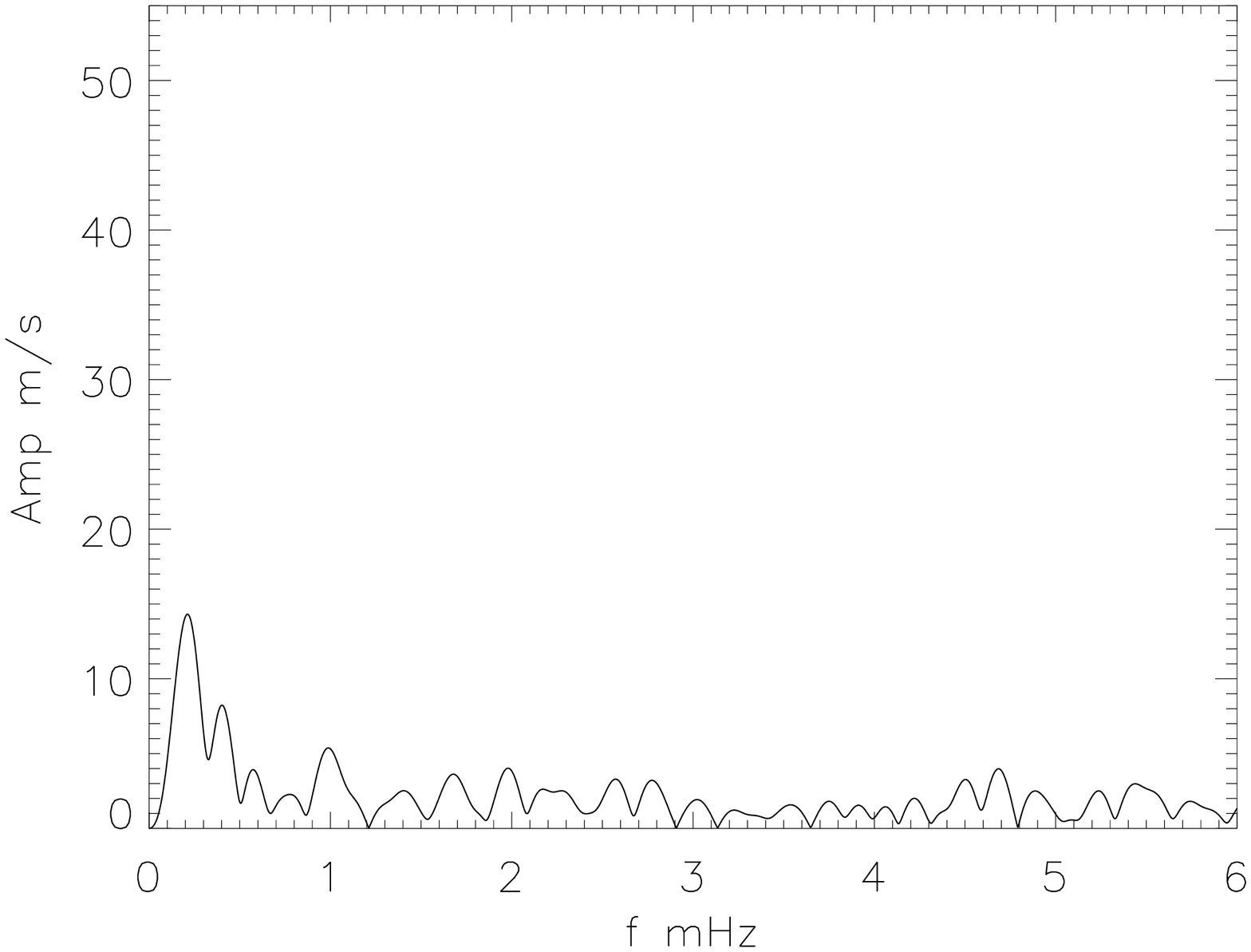}
  \includegraphics[width=60mm,height=82mm,
  angle=270]{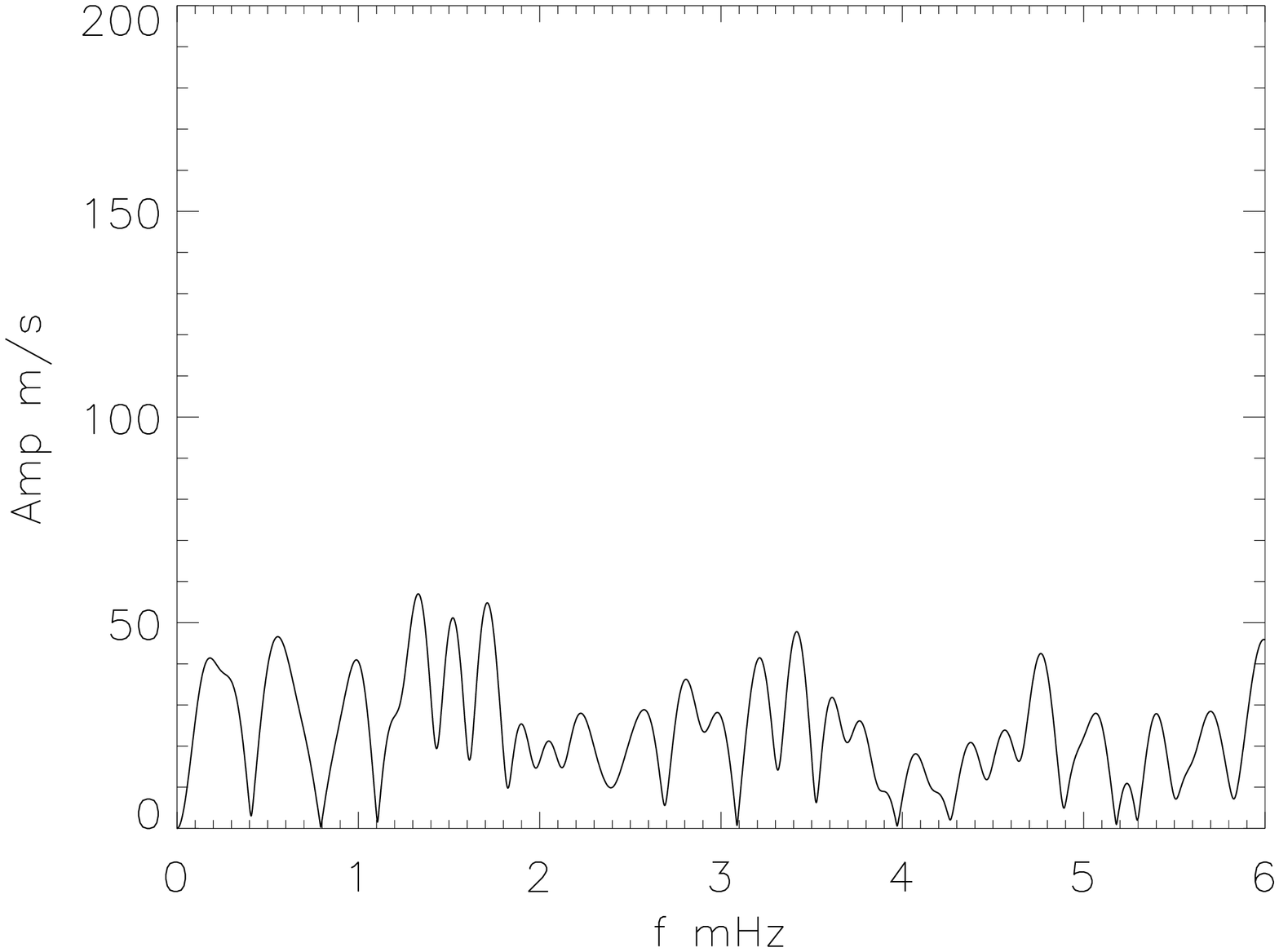}
  \caption{\label{fig:204367cc}Amplitude spectra from
    cross-correlation for the HD\,204367 spectra. Top: for the
    wavelength region $\lambda\lambda\,5150-5800$\,\AA; Bottom: for
    the wavelength region $\lambda\lambda\,6350-6700$\,\AA.}
\end{figure}
\begin{figure}
  \vspace{3pt}
  \includegraphics[width=65mm, angle=270]{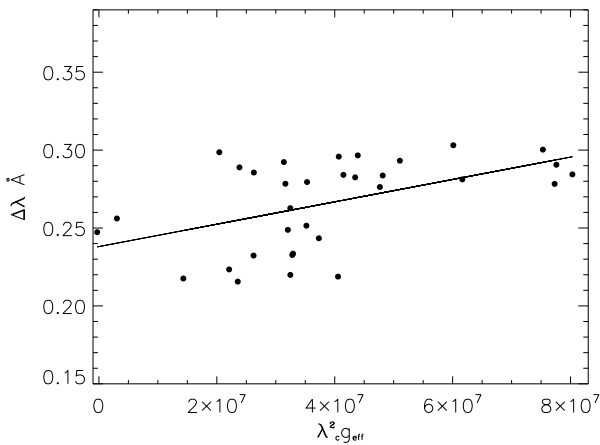}
  \caption{\label{fig:204367mag2}Magnetic broadening measurements of
    33 Fe lines of HD\,204367.  Measured Gaussian FWHM
    ($\Delta\lambda$) vs product of laboratory wavelength squared and
    Land\'e factor ($\lambda^2_c\,g_{\rm eff}$).  A weak relation
    ($r=0.50$) fitted with a least-squares linear fit, is indicated
    with a line.}
\end{figure}

Radial velocity shifts of 38 stellar lines were measured. The noise is
considerable, but when combining all lines
(Fig.\,\ref{fig:204367cog}), we can exclude rapid pulsation with
amplitudes above 36\,\ms\ ($\sigma=12$\,\ms), and 46\,\ms\
($\sigma=14$\,\ms) for the \halpha\ line core alone
(Table\,\ref{tab:cogpow}). There are no significant peaks in the
periodograms.  Cross-correlations produce similar `flat' amplitude
spectra (Table\,\ref{tab:ccpow} and Fig.\,\ref{fig:204367cc}).
An upper limit of the mean quadratic field is found with 27 Fe lines
to $\left<B_{\rm q}\right>\le1.4$\,kG, while
{\small SYNTHMAG} models compared to 16 Cr and Fe
lines gave an upper limit of $\langle B_{\rm synth}\rangle\le1$\,kG.
Further, line width measurements of 33 iron lines showed a weak
relation (Fig.\,\ref{fig:204367mag2}) that would indicate a rather
weak field of $0.77\pm0.24$\,kG (1\,$\sigma$ error).  This fit is,
however, barely significant as a significance test only gave $t=2.9$
while $t>3.0$ is required. Polarimetry is therefore needed for
verification.

This star shows no detectable periodic variability down to 72\,\ms\
($\sigma=25$\,\ms) for all Nd and Pr lines combined, and 5\,\ms\
($\sigma=2$\,\ms) for cross-correlations.  Rotation is slow
(\vsini\,=\,10.8\,\kms), the spectrum is nearly devoid of REEs, and
the star may have a small magnetic field of $\sim$0.8\,kG.

\subsection{HD\,208217}
\label{sec:208217}

This known magnetic and peculiar star, classified as Ap SrEuCr, A1
\citep{houketal75} has been previously examined in several
studies. \citet{martinez93} spent nearly 13\,h on the star on 8
separate nights. With $\delta c_1 = -0.19$ it is strongly
peculiar. The high-speed photometry excludes periodicities above
0.4\,mmag (frequencies above 0.8\,mHz) and 0.8\,mmag (frequencies
above 0.5\,mHz).  With spectroscopic observations,
\citet{mathysetal97} detected a mean magnetic field modulus varying
with a semi-amplitude of nearly 1000\,G about a mean value of 7958\,G
($\sigma=588$ G) over a rotation period of 8.44475\,d
\citep{manfroidetal97}. For this period, the epoch of our
spectroscopic observations occurs near a negative extremum of the
longitudinal field (Mathys, unpublished observations). For roAp
stars, maximum pulsation amplitude occurs when one of the stellar
magnetic poles come into sight. In this case the negative magnetic
pole, and our chances for detecting roAp pulsations should therefore
be optimal. Mathys et al. furthermore found the star to be a
single-lined binary with a most likely orbital period of at least
2\,yr.

\begin{figure}
  \vspace{3pt}
  \includegraphics[width=60mm,height=82mm,
  angle=270]{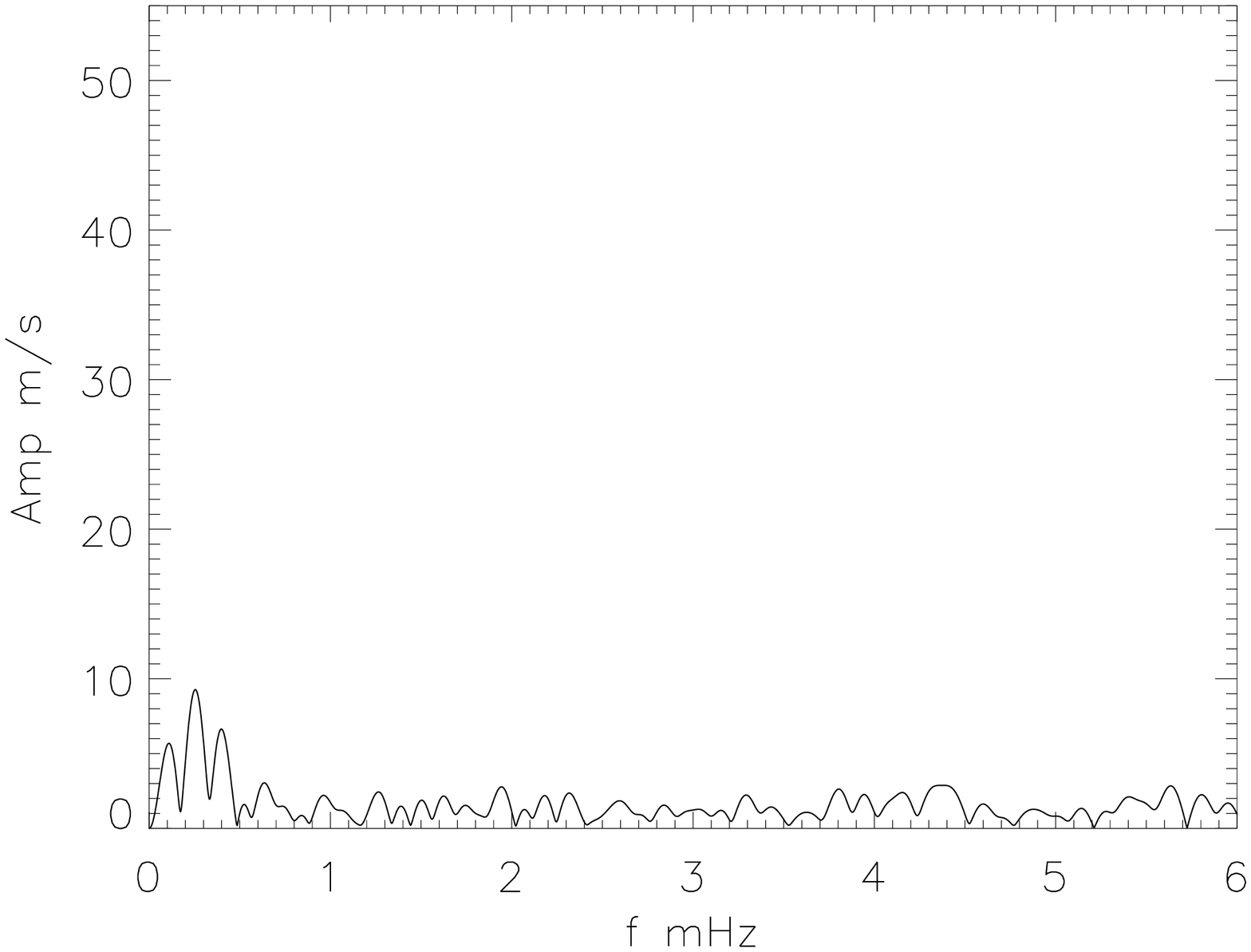}
  \includegraphics[width=60mm,height=82mm,
  angle=270]{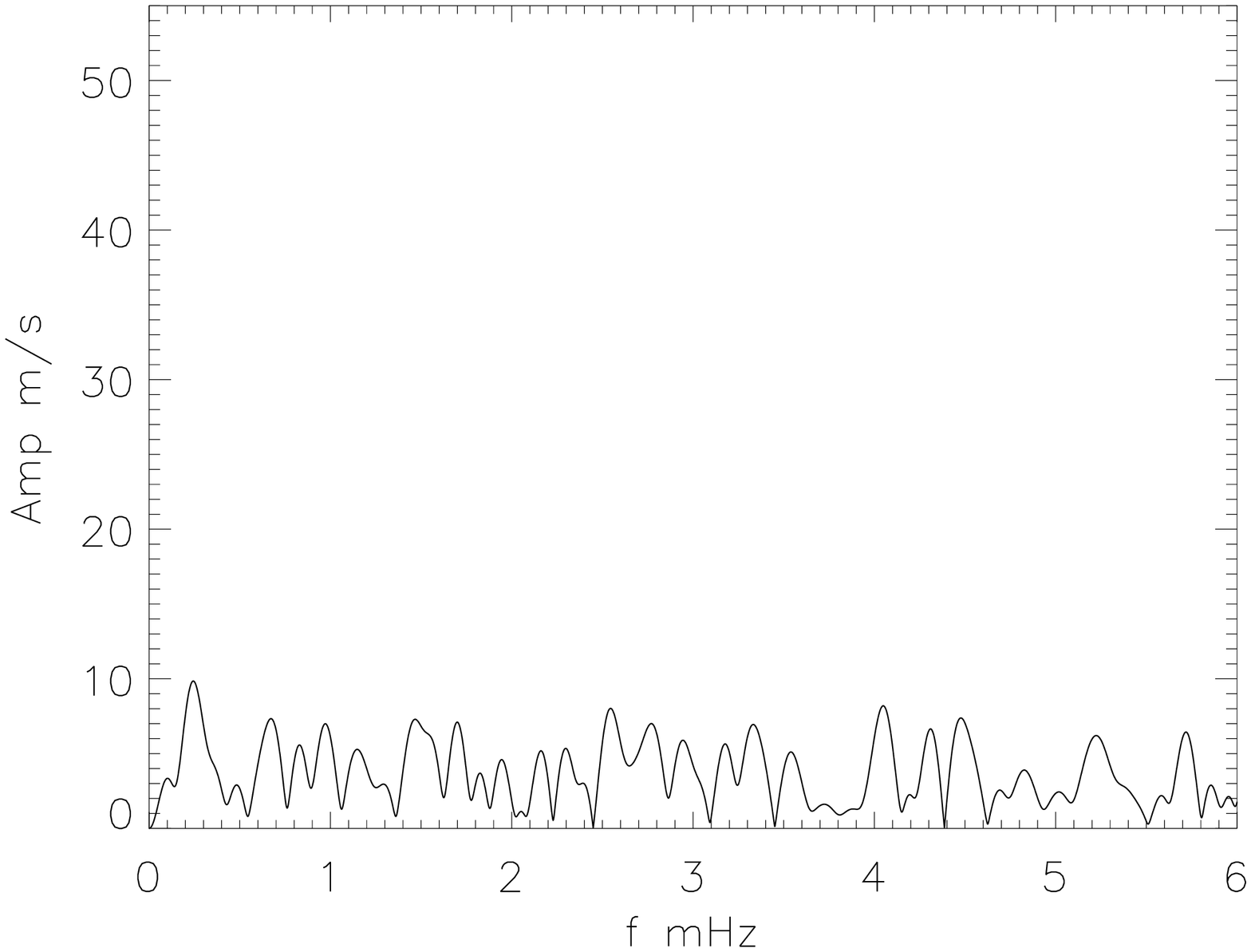}
  \caption{\label{fig:208217cc}Amplitude spectra from
    cross-correlation for the HD\,208217 spectra. Top: for the
    wavelength region $\lambda\lambda\,5150-5800$\,\AA; Bottom: for
    the wavelength region $\lambda\lambda\,6350-6700$\,\AA.}
\end{figure}

\begin{figure}
  \vspace{3pt} \hspace{-3pt}
  \includegraphics[width=8.5cm,height=13.5cm, angle=0]{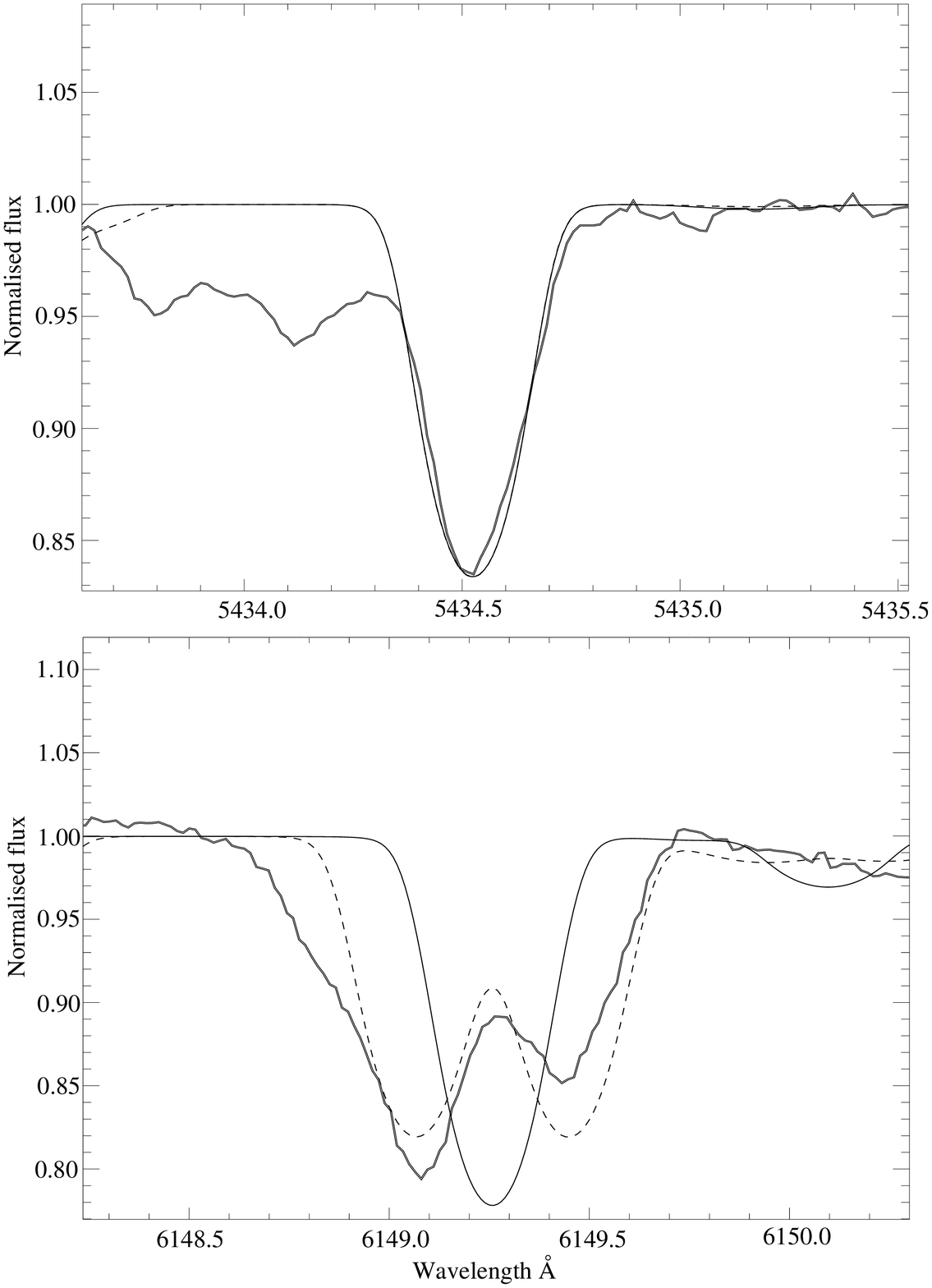}
  \caption{\label{fig:208217mag1}Magnetic field strength of HD\,208217
    demonstrated with the average UVES spectrum (thick line) for the
    absorption lines \ion{Fe}{i} 5434.52 (top panel) and \ion{Fe}{ii}
    6149.25\,\AA\ (bottom) which respectively have a low and a high
    Land\'e factor.  Two {\small SYNTHMAG} models are superposed for
    mean field moduli of 0.0 (full line) and 8.0\,kG (dashed
    line). The models use \vsini\,=\,9.4\kms, $[$Fe/N$]=-5.4$
    (\ion{Fe}{i}\,5434.52) and $-4.4$ (\ion{Fe}{ii}\,6149.25).  }
\end{figure}

With a typical $S/N$ of 150 per spectrum, this set of 138 UVES spectra
obtained in 2.52\,h with a 66-s time resolution has the highest
quality of this study. The spectra show strong lines of Nd, Pr and
\euii, while \ion{Ca}{i} is weak which could indicate a higher
temperature than the $T_{\rm eff} = 7700$\,K from photometry, but the
wings of \halpha\ agree well with models for $T_{\rm eff}$ in the
range $7500-8000$\,K.  Lines of REEs such as \ion{Nd}{ii}\,6549.52 are
mostly single; \euii\,6437.64\,\AA\ is double while the single
\euii\,6437.64 appears broadened. The lines of \ion{Ce}{ii} are
absent.

Radial velocity shifts of 52 lines were measured
(Table\,\ref{tab:cogpow}) and show for all lines combined no rapid
oscillations above 14\,\ms\ ($\sigma=4$\,\ms). The core of \halpha\ is
stable to 40\,\ms\ ($\sigma=13$\,\ms).  However, lines of \priii\ and
\euii\ show peaks of $3.9\,\sigma$ and $4.9\,\sigma$ respectively but
for different frequencies and are rejected.  The cross-correlations
produce flat amplitude spectra reaching the lowest noise level in this
study: the integrated radial-velocity noise is only $1.4-2.8$\,\ms\
and no peaks appear above $3\,\sigma$.
Lines of REEs are in general single, however \euii\,6437.64 is
broadened while \euii\,6437.64 is double. Also \ion{Fe}{ii}\,6149.25
and \ion{Fe}{ii}\,6432.68\,\AA\ are double, and iron lines are
generally broadened, consistently with the strong magnetic field known
for 
this star.  The rotation is low (\vsini\,=\,10.8\,\kms), but the
spectrum is distorted by the magnetic field and blends from the
abundant peculiar elements, so line identification and analysis is
complicated.  Using 16 Fe lines, the mean quadratic field is found to
$\left<B_{\rm q}\right>=8.0\pm0.5$\,kG.  
With {\small SYNTHMAG} fitting of 17 lines of different Land\'e factors (see examples in
Fig.\,\ref{fig:208217mag1})
the field strength is found to be $\langle B_{\rm synth}\rangle=7.5\pm0.5$\,kG,
while  the Zeeman splitting of \ion{Fe}{ii}\,6149.25 and \ion{Cr}{ii}\,5318\,\AA\
gives a magnetic field modulus of $8.0\pm0.3$\,kG.

Summarising, this star has strongly peculiar REE abundances, a (known)
strong magnetic field of $\left<B_{\rm q}\right>=8.0\pm0.5$\,kG and is
stable to 15\,\ms\ ($\sigma=4$\,\ms) for all Nd and Pr lines combined,
and 3\,\ms\ ($\sigma=1$\,\ms) for cross-correlations.

\section{Discussion}

The class of roAp stars is notoriously difficult to supplement with
new members, as demonstrated by several photometric and spectroscopic
studies before this one.  But motivated by the recent discovery of the
luminous roAp star HD\,116114 \citep{elkinetal05}, we have
spectroscopically tested a sample of 9 luminous Ap stars for rapid
pulsations. Using lines known to show pulsations in roAp stars, we
reach typical upper amplitude limits in radial velocity of:
$40-75$\,\ms\ ($\sigma=13-24$\,\ms) for the line core of \halpha,
$20-65$\,\ms\ ($\sigma=7-20$\,\ms) when combining all measured Nd and
Pr lines, and $20-40$\,\ms\ ($\sigma=7-11$\,\ms) when combining all
measured lines.  With cross-correlations, using large wavelength
regions, we typically reach upper amplitude limits of $4-10$\,\ms\
($\sigma=1-4$\,\ms). In spite of a clear theoretical prediction
\citep{cunha02} and empirical (HD\,116114) evidence for roAp
pulsations in this part of the Hertzsprung--Russell diagram, we end up
with 9 null-results, or noAp stars.  A number of questions are
therefore pertinent to discuss.

\vspace{2mm}
\noindent{\em How well does our test sample resemble known roAp
  stars?}
\vspace{1mm}

\noindent
All studied stars have strong REE lines, except for HD\,204367 and
possibly also HD\,197417. They have the core-wing anomaly typical for
roAp stars, and {\it Hipparcos} luminosities with our temperature
estimates place them inside the predicted roAp instability strip.
Several appear to have spotted surface distributions of REEs (such as
HD\,170565, HD\,151301, HD\,132322 and perhaps also HD\,197417) which
is typically associated with the strong magnetic fields common in
known roAp stars.  Indeed most of the stars are magnetic, and cover a
range in magnetic field strengths of $0.4-8.0$\,kG, comparable to that
of known roAp stars \citep{kurtzetal06b}. In the case of the
sharp-lined HD\,110072, we compared its spectrum in detail with those
of two known roAp stars, and found remarkable similarities for the
REEs.  However, HD\,110072 and HD\,208217 are double-lined and
single-lined binaries, respectively, which might indirectly influence
their stability to high-frequency pulsations by reducing the magnetic
field intensity (see \citealt{cunha02} and references therein).
Still, the orbits are probably too wide in both cases for tidal
interaction to occur and HD\,208217 has a known strong magnetic field
($\left<B\right>=8$ kG). The known roAp stars have cases of wide
binaries, such as $\beta\,$CrB (spectroscopic binary), HR\,3831,
$\alpha$\,Cir, $\gamma$\,Equ and HD\,99563 (visual binaries).  In
these regards, the studied sample has the characteristics of roAp
stars.

\vspace{2mm}
\noindent{\em Could pulsations have been overlooked?}
\vspace{1mm}

\noindent\citet{kurtzetal06b} published radial velocity amplitudes for the
\halpha\ cores of 16 roAp stars.  Of these, only 3 have amplitudes
below 75\,\ms\ (HD\,116114, HD\,154708 and HD\,166473, of which two
have amplitudes above 3\,$\sigma$), while the rest range from
$148-2528$\,\ms.  Further, radial velocity series for Pr and Nd line
measurements in UVES spectra of these 16 roAp stars (Kurtz et al.,
partly unpublished), show that about 75 per cent of the measured and
significant (3\,$\sigma$) amplitudes are within the range
$350-1600$\,\ms.  Five of these stars have very small Nd and Pr
amplitudes ($60-90$\,\ms): HD\,166473, HD\,116114, $\beta$\,CrB,
33\,Lib and HD\,154708. In such difficult cases, other lines or
combinations of several lines makes detection of pulsations possible.
We successfully tested our procedures on the two latter roAp stars in
Sect.\,\ref{sect-rvshift}, and also used combinations of several
lines, including of different elements. More of the other 19 known
roAp stars have low amplitudes, such as 10\,Aql, but our tests and
analyses show that we reach these amplitude levels and should have
detected such rapid pulsations if present in the studied sample.

A complication for our analyses is typical \vsini\,$\sim10-30$\kms
combined with double lines due to either spots on the stellar surface
and/or magnetic splitting, which results in considerable line blending
and makes line identification and analysis more difficult.  However,
our radial-velocity analysis was based partly on cross-correlations
that are more robust than line measurements (as shown by our tests for
two roAp stars) and this method similarly results in flat amplitude
spectra that exclude rapid oscillations to relatively small
roAp-amplitude levels.  It also seems improbable that, e.g.,
unfavourable viewing angles of the global pulsations or short mode
lifetimes could explain a momentary lapse of detectable pulsation
amplitudes simultaneously in all nine stars.  In fact, we know
independently that HD\,208217 was observed near its magnetic negative
extremum where roAp pulsations are expected to have maximum amplitudes.
Future surveys like this one may benefit from being repeated at different
rotation phases.  We also note that the near-normal REE abundances of
HD\,197417 and HD\,204367 reduce the probability that they are roAp
stars, given the strong peculiarity of all known roAp stars.

\vspace{2mm}
\noindent {\em Do these stars really not pulsate?}
\vspace{1mm}

\noindent
In pulsators such as roAp stars oscillations are intrinsically
unstable. Their excitation depends on the balance between the driving
and damping of the oscillations over each pulsation cycle.  In roAp
stars this balance is thought to be particularly delicate. On one
hand, the amount of energy input through the opacity mechanism acting
on the hydrogen ionisation region depends strongly on the interaction
between the magnetic field and envelope convection, being maximal in
the regions where envelope convection is suppressed
\citep{balmforthetal01,cunha02}.  On the other hand, the direct effect
of the magnetic field on pulsations can introduce significant energy
losses, through slow Alfv\' en waves in the interior and through
acoustic waves in the atmosphere, both resulting from mode conversion
in the magnetic boundary layer \citep{cunha00,saio05}.  Due to this
delicate balance, it is not too surprising that roAp and noAp stars
occupy the same locus in the HR diagram. Despite the developments in
theoretical studies of linear non-adiabatic pulsations in models of
roAp stars, we still lack a theoretical study that takes into account
all these phenomena simultaneously and, thus, cannot firmly predict
the conditions under which pulsations should be expected in roAp
stars. In fact, both studies of \cite{cunha02} and \cite{saio05}
considered the extreme case in which envelope convection is fully
suppressed. Moreover, the first of these studies did not consider the
direct effect of the magnetic field on pulsations and neglected the
energy losses as a result of mode conversion, and the second study,
while considering mode conversion, assumed the waves are fully
reflected at the surface, hence neglecting energy losses through
acoustic running waves in the atmosphere.

As discussed by \cite{cunha02}, the condition for suppression of
envelope convection, which seems necessary to make the high frequency
modes unstable, is in principle harder to fulfil in evolved stars due
to the increase with age of the absolute value of the buoyancy
frequency in the region where hydrogen is ionised. Hence, it is likely
that in evolved stars oscillations are excited only if the magnetic
field is relatively strong.  Unfortunately, the complexity of the
interaction between magnetic field and convection makes it impossible
to derive a global convective stability criterion, even if local
criteria for convective stability may be established
\citep{Gough66,Moss69} \citep[see also][for an extensive discussion on
this subject]{theado05}. Thus, the magnetic intensity needed to
suppress convection at a given age, for a given mass, is very hard to
establish.  The more evolved roAp star HD\,116114, in which a
relatively low frequency oscillation was found well in agreement with
theoretical predictions, has a magnetic field modulus of $\approx
6$~kG. In contrast with this, most stars in our sample have estimated
mean magnetic field moduli around or below 2~kG. The clear exceptions
are HD\,107107, HD\,131750 and HD\,208217. Of these three, the latter
is clearly an important test case to check the theoretical
predictions. It has the strongest confirmed magnetic field in our
sample and is strongly peculiar. However, we observed HD\,208217 when
one of its magnetic poles were almost visible, so pulsation should
have been near its maximum amplitude.

From the observational point of view, one way to investigate the
conditions under which roAp star oscillations are excited, and thus
test theoretical models, is by identifying systematic differences
between roAp and noAp stars.  This study would have been able to
detect pulsations in all the known roAp stars, and any missed rapid
pulsations must have amplitudes lower than these.  Hence we conclude
that based on the obtained data, most stars in our sample are indeed
noAp stars, and also excellent roAp candidates.  Despite this
conclusion, it is premature to state that we can confirm the evidence
that noAp stars are in average more luminous and more evolved than
roAp stars, as indicated by earlier studies based on photometric
surveys for pulsations in roAp candidates
\citep{north97,handleretal99,hubrig00}. To conclude that, we would
also need to search for rapid pulsations in a control group of less
evolved roAp stars with lower luminosity and compare the frequency of
null-results in the two cases.  Such a survey has been started and
results are expected in the near future.  The next step is to
analyse the noAp stellar atmospheres in detail, taking temperature
gradients and the abundance stratification into account.

New spectra are needed of HD\,132322 to clarify the origin of its
double line structures, and of HD\,110072 to verify its secondary
spectrum and to test for the `spurious' absorption lines in the recent
FEROS spectrum. Polarimetry of HD\,110072 and HD\,204367 is needed to
confirm the detected magnetic fields.

\section*{Acknowledgments}

LMF, DWK, JDR and VGE acknowledge support for this work from the
Particle Physics and Astronomy Research Council (PPARC). MSC is
supported by by the EC's FP6, FCT and FEDER (POCI2010) through the
HELAS international collaboration and through the project
POCI/CTE-AST/57610/2004FCT-Portugal. LMF received support from the
Danish National Science Research Council through the project ``Stellar
structure and evolution: new challenges from ground and space
observations'' carried out at Aarhus University and Copenhagen
University.  We acknowledge resources provided by the electronic data
bases (VALD, VizieR, SIMBAD, NASA's ADS)

\clearpage
\appendix
\begin{flushleft}
  \section{Supplementary material}
  \subsection{Additional tables}
\end{flushleft}
\begin{table}
  \centering
  \begin{minipage}{80mm}
    \caption{\label{tab:model} Abundances in $\log{N/N_{\rm{tot}}}$
      for our model spectrum used for line identification. Effective
      temperature $\mteff=8750$\,K and surface gravity is $\log\,g
      =4.0$. }
    \begin{tabular}{@{}l@{\,\,}rl@{\,\,}rl@{\,\,}rl@{\,\,}rl@{\,\,}r@{}}
      \hline
      H &  0.91& He& -1.05& Li&-10.88& Be&-10.89& B & -9.44 \\
      C & -3.48 & N & -3.99& O & -3.61& F & -7.48& Ne& -3.95 \\
      Na& -5.71& Mg& -4.46 & Al& -5.57& Si& -4.89& P & -6.59 \\
      S & -4.83& Cl& -6.54& Ar& -5.48 & K & -6.82& Ca& -5.88 \\
      Sc& -8.94& Ti& -6.50& V & -8.04& Cr& -3.57 & Mn& -6.65 \\
      Fe& -3.57& Co& -5.52& Ni& -5.59& Cu& -7.83& Zn& -7.84  \\
      Ga& -9.16& Ge& -8.63& As& -9.67& Se& -8.69& Br& -9.41 \\
      Kr& -8.81 & Rb& -9.44& Sr& -9.14& Y & -9.80& Zr& -9.54 \\
      Nb& -9.62& Mo& -9.12 & Tc&-20.00& Ru& -9.20& Rh& -9.92 \\
      Pd& -9.35& Ag&-11.10& Cd& -9.18 & In& -9.58& Sn& -9.04 \\
      Sb&-11.04& Te& -9.80& I &-10.53& Xe& -9.81 & Cs& -9.92 \\
      Ba& -8.91& La& -9.82& Ce&-10.49& Pr& -8.83& Nd& -8.94  \\
      Pm&-20.00& Sm&-10.04& Eu& -9.93& Gd&-10.92& Tb& -9.94 \\
      Dy& -9.34 & Ho&-11.78& Er& -9.01& Tm&-12.04& Yb&-10.06 \\
      Lu&-10.28& Hf& -9.16 & Ta&-11.91& W &-10.93& Re&-11.77 \\
      Os&-10.59& Ir&-10.69& Pt&-10.24 & Au&-11.03& Hg&-10.95 \\
      Tl&-11.14& Pb&-10.19& Bi&-11.33& Po&-19.00 & At&-19.00 \\
      Rn&-19.00& Fr&-19.00& Ra&-19.00& Ac&-19.00& Th&-10.92  \\
      Pa&-19.00& U &-10.51& Np&-19.00& Pu&-19.00& Am&-19.00 \\
      Cm&-19.00 & Bk&-19.00& Cf&-19.00& Es&-19.00&   &       \\
      \hline
      \hline
    \end{tabular}
  \end{minipage}
\end{table}
\newpage
\begin{figure*}
  \flushleft\subsection{Selected regions of average spectra}
  \vspace{3pt}
  \includegraphics[height=0.98\textwidth,
  angle=270]{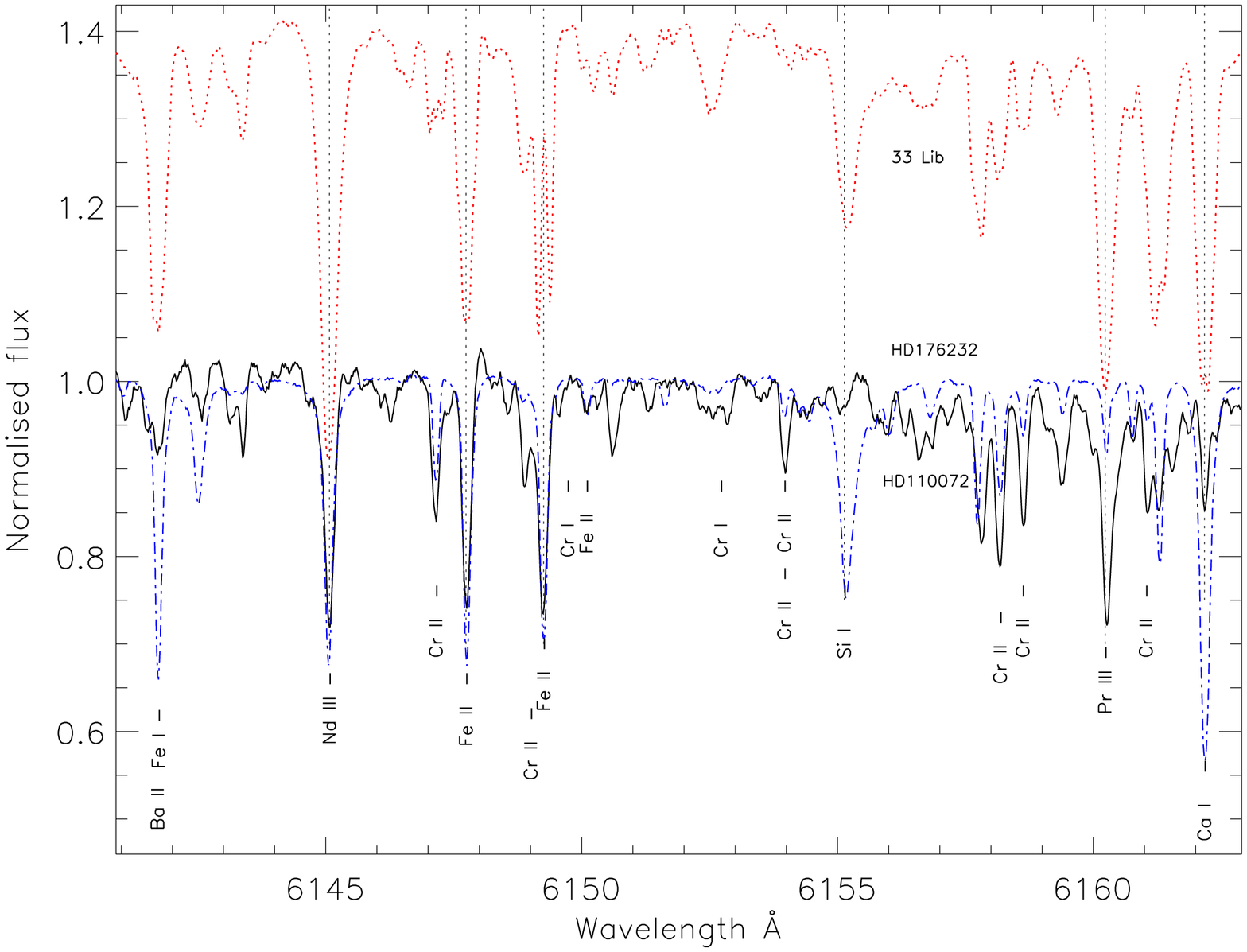}
  \caption{\label{fig:110072cmproap} Comparison of the averaged
    spectrum of HD\,110072 (full line) with averaged UVES spectra of
    the two known roAp stars HD\,176232 (dot-dashed) and 33\,Lib
    (offset and dotted). Note the similarities in strong REEs
    (\ion{Nd}{iii} and \ion{Pr}{iii}), while \ion{Ba}{ii}, \ion{Si}{i}
    and \ion{Ca}{i} all are considerably weaker in HD\,110072. The
    spectrum of the fainter HD\,110072 was smoothed over every 4
    pixels. Note the absence of lines from a secondary at this
    wavelength region of HD\,110072.  }
\end{figure*}
\begin{figure*}
  \includegraphics[height=18cm, width=0.95\textwidth, angle=0]{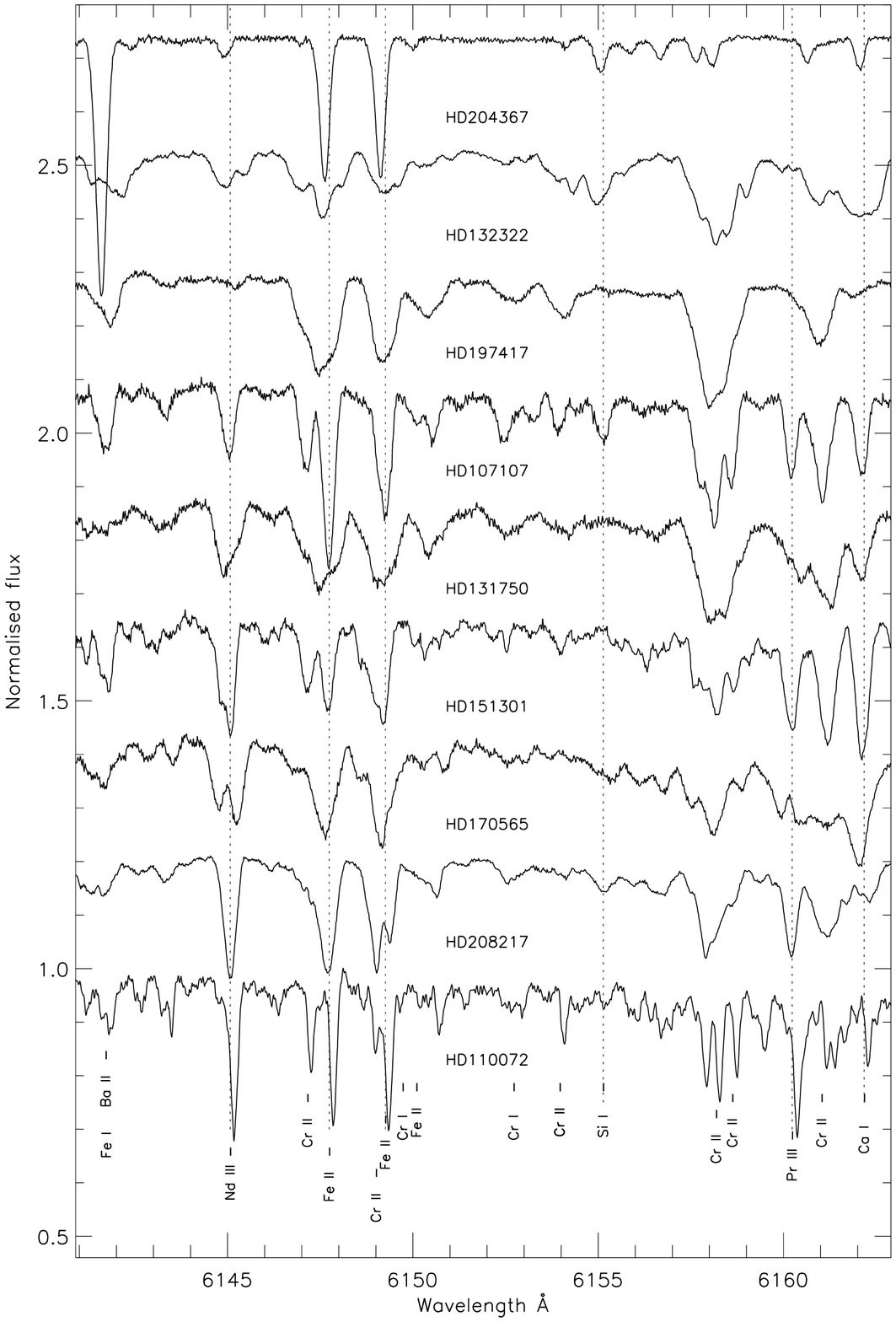}
  \caption{\label{fig:spot}Comparison of the profiles of all examined
    Ap stars. To increase the readability, the individual spectra have
    been offset in flux, and appear with increasing temperature 
    upwards.  The selected region shows examples of double lines due to 
    magnetic splitting, such as HD\,208217, or spots on the surface of 
    the stars.  The spectrum of
    HD\,110072 was smoothed over every two pixels.  }
\end{figure*}
\clearpage \flushleft\subsection[]{Amplitude spectra for combined line
  measurements} These figures show star-by-star amplitude spectra
calculated for combined radial-velocity series for elements as
indicated by top labels: `all' indicates all available lines, `NdPr'
all lines for neodymium and praseodymium. For centre-of-gravity line
measurements. Note that for double lines, both components were
measured separately and included. Compare with
Table\,\ref{tab:cogpow}.

\begin{figure*}
  \vspace{3pt}
  \includegraphics[height=5.6cm,
  angle=270]{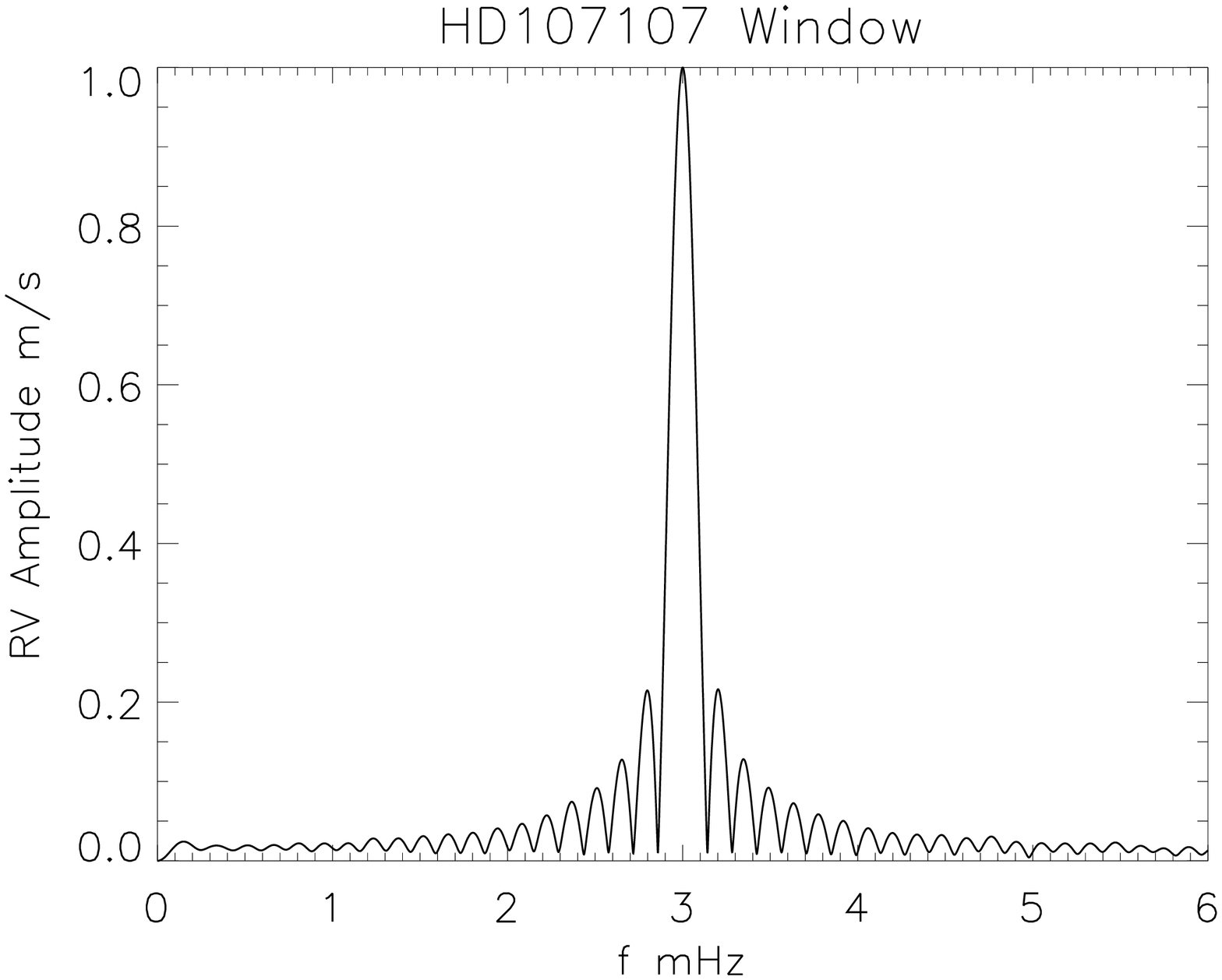}
  \includegraphics[height=5.6cm, angle=270]{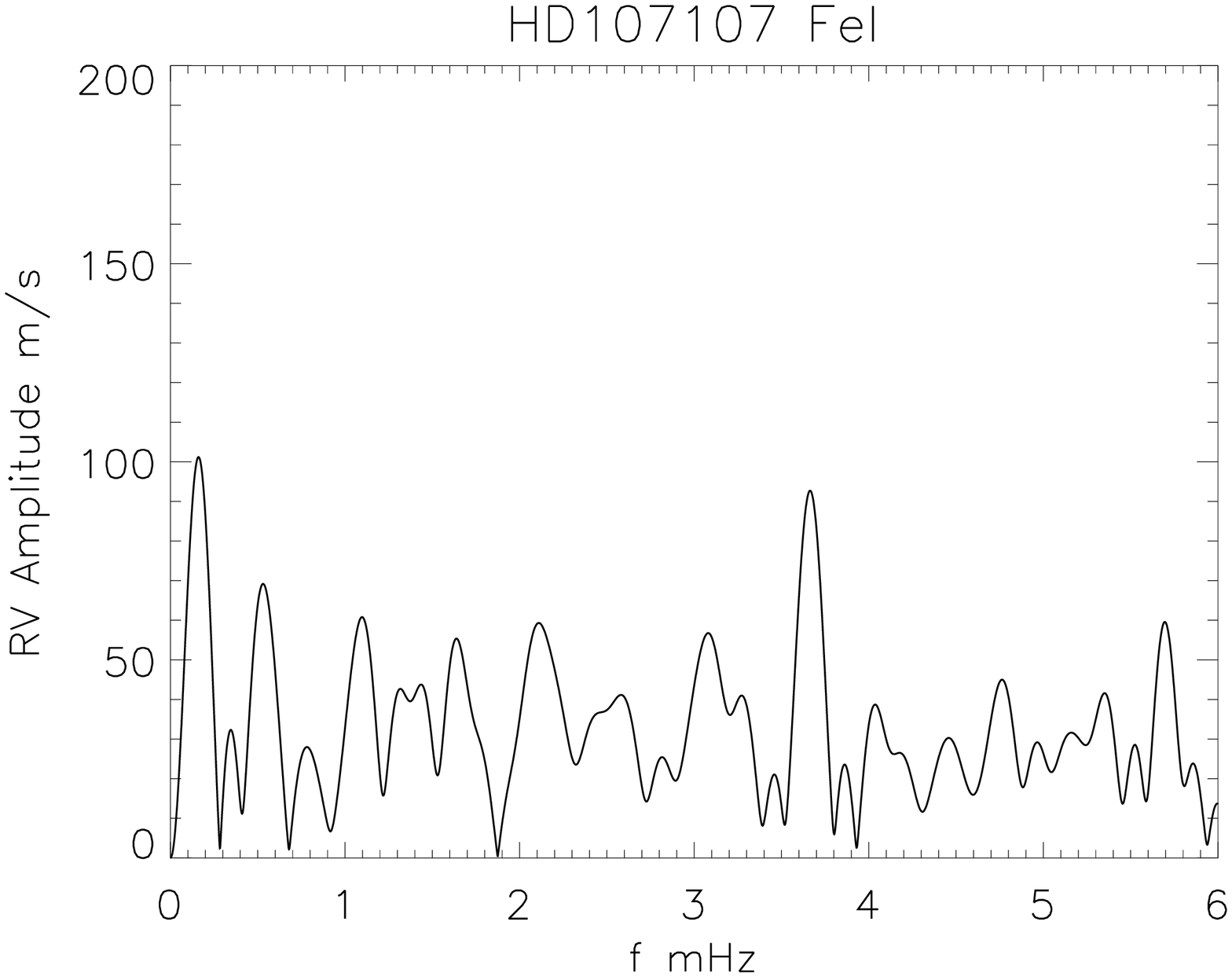}
  \includegraphics[height=5.6cm, angle=270]{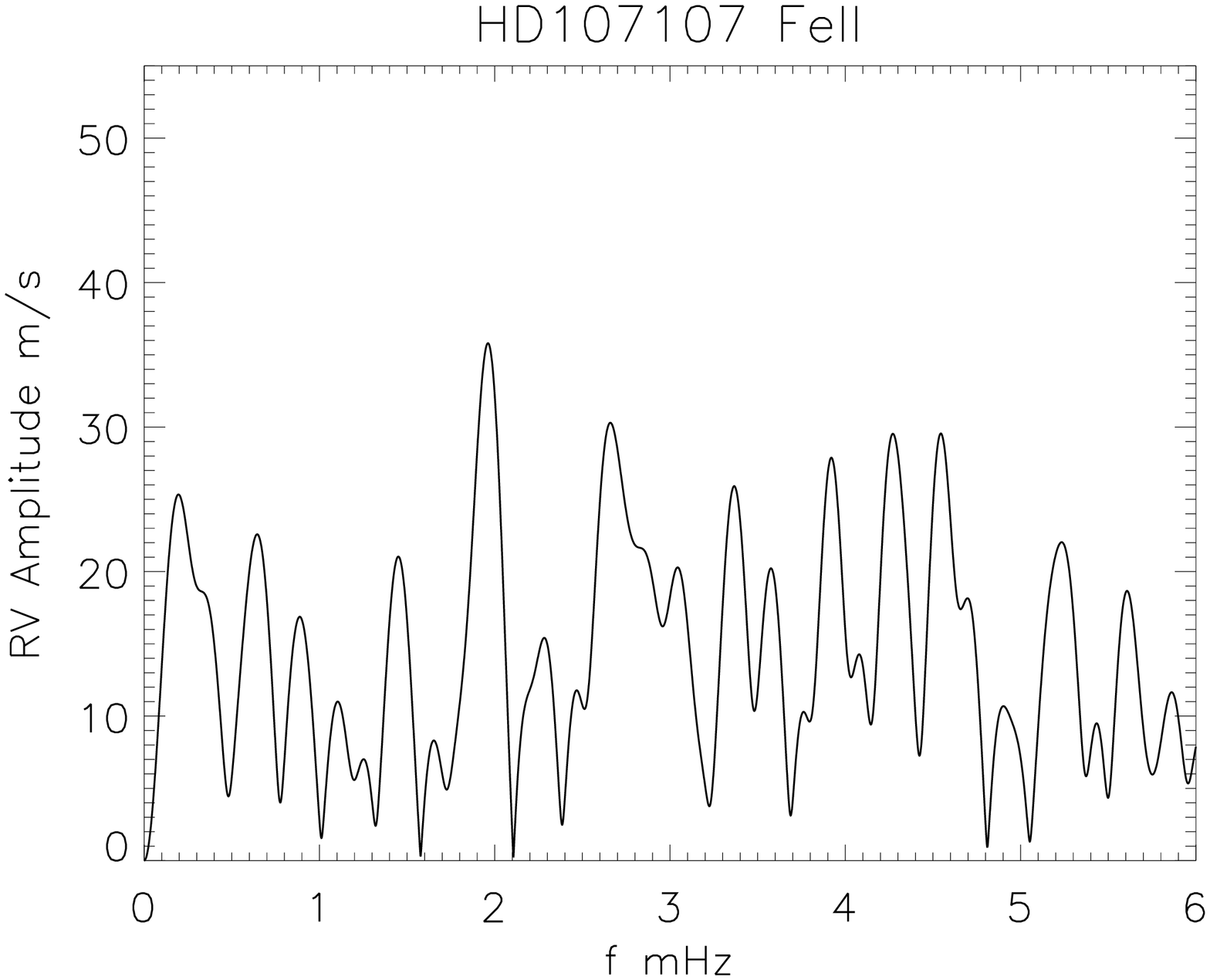}
  \includegraphics[height=5.6cm, angle=270]{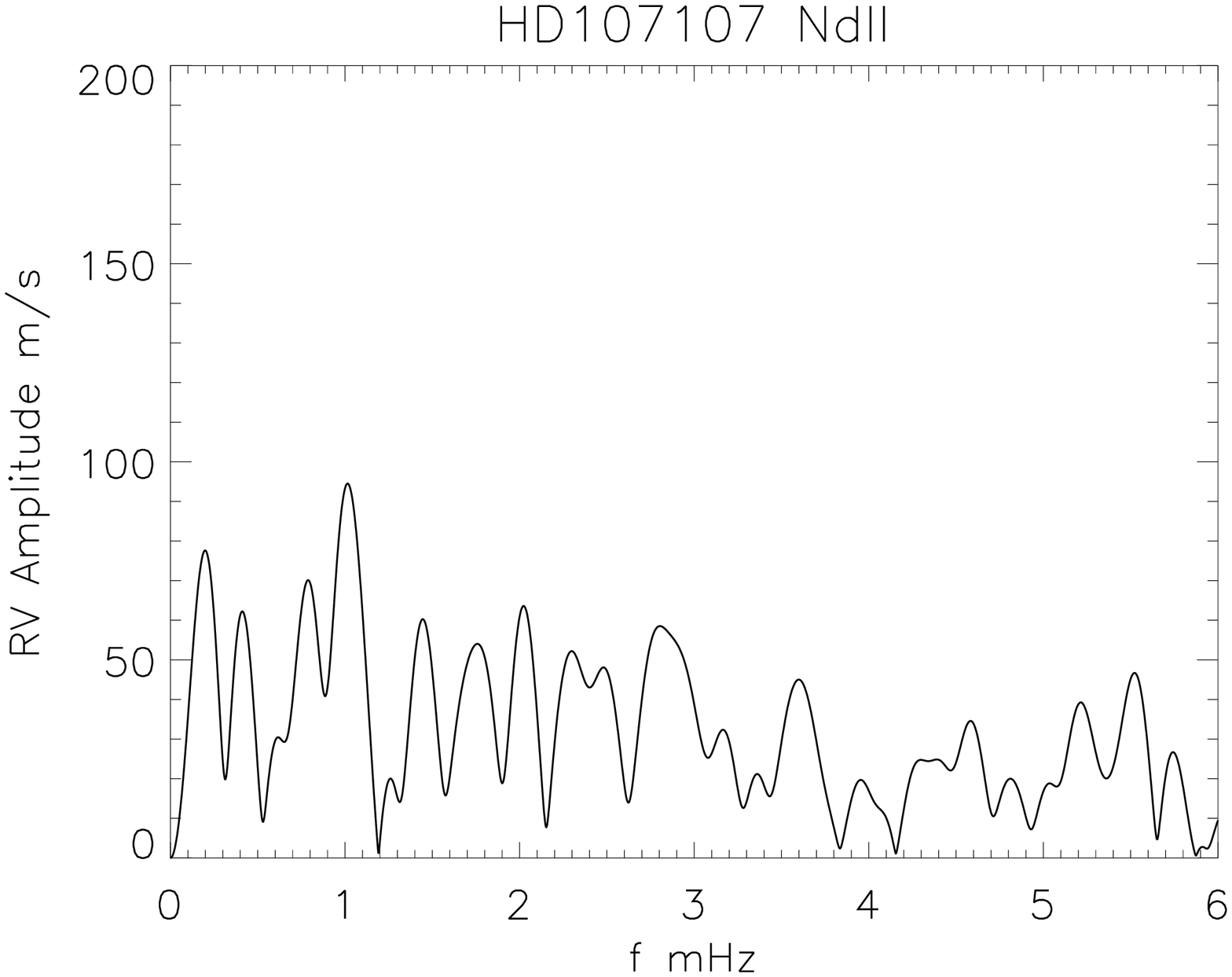}
  \includegraphics[height=5.6cm,
  angle=270]{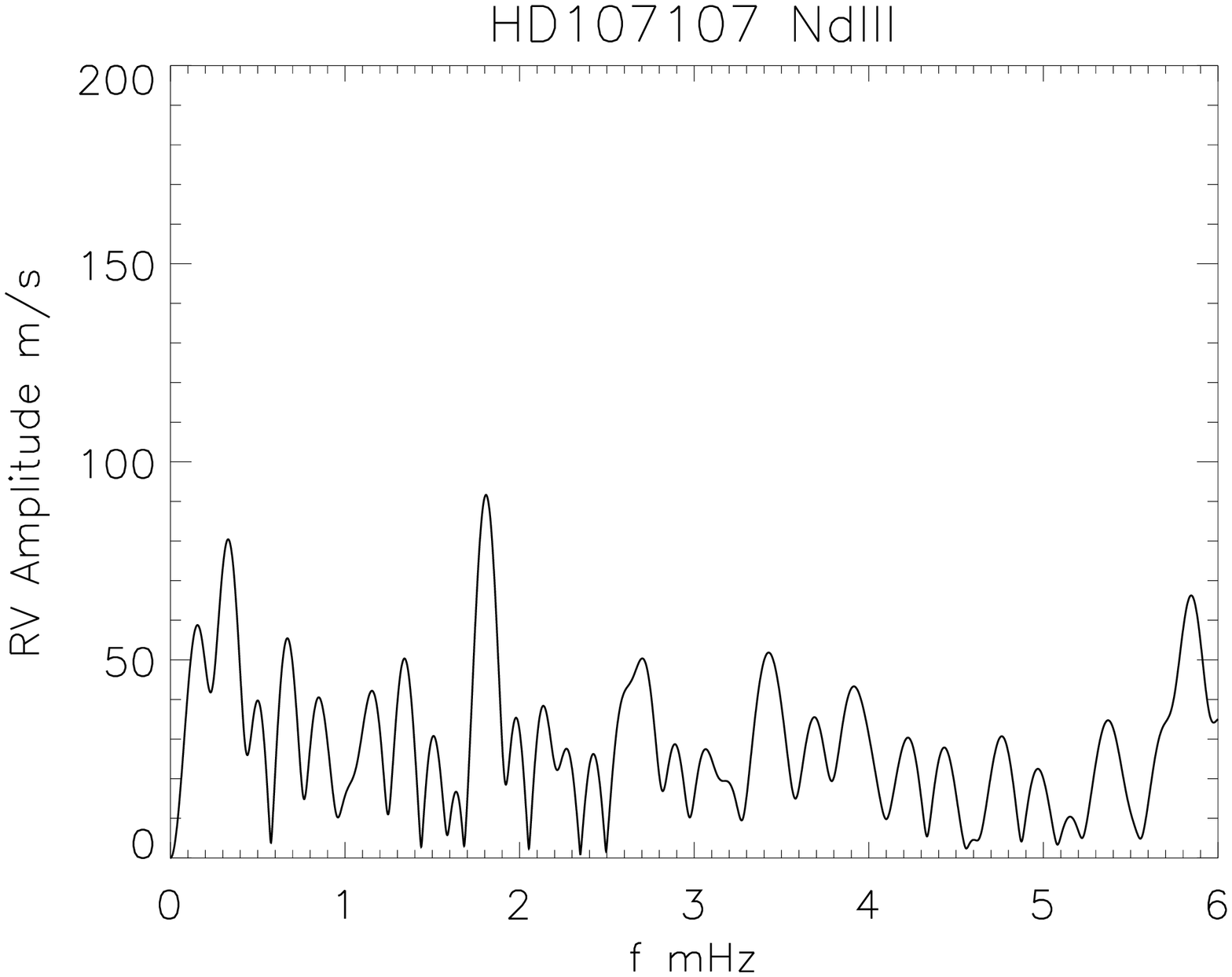}
  \includegraphics[height=5.6cm, angle=270]{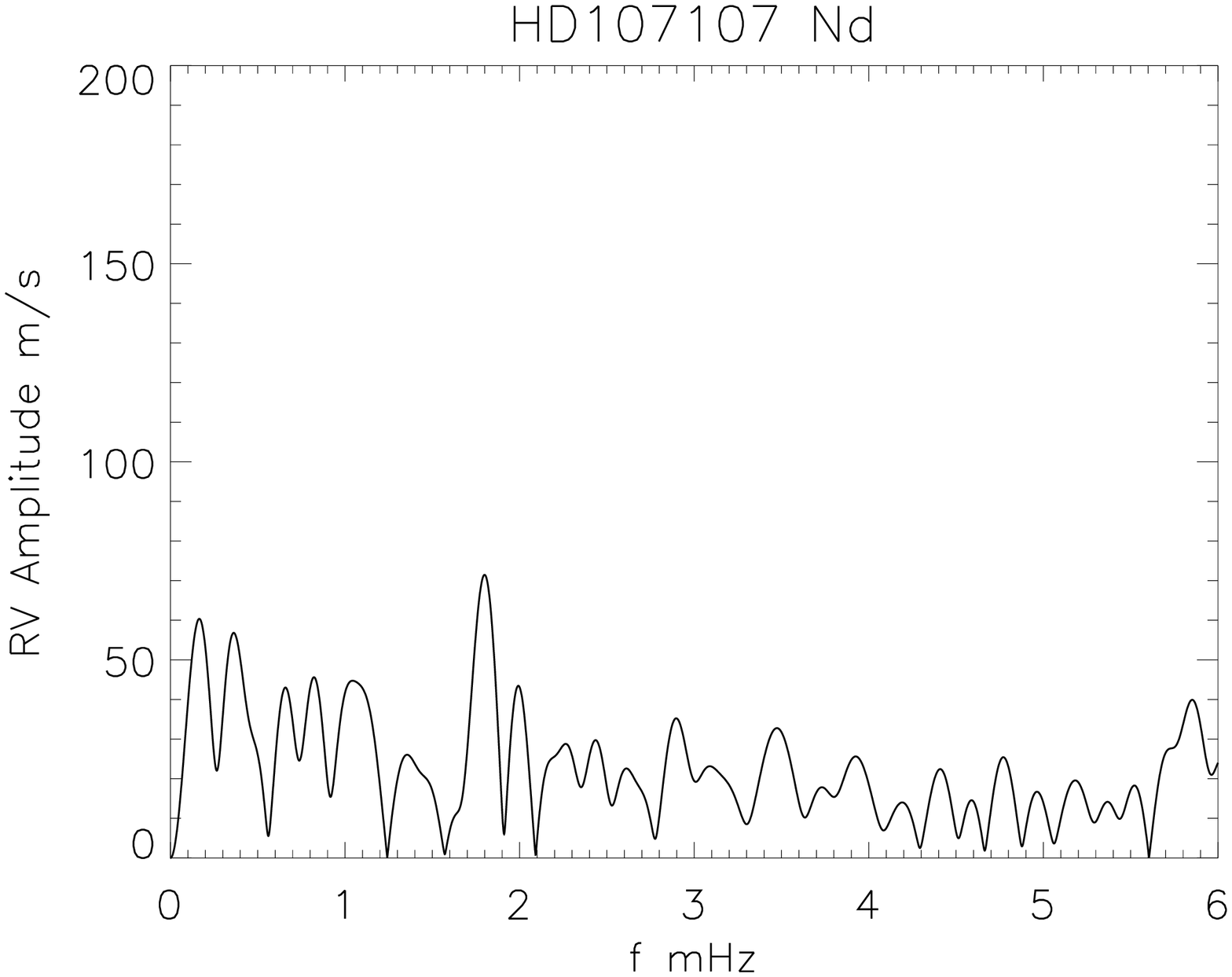}
  \includegraphics[height=5.6cm, angle=270]{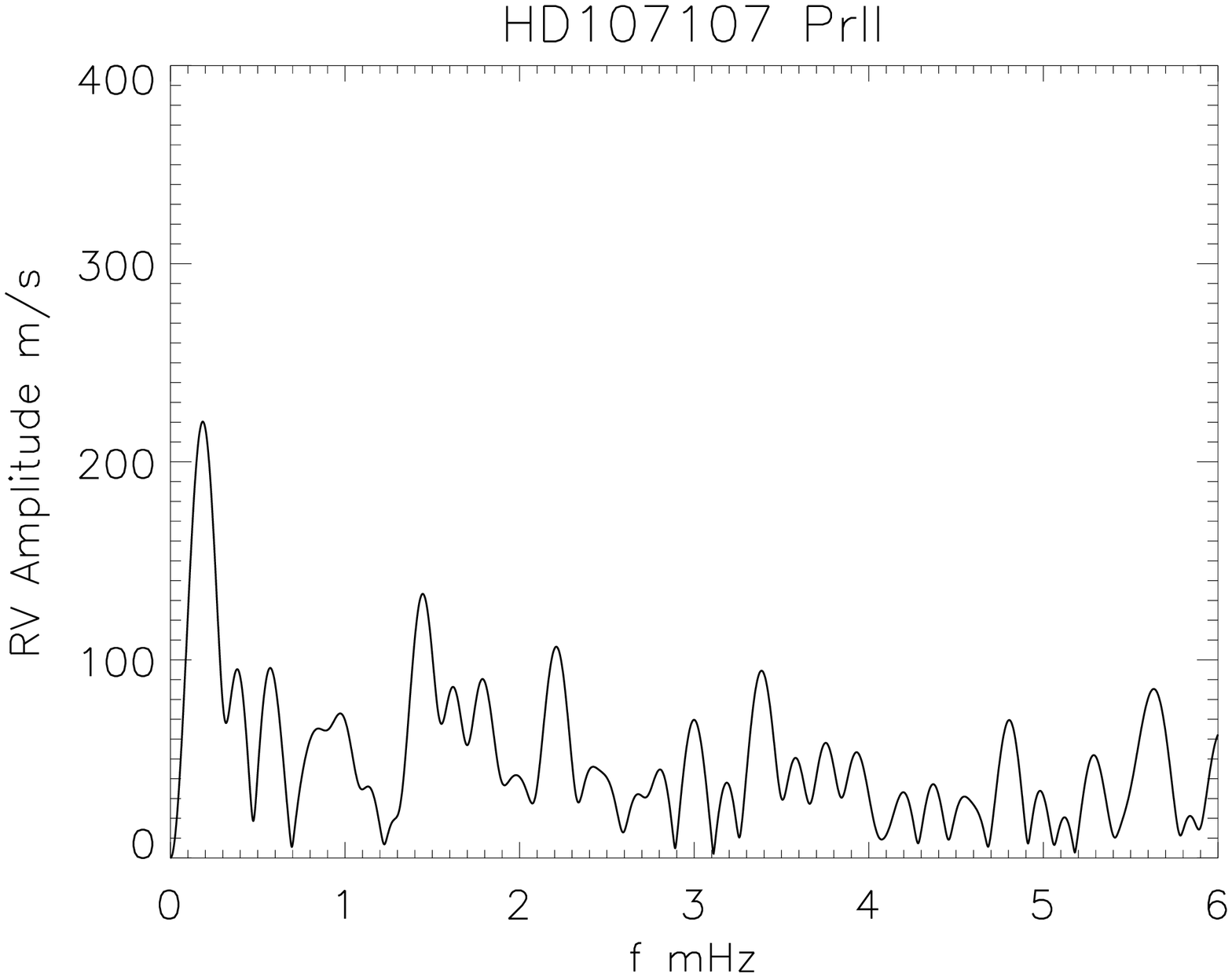}
  \includegraphics[height=5.6cm,
  angle=270]{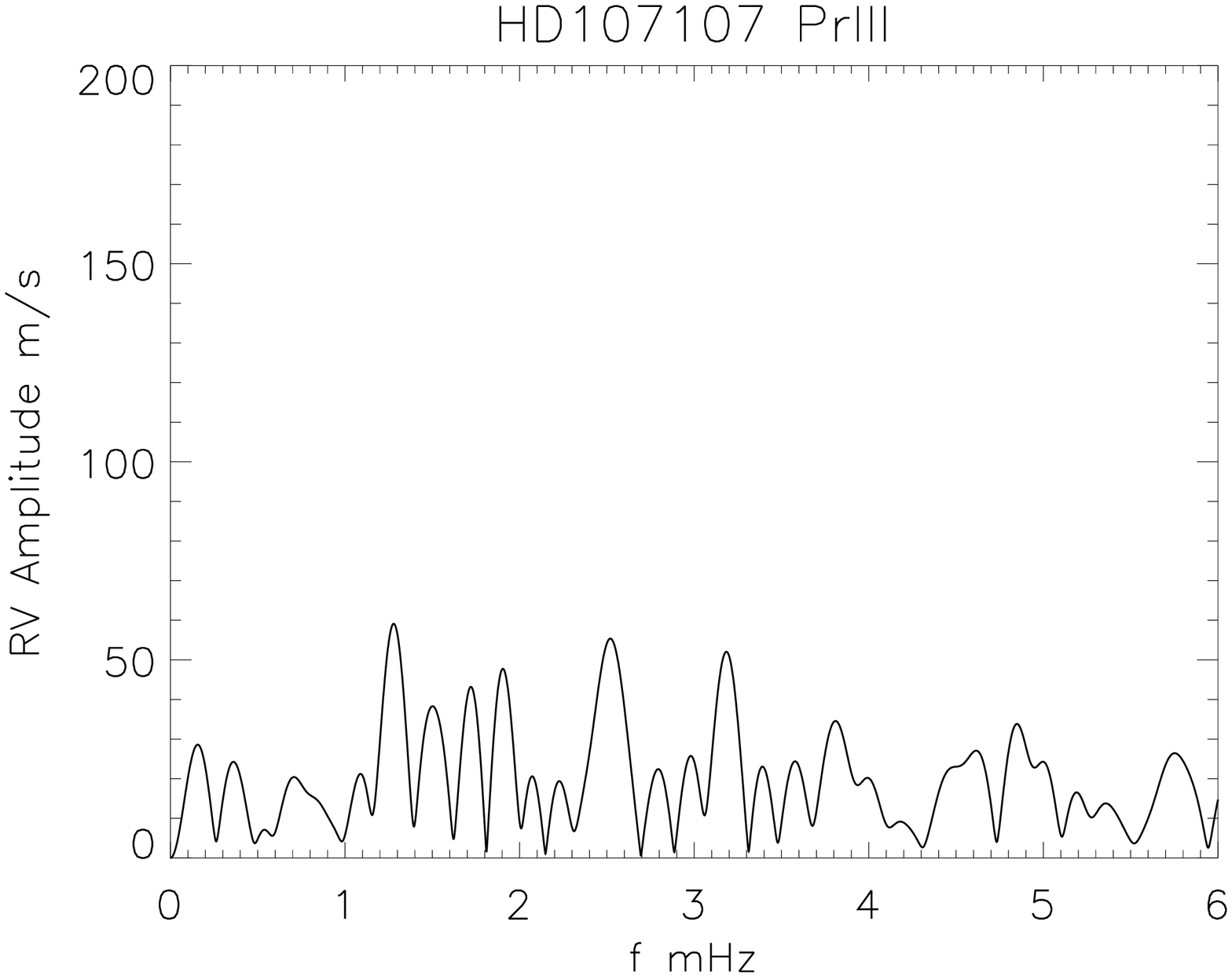}
  \includegraphics[height=5.6cm, angle=270]{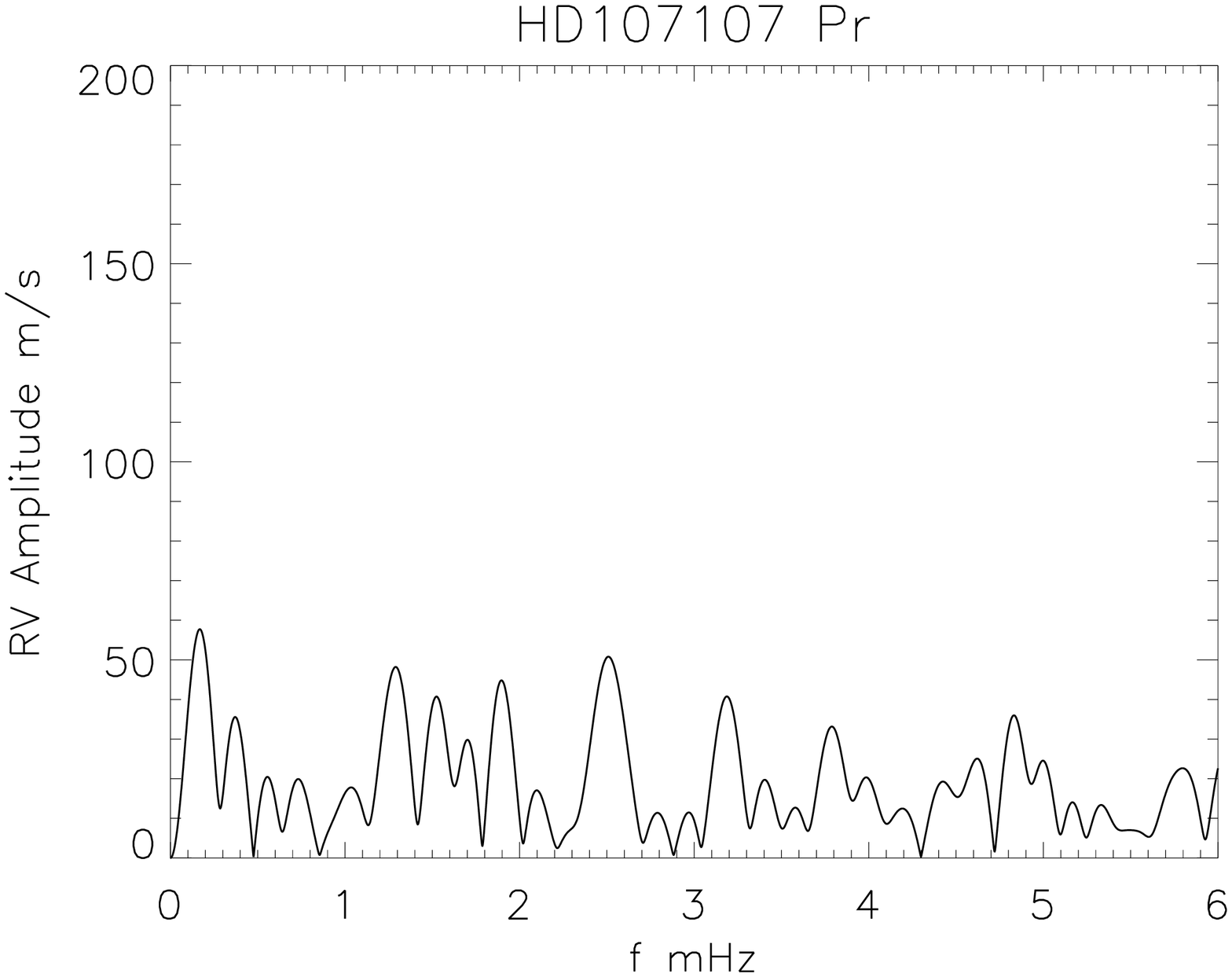}
  \includegraphics[height=5.6cm, angle=270]{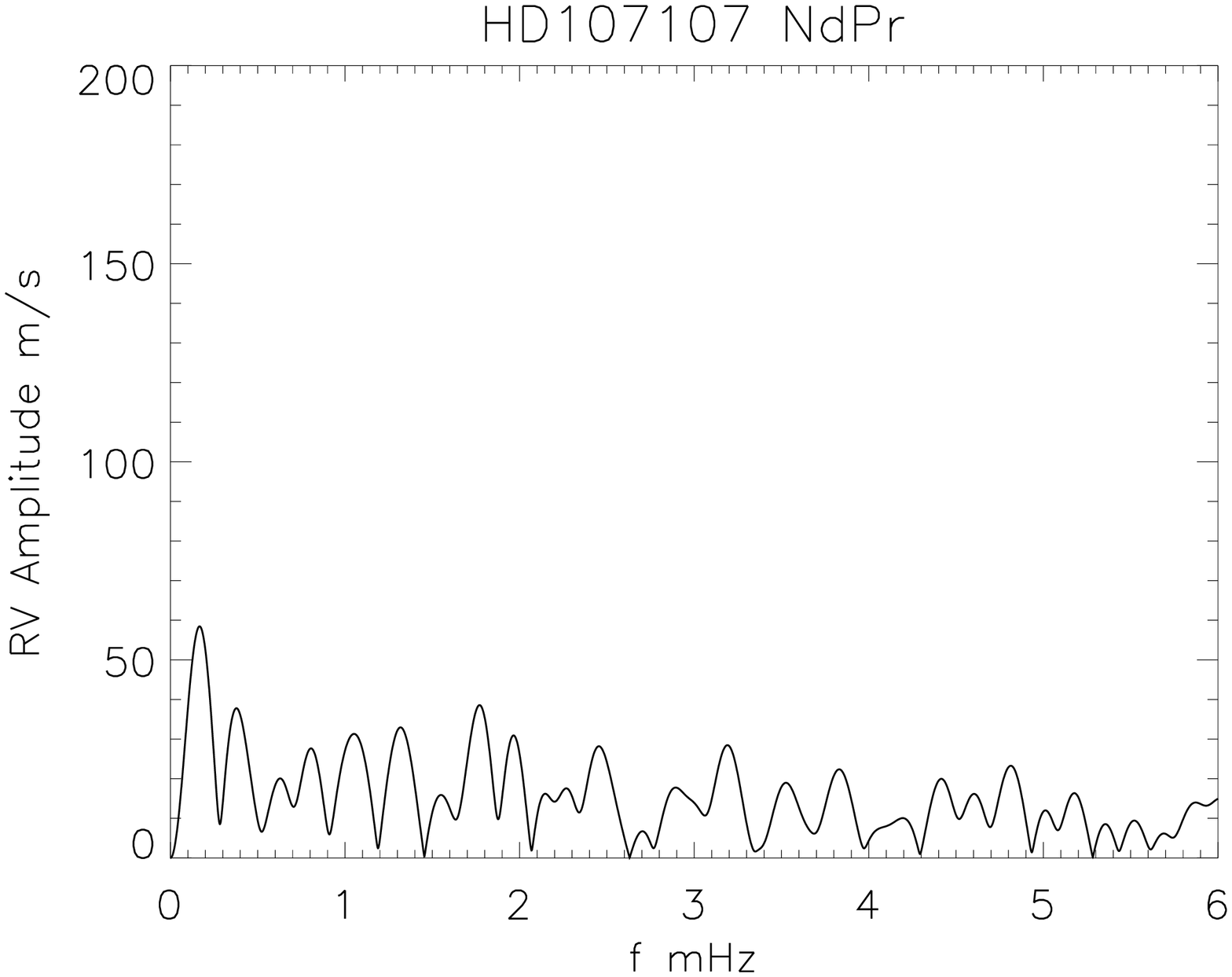}
  \includegraphics[height=5.6cm, angle=270]{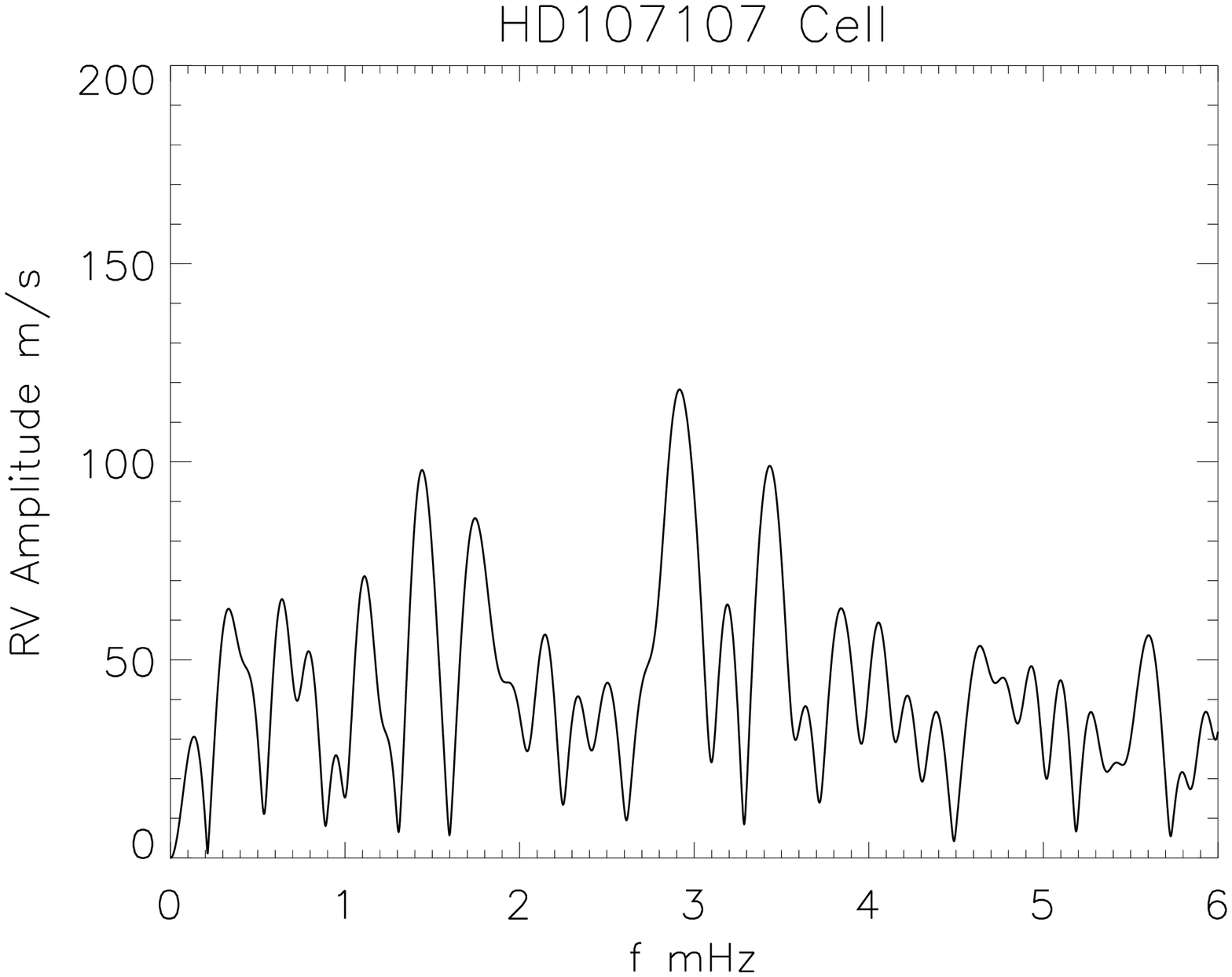}
  \includegraphics[height=5.6cm, angle=270]{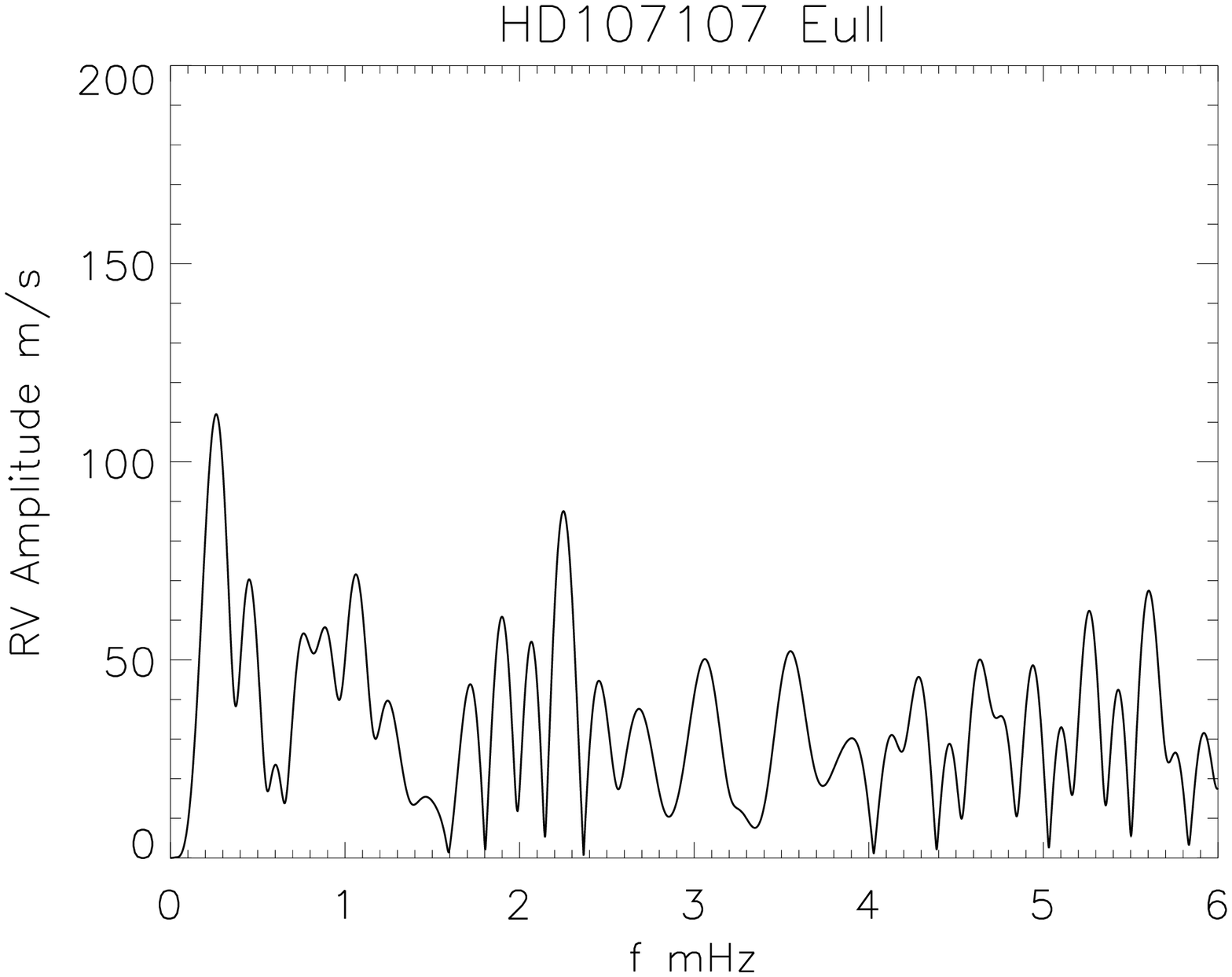}
  \includegraphics[height=5.6cm, angle=270]{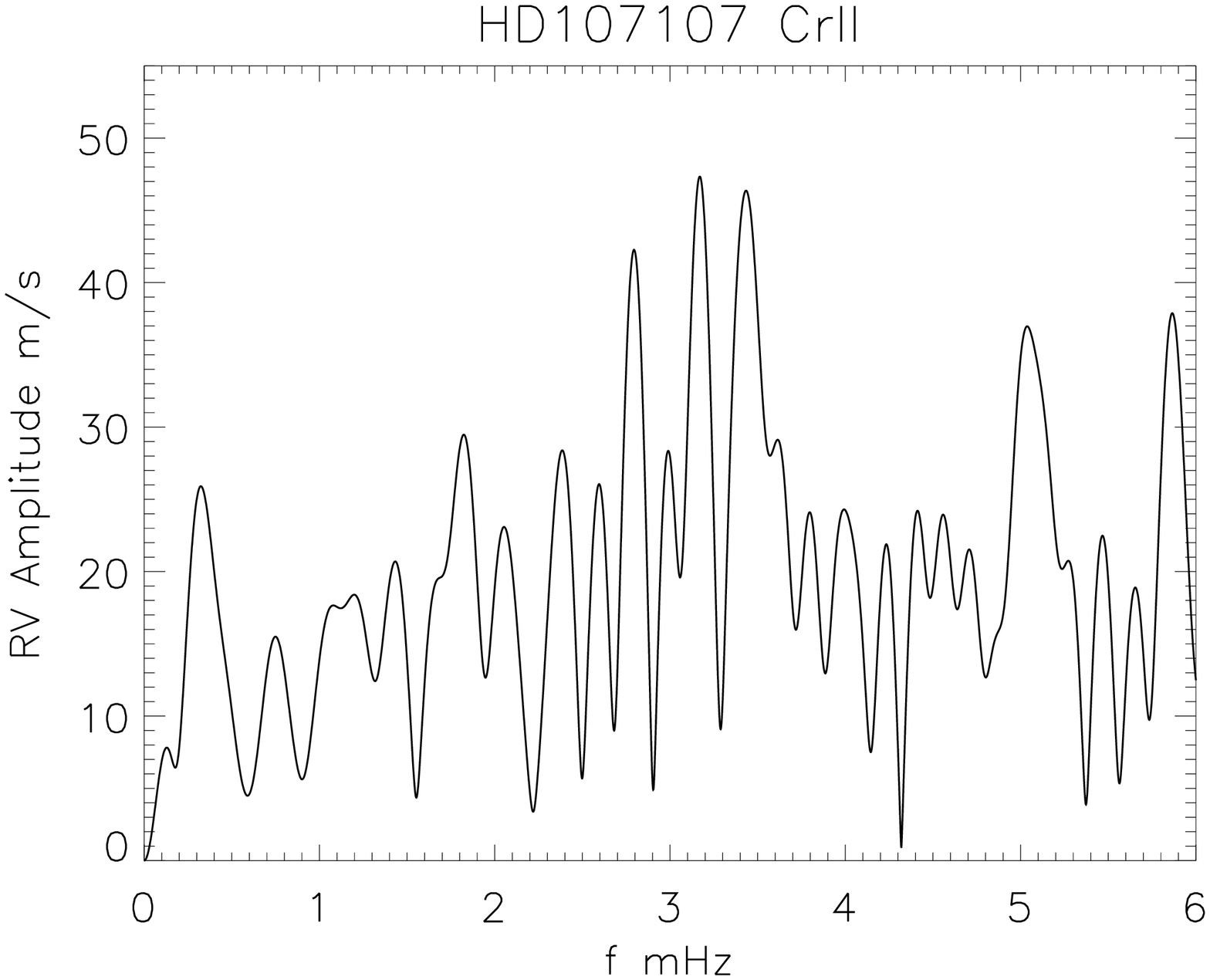}
  \includegraphics[height=5.6cm, angle=270]{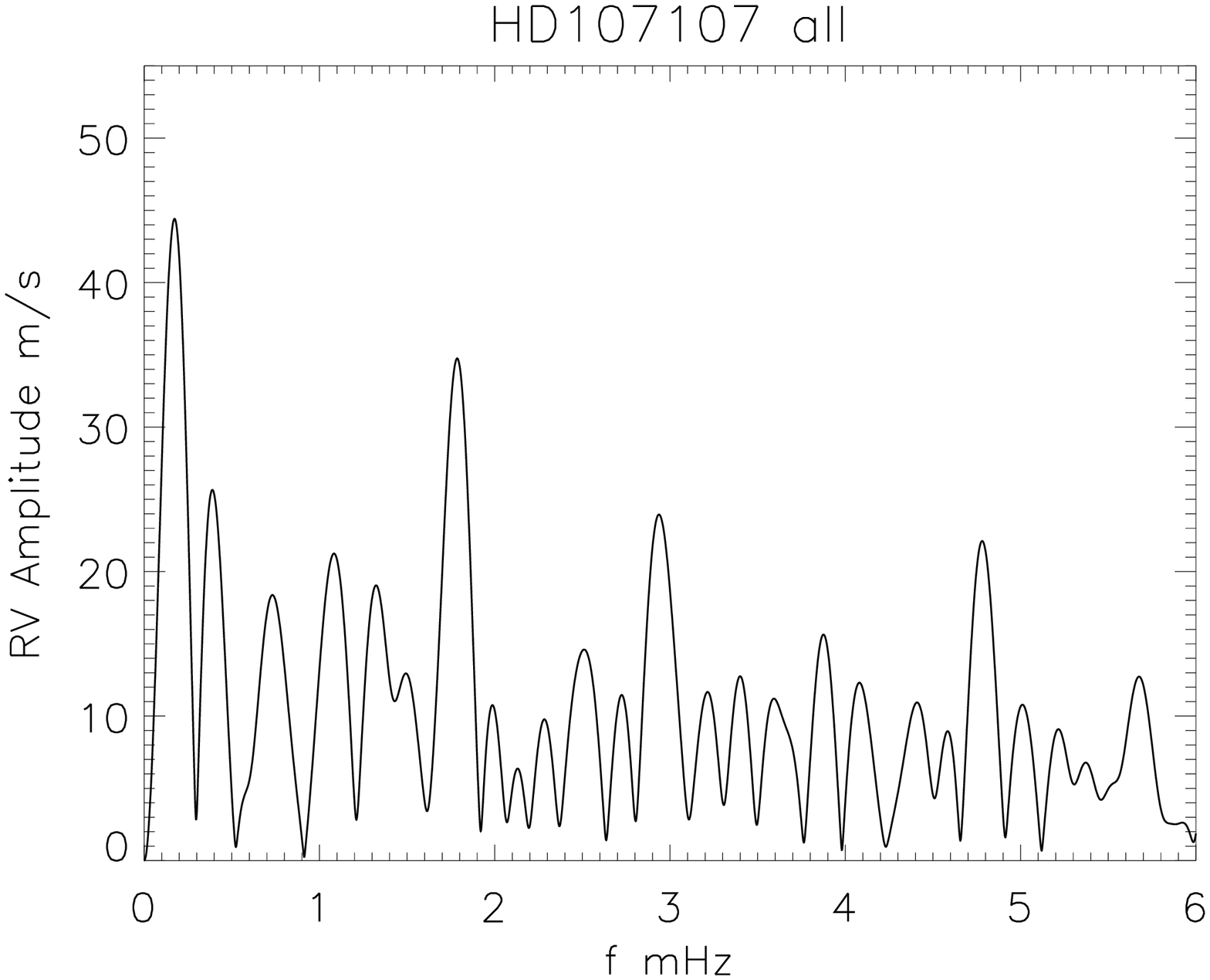}
  \includegraphics[height=5.6cm, angle=270]{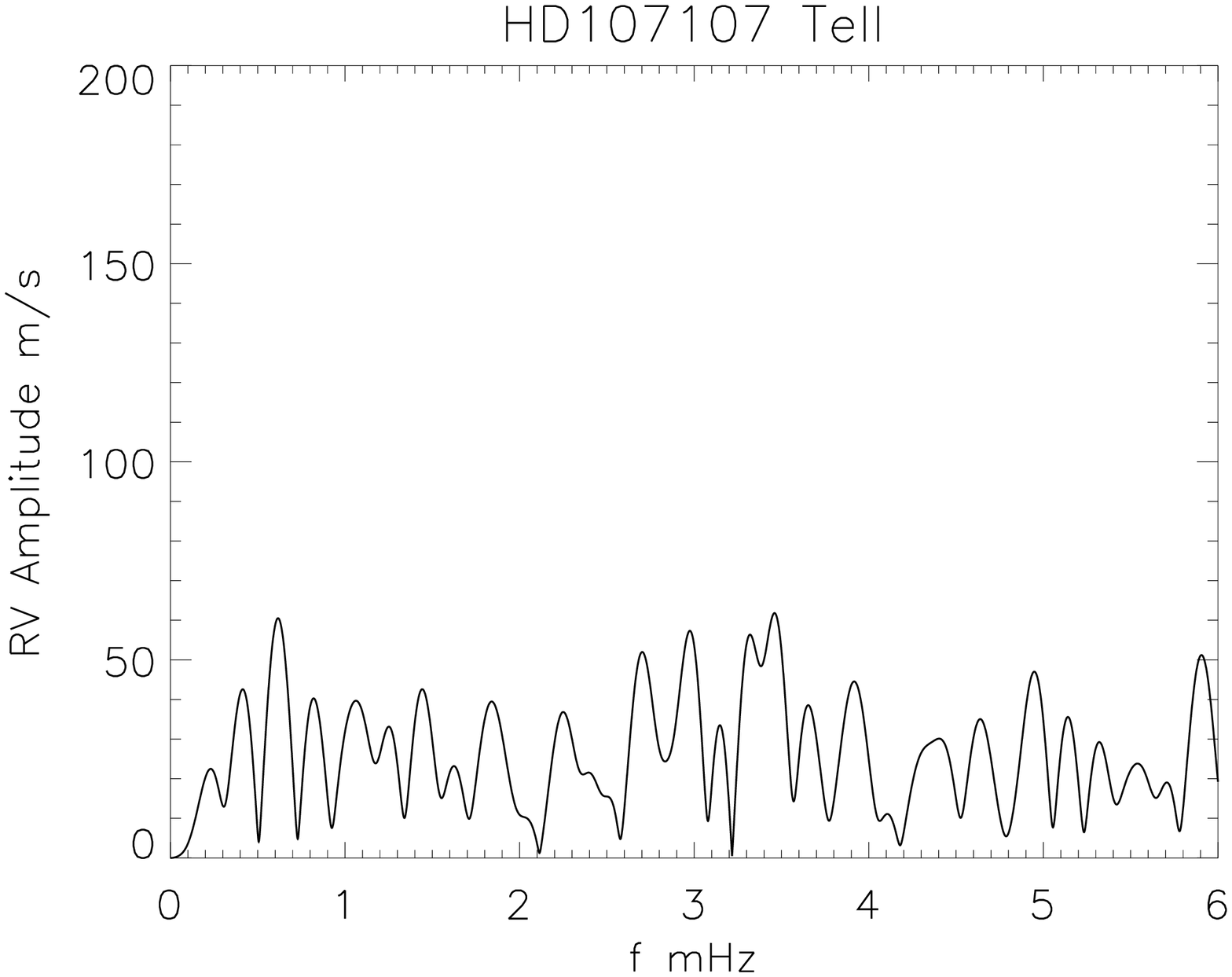}
  \caption{\label{fig:107107cog}Amplitude spectra for HD\,107107.
    Panel labels indicate the element lines(s) in the combined RV
    series.  Top left panel is the window function for a 3\,mHz
    signal.  Panels `all' and `Tell' are, respectively, for all
    available lines combined and for all telluric lines combined.  }
\end{figure*}

\begin{figure*}
  \vspace{3pt}
  \includegraphics[height=5.6cm,
  angle=270]{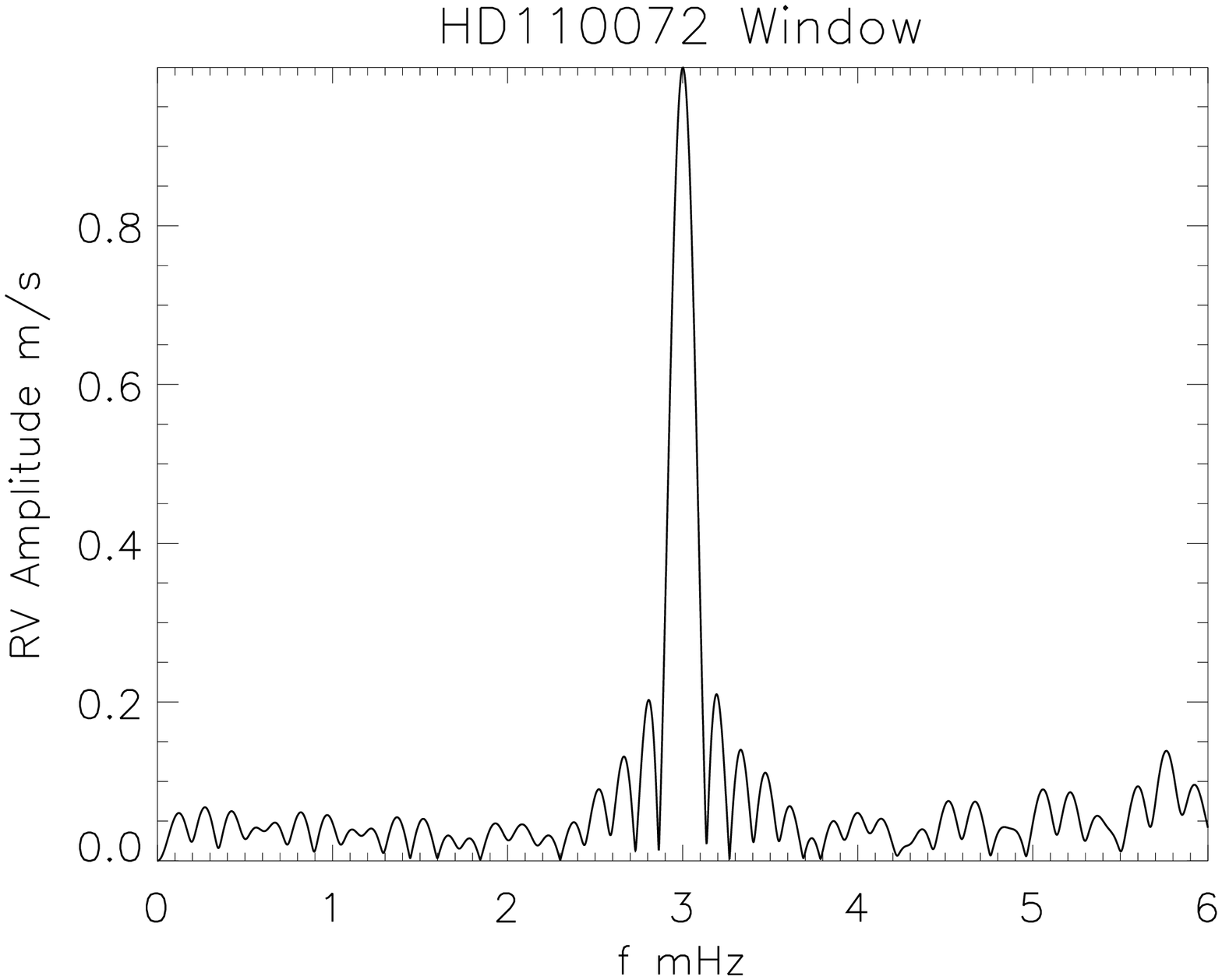}
  \includegraphics[height=5.6cm, angle=270]{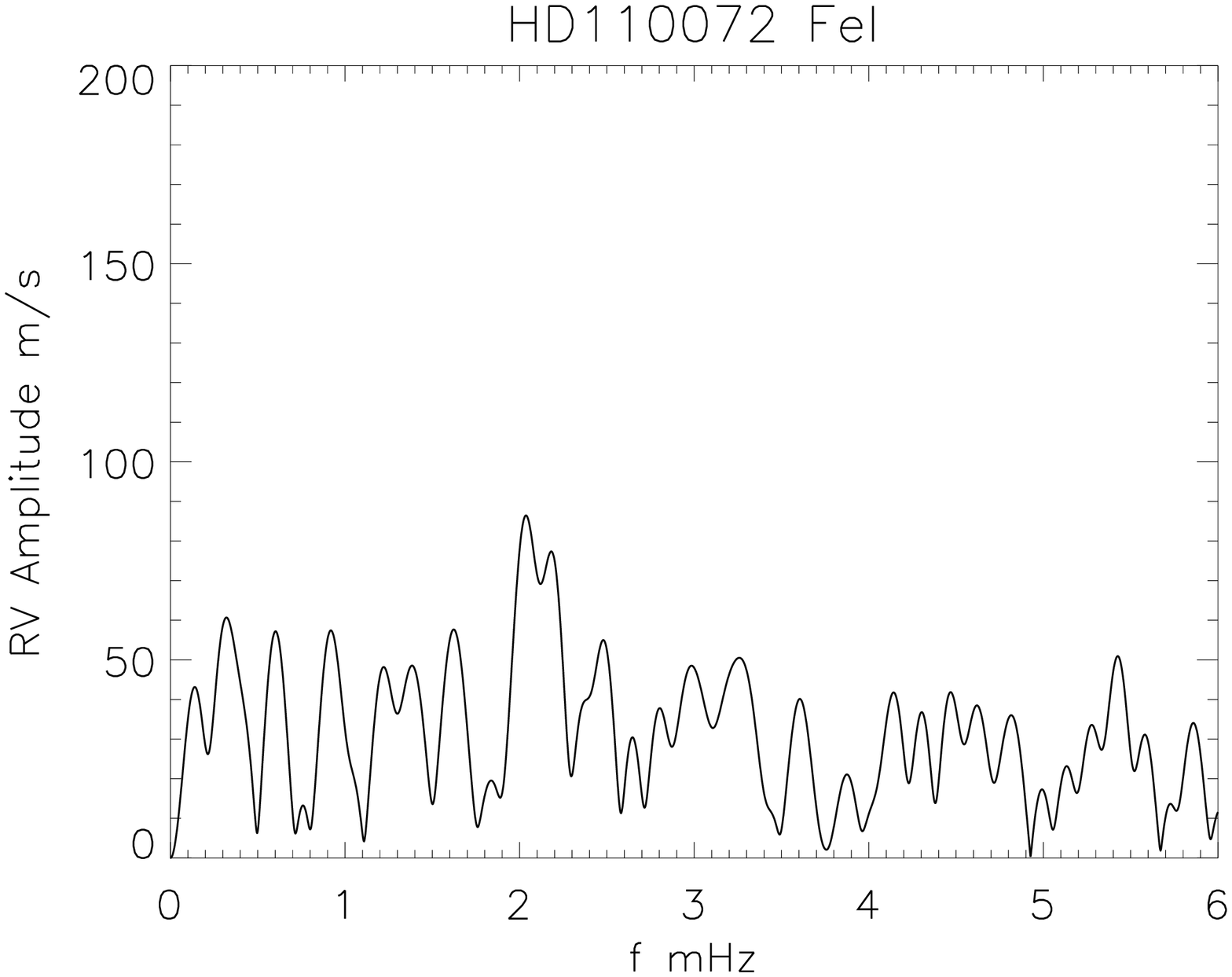}
  \includegraphics[height=5.6cm, angle=270]{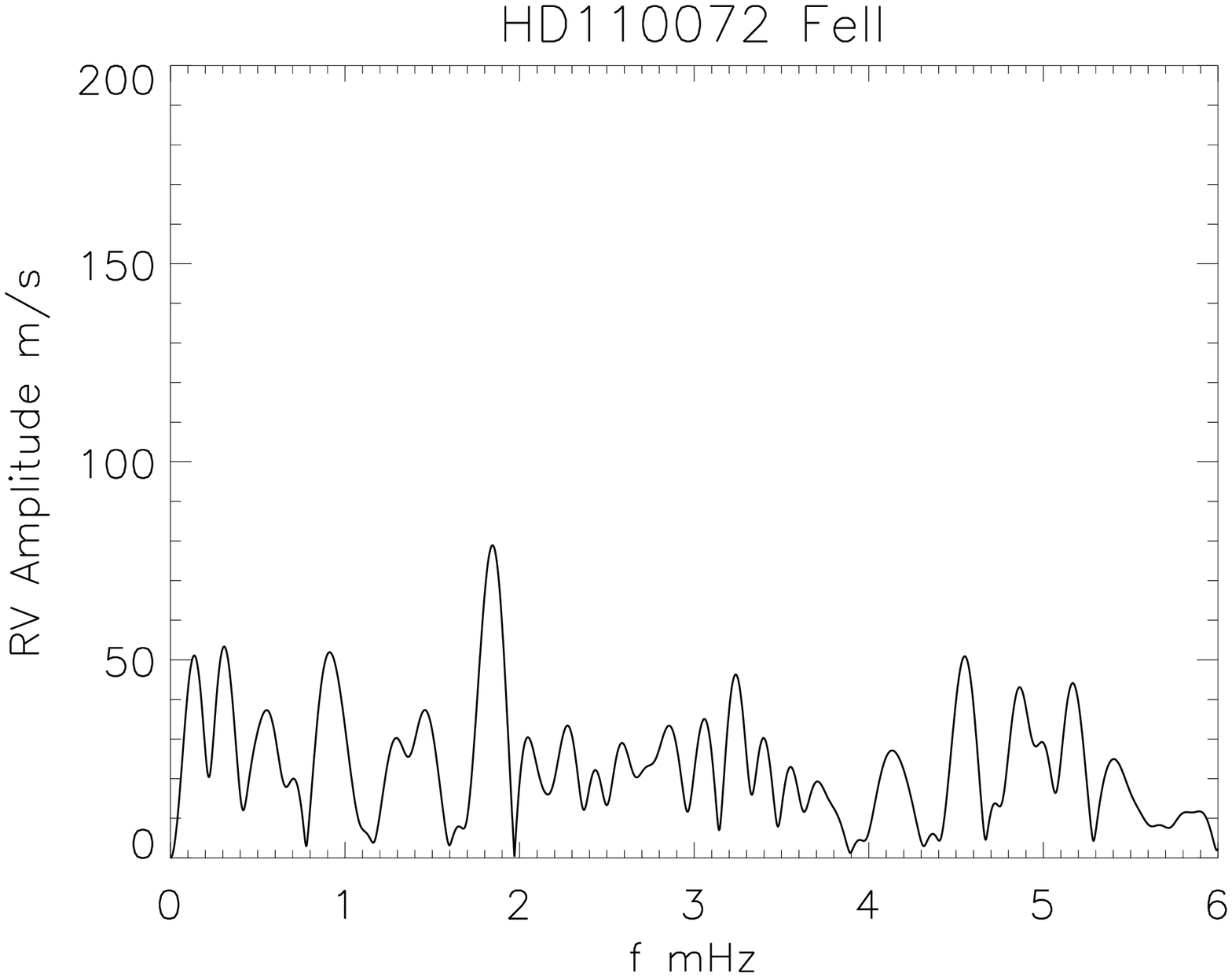}
  \includegraphics[height=5.6cm, angle=270]{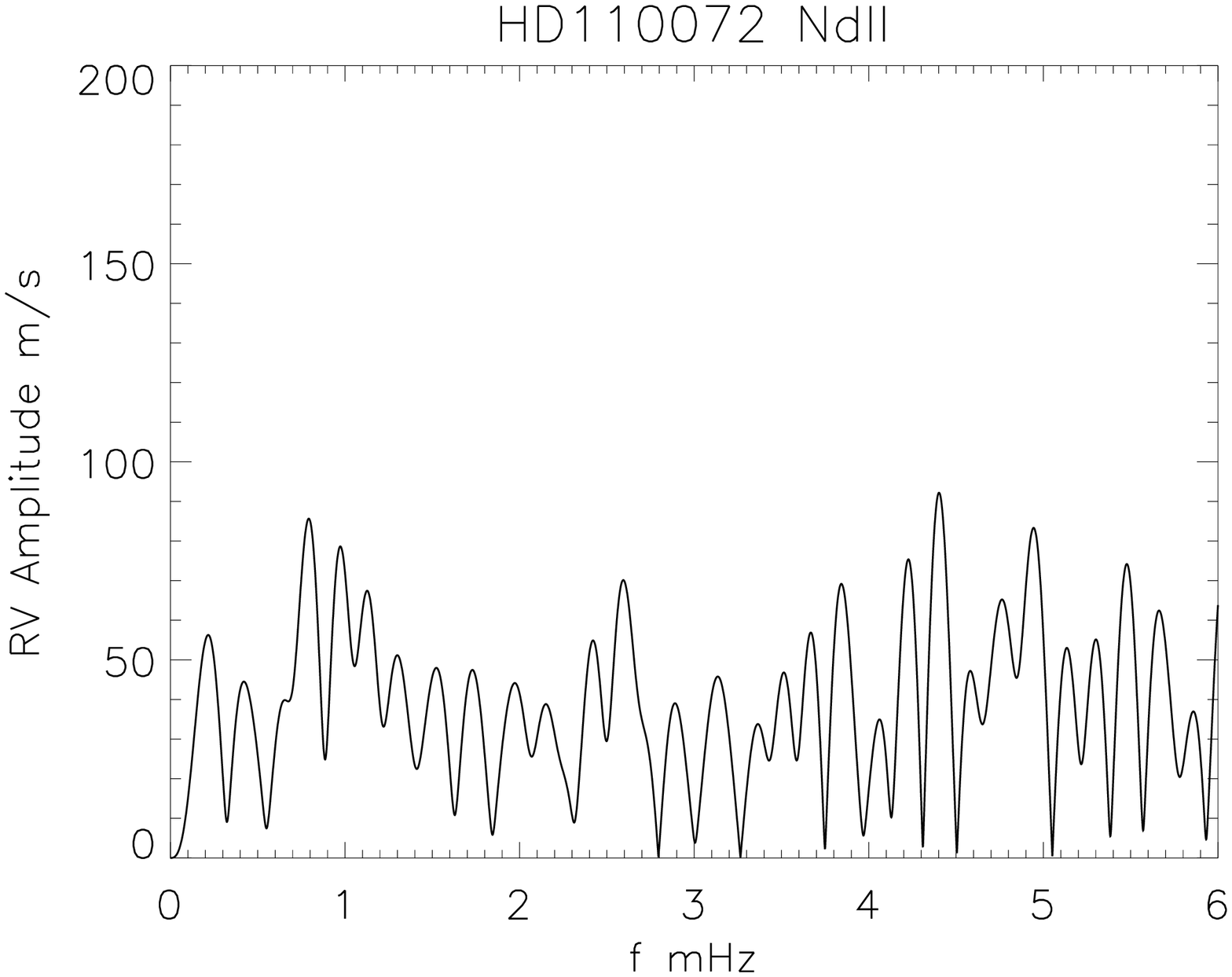}
  \includegraphics[height=5.6cm,
  angle=270]{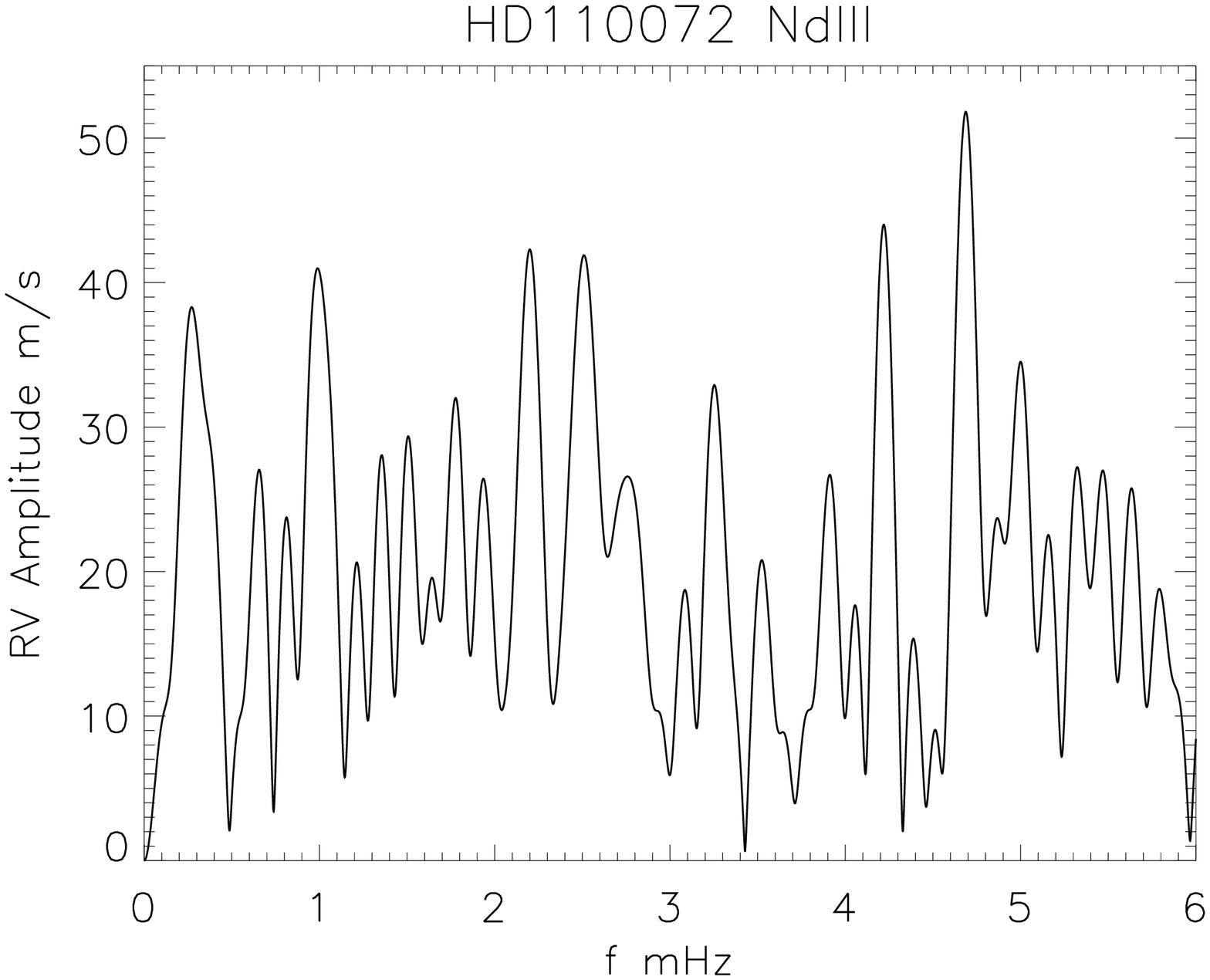}
  \includegraphics[height=5.6cm, angle=270]{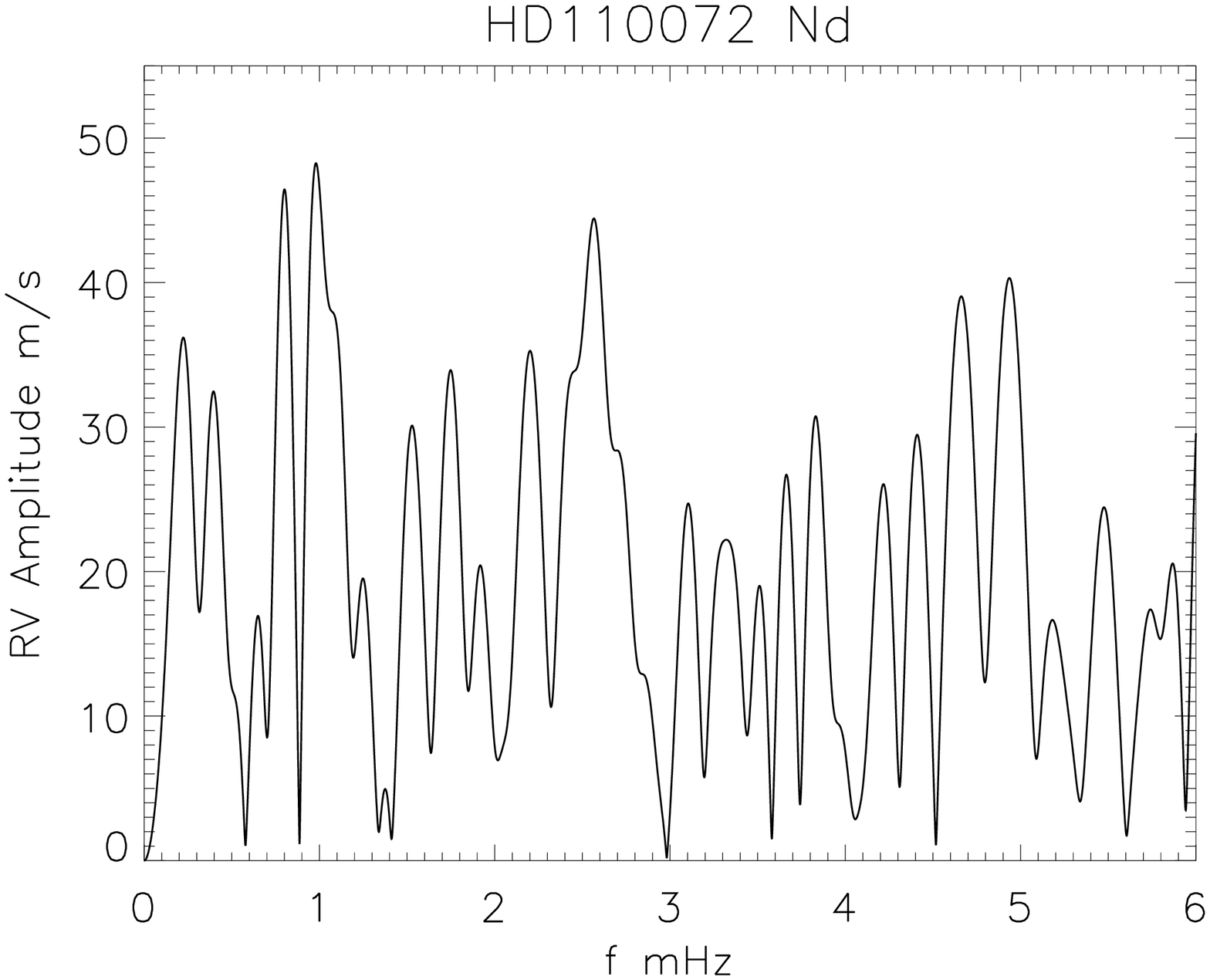}
  \includegraphics[height=5.6cm, angle=270]{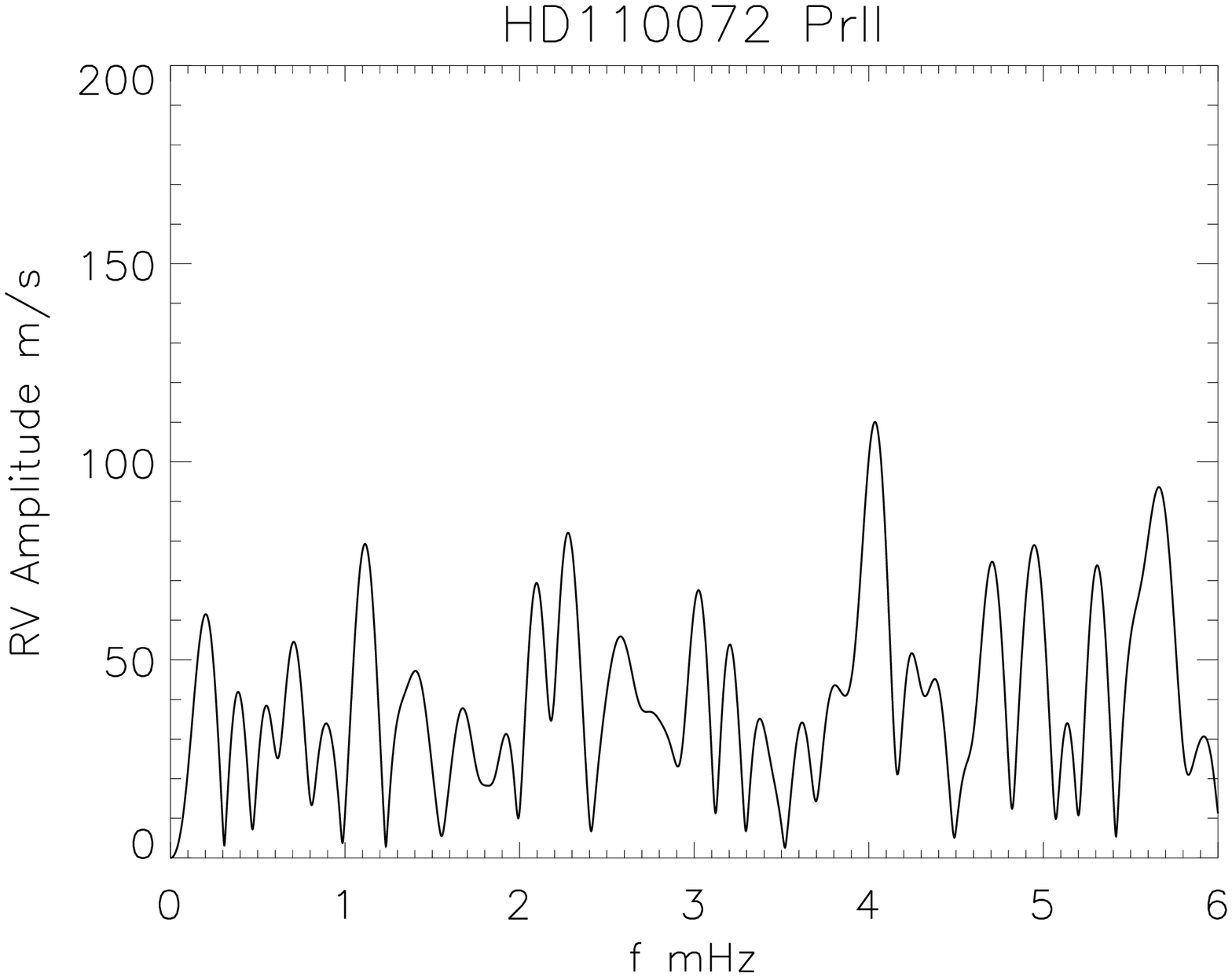}
  \includegraphics[height=5.6cm,
  angle=270]{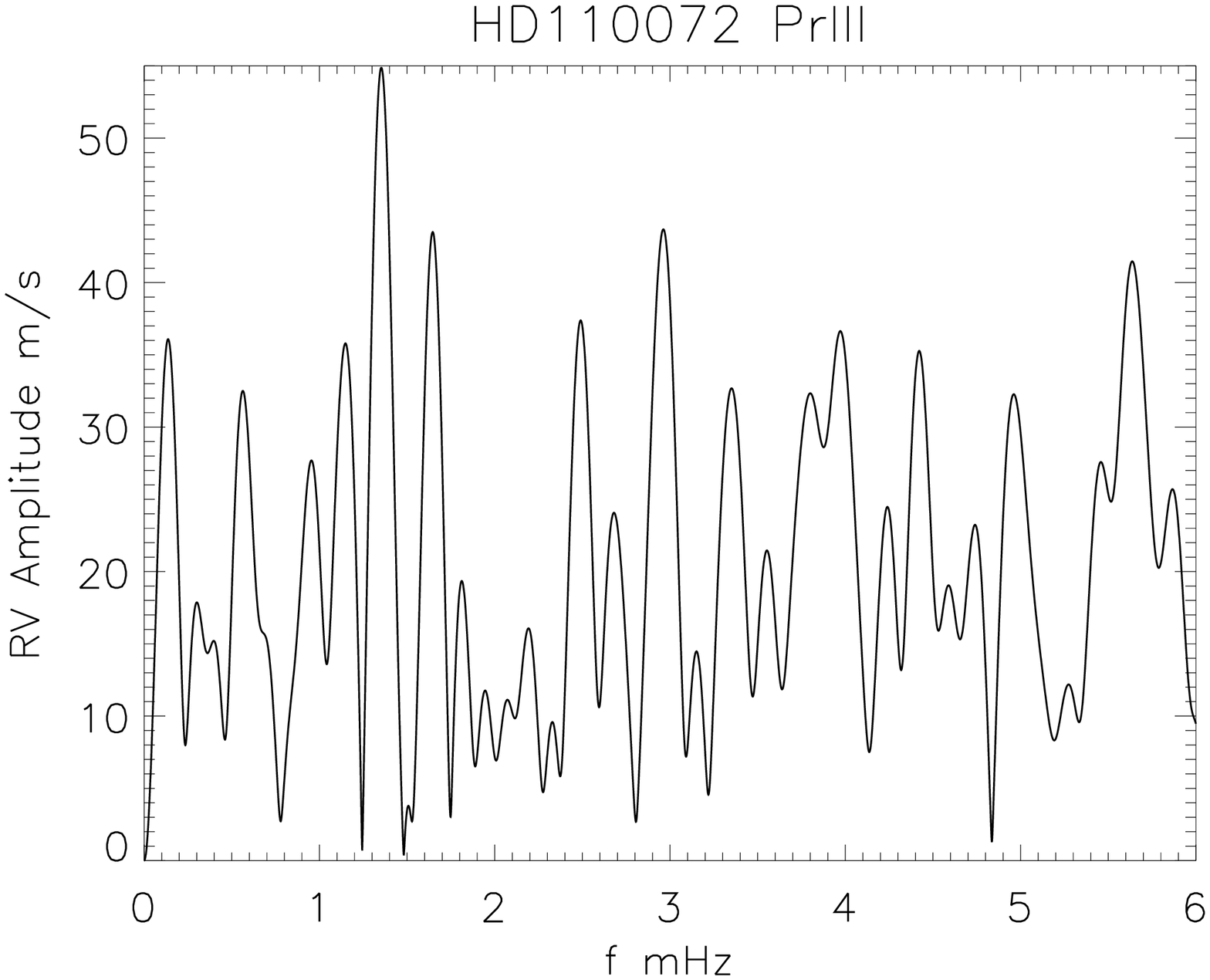}
  \includegraphics[height=5.6cm, angle=270]{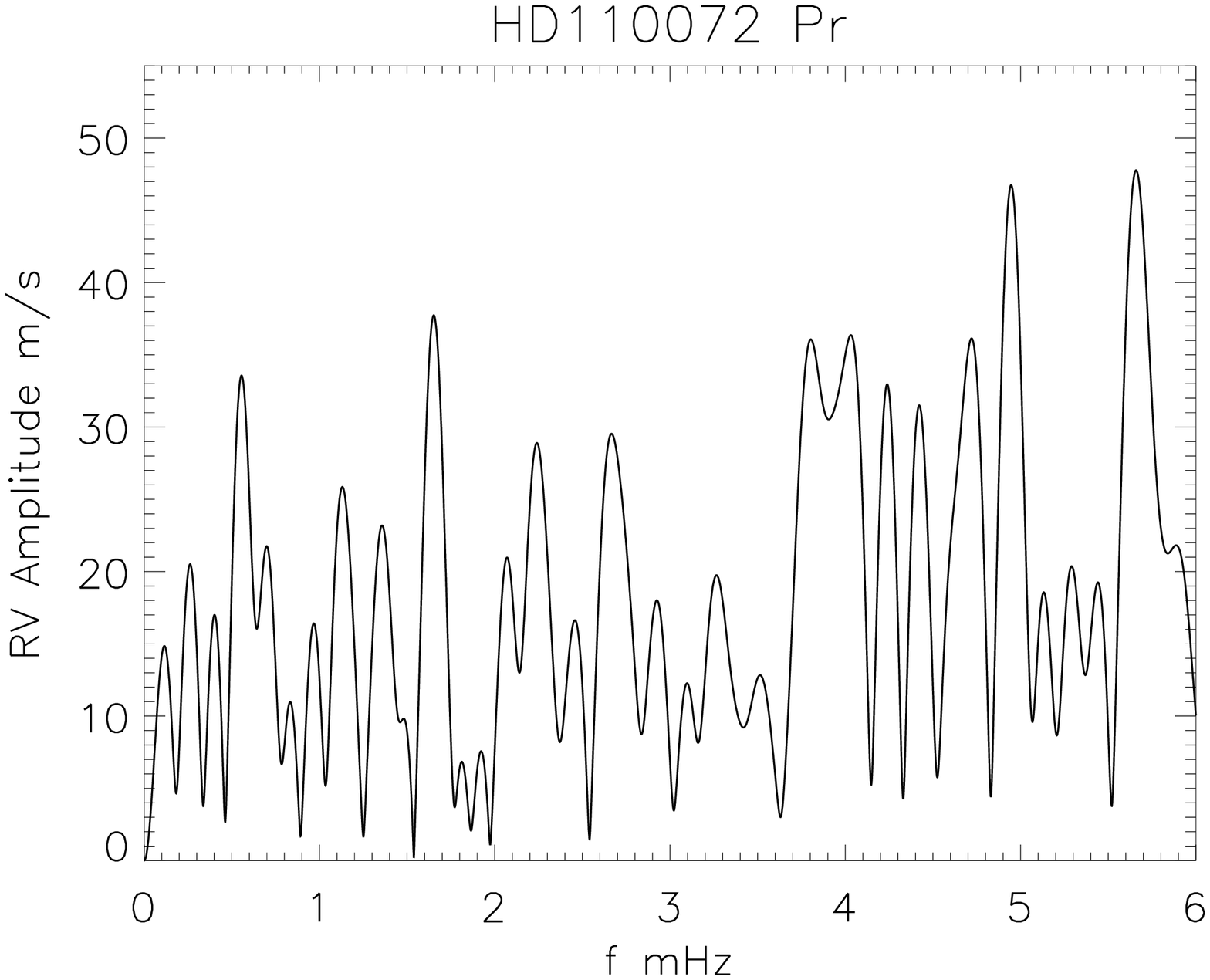}
  \includegraphics[height=5.6cm, angle=270]{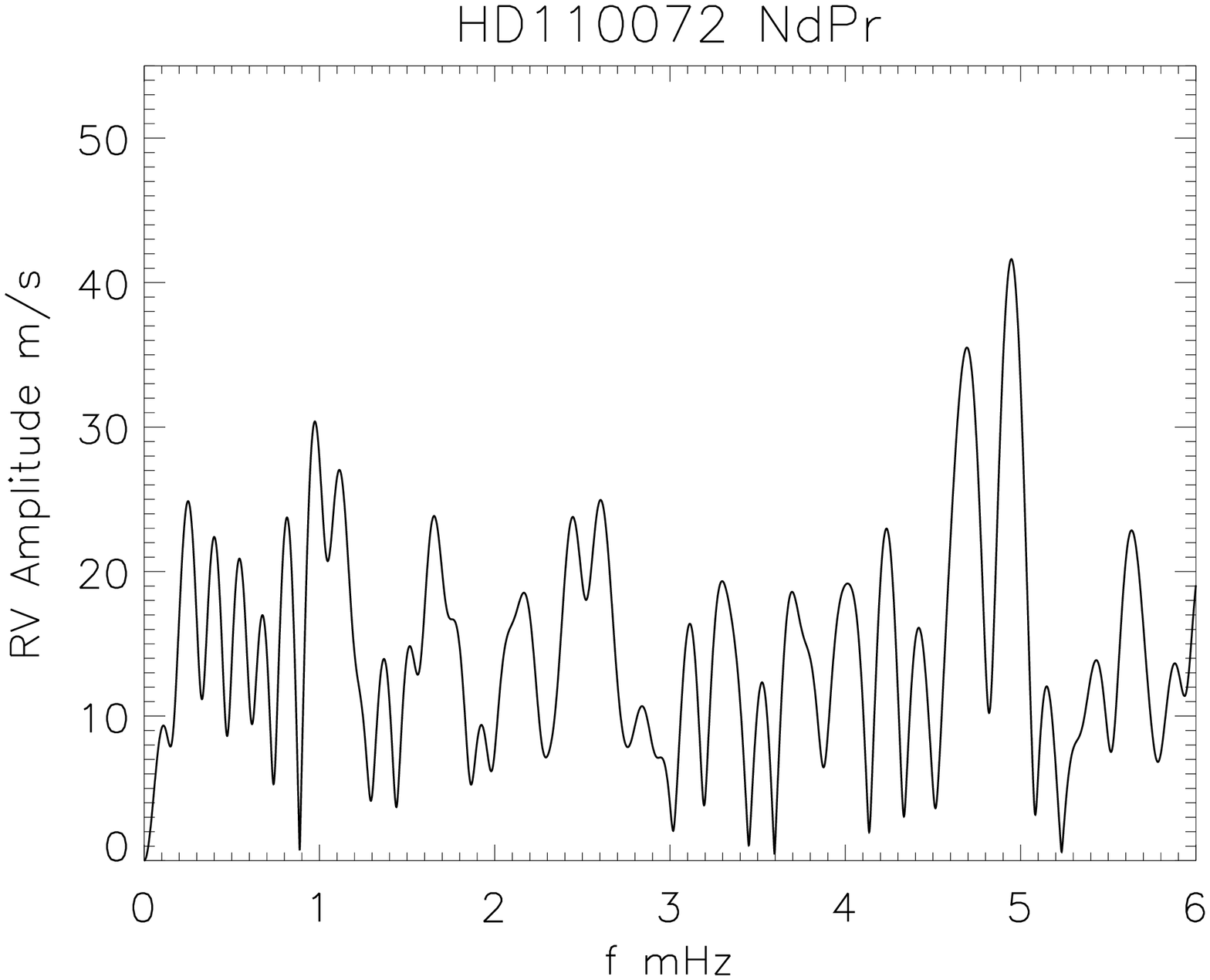}
  \includegraphics[height=5.6cm, angle=270]{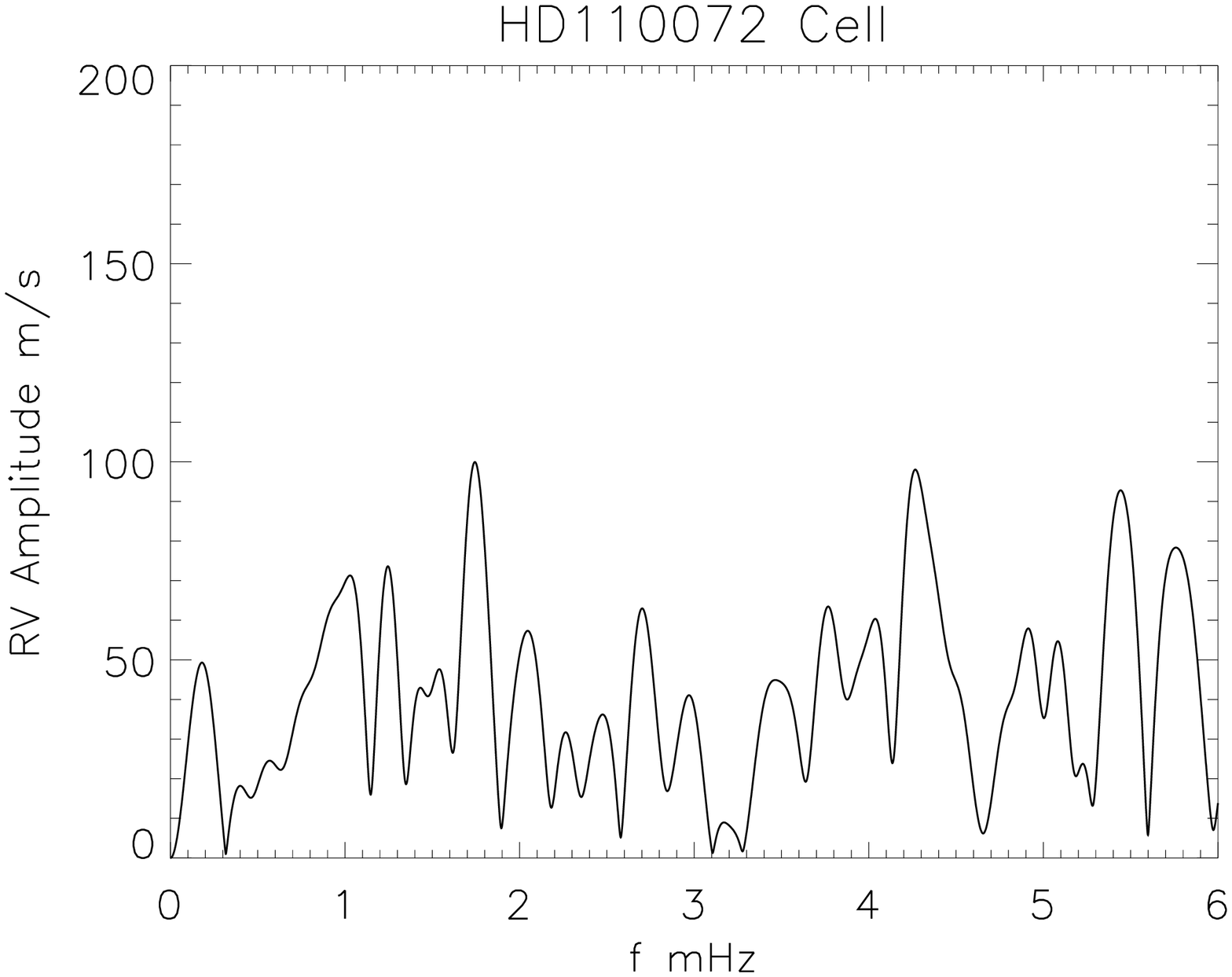}
  \includegraphics[height=5.6cm, angle=270]{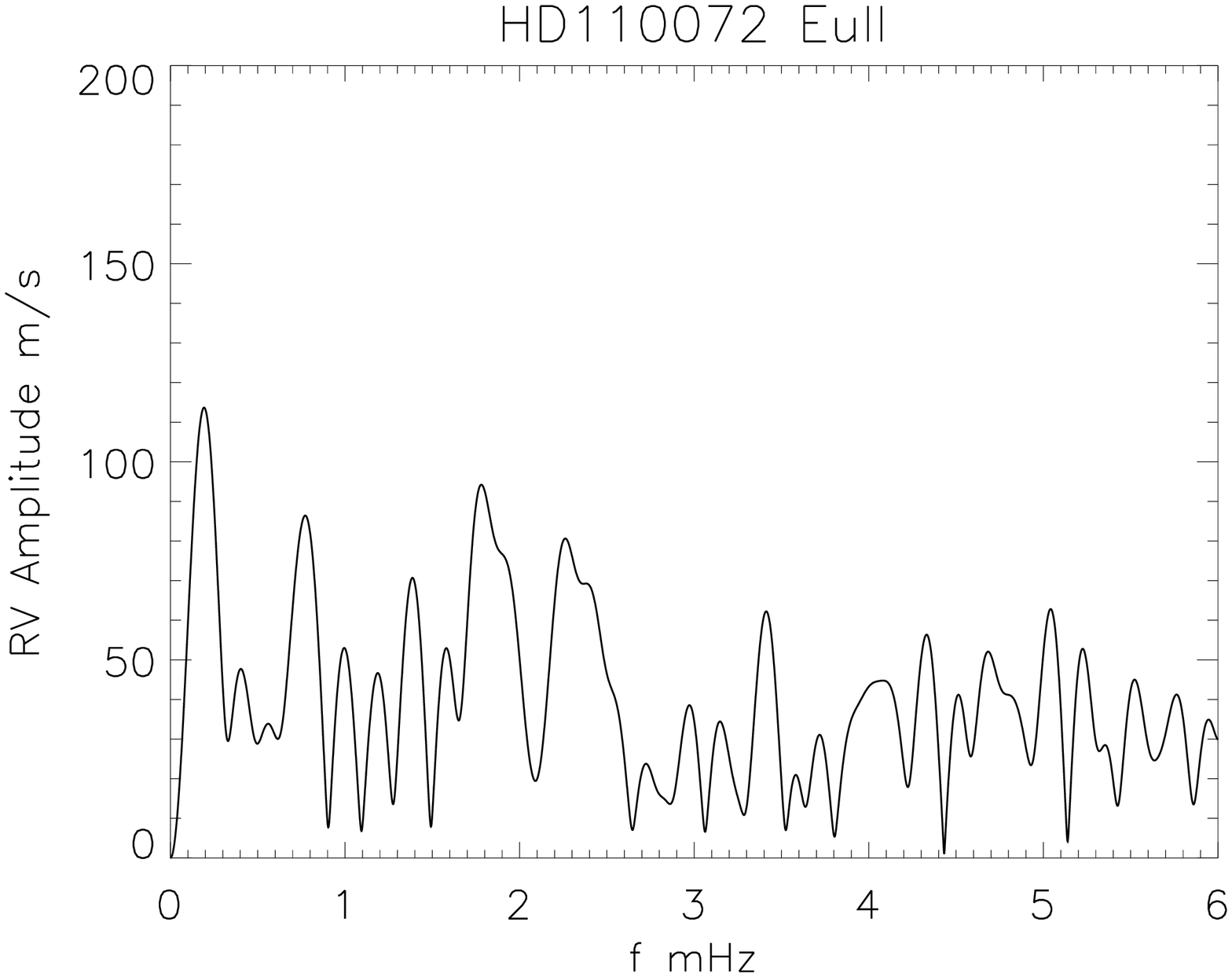}
  \includegraphics[height=5.6cm, angle=270]{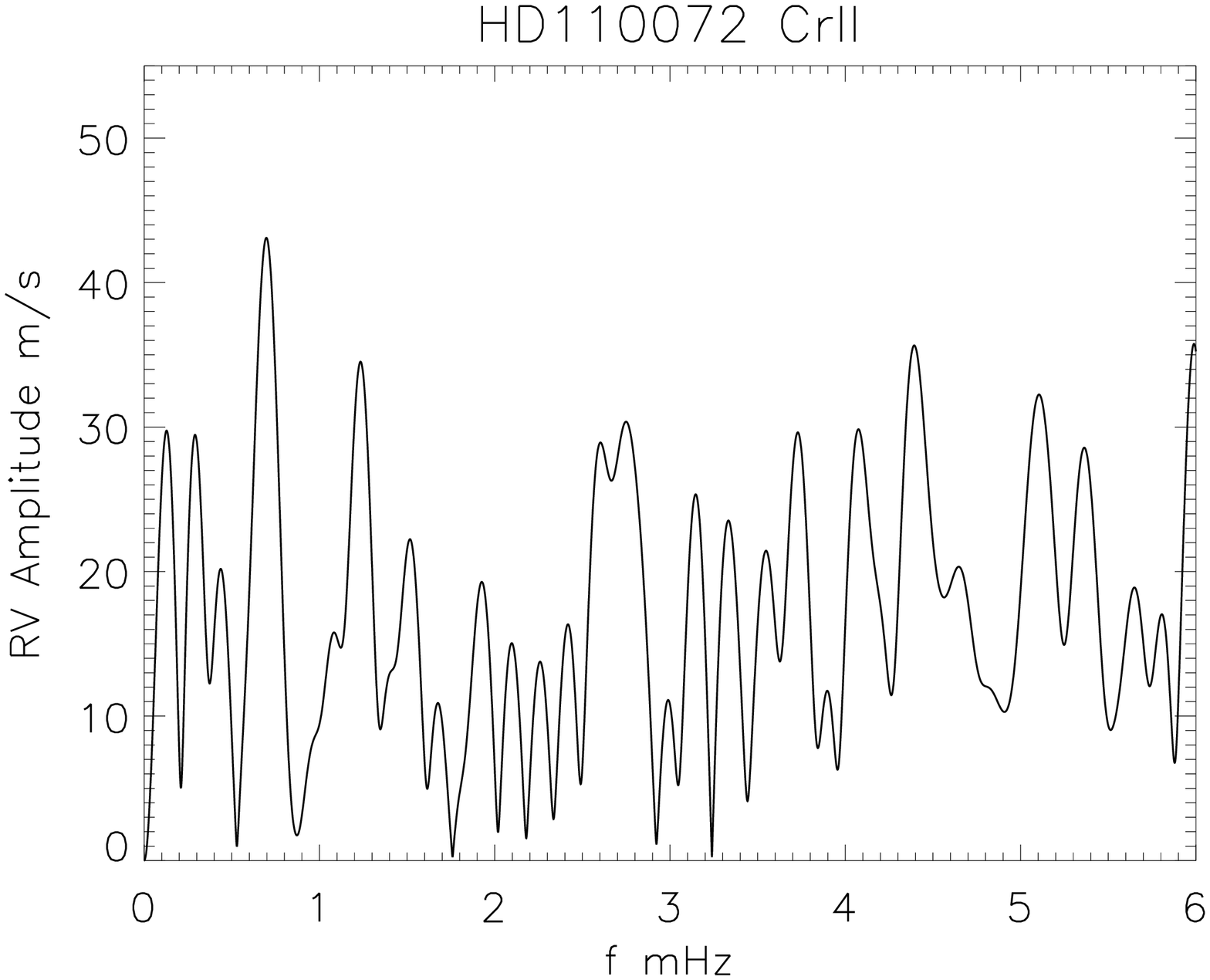}
  \includegraphics[height=5.6cm, angle=270]{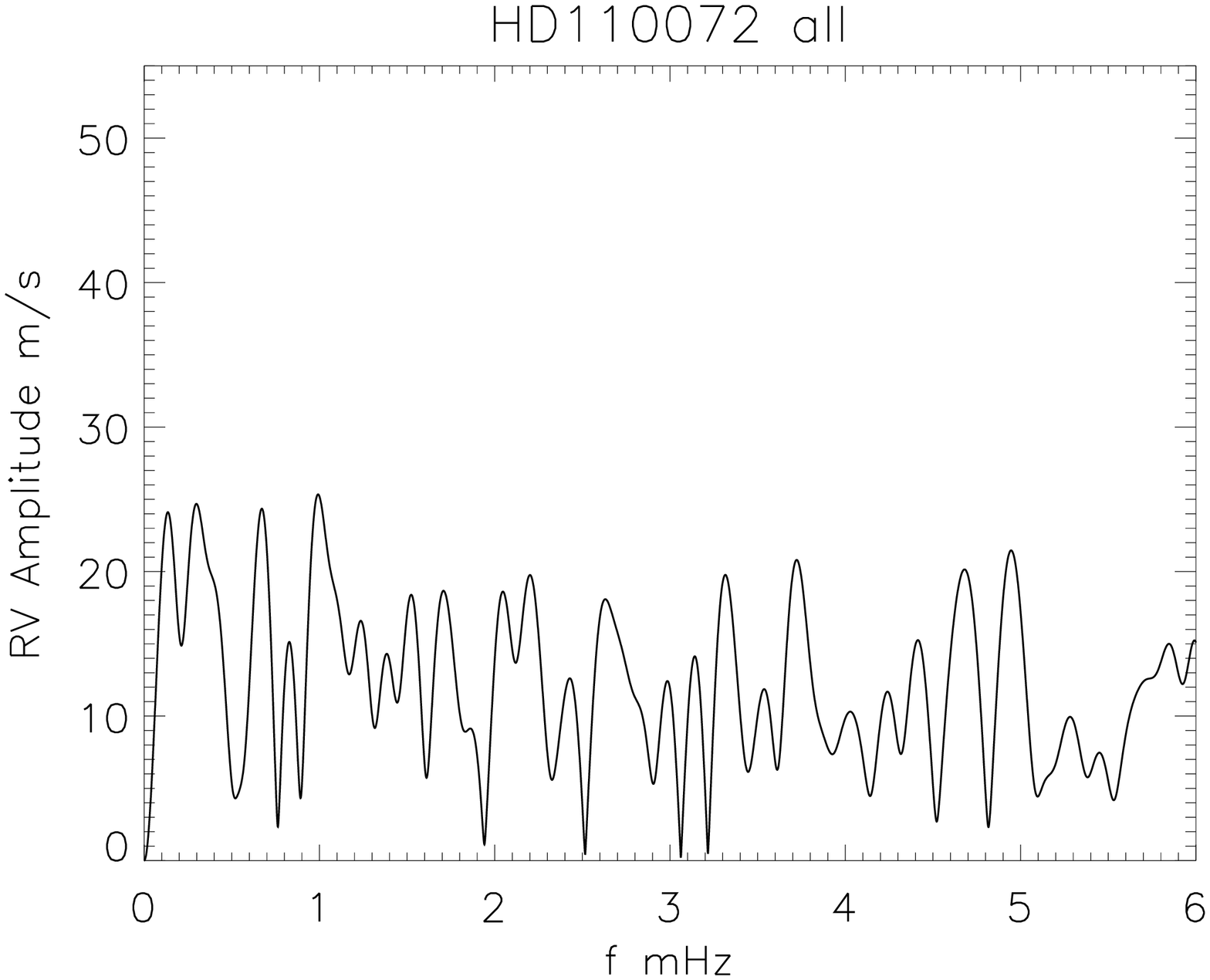}
  \includegraphics[height=5.6cm, angle=270]{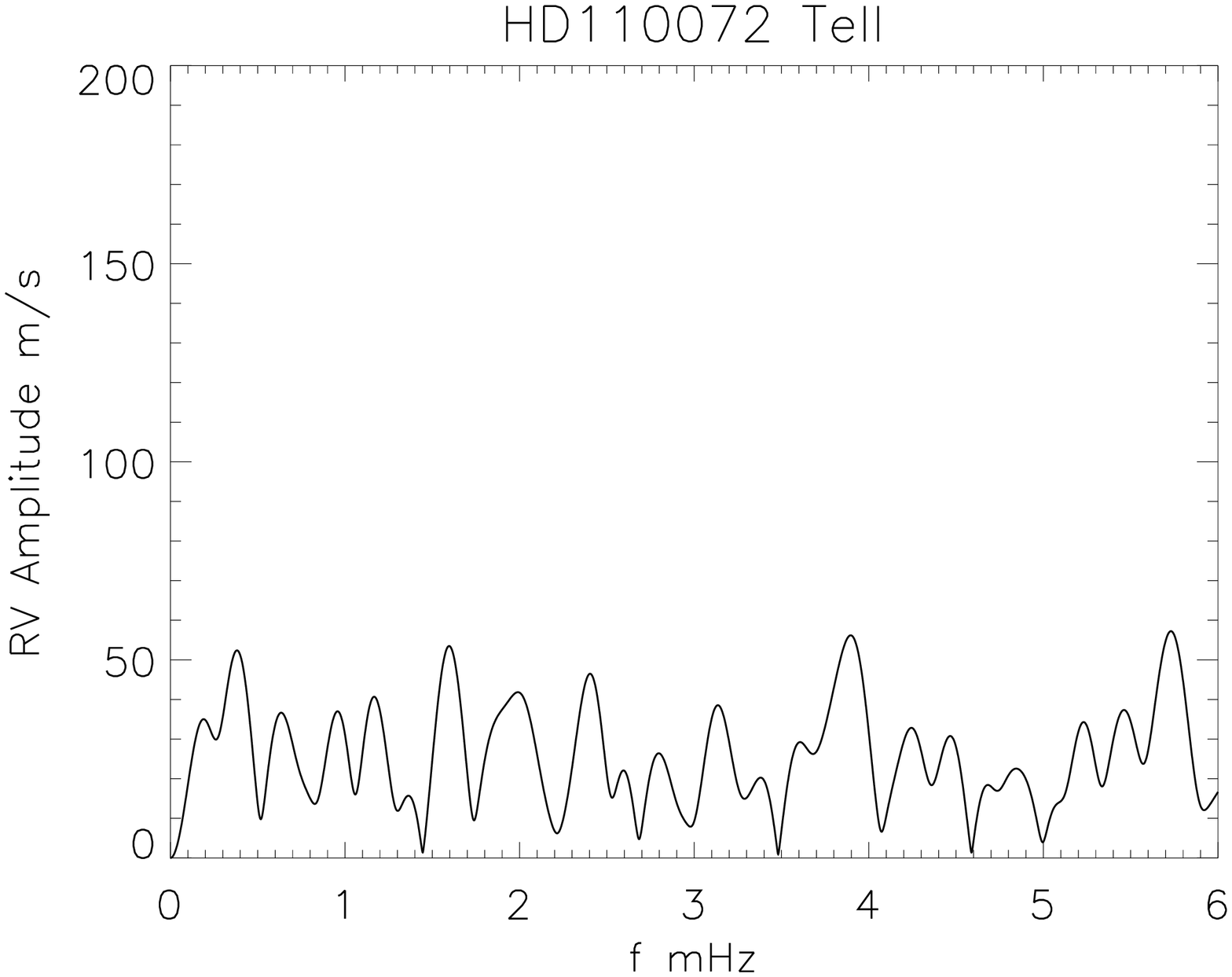}
  \caption{\label{fig:110072cog}Same as Fig.\,\ref{fig:107107cog} but
    for HD\,110072.  The Nyquist frequency is 4.6\,mHz.}
\end{figure*}

\begin{figure*}
  \vspace{3pt}
  \includegraphics[height=5.6cm,
  angle=270]{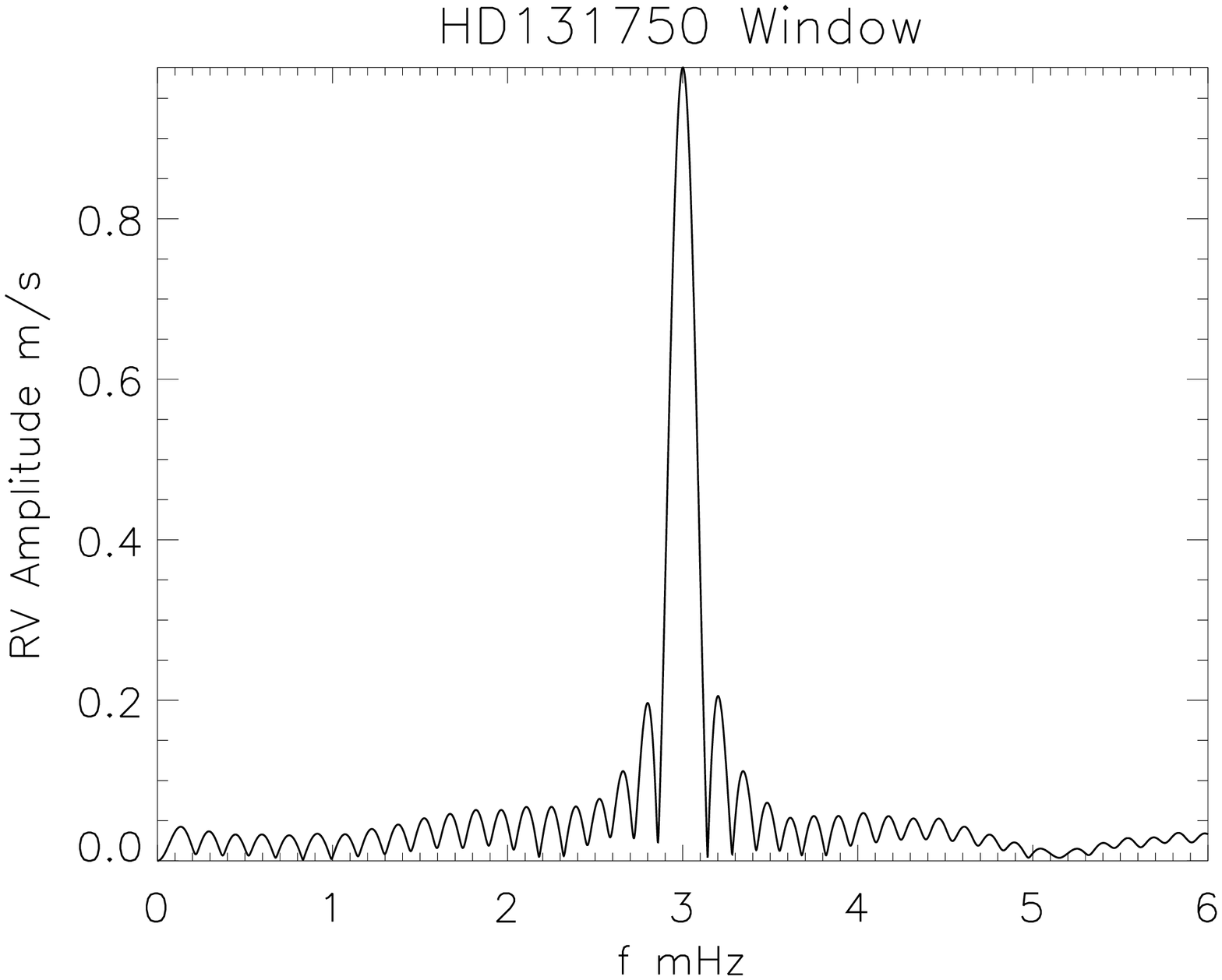}
  \includegraphics[height=5.6cm, angle=270]{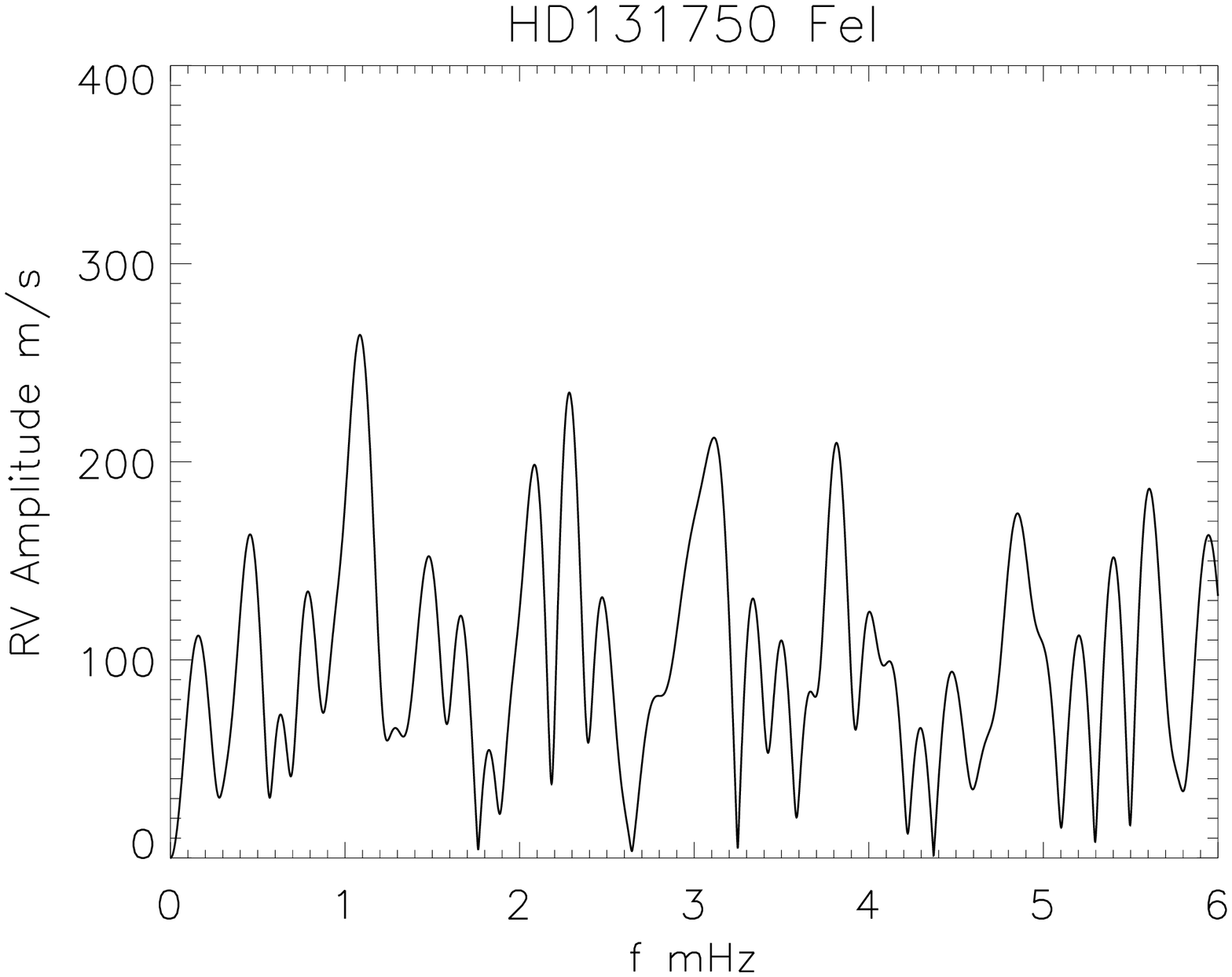}
  \includegraphics[height=5.6cm, angle=270]{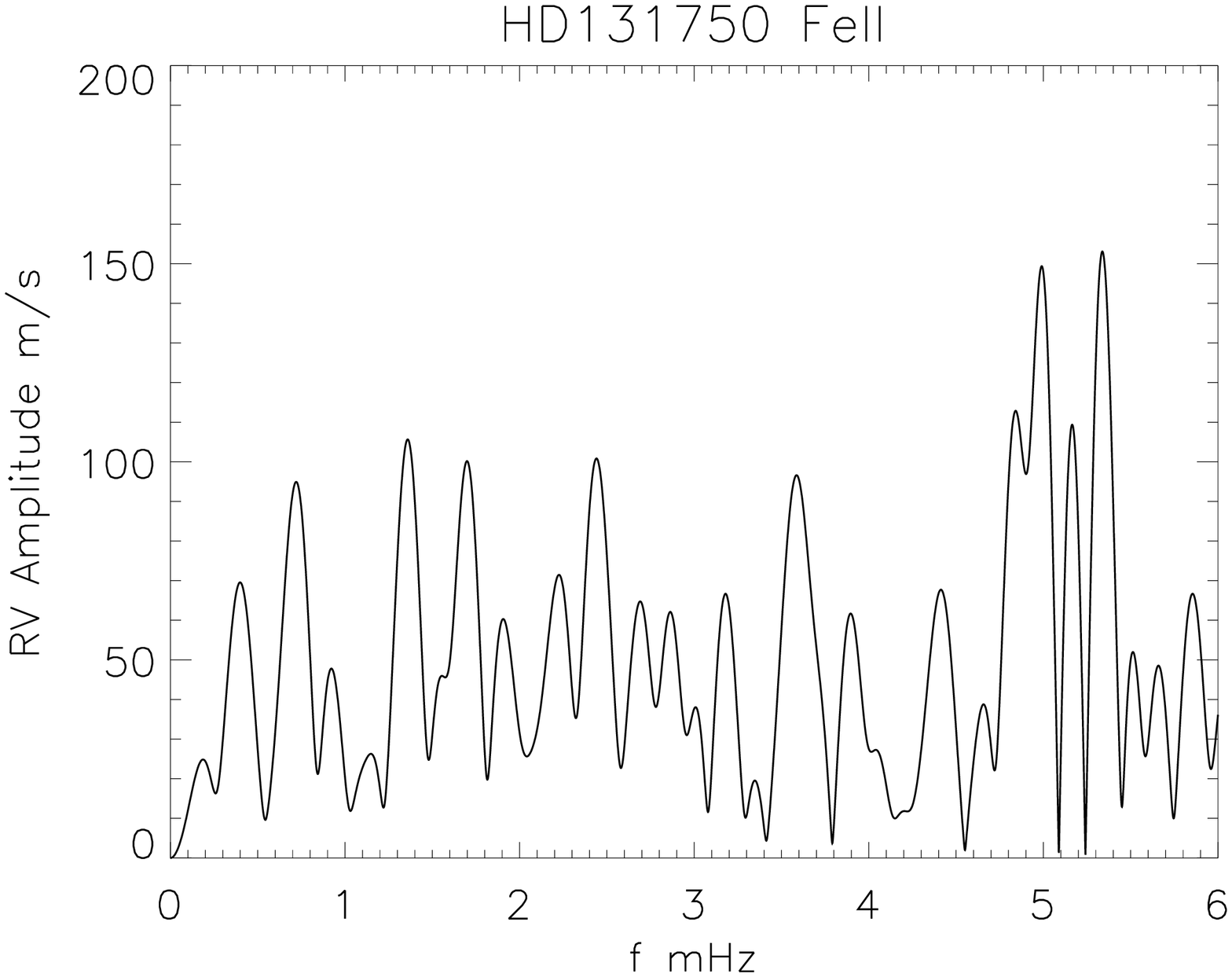}
  \includegraphics[height=5.6cm, angle=270]{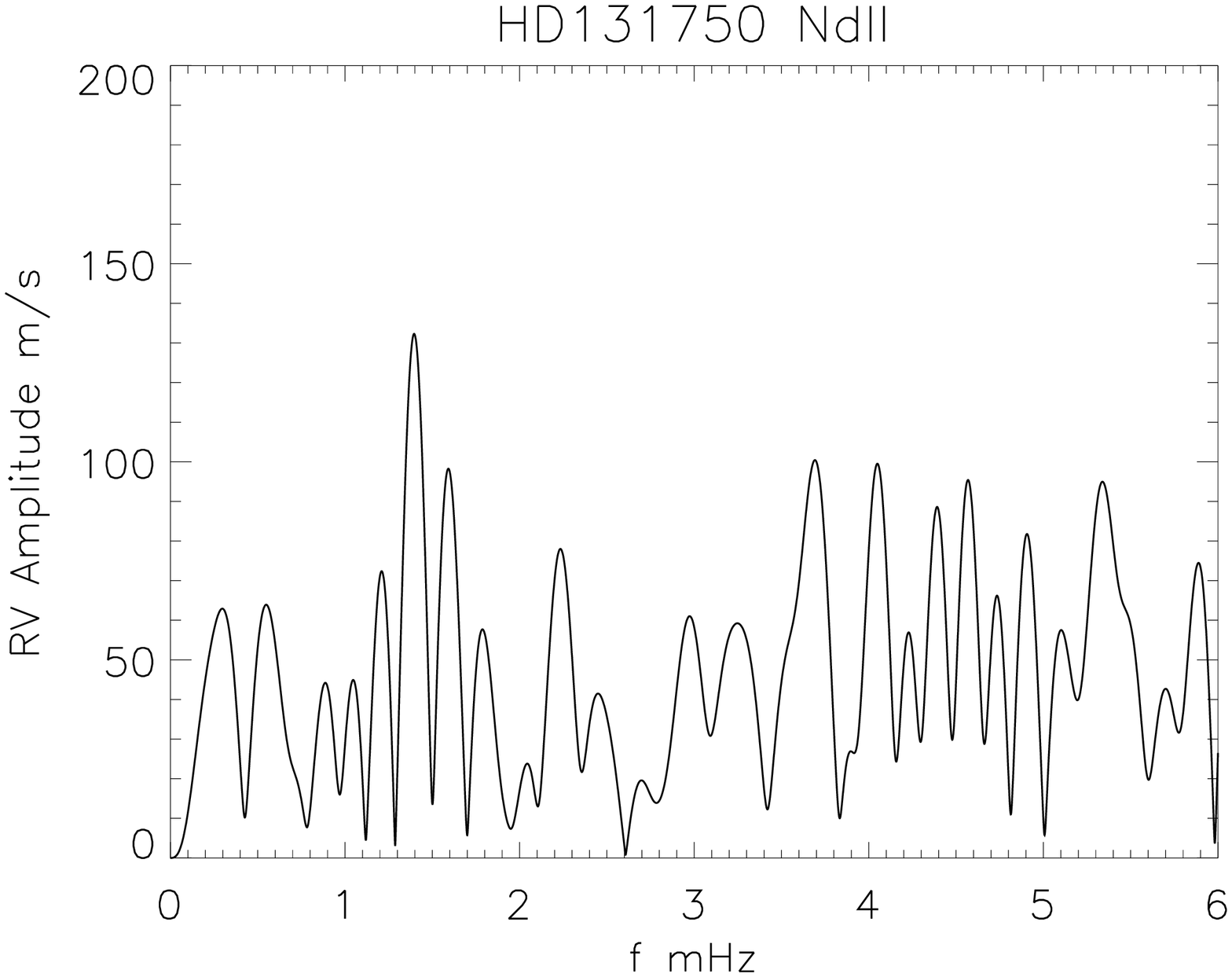}
  \includegraphics[height=5.6cm,
  angle=270]{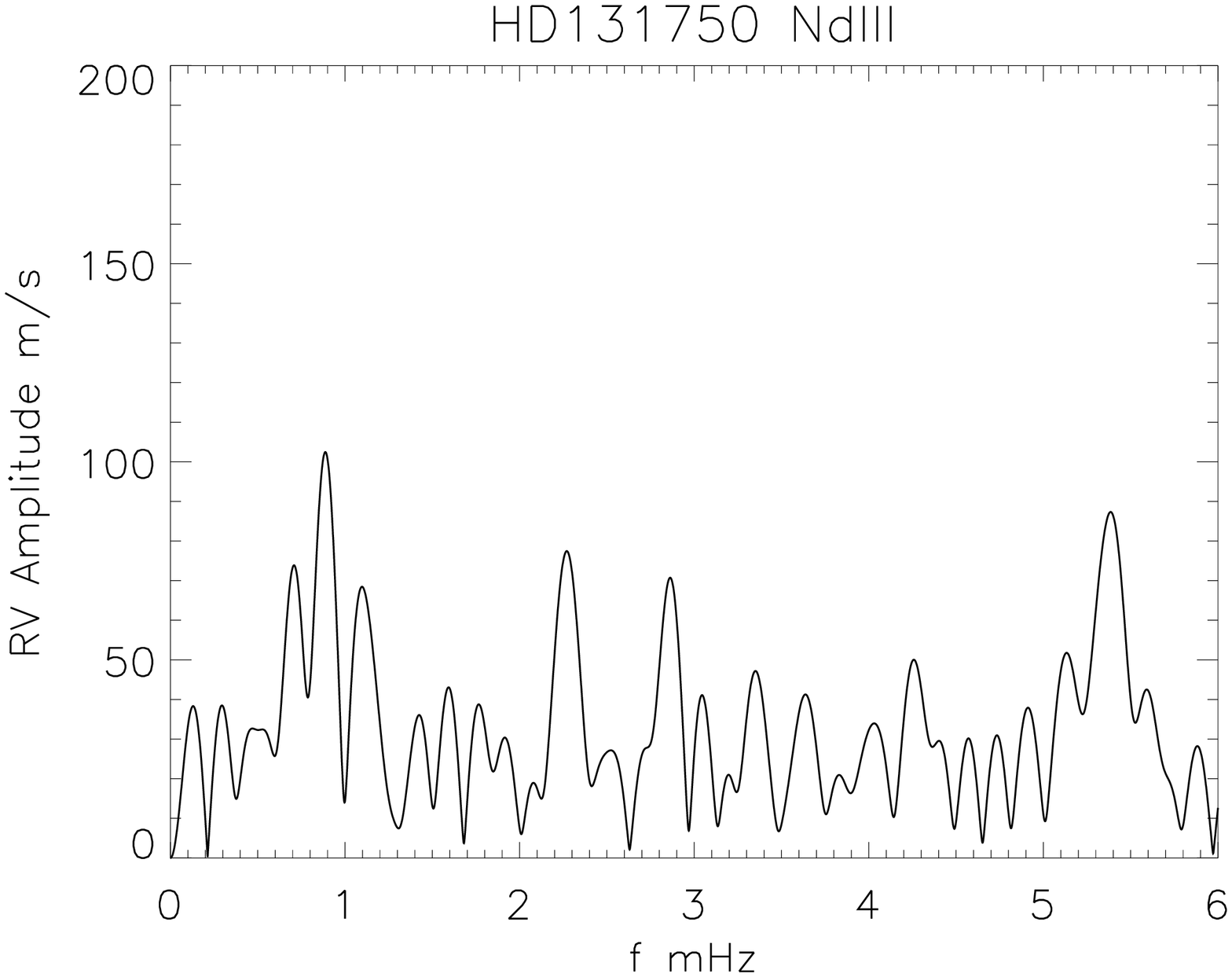}
  \includegraphics[height=5.6cm, angle=270]{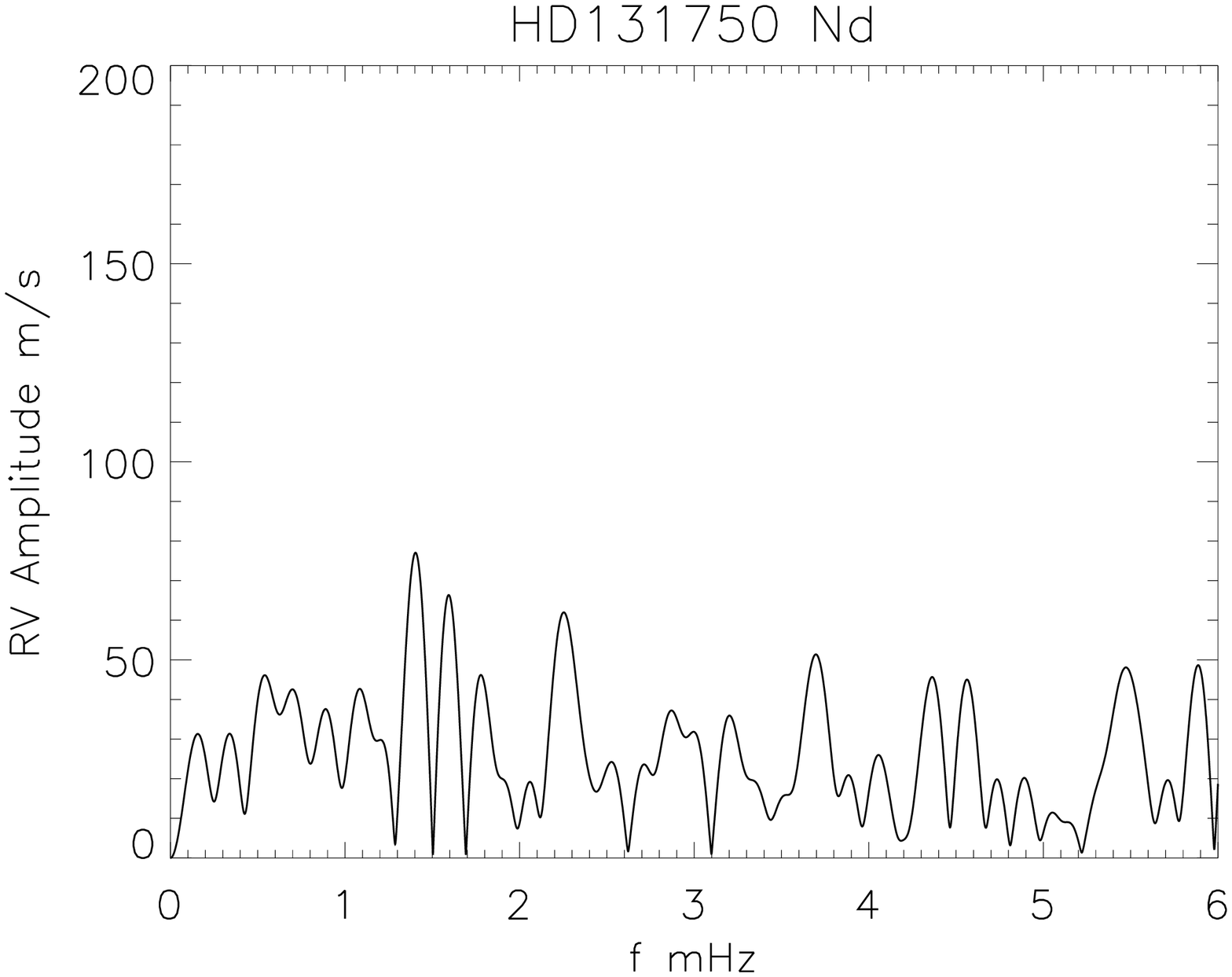}
  \includegraphics[height=5.6cm, angle=270]{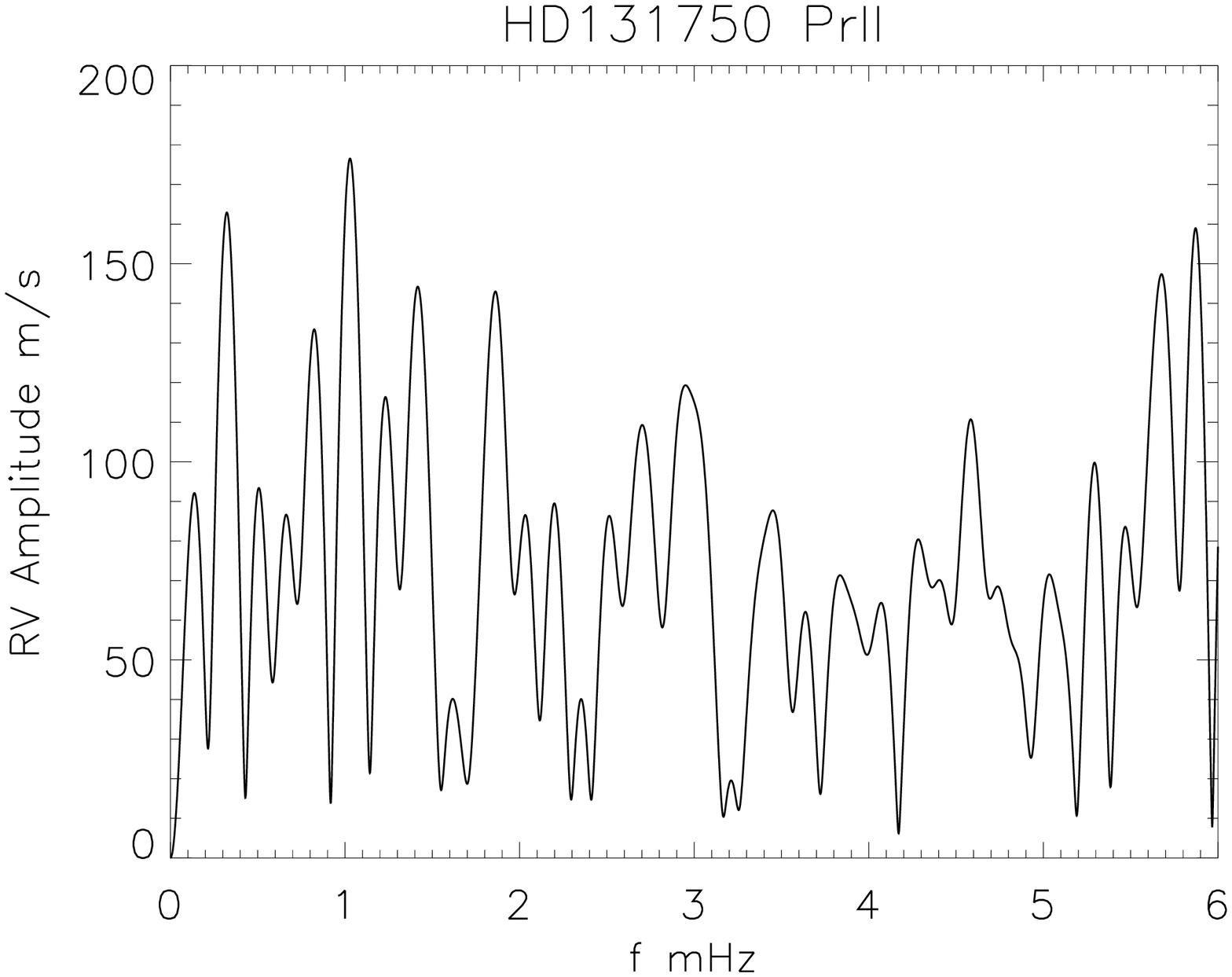}
  \includegraphics[height=5.6cm,
  angle=270]{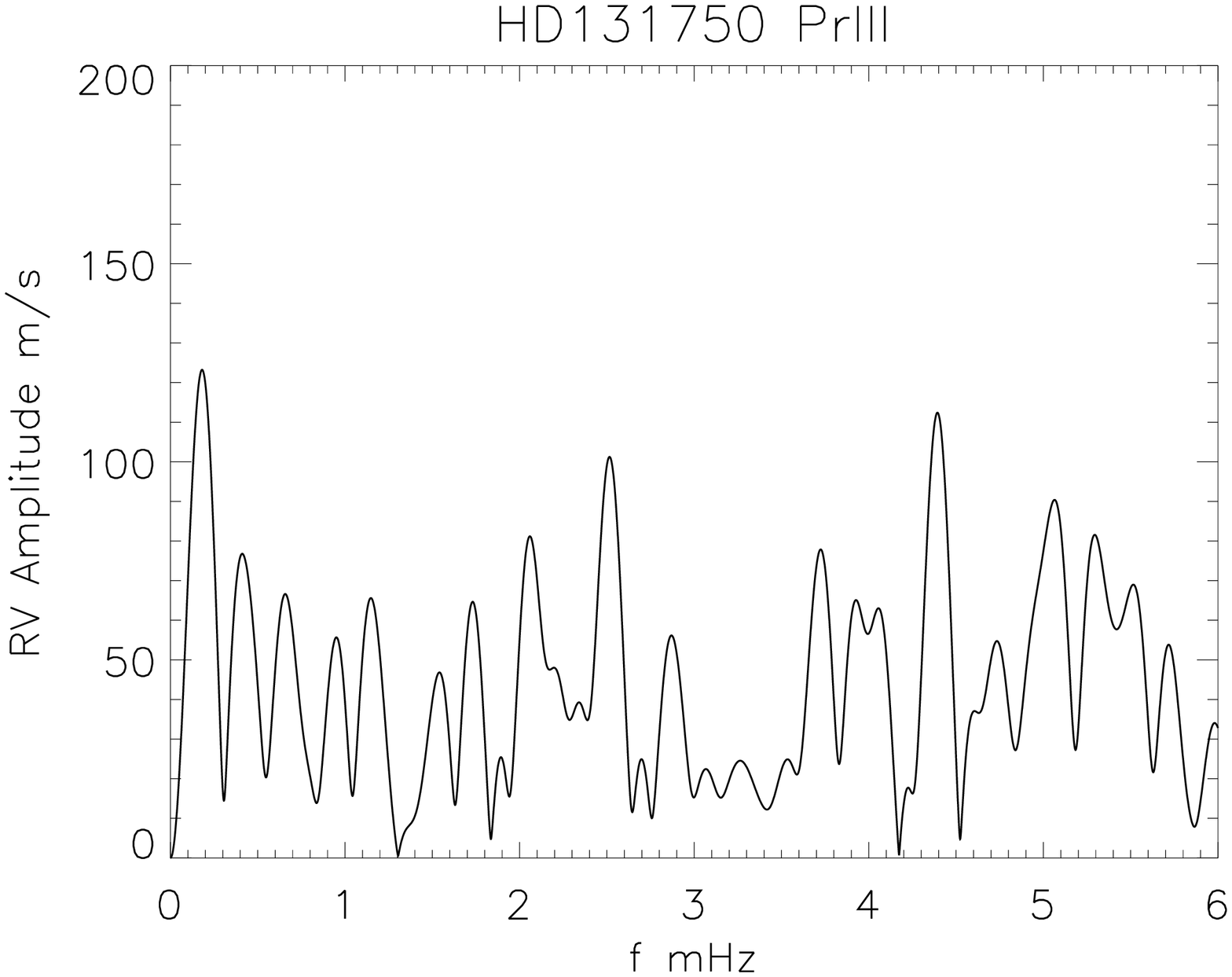}
  \includegraphics[height=5.6cm, angle=270]{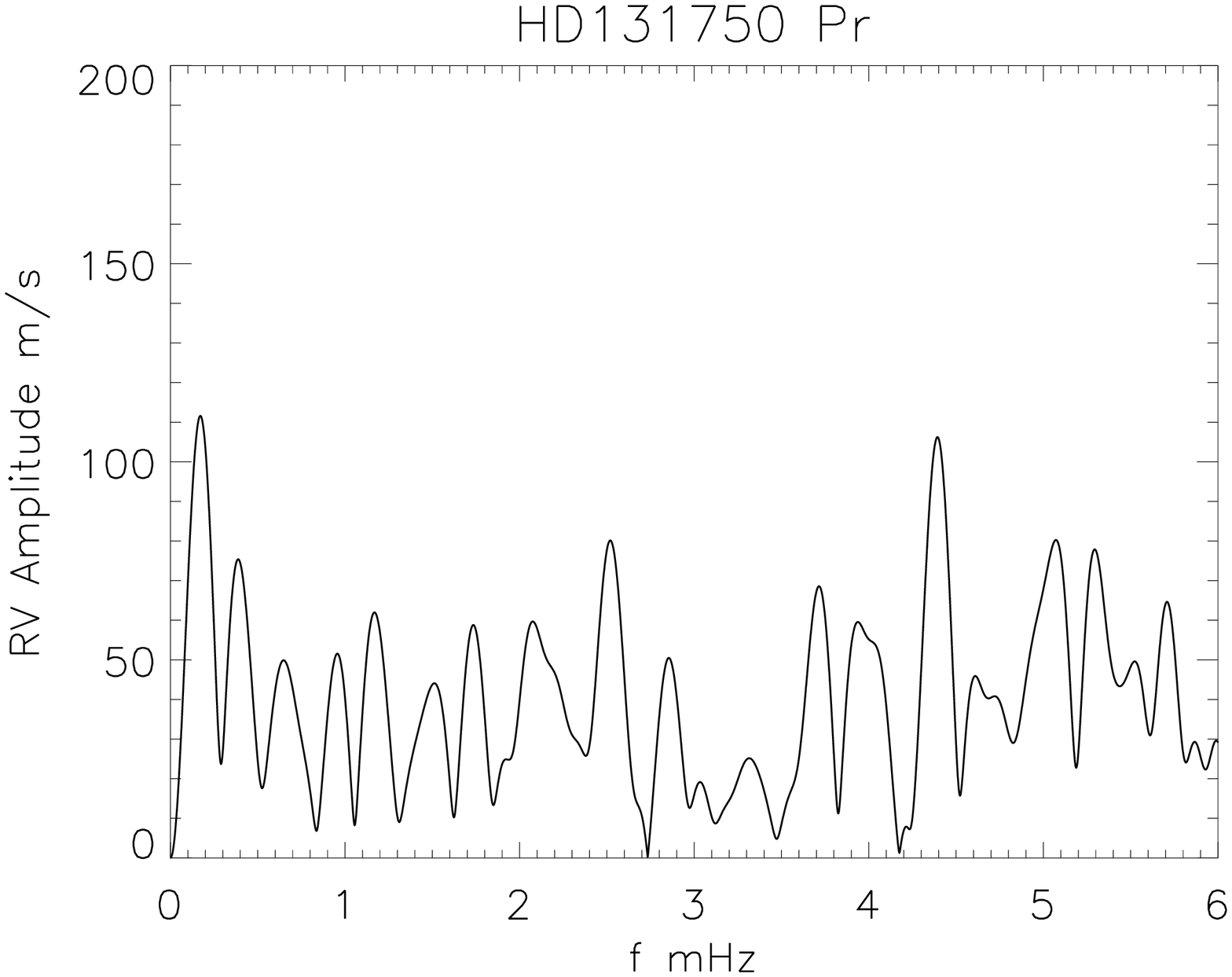}
  \includegraphics[height=5.6cm, angle=270]{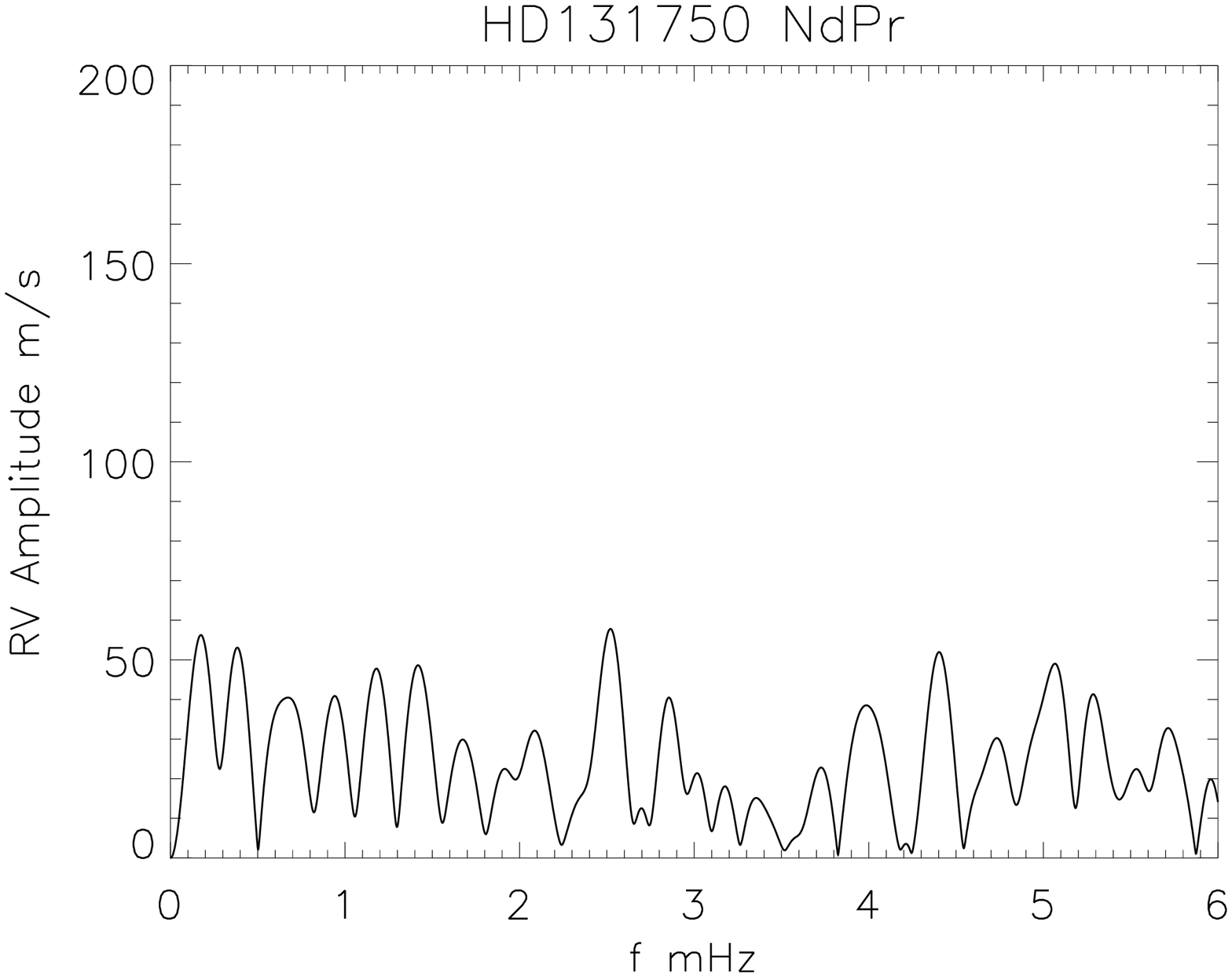}
  \includegraphics[height=5.6cm, angle=270]{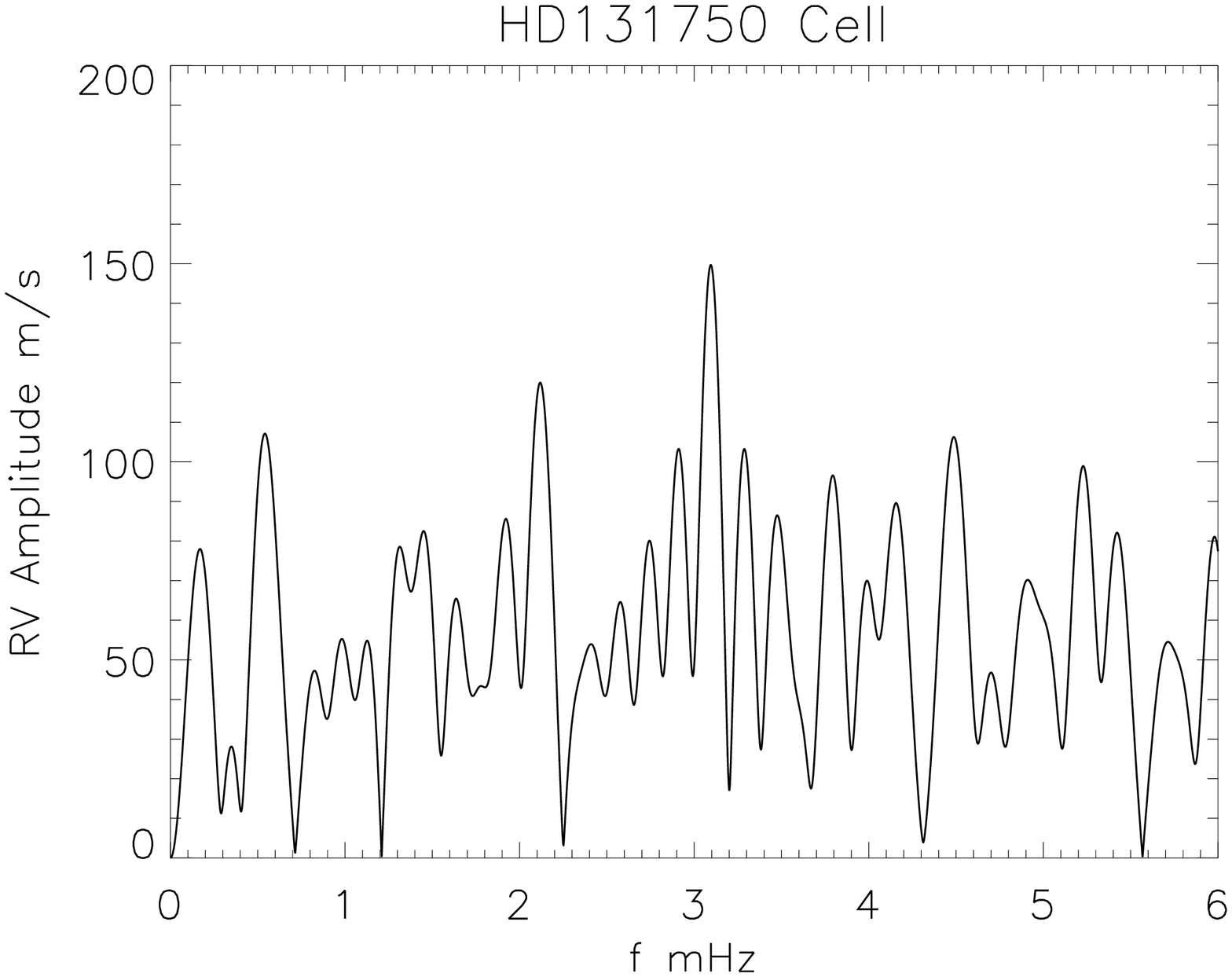}
  \includegraphics[height=5.6cm, angle=270]{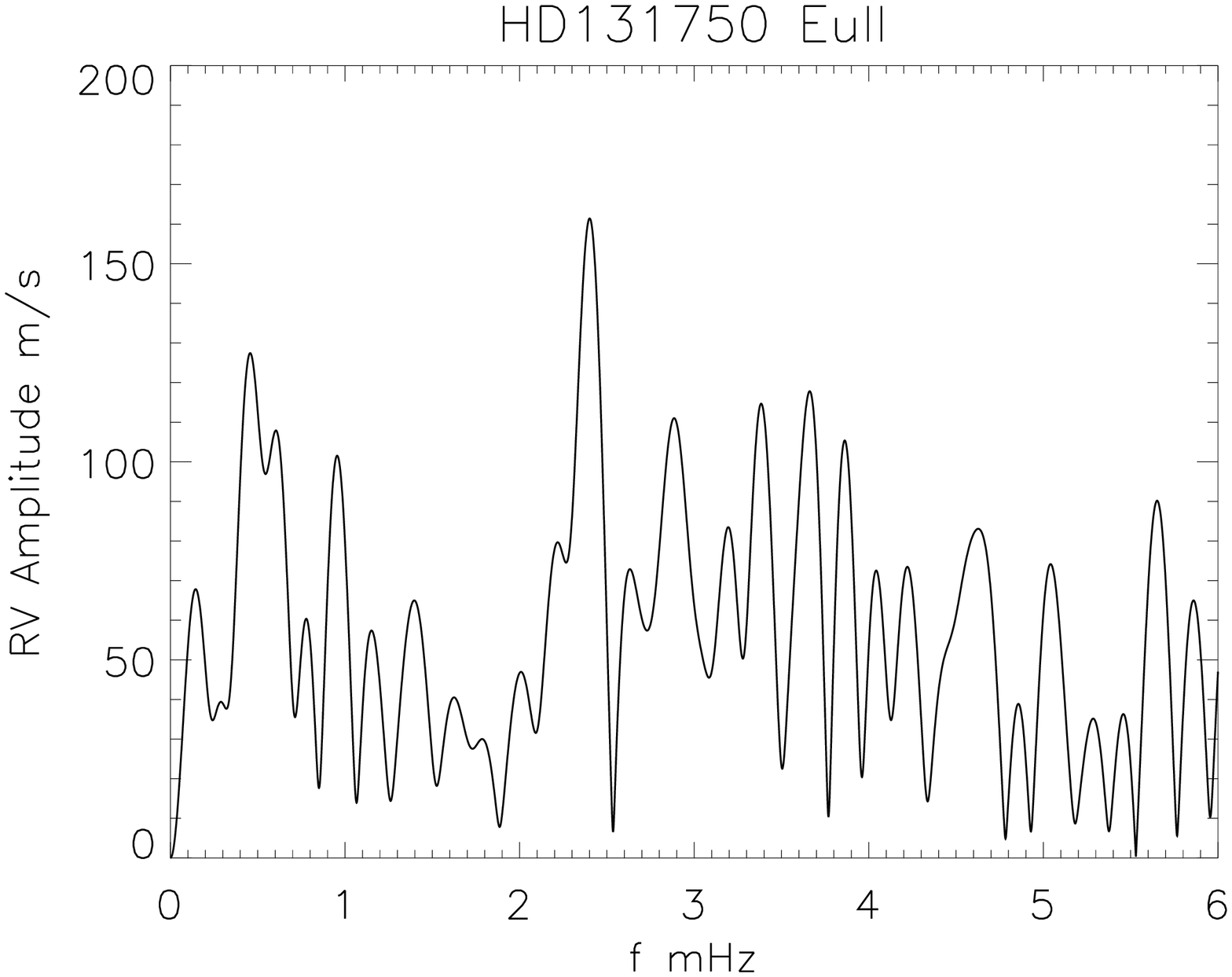}
  \includegraphics[height=5.6cm, angle=270]{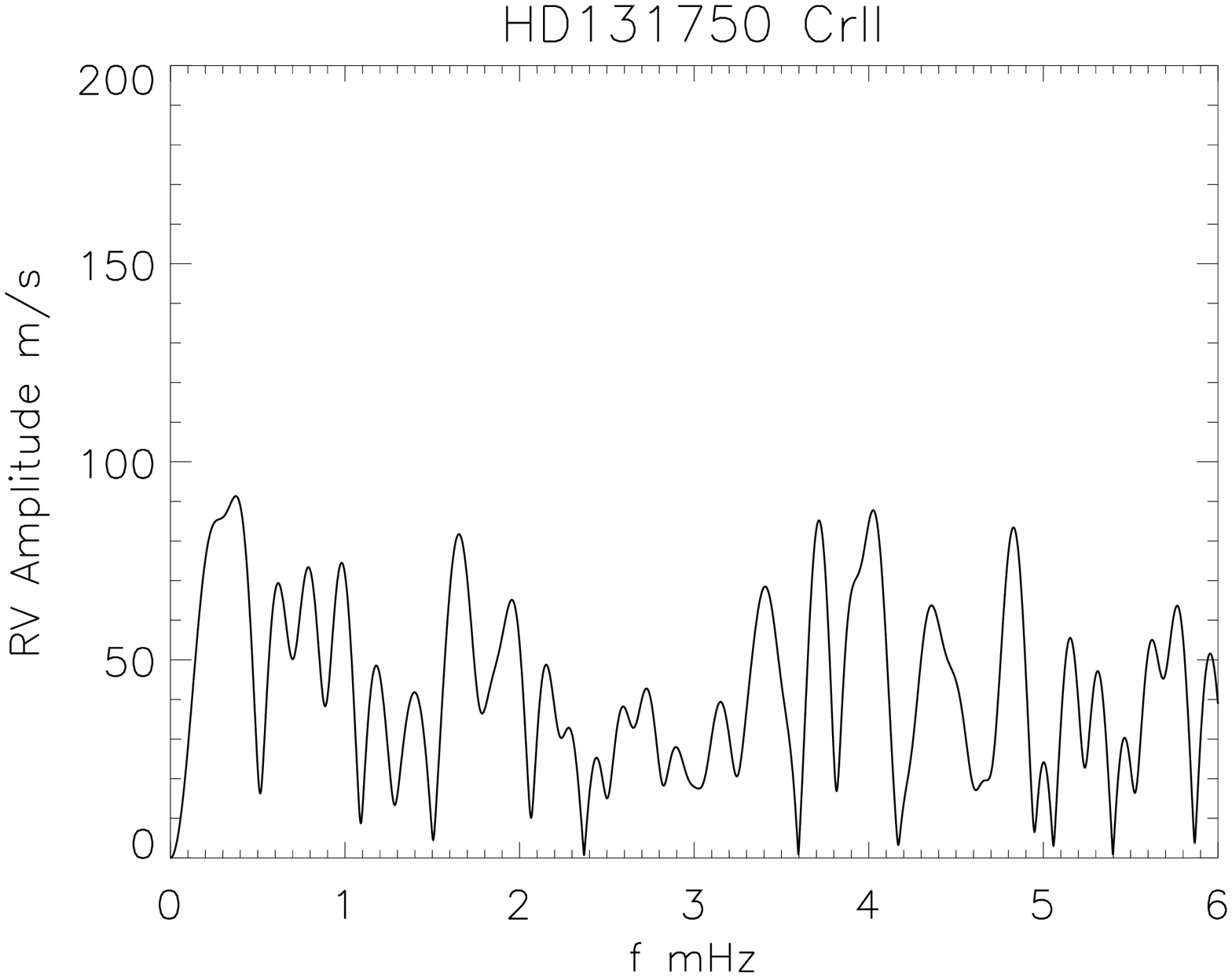}
  \includegraphics[height=5.6cm, angle=270]{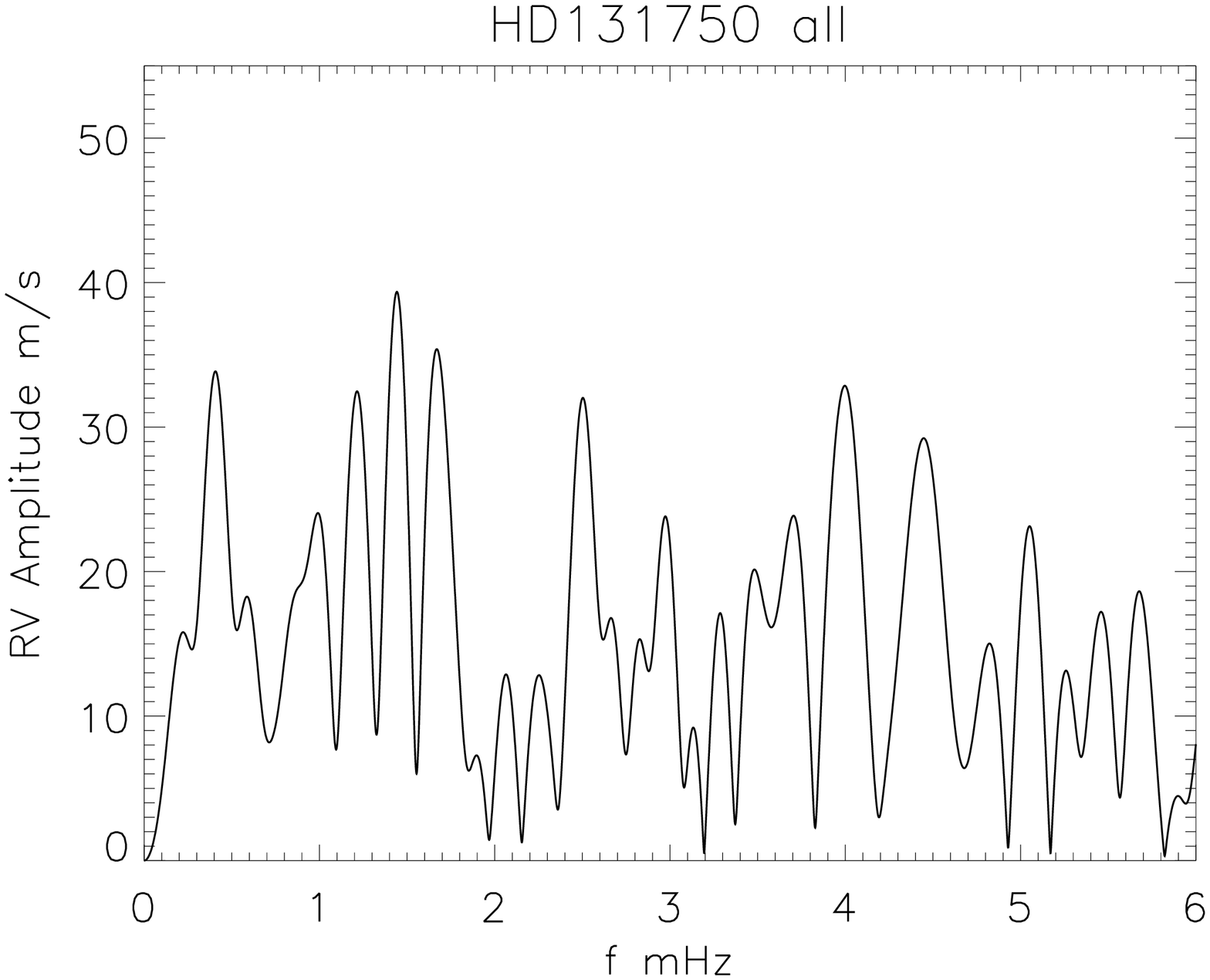}
  \includegraphics[height=5.6cm, angle=270]{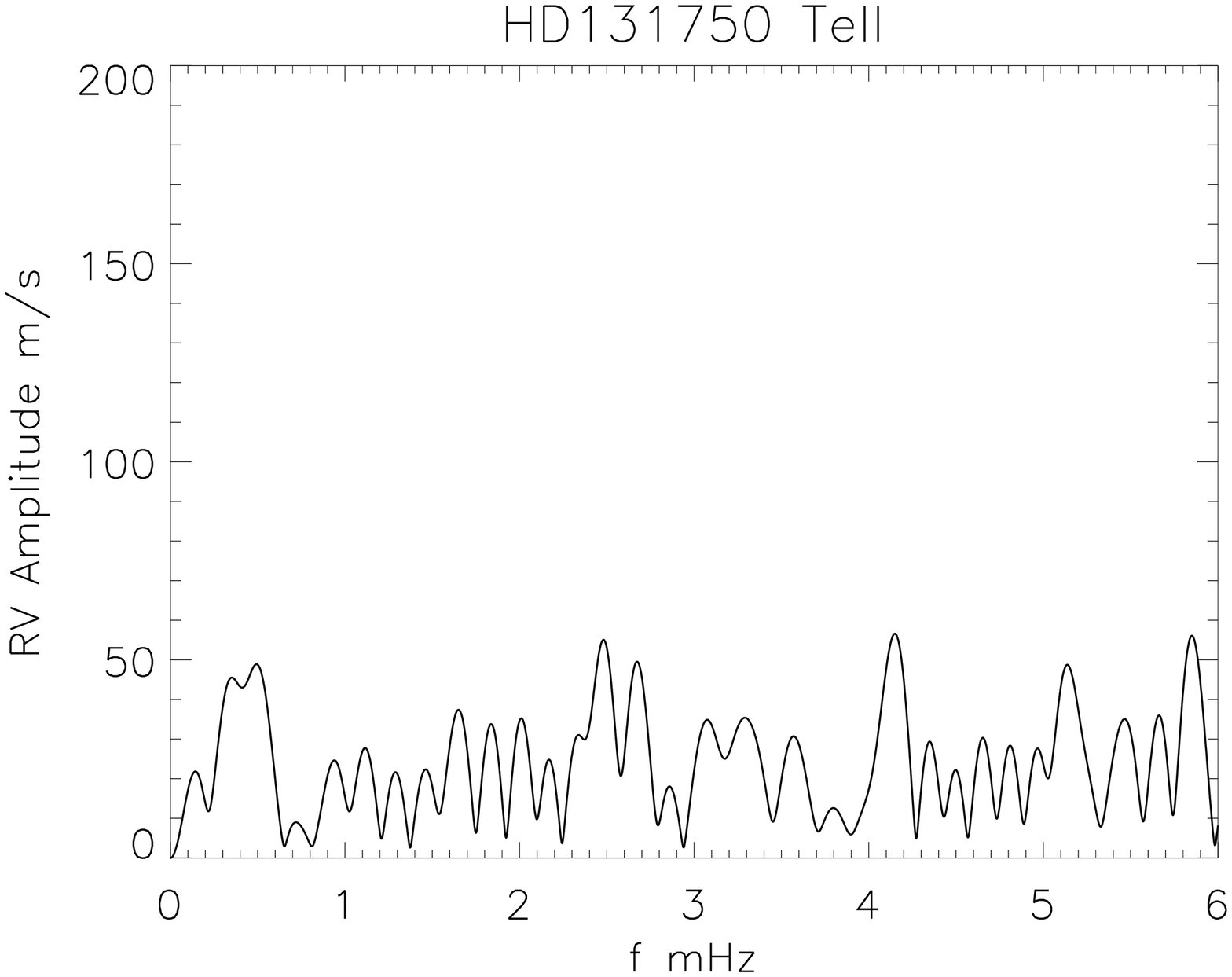}
  \caption{\label{fig:131750cog}Same as Fig.\,\ref{fig:107107cog} but
    for HD\,131750.  }
\end{figure*}

\begin{figure*}
  \vspace{3pt}
  \includegraphics[height=5.6cm,
  angle=270]{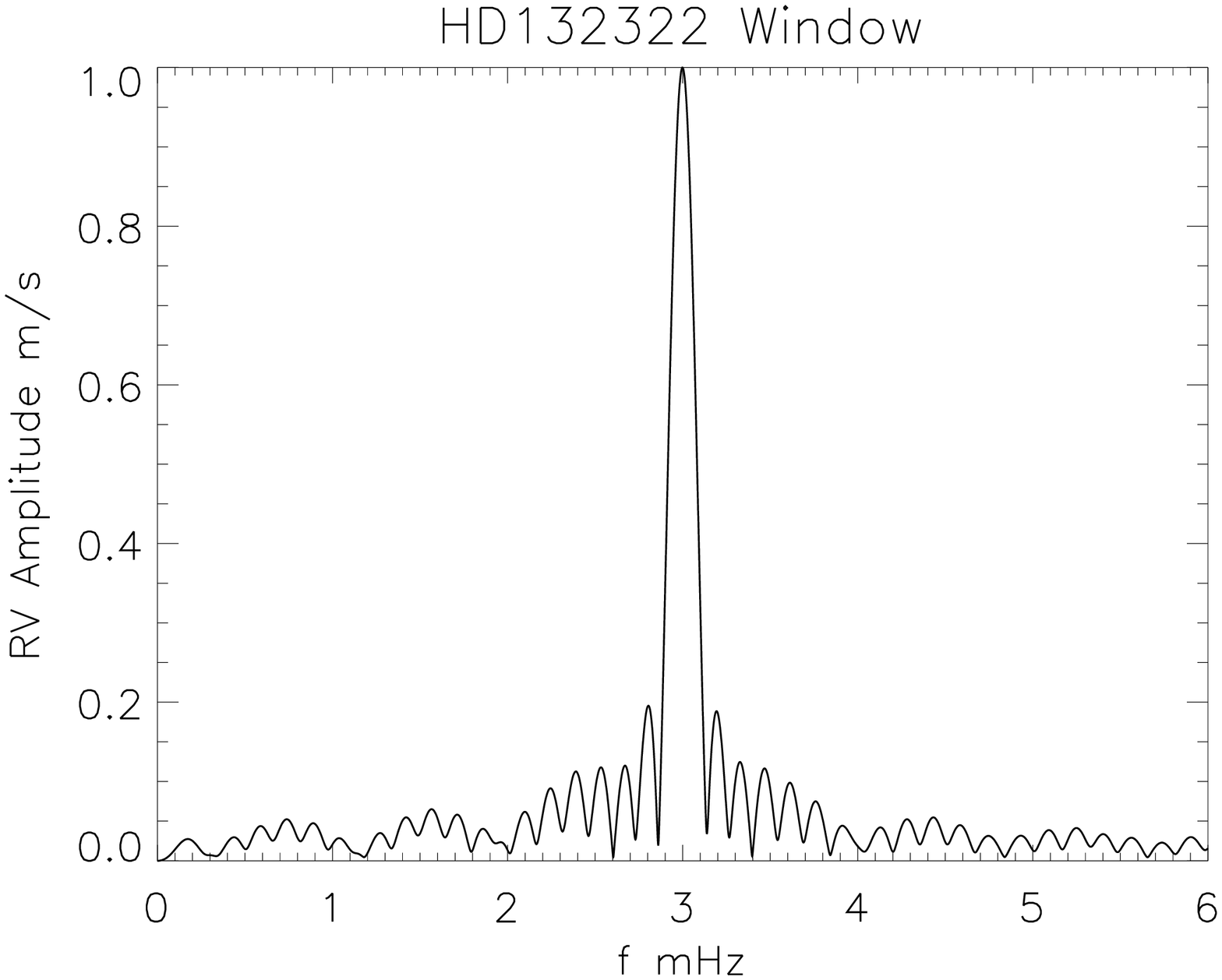}
  \includegraphics[height=5.6cm, angle=270]{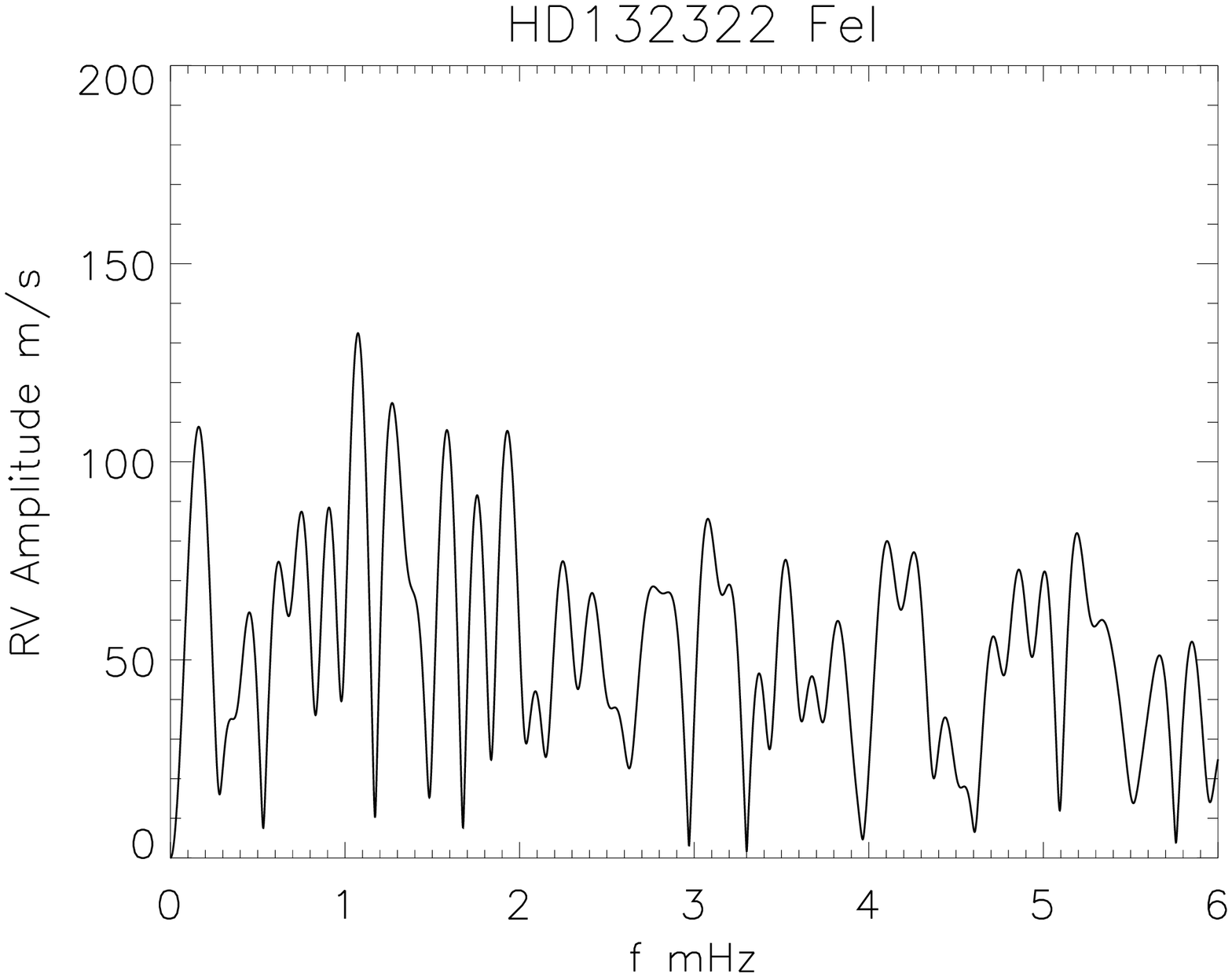}
  \includegraphics[height=5.6cm, angle=270]{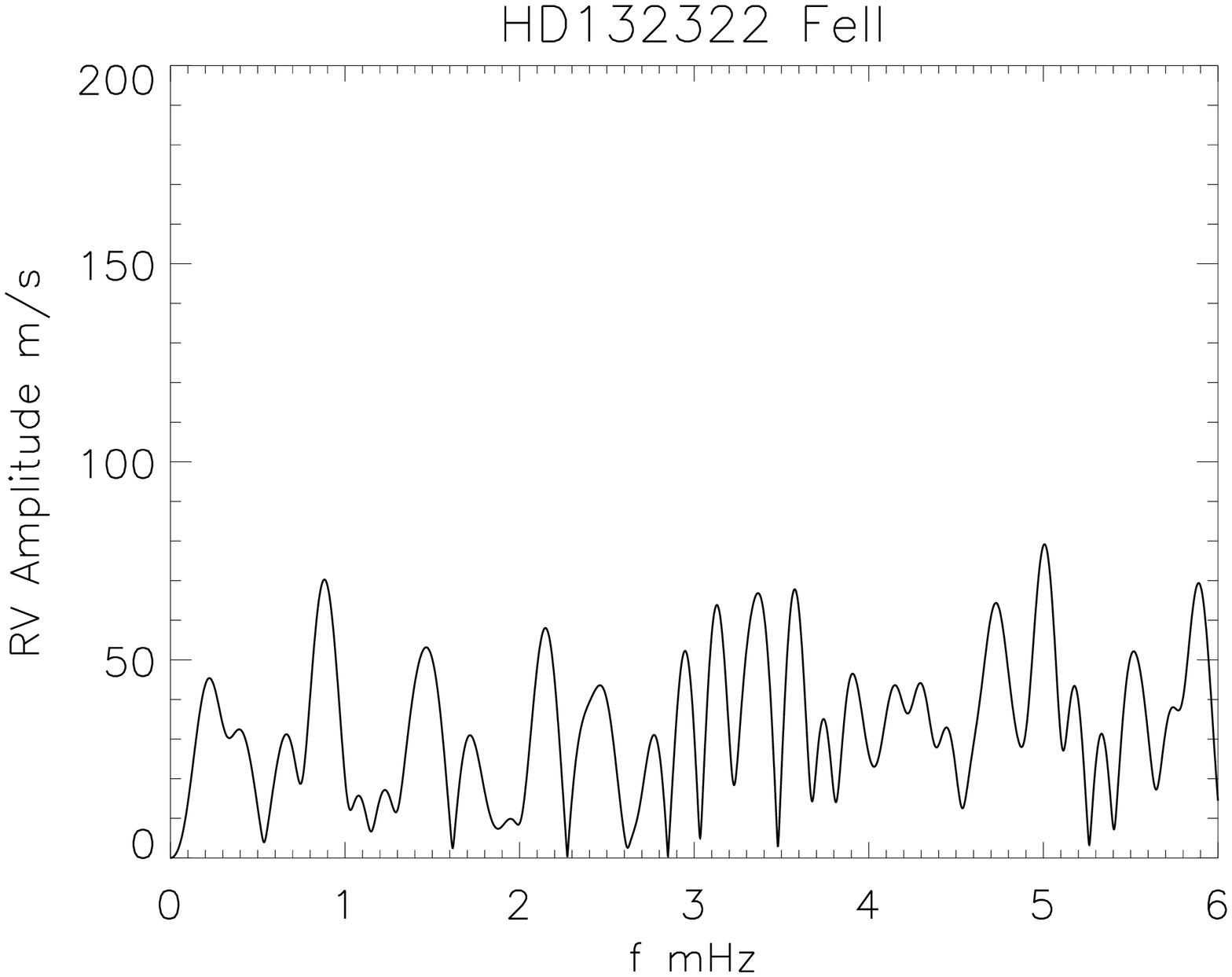}
  \includegraphics[height=5.6cm, angle=270]{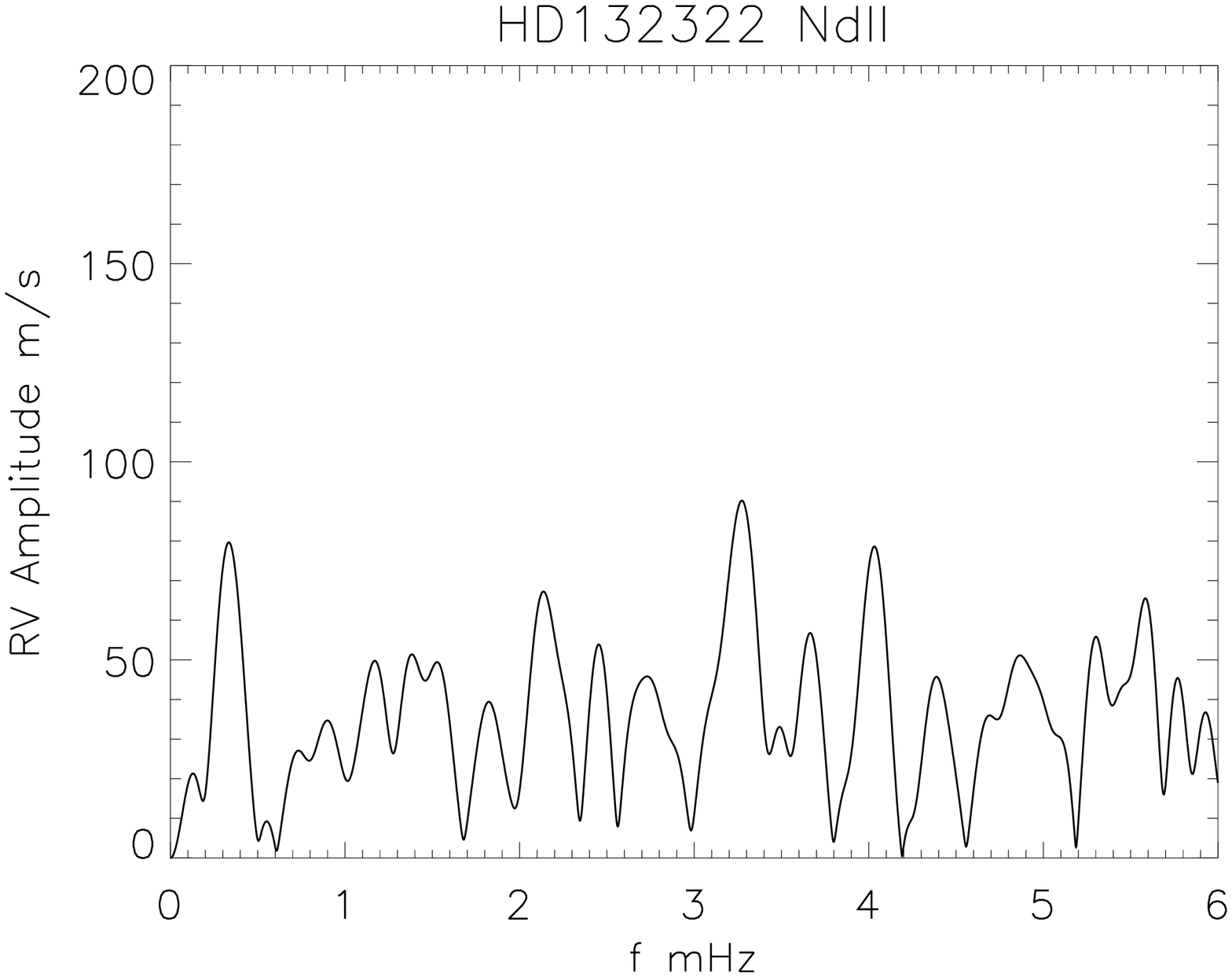}
  \includegraphics[height=5.6cm,
  angle=270]{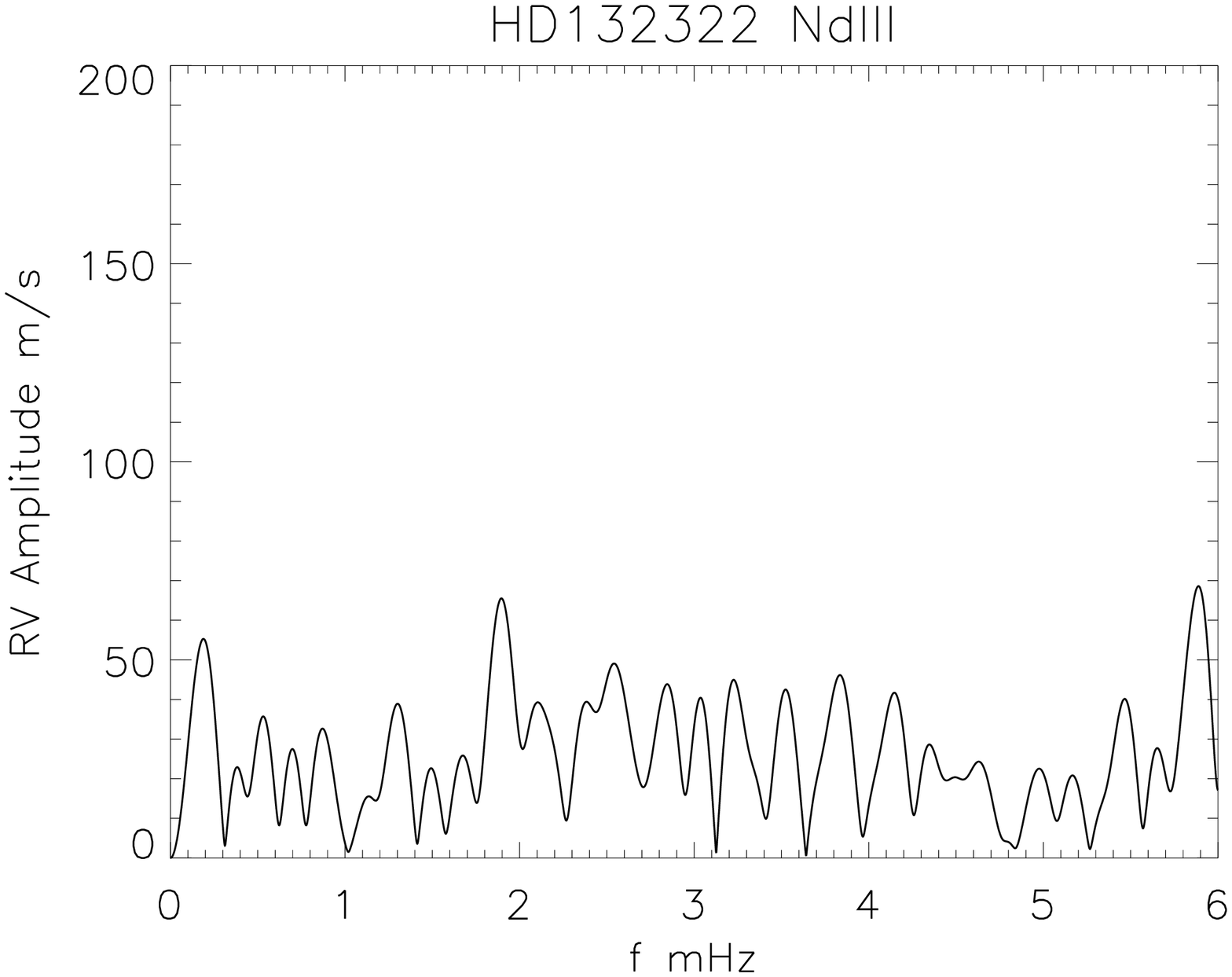}
  \includegraphics[height=5.6cm, angle=270]{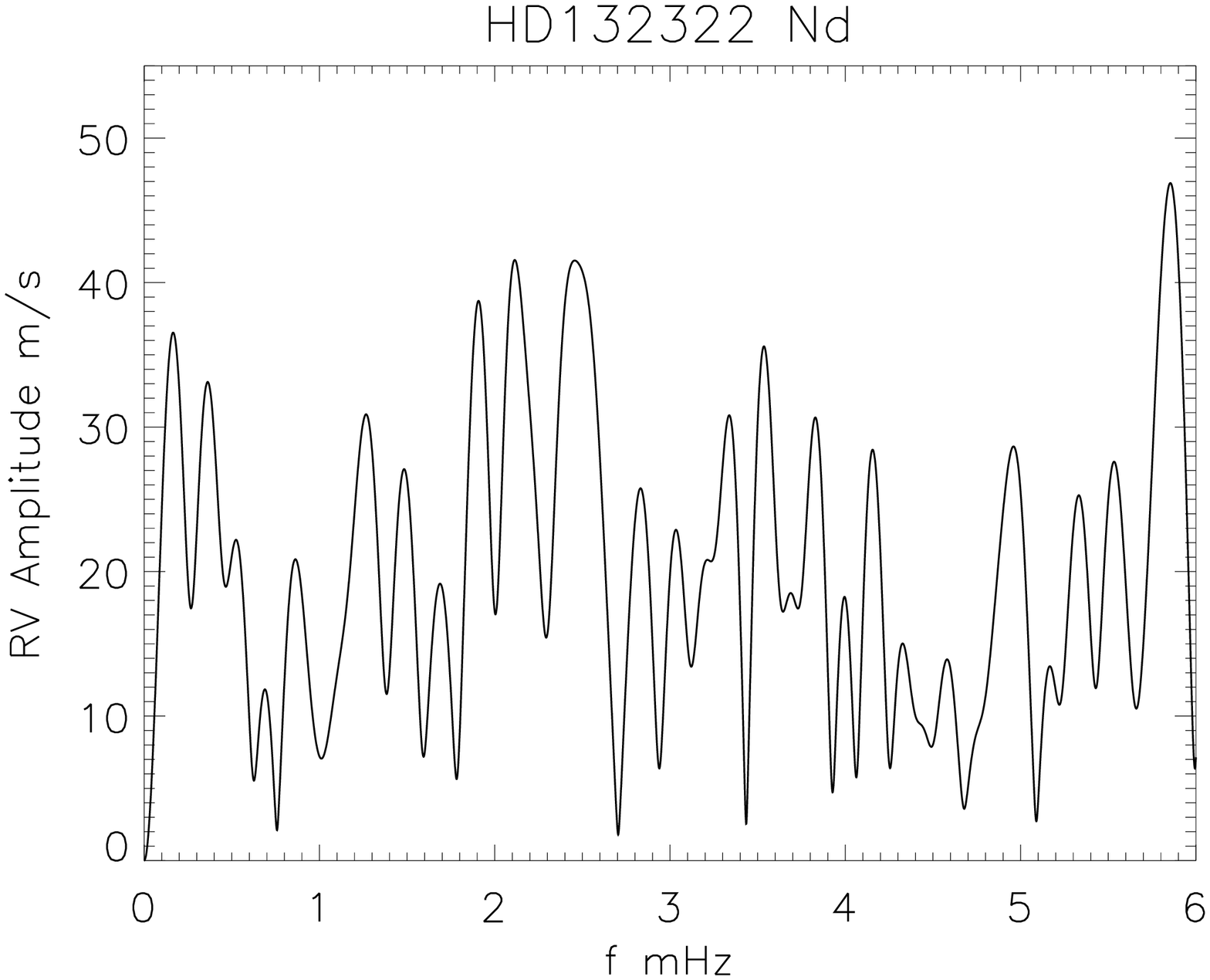}
  \includegraphics[height=5.6cm, angle=270]{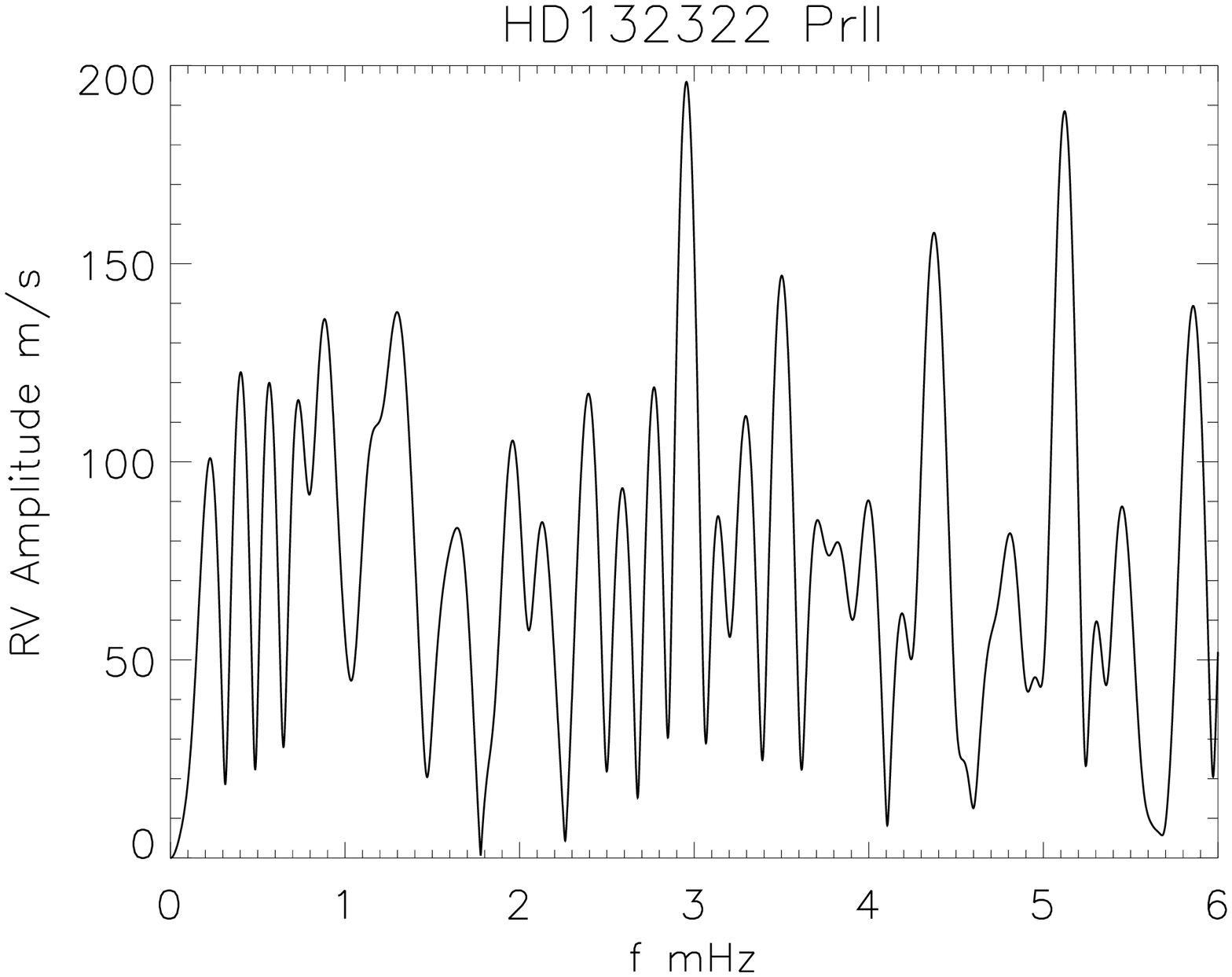}
  \includegraphics[height=5.6cm,
  angle=270]{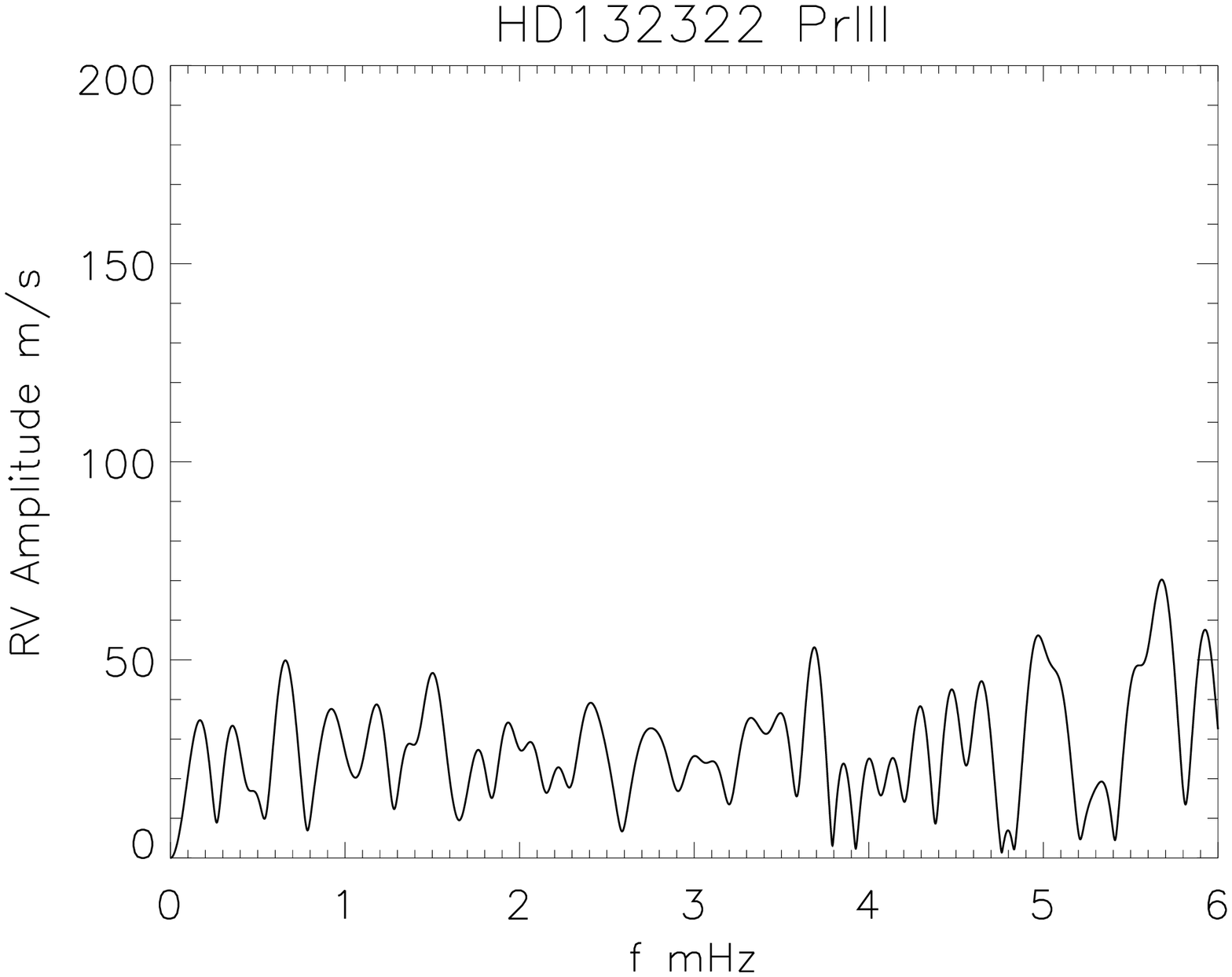}
  \includegraphics[height=5.6cm, angle=270]{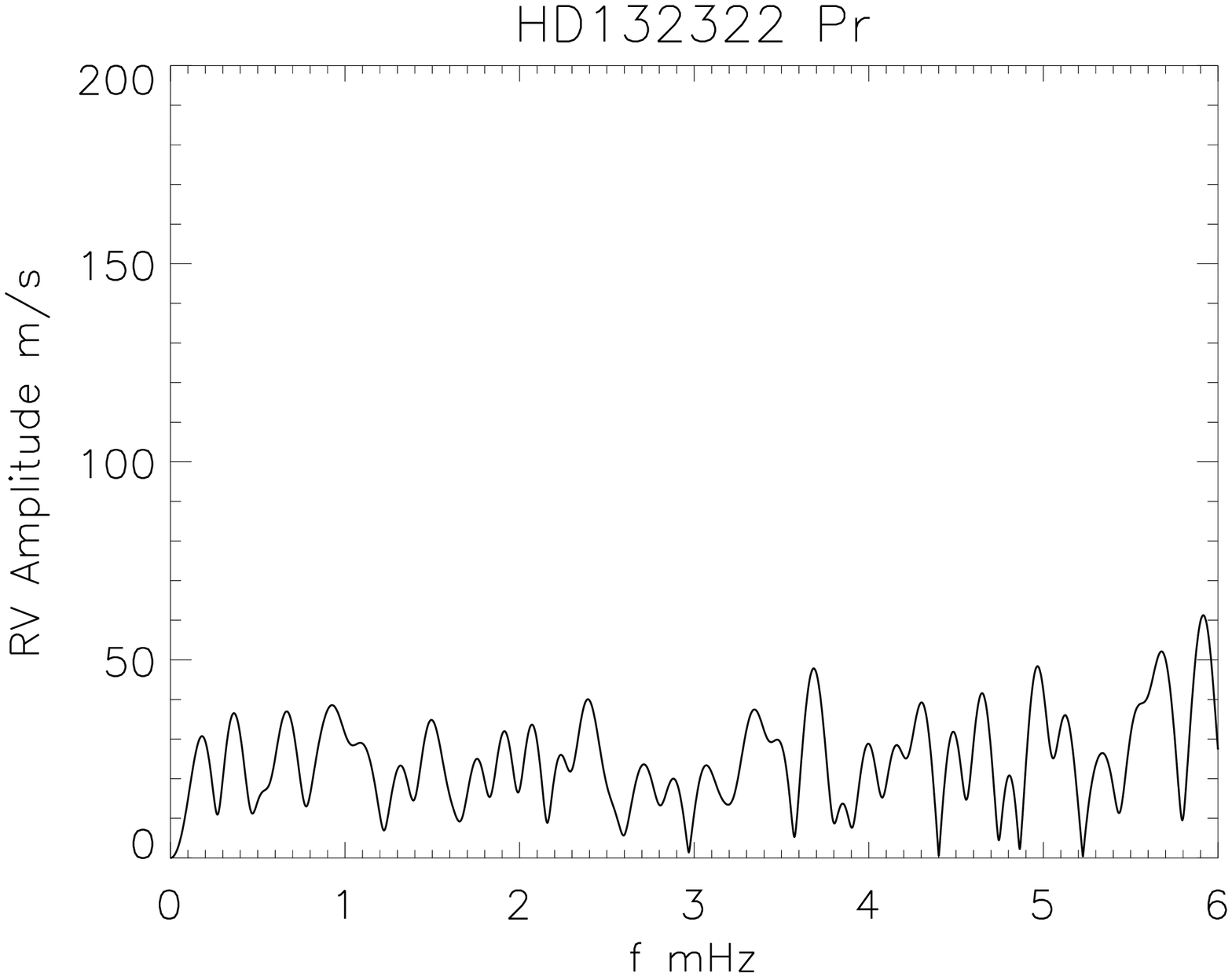}
  \includegraphics[height=5.6cm, angle=270]{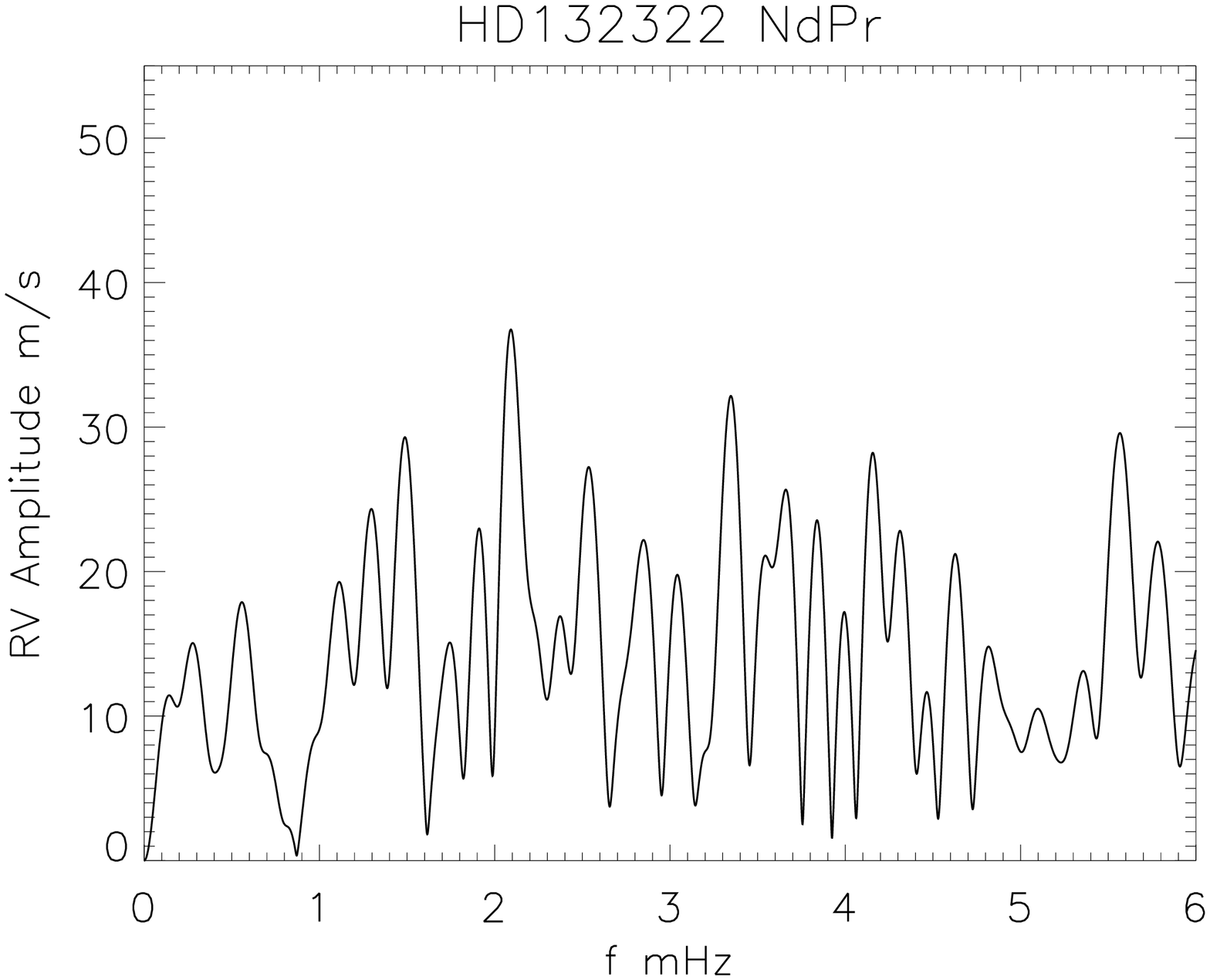}
  \includegraphics[height=5.6cm, angle=270]{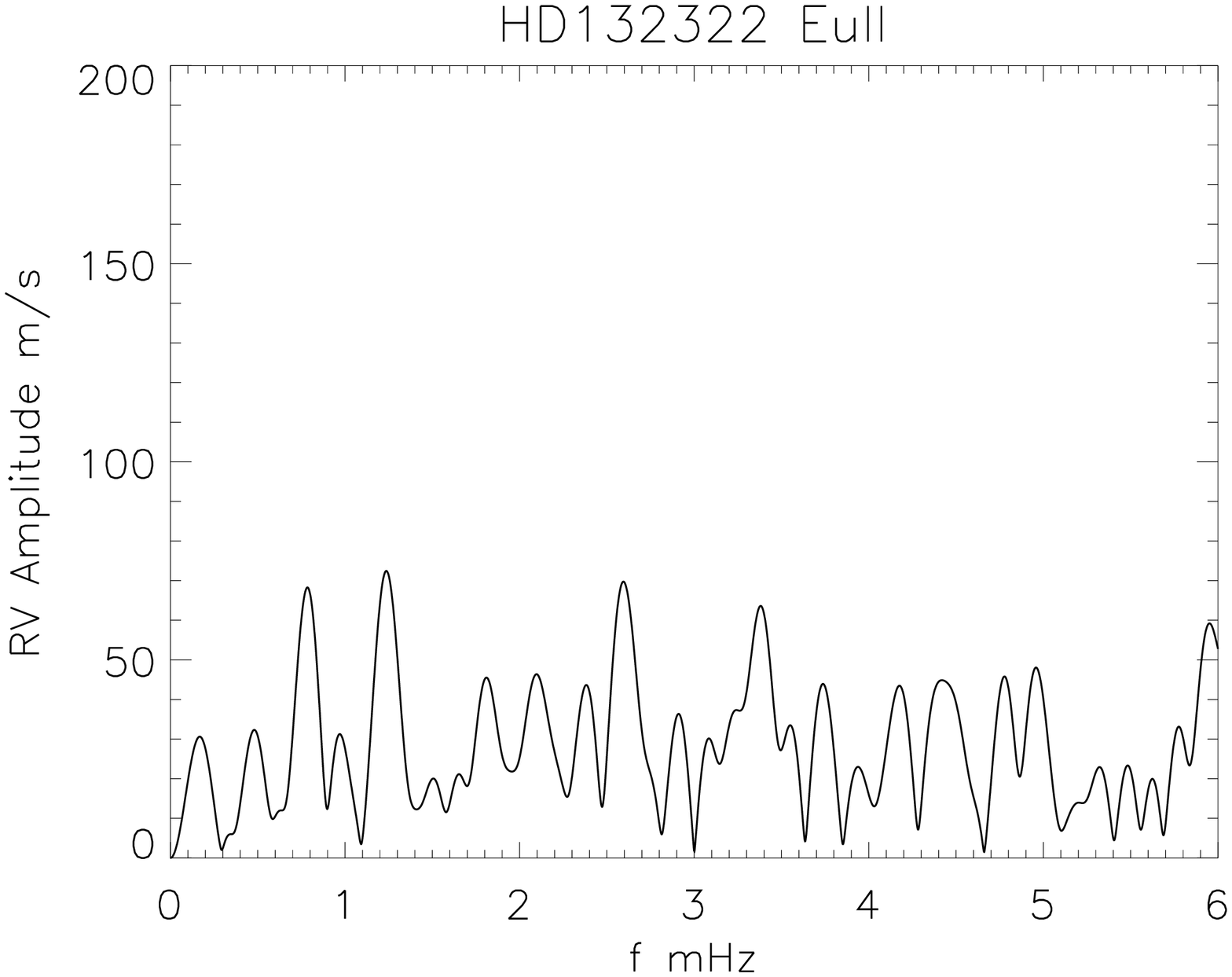}
  \includegraphics[height=5.6cm, angle=270]{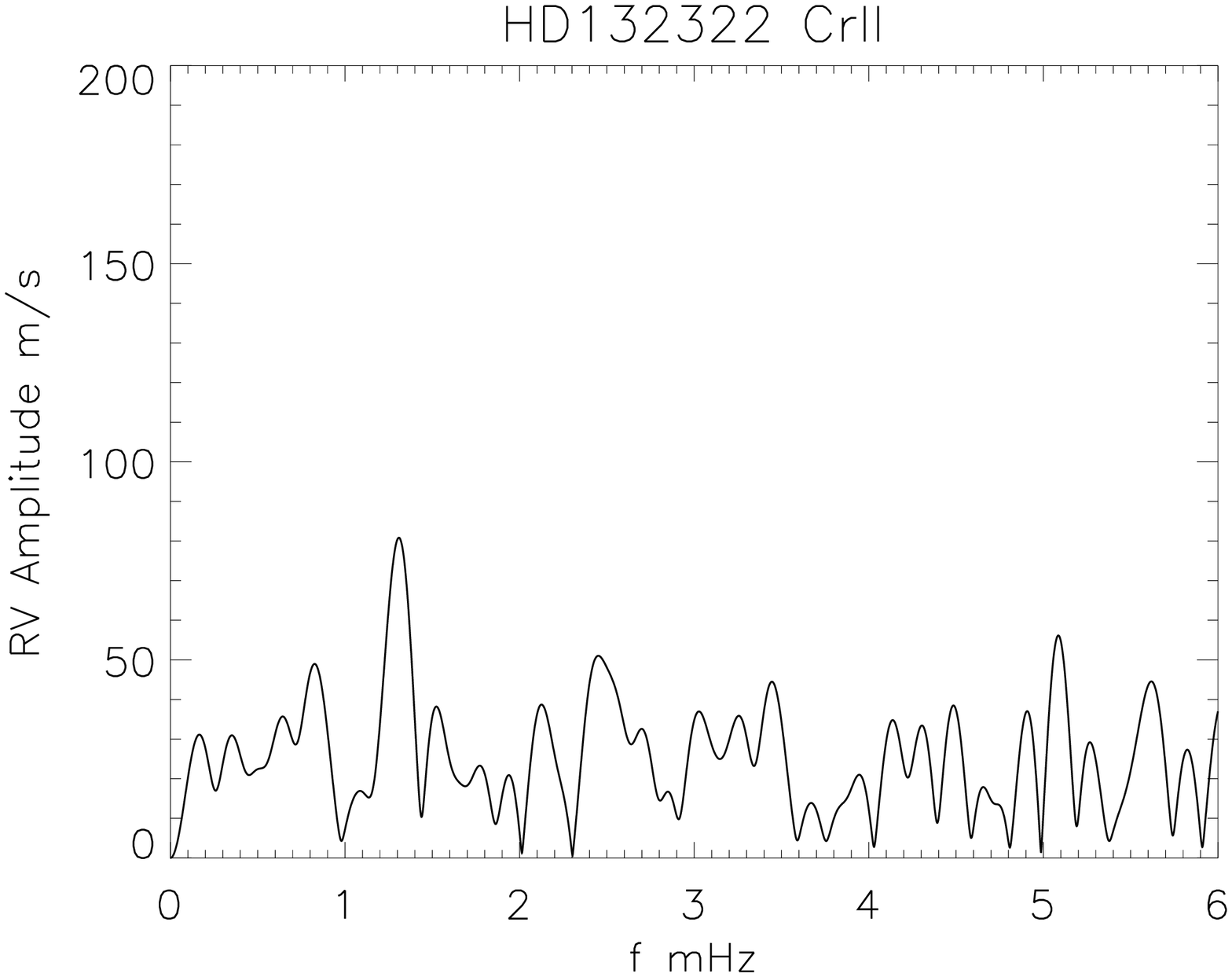}
  \includegraphics[height=5.6cm, angle=270]{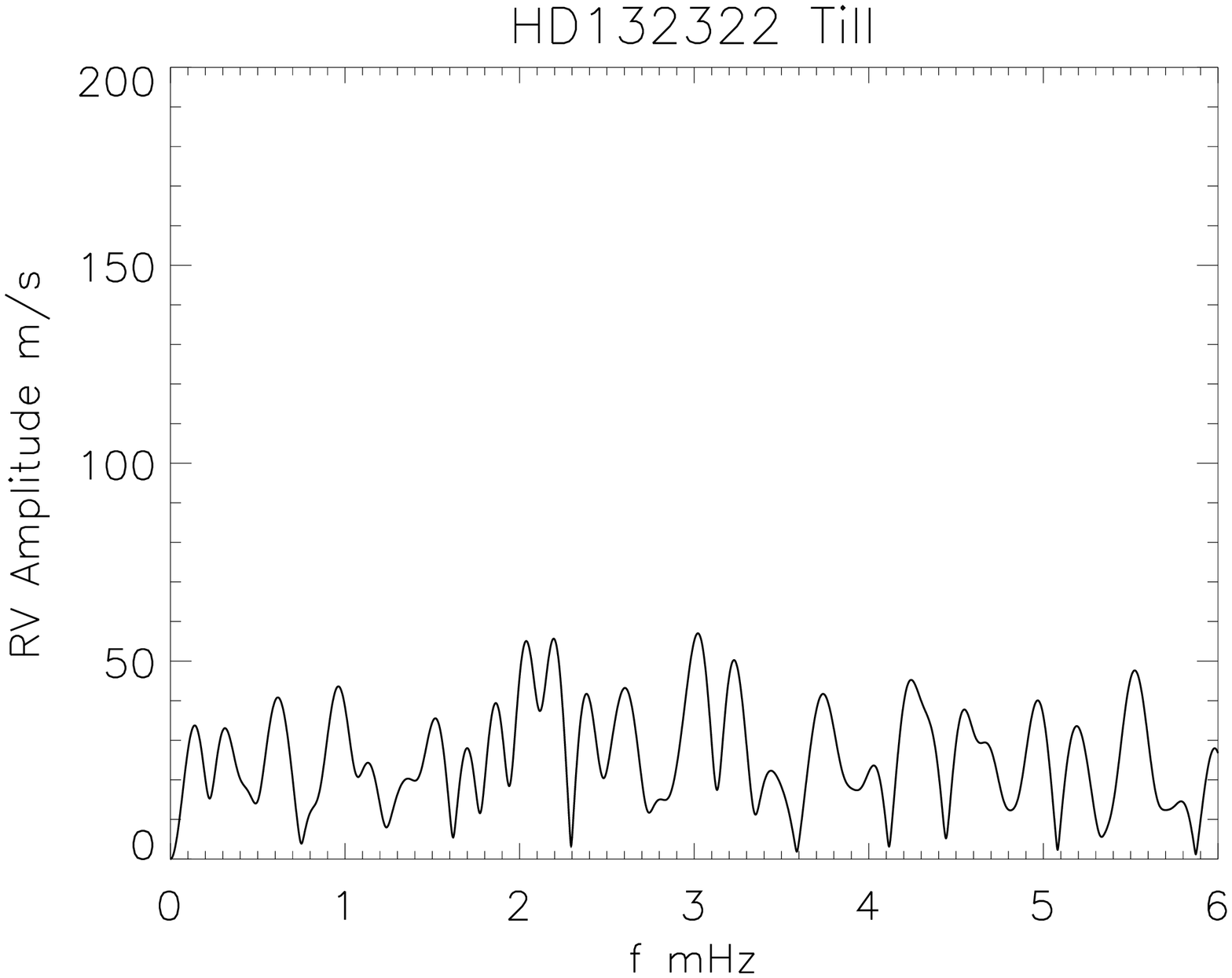}
  \includegraphics[height=5.6cm, angle=270]{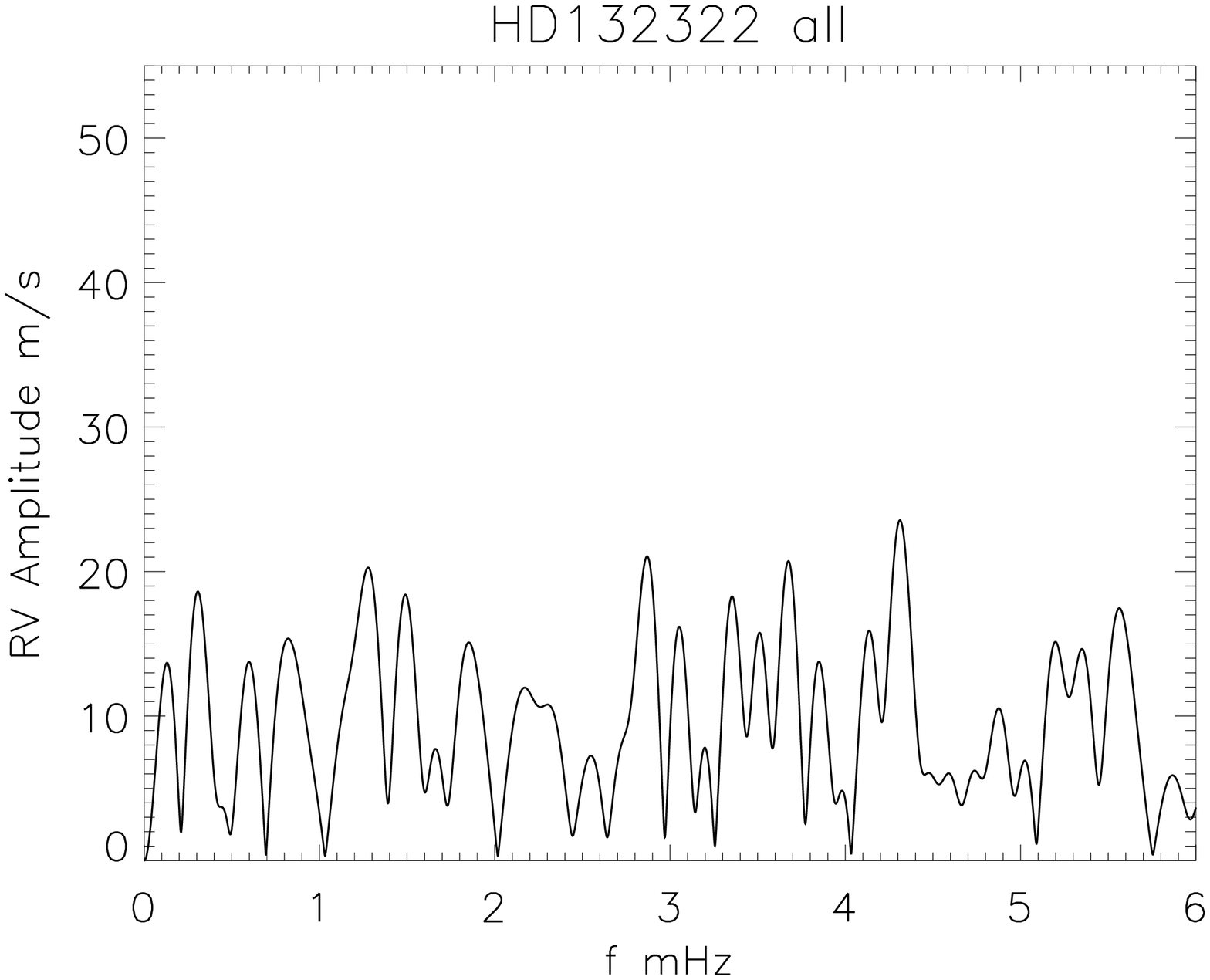}
  \includegraphics[height=5.6cm, angle=270]{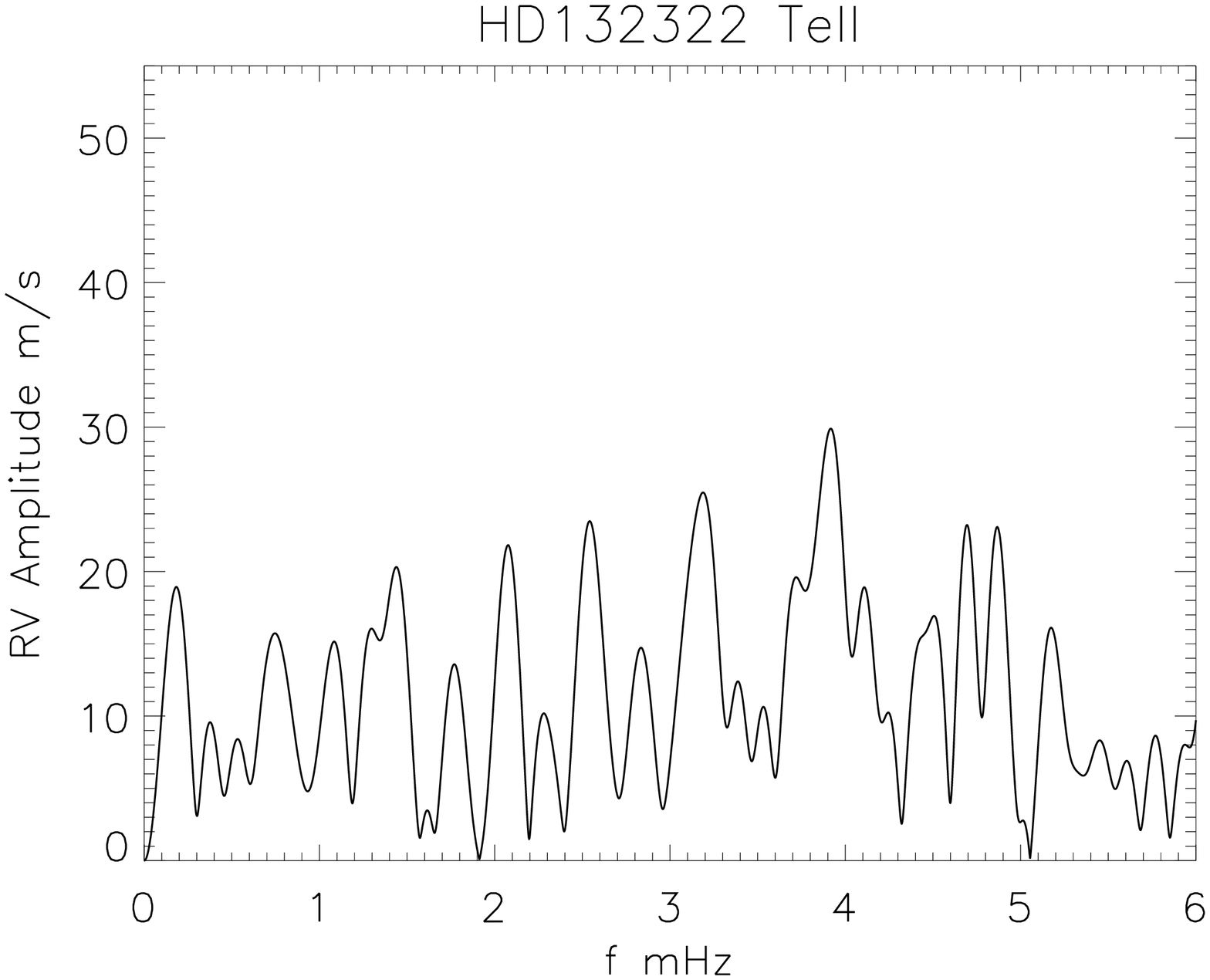}
  \caption{\label{fig:132322cog}Same as Fig.\,\ref{fig:107107cog} but
    for HD\,132322.  The Nyquist frequency is 4.7\,mHz.}
\end{figure*}

\begin{figure*}
  \vspace{3pt}
  \includegraphics[height=5.6cm,
  angle=270]{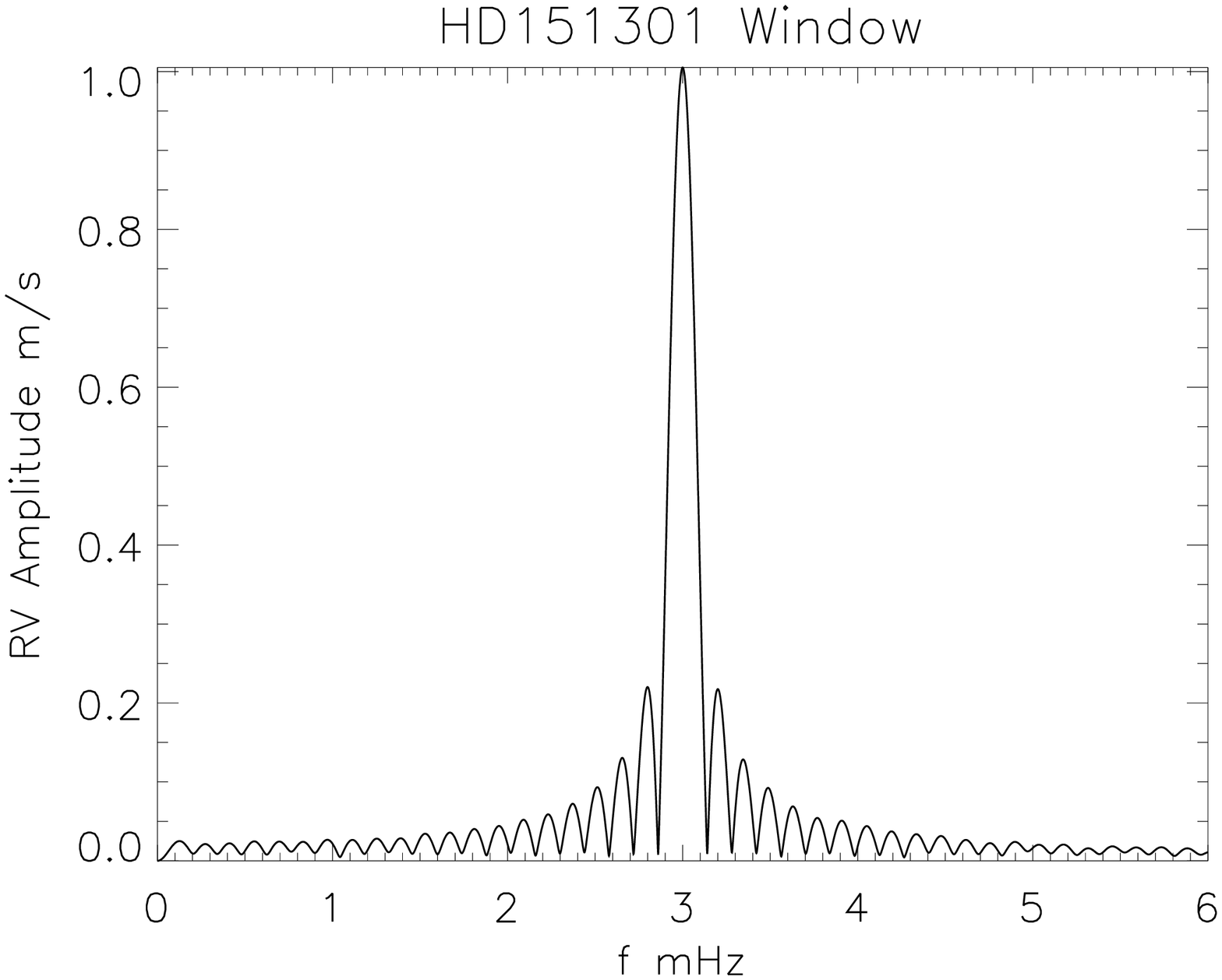}
  \includegraphics[height=5.6cm, angle=270]{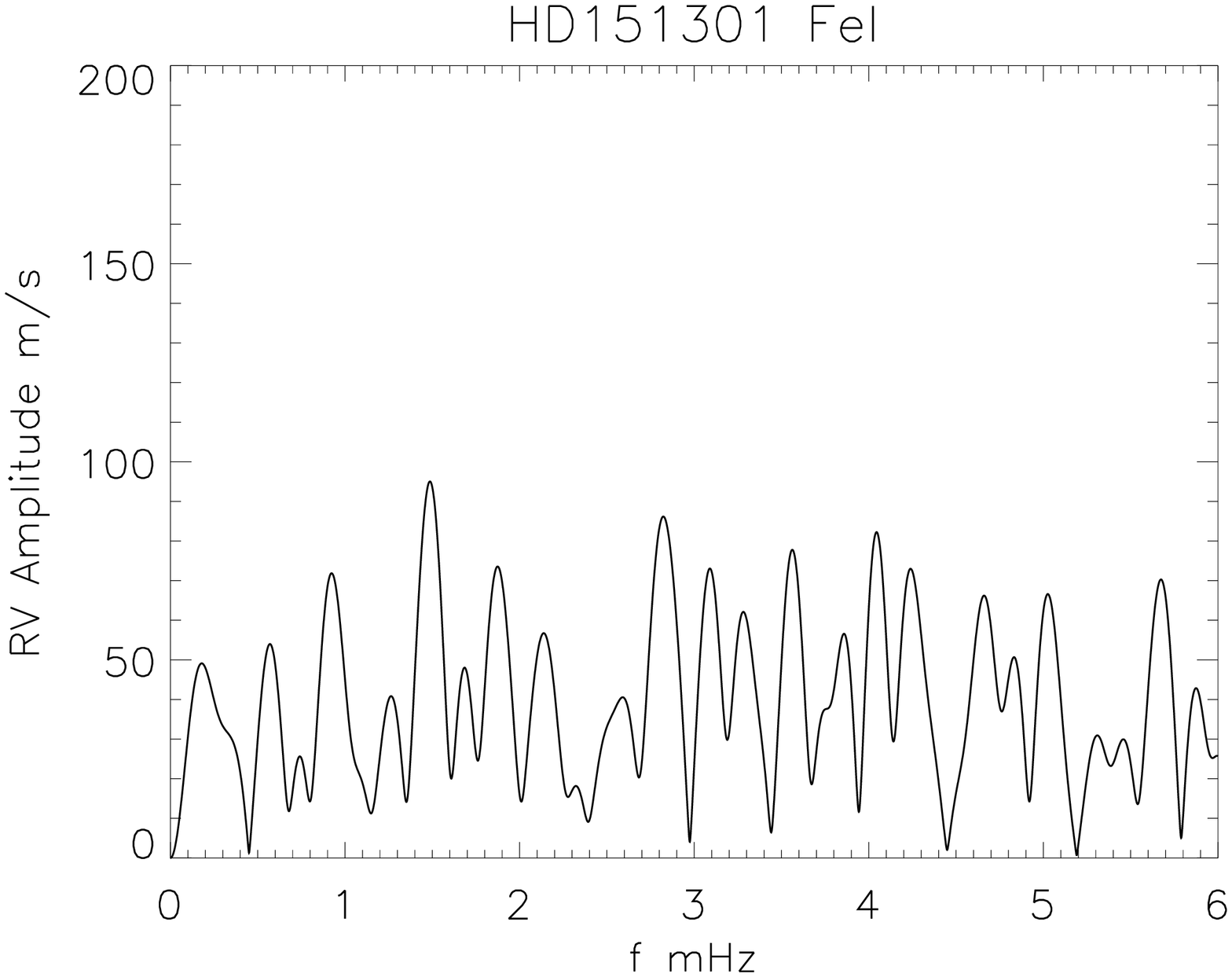}
  \includegraphics[height=5.6cm, angle=270]{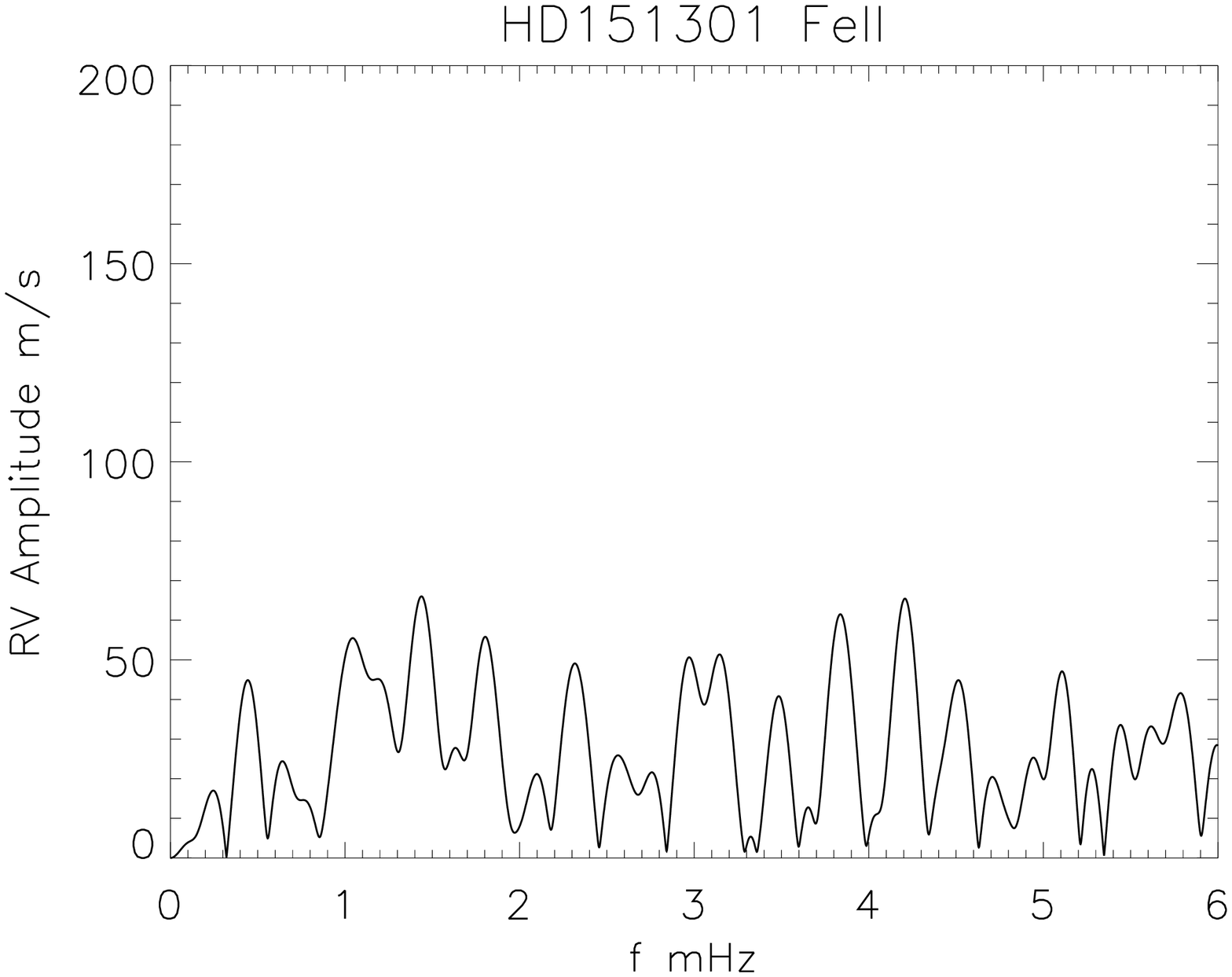}
  \includegraphics[height=5.6cm, angle=270]{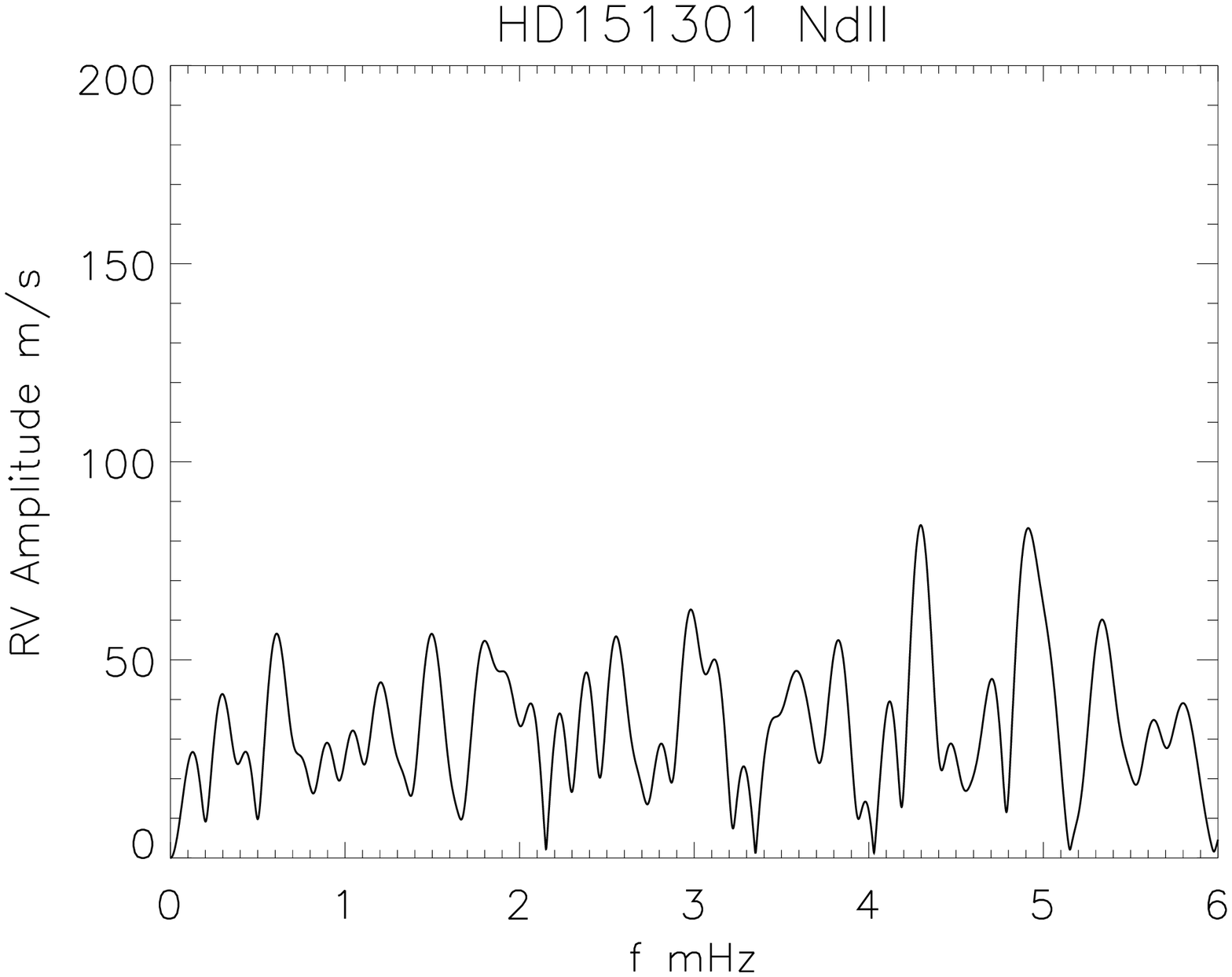}
  \includegraphics[height=5.6cm,
  angle=270]{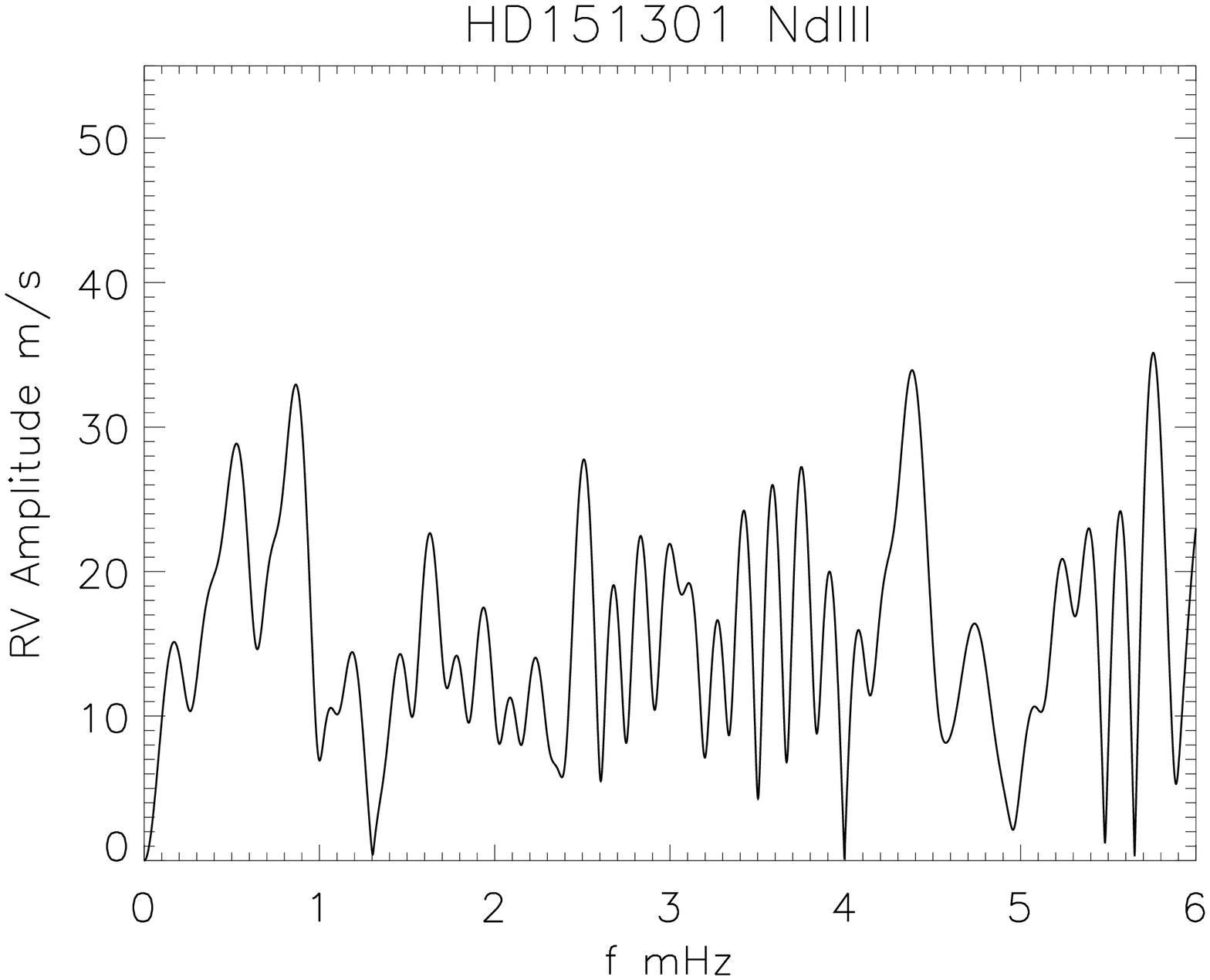}
  \includegraphics[height=5.6cm, angle=270]{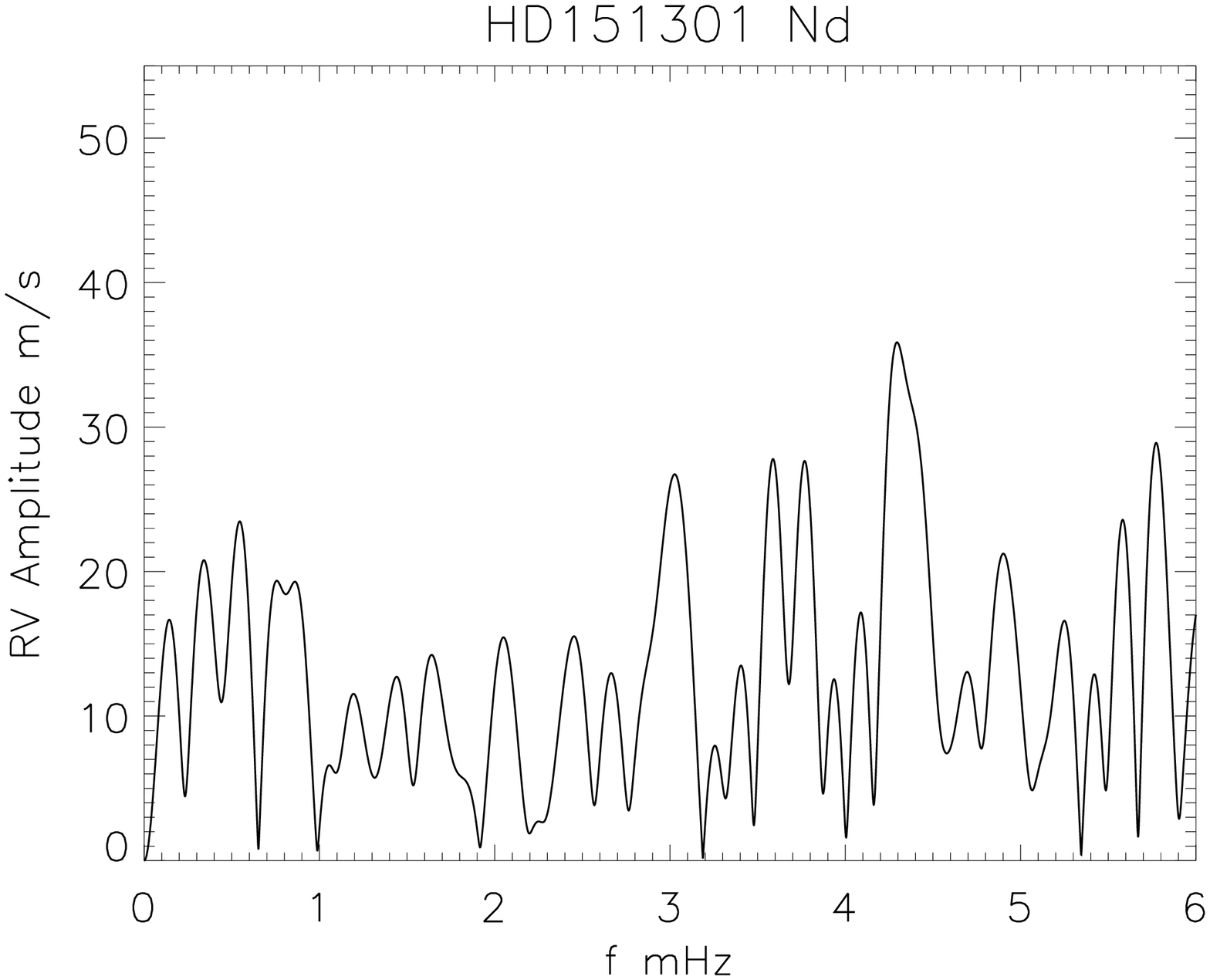}
  \includegraphics[height=5.6cm, angle=270]{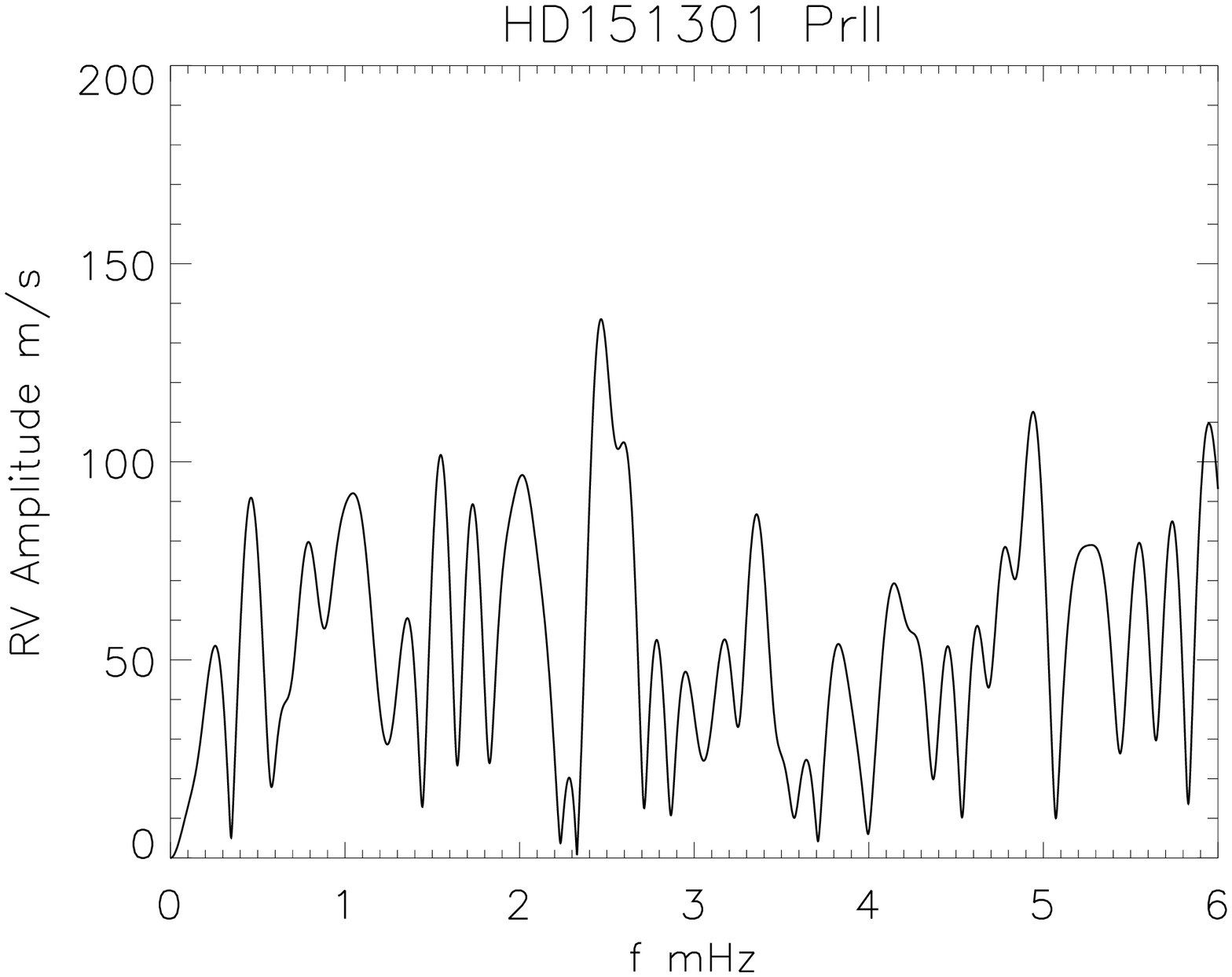}
  \includegraphics[height=5.6cm,
  angle=270]{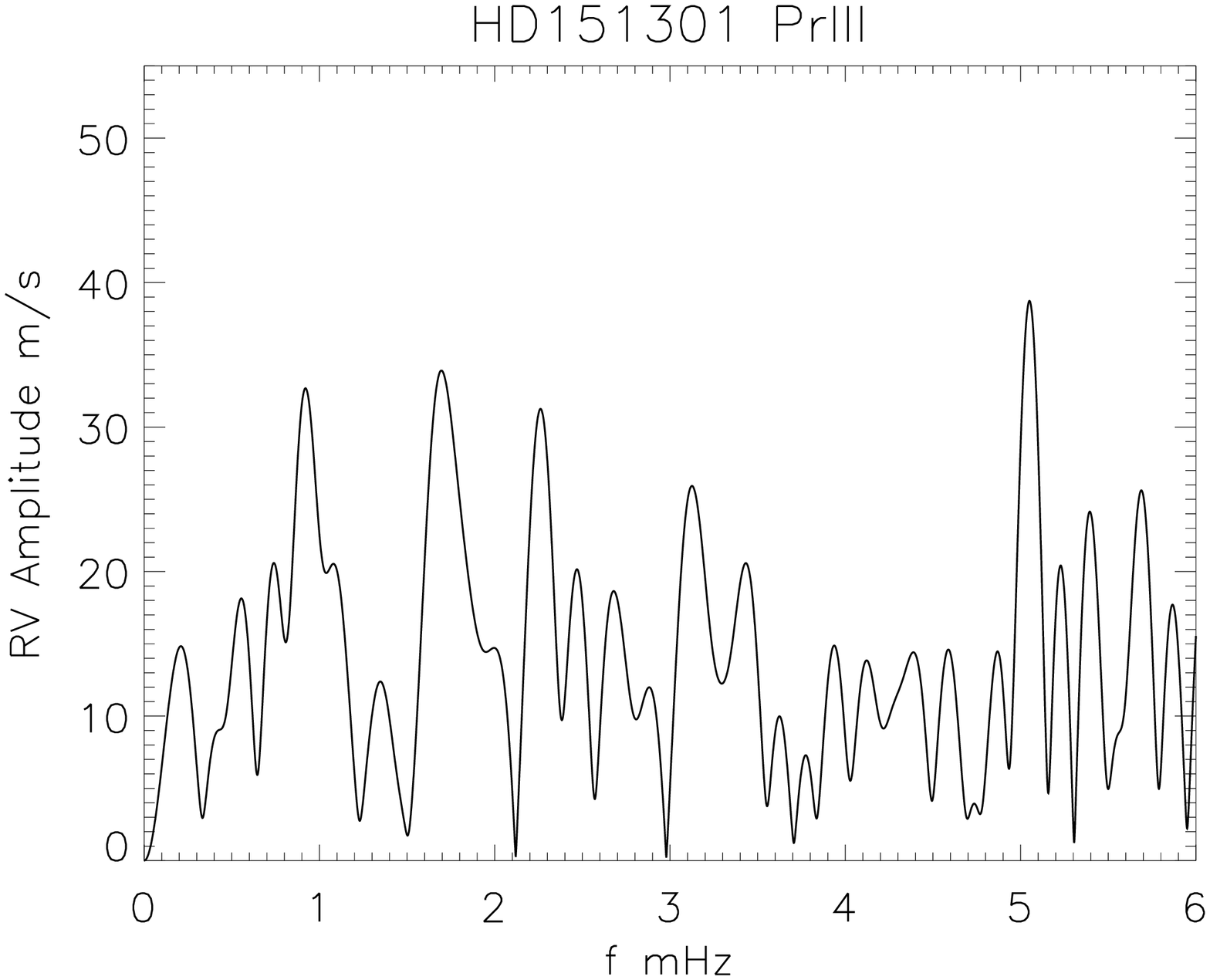}
  \includegraphics[height=5.6cm, angle=270]{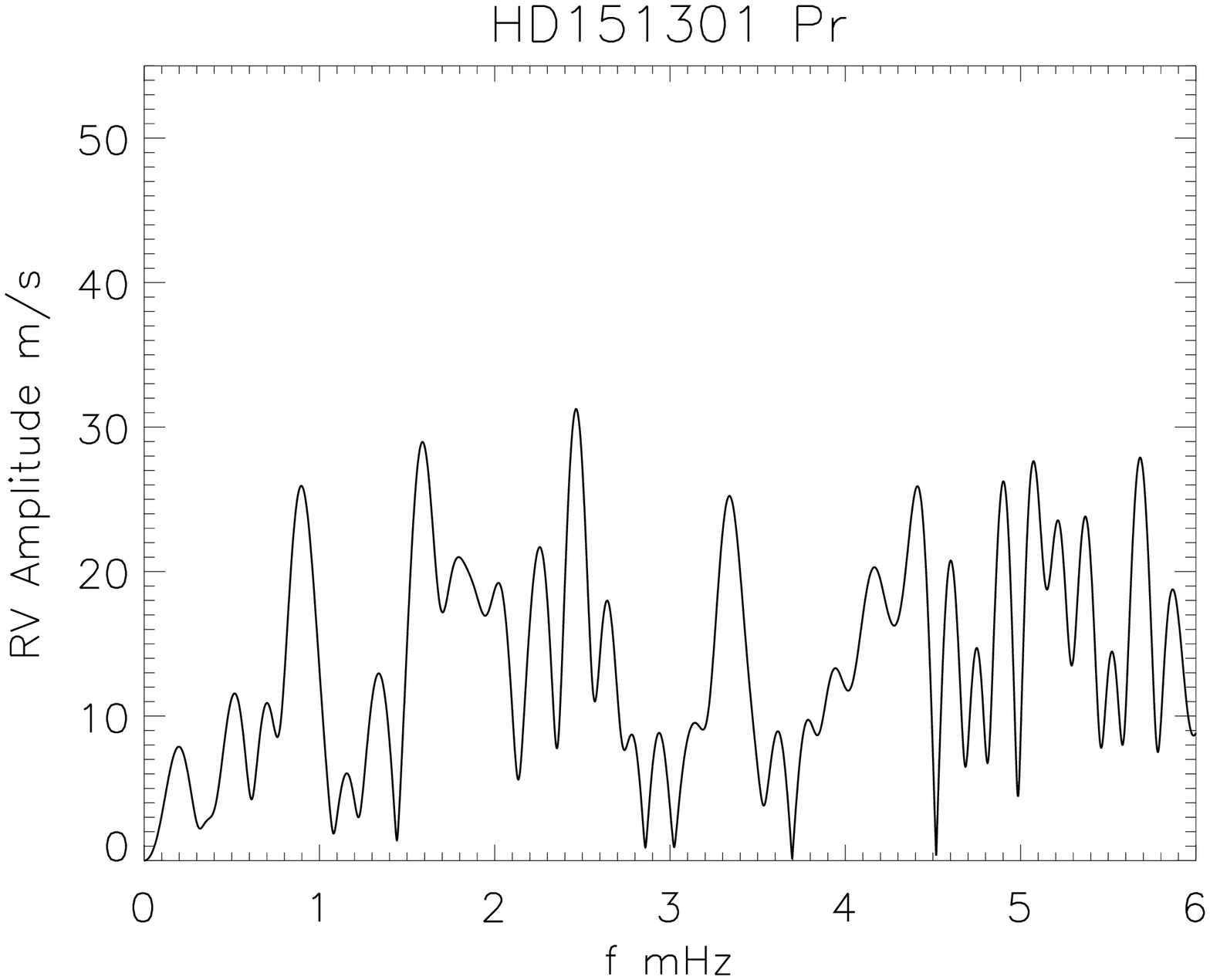}
  \includegraphics[height=5.6cm, angle=270]{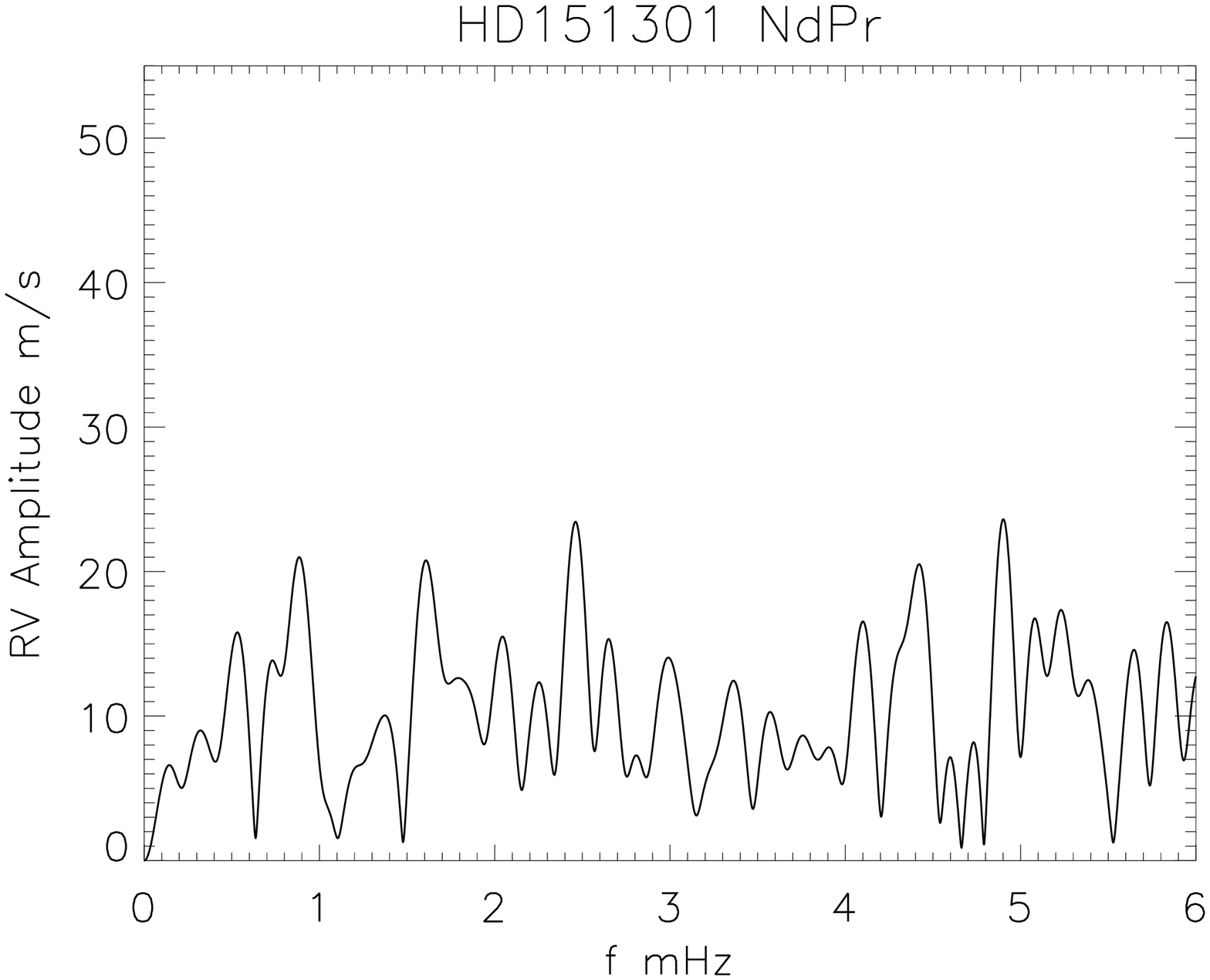}
  \includegraphics[height=5.6cm, angle=270]{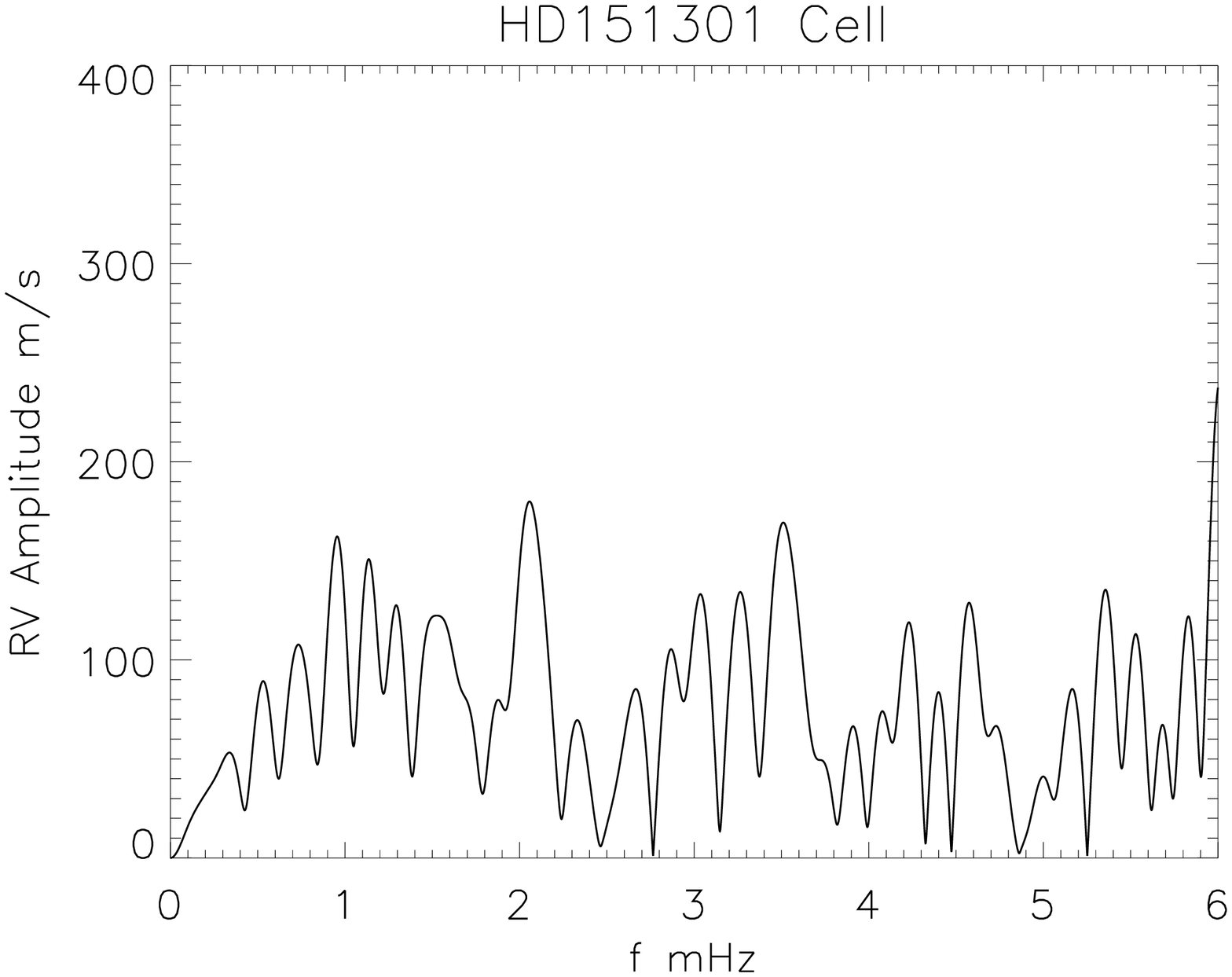}
  \includegraphics[height=5.6cm, angle=270]{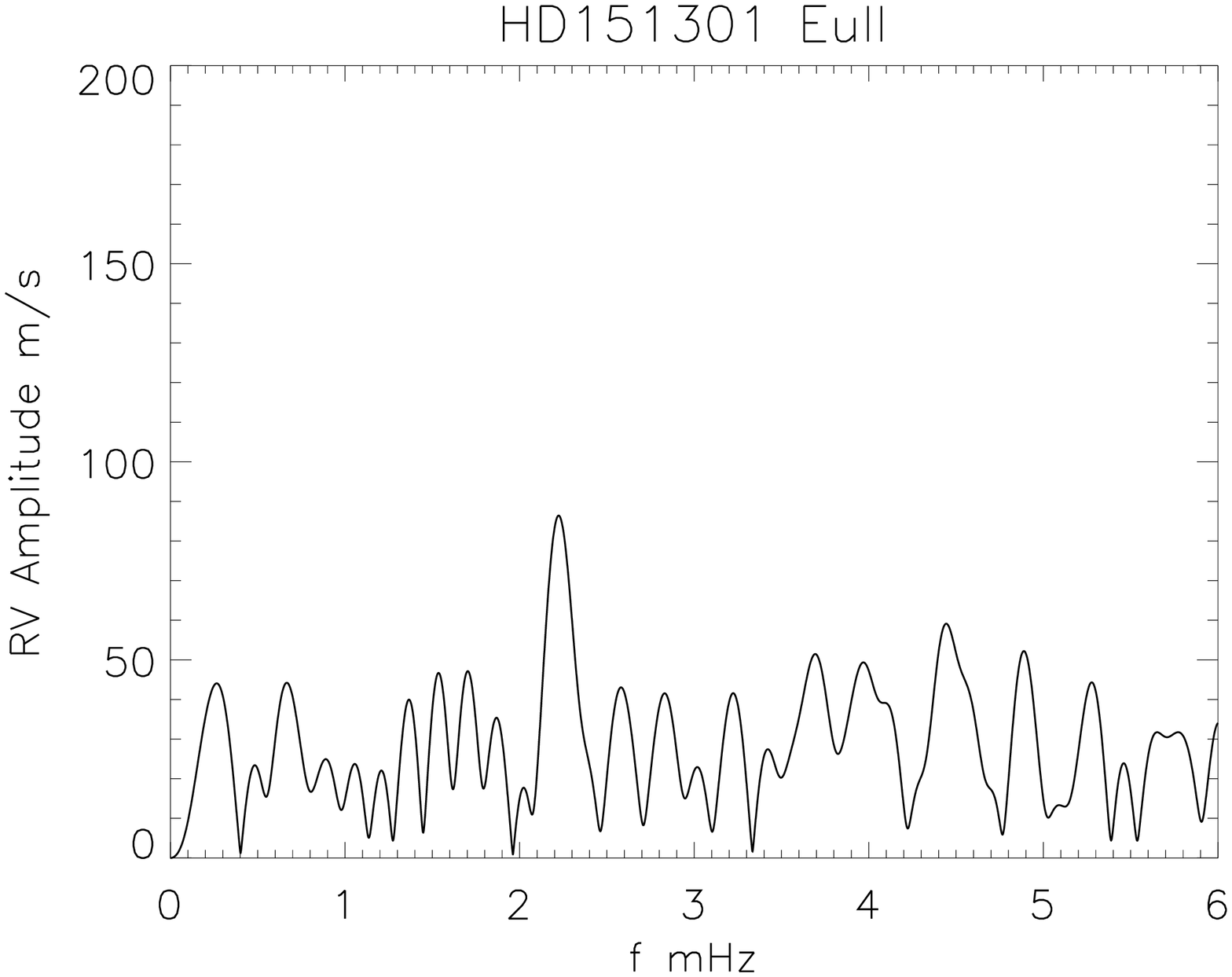}
  \includegraphics[height=5.6cm, angle=270]{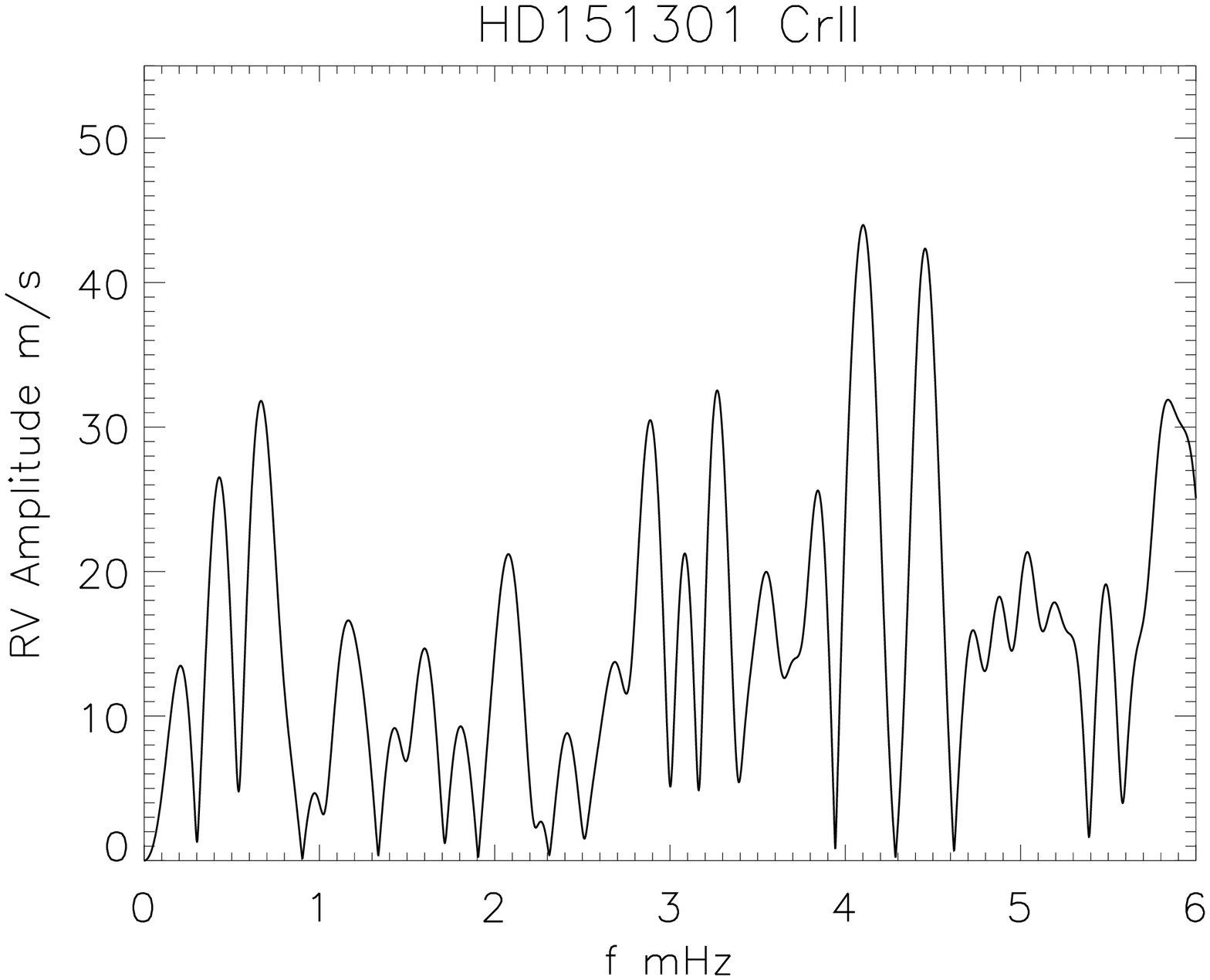}
  \includegraphics[height=5.6cm, angle=270]{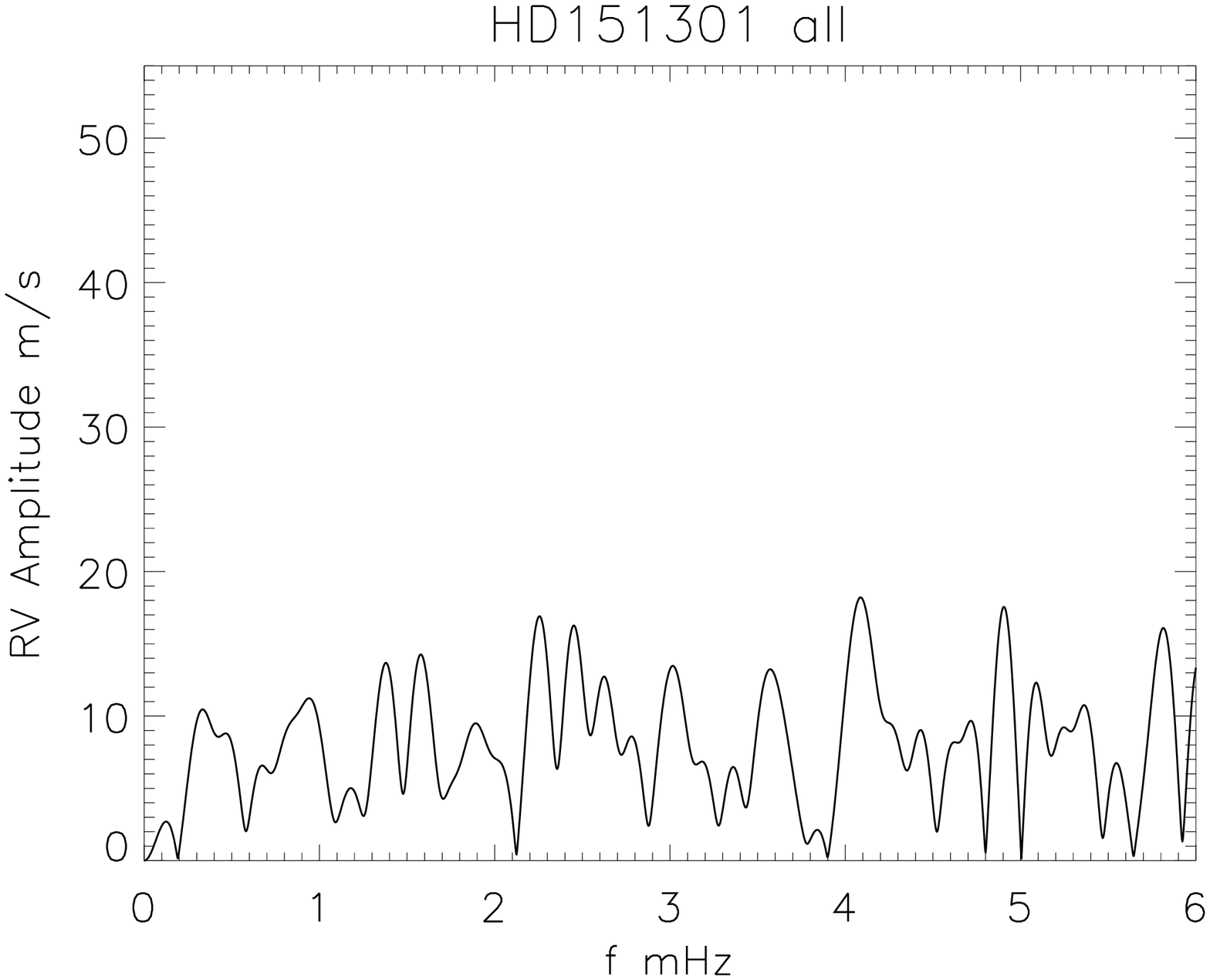}
  \includegraphics[height=5.6cm, angle=270]{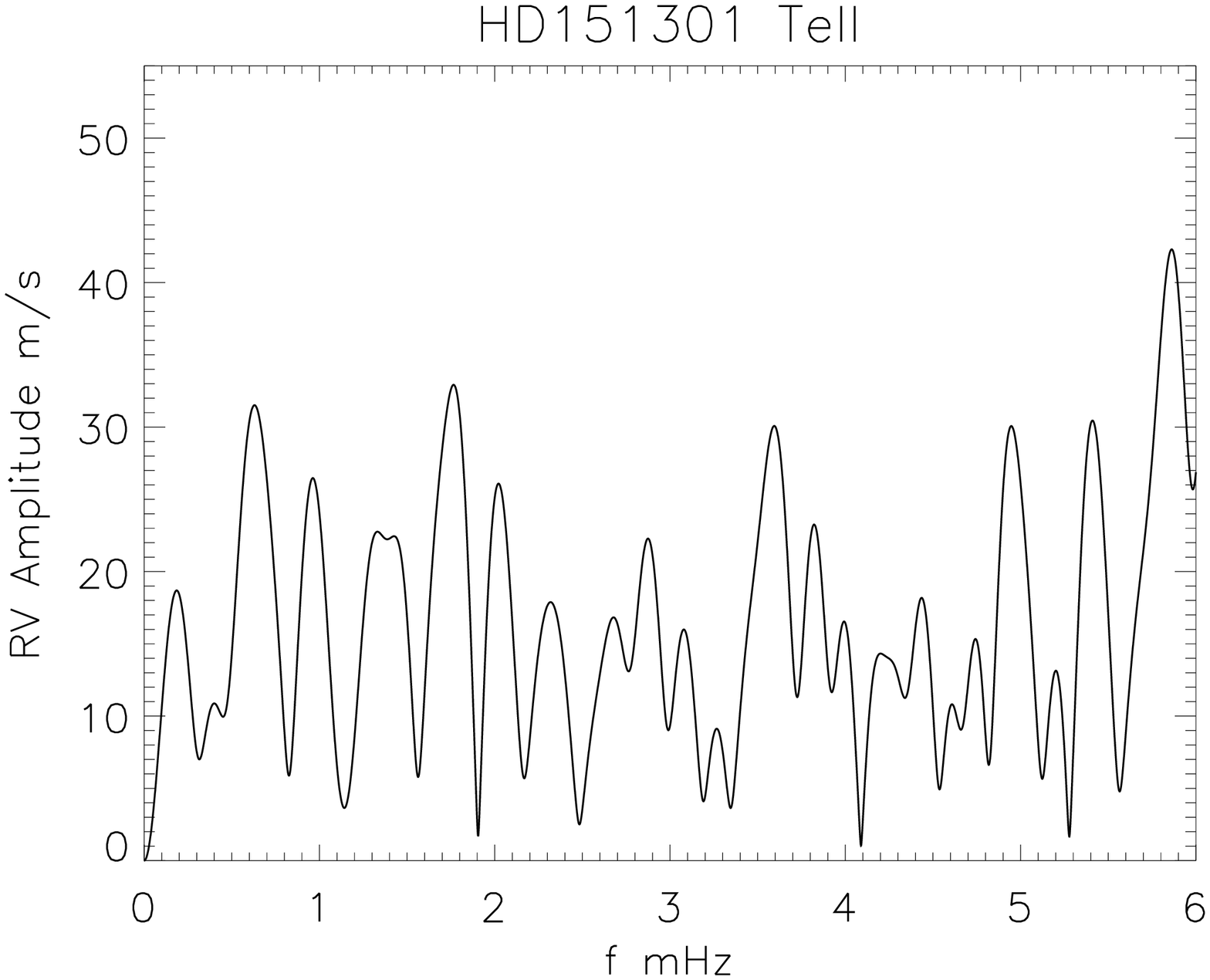}
  \caption{\label{fig:151301cog}Same as Fig.\,\ref{fig:107107cog} but
    for HD\,151301.  }
\end{figure*}

\begin{figure*}
  \vspace{3pt}
  \includegraphics[height=5.6cm,
  angle=270]{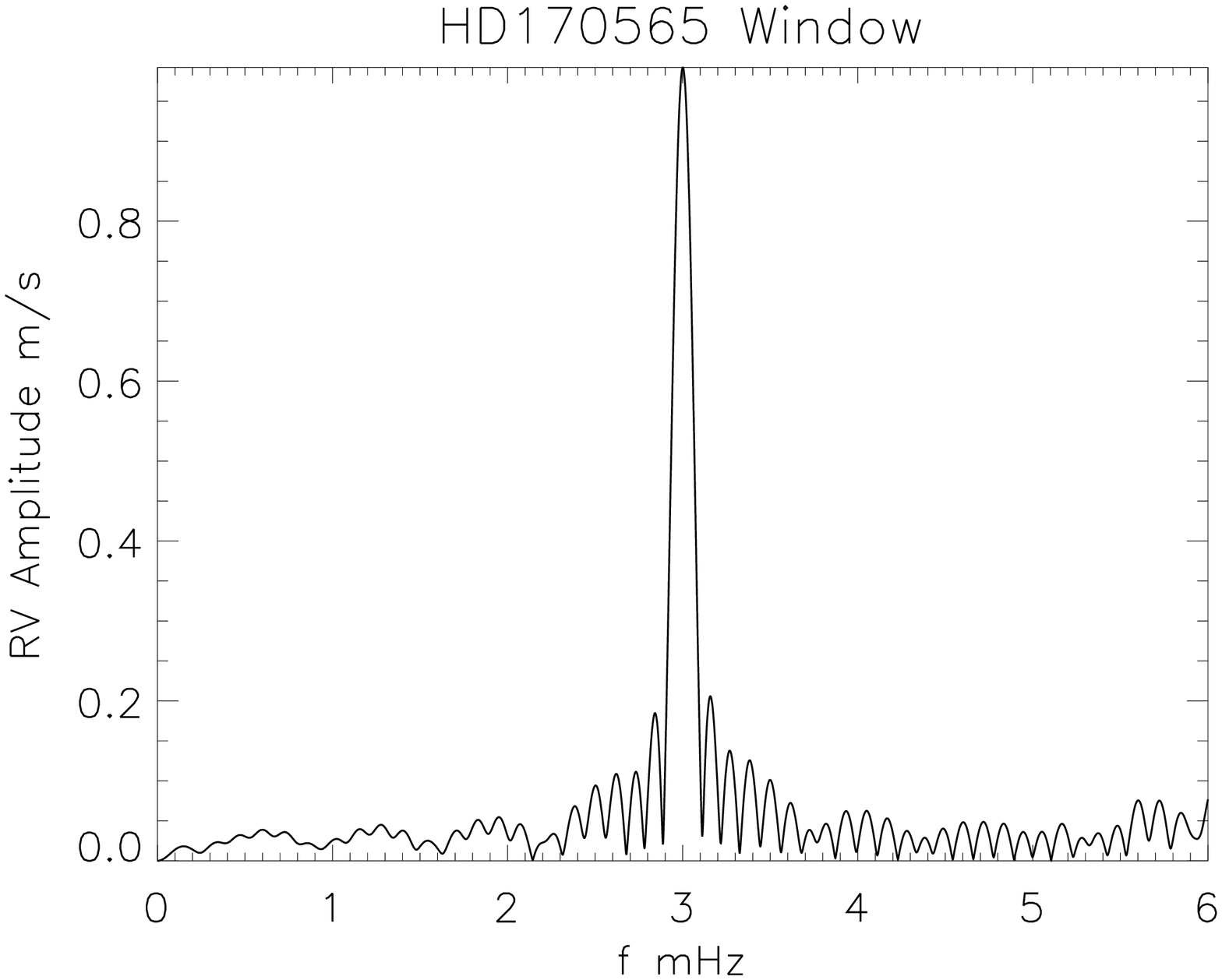}
  \includegraphics[height=5.6cm, angle=270]{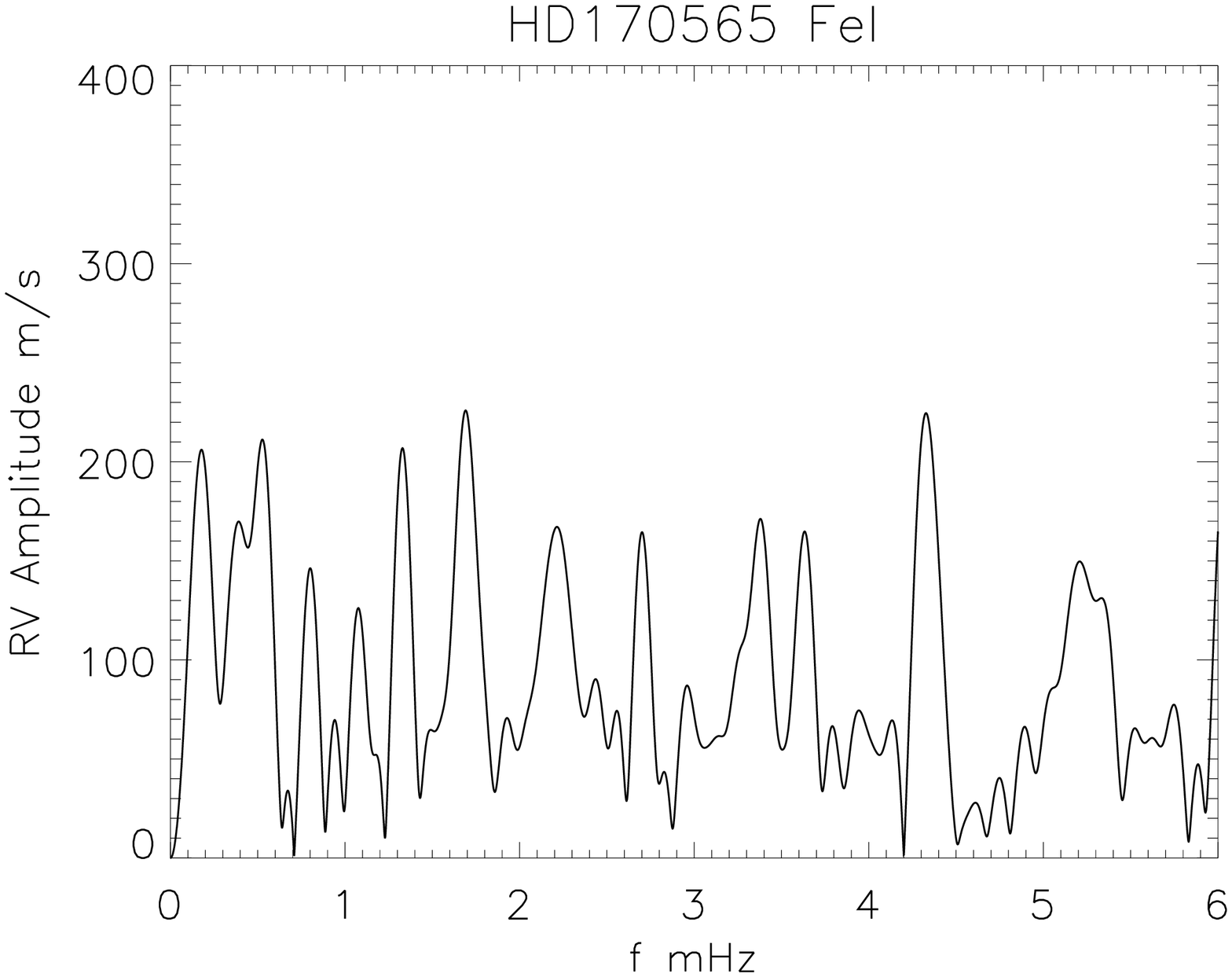}
  \includegraphics[height=5.6cm, angle=270]{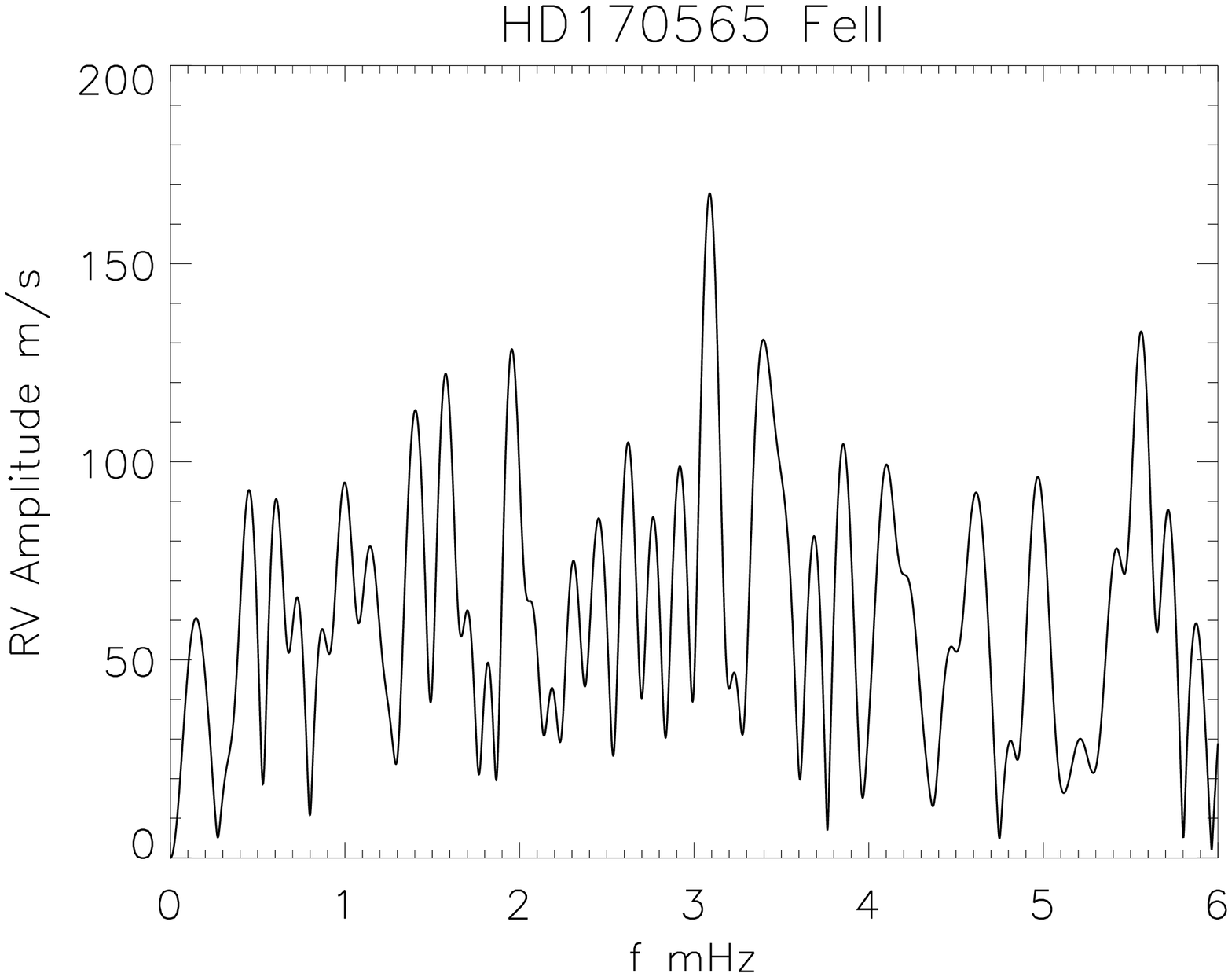}
  \includegraphics[height=5.6cm,
  angle=270]{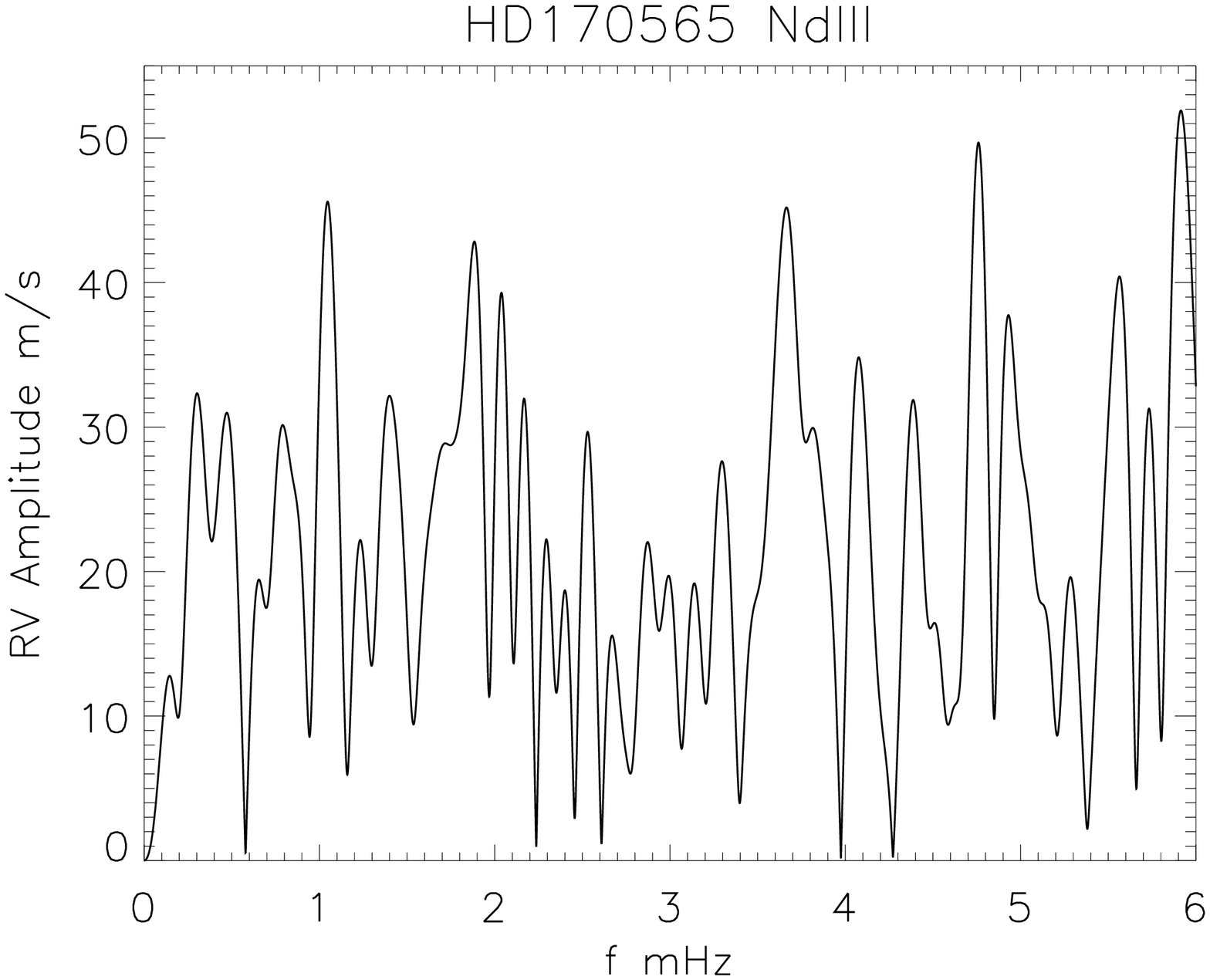}
  \includegraphics[height=5.6cm, angle=270]{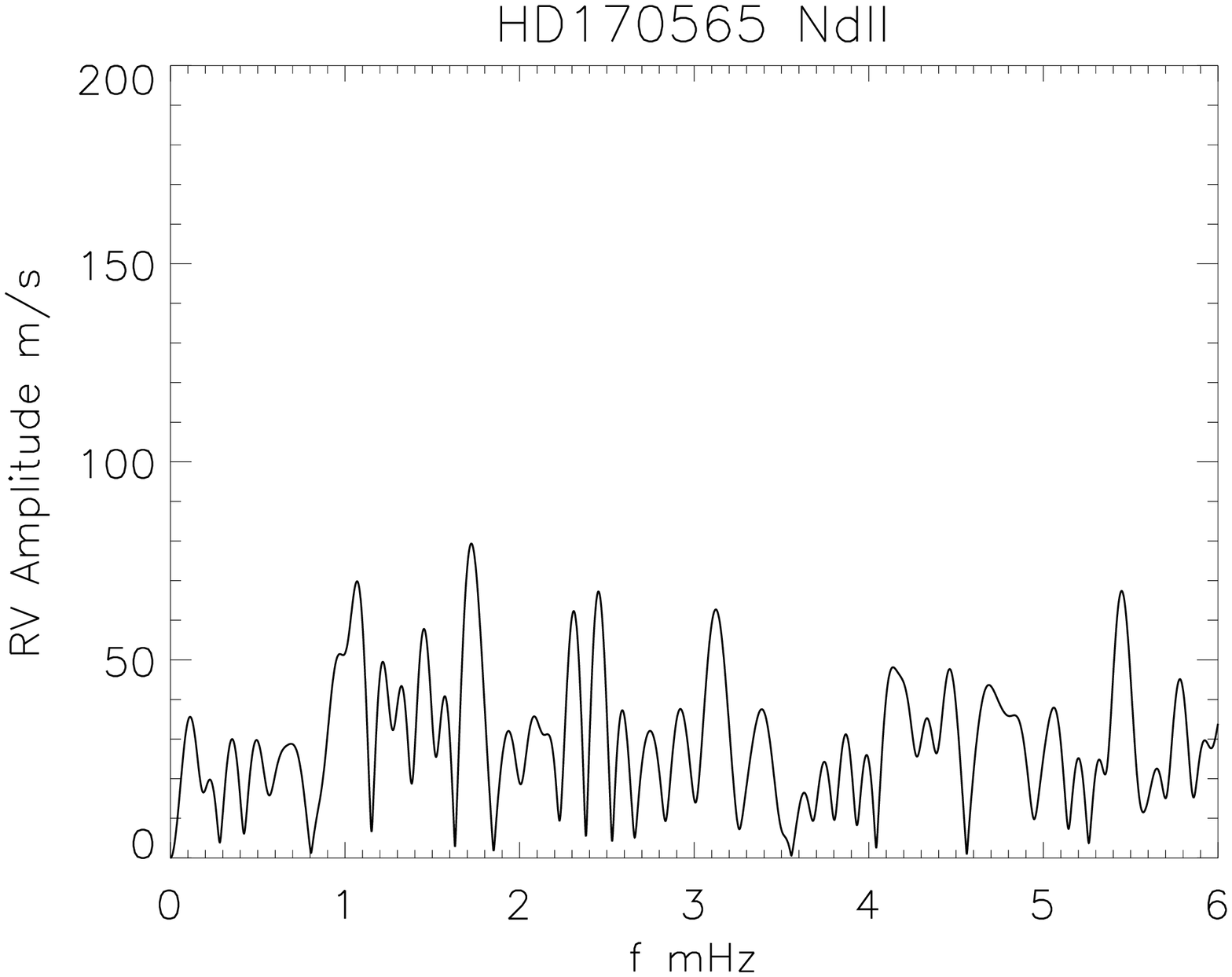}
  \includegraphics[height=5.6cm, angle=270]{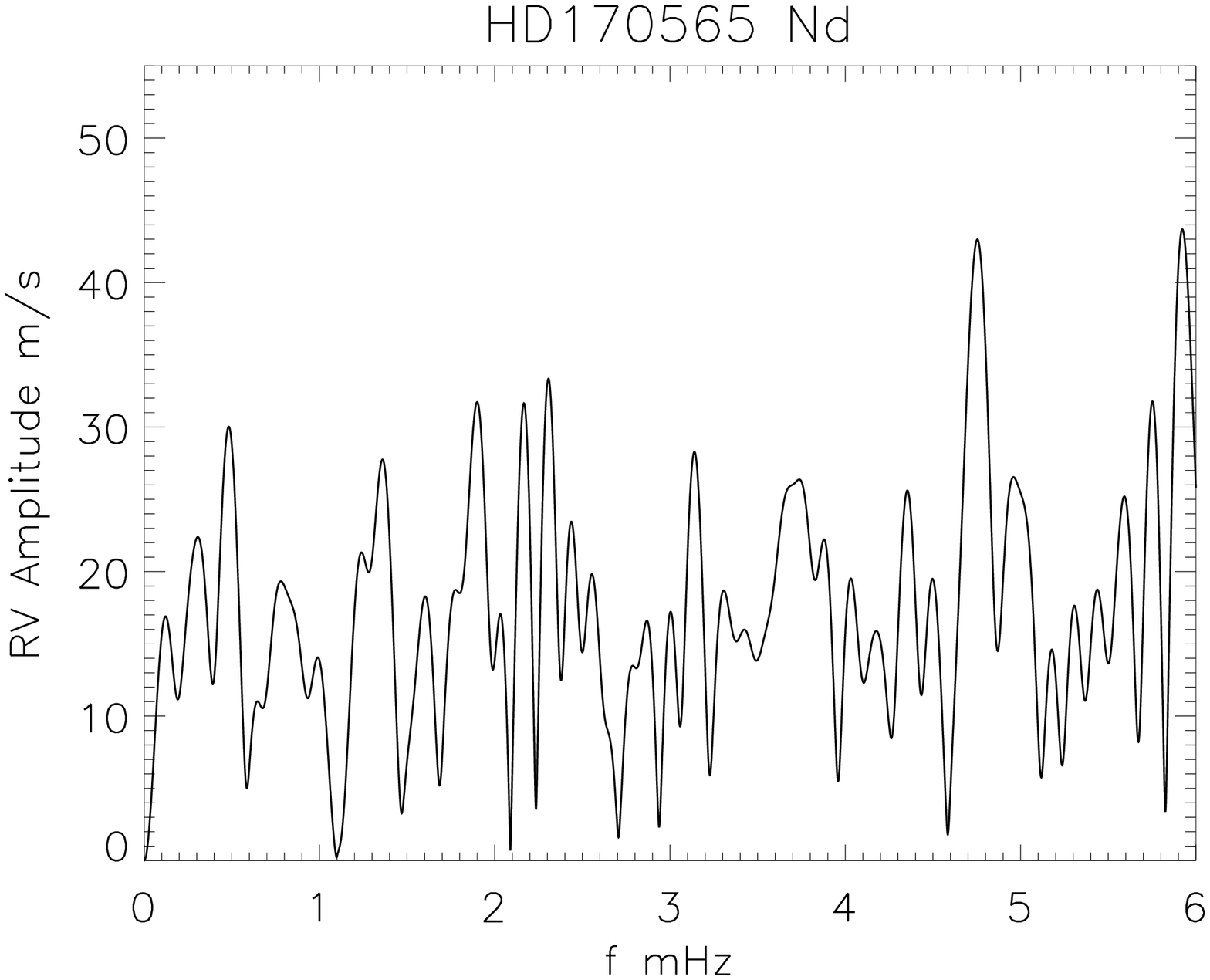}
  \includegraphics[height=5.6cm,
  angle=270]{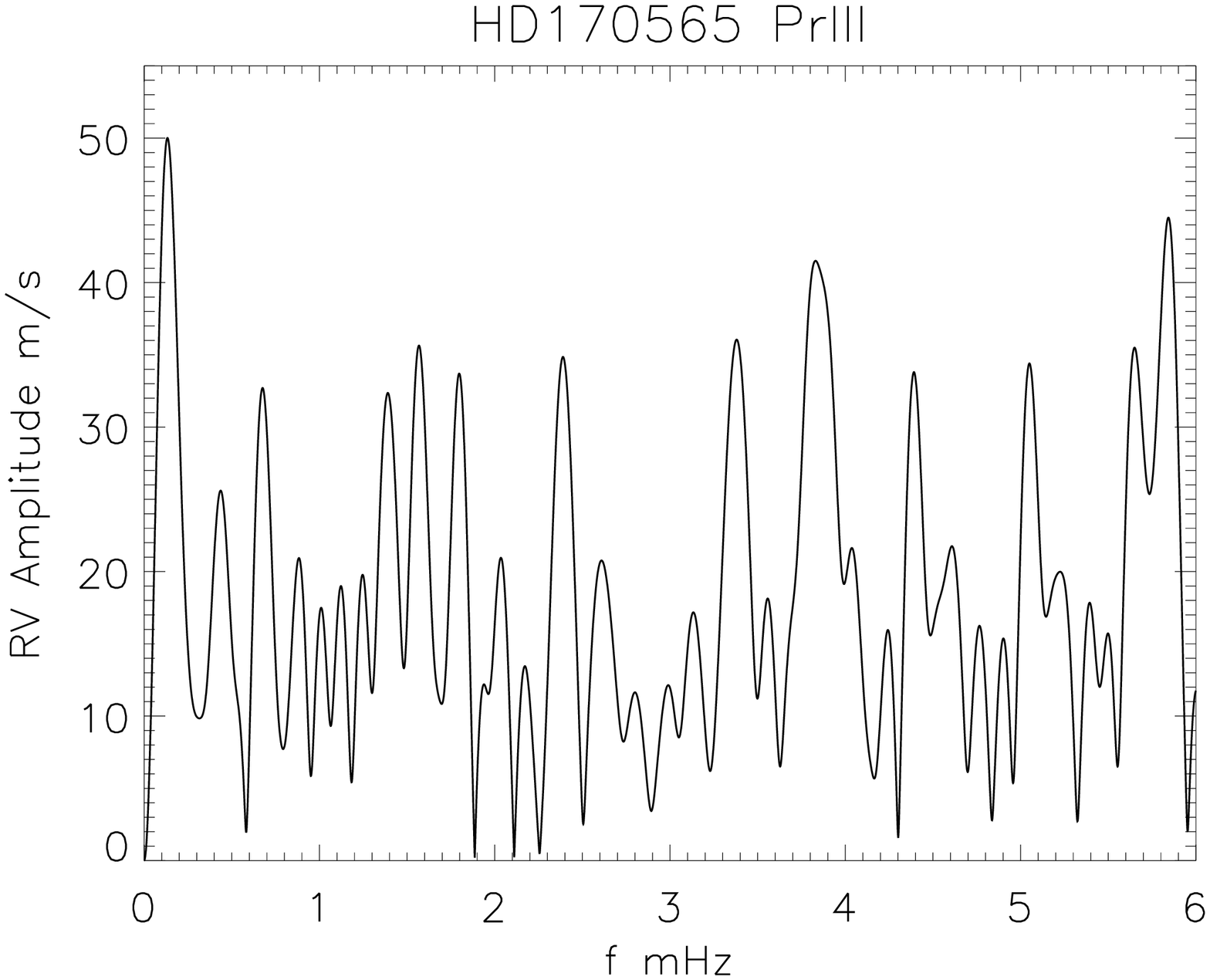}
  \includegraphics[height=5.6cm, angle=270]{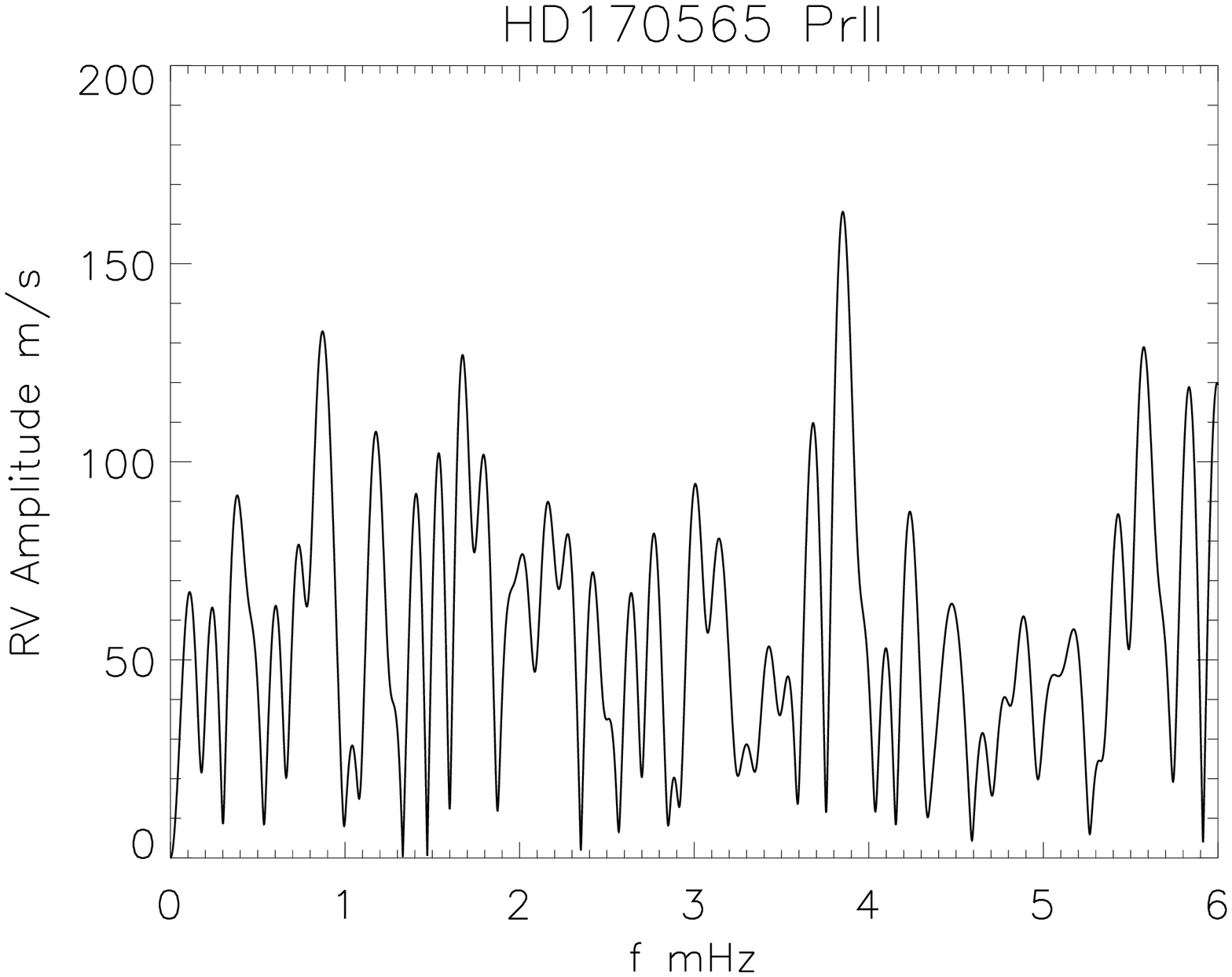}
  \includegraphics[height=5.6cm, angle=270]{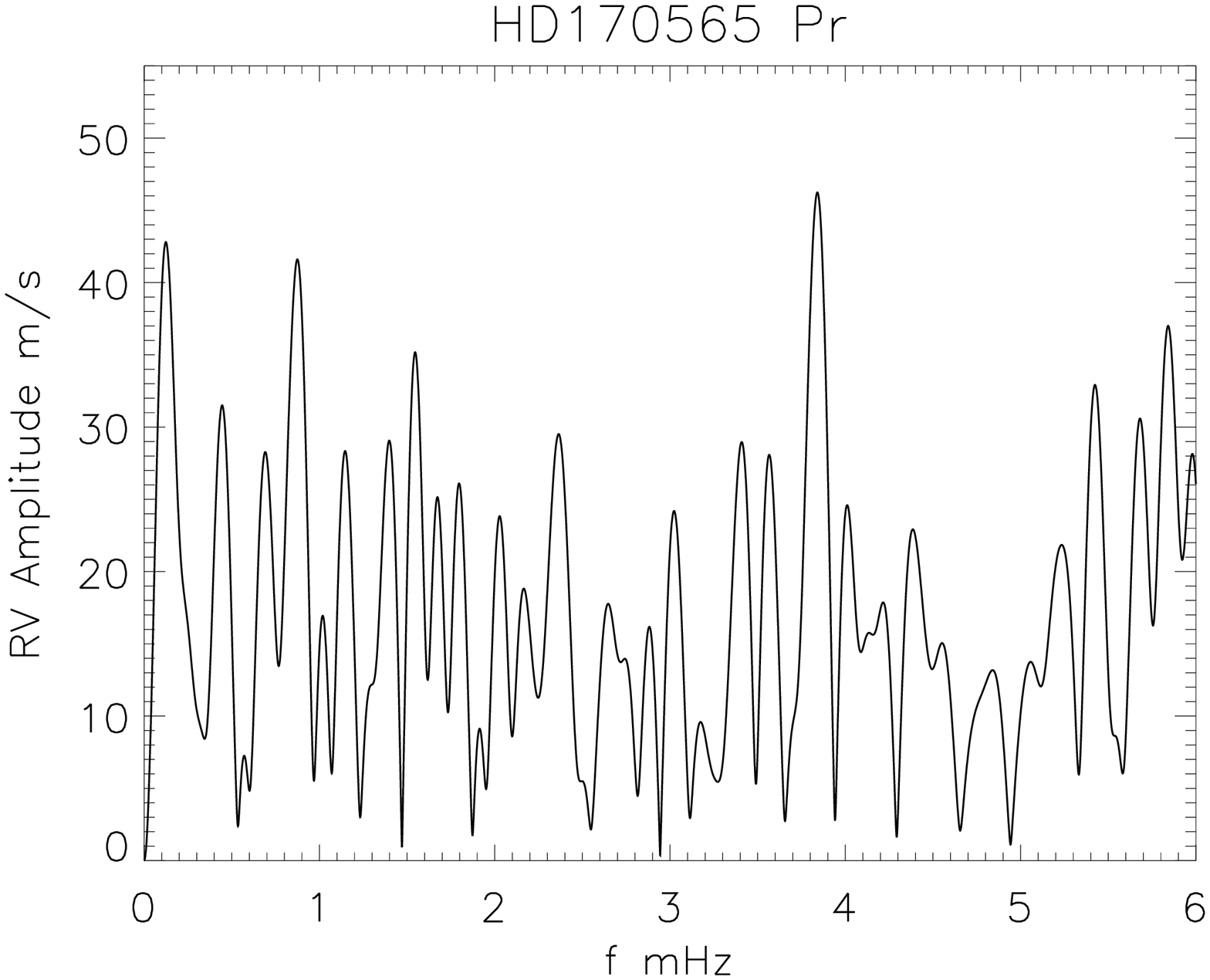}
  \includegraphics[height=5.6cm, angle=270]{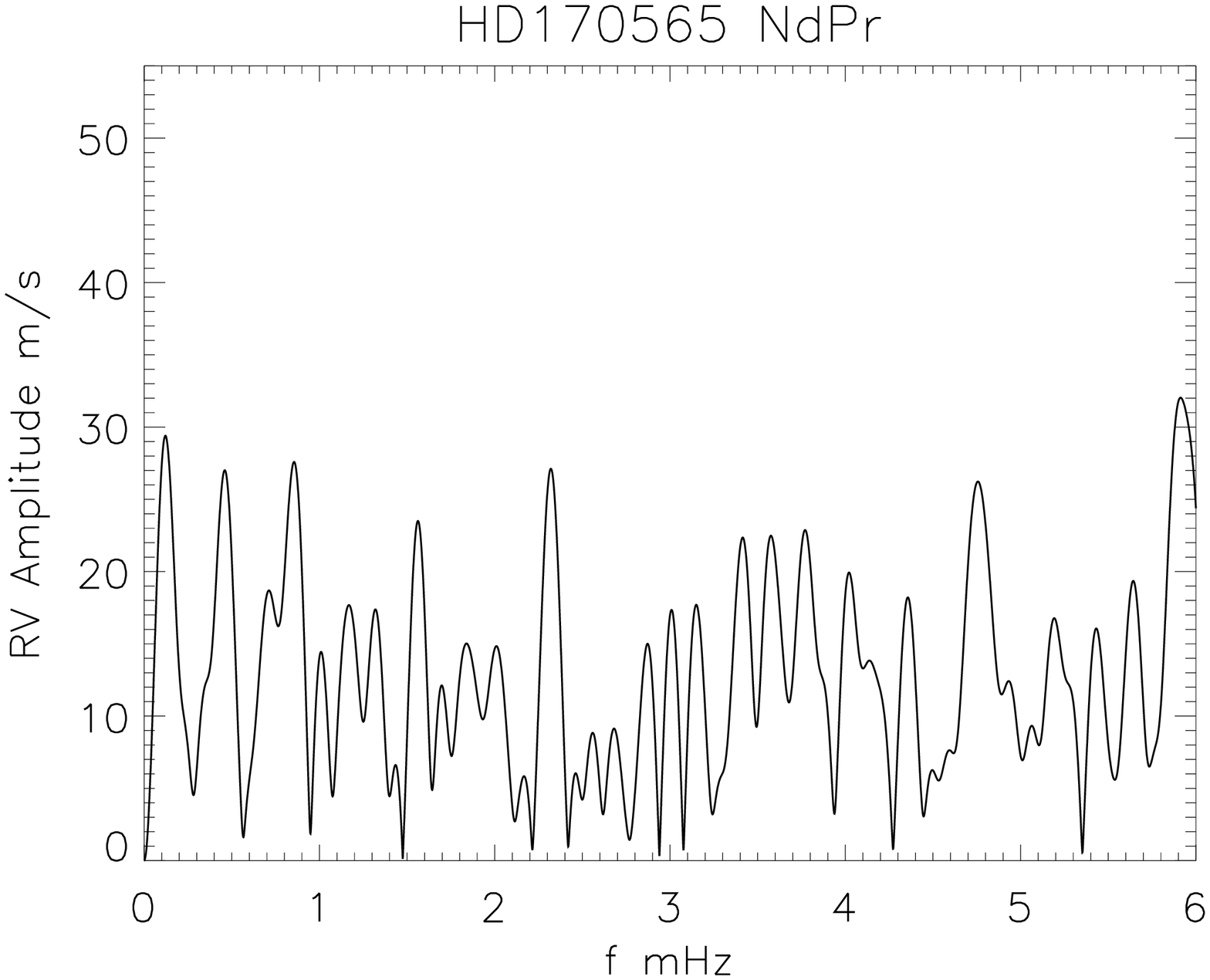}
  \includegraphics[height=5.6cm, angle=270]{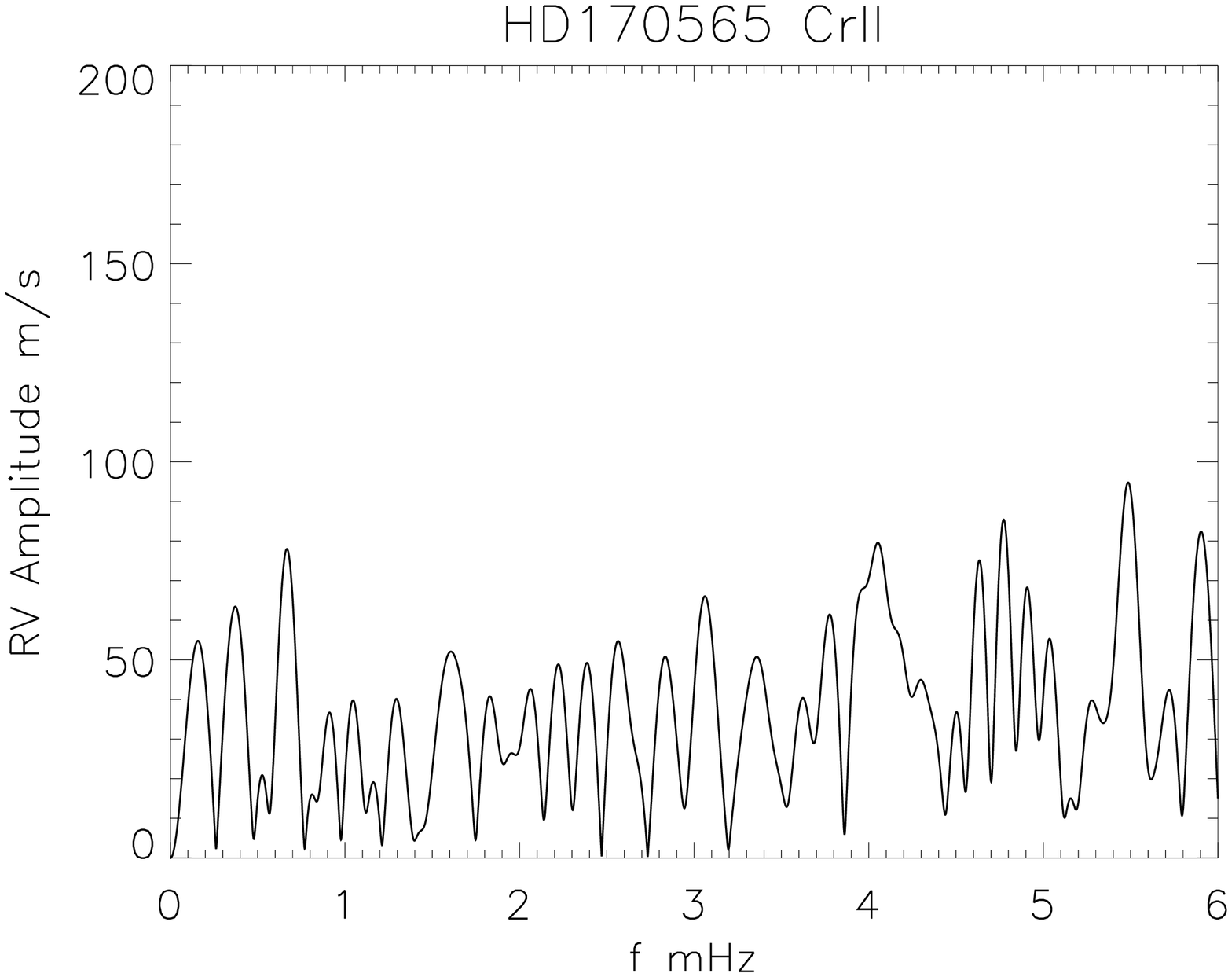}
  \includegraphics[height=5.6cm, angle=270]{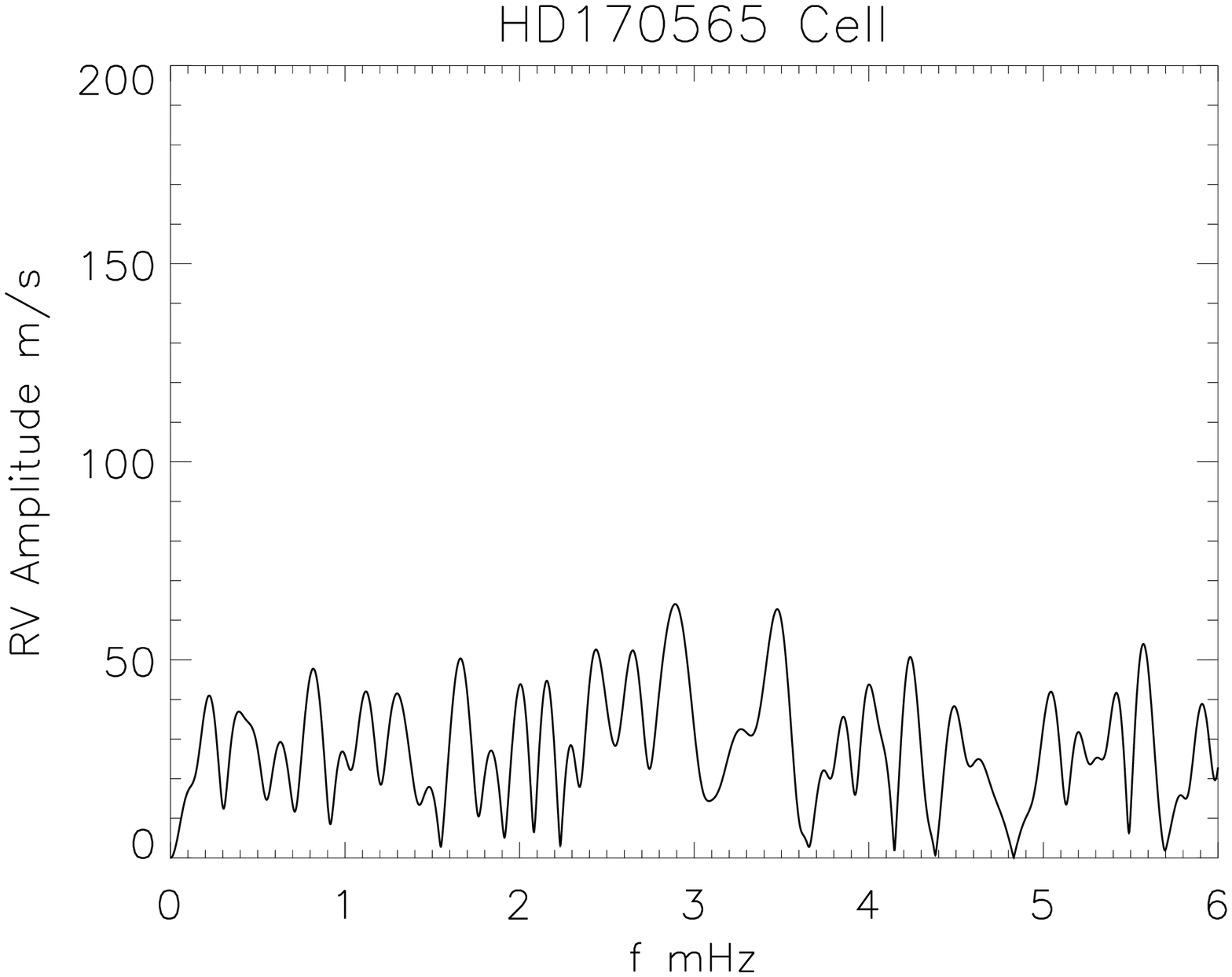}
  \includegraphics[height=5.6cm, angle=270]{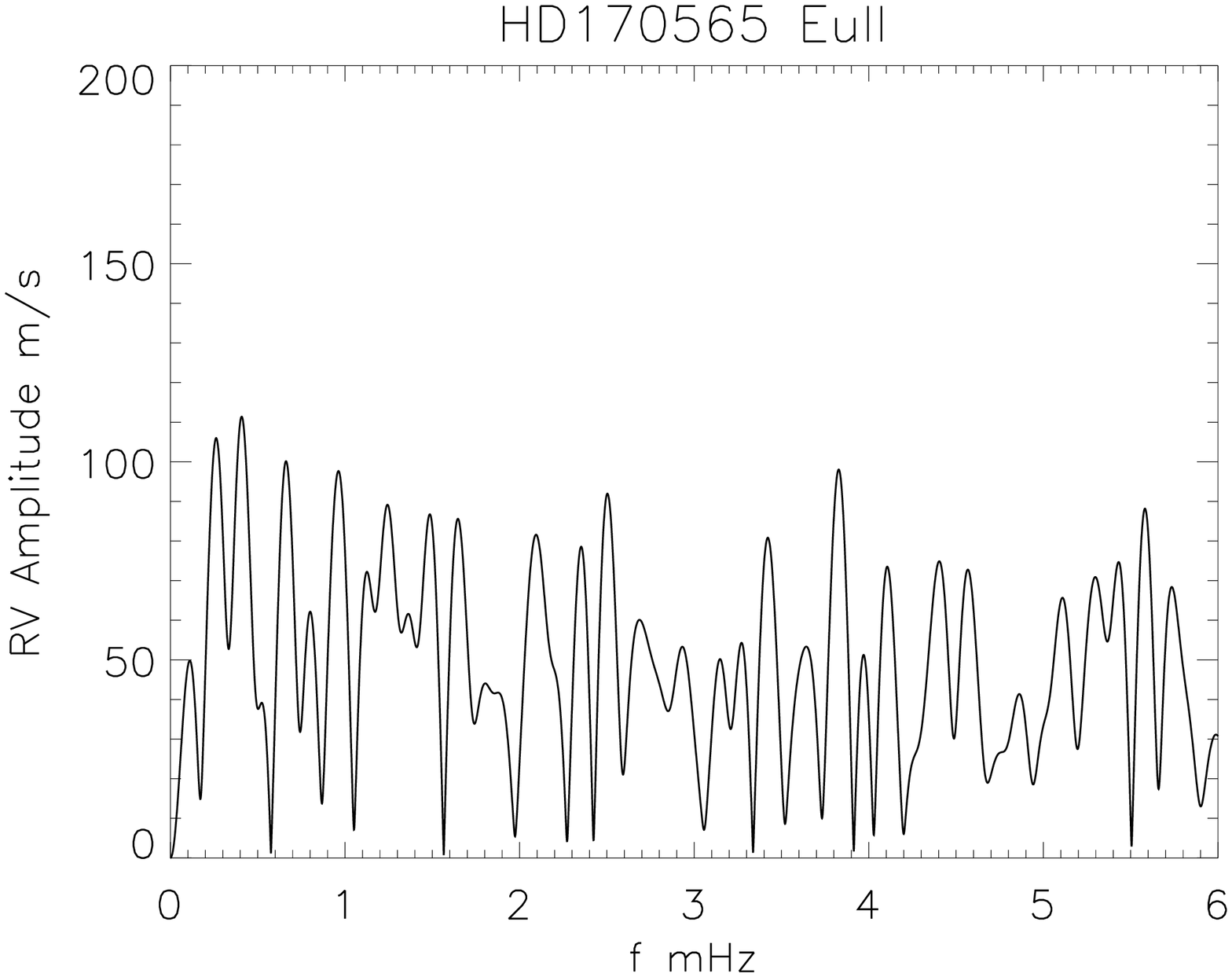}
  \includegraphics[height=5.6cm, angle=270]{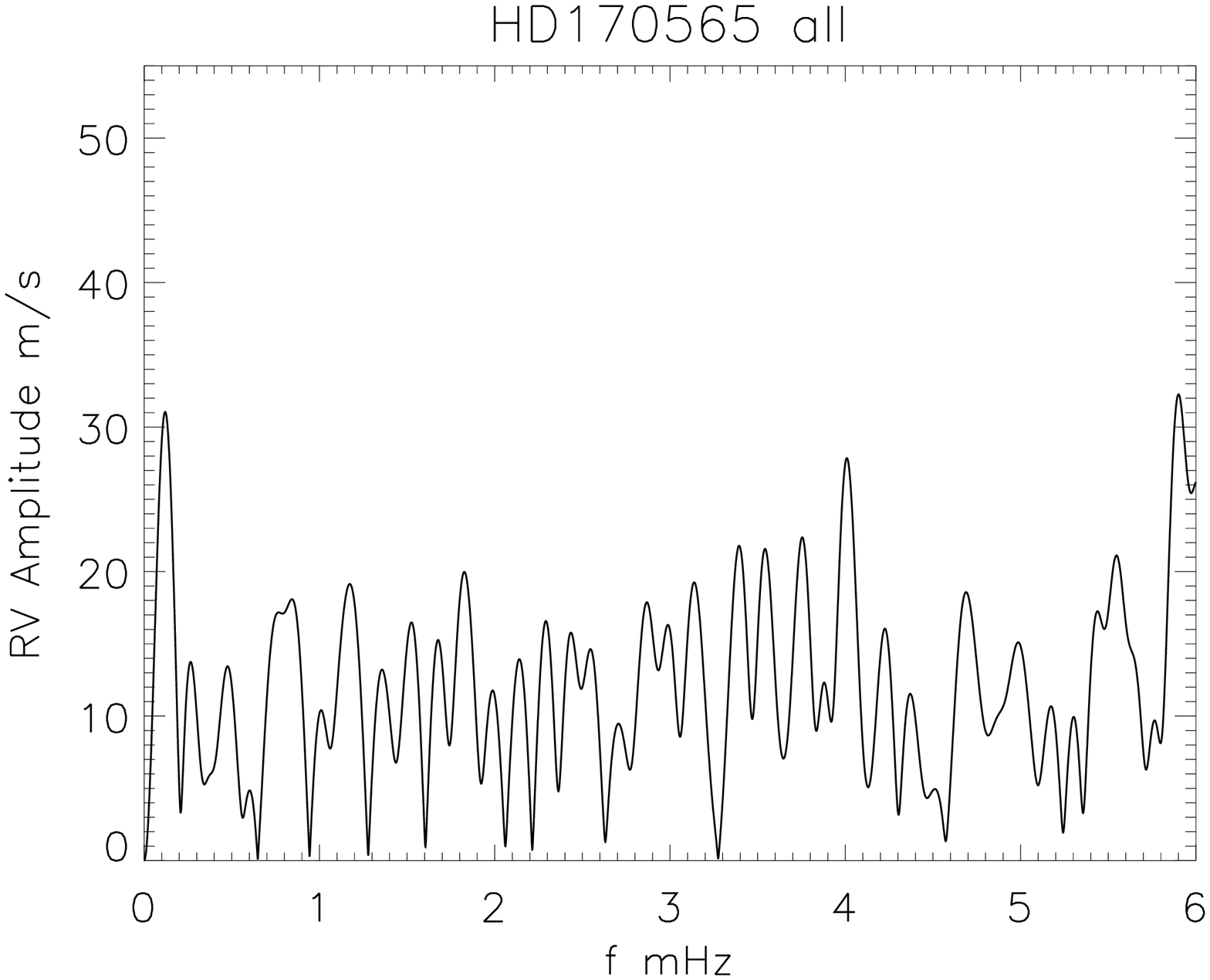}
  \includegraphics[height=5.6cm, angle=270]{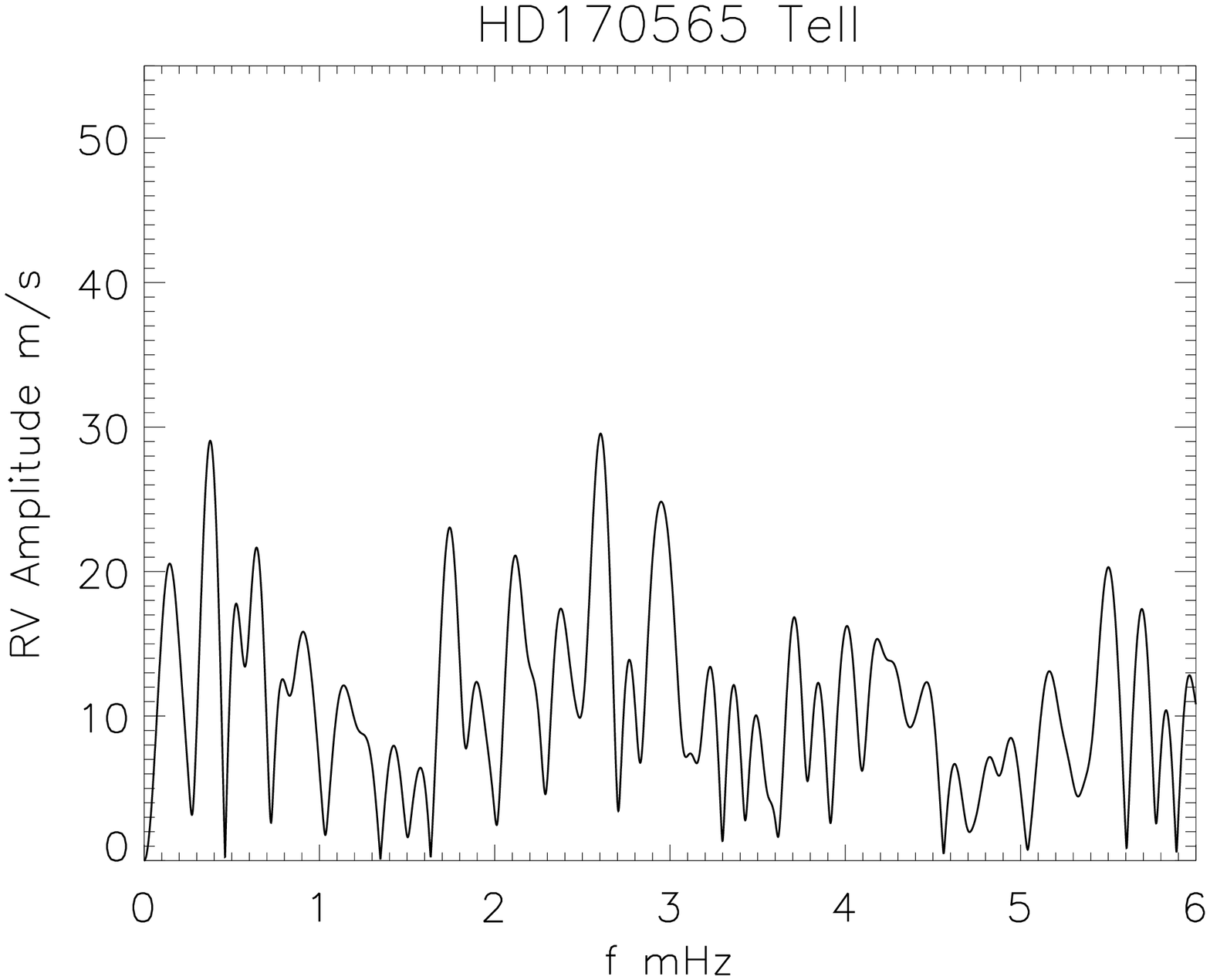}
  \caption{\label{fig:170565cog}Same as Fig.\,\ref{fig:107107cog} but
    for HD\,170565.  The Nyquist frequency is 4.7\,mHz.}
\end{figure*}

\begin{figure*}
  \vspace{3pt}
  \includegraphics[height=5.6cm,
  angle=270]{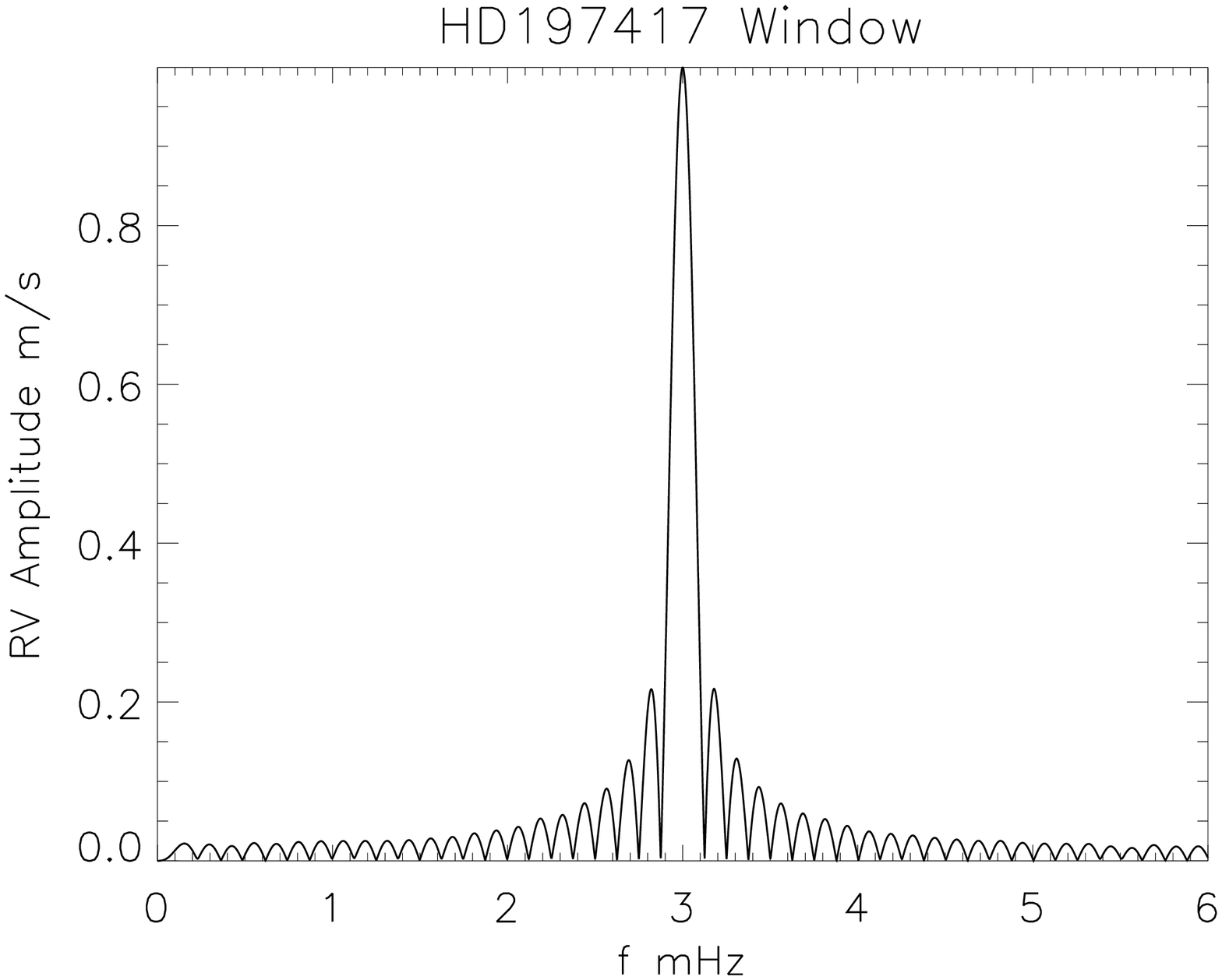}
  \includegraphics[height=5.6cm, angle=270]{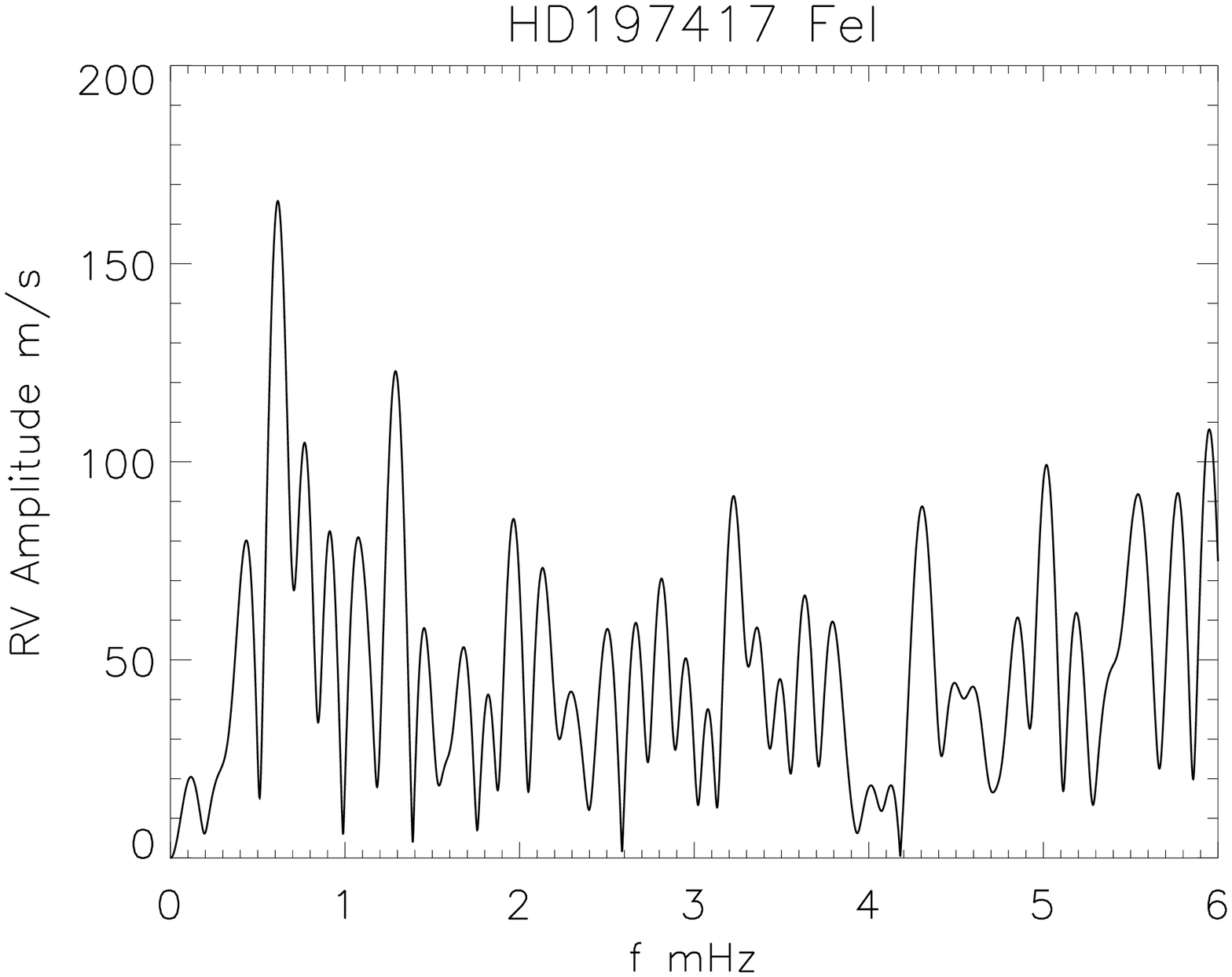}
  \includegraphics[height=5.6cm, angle=270]{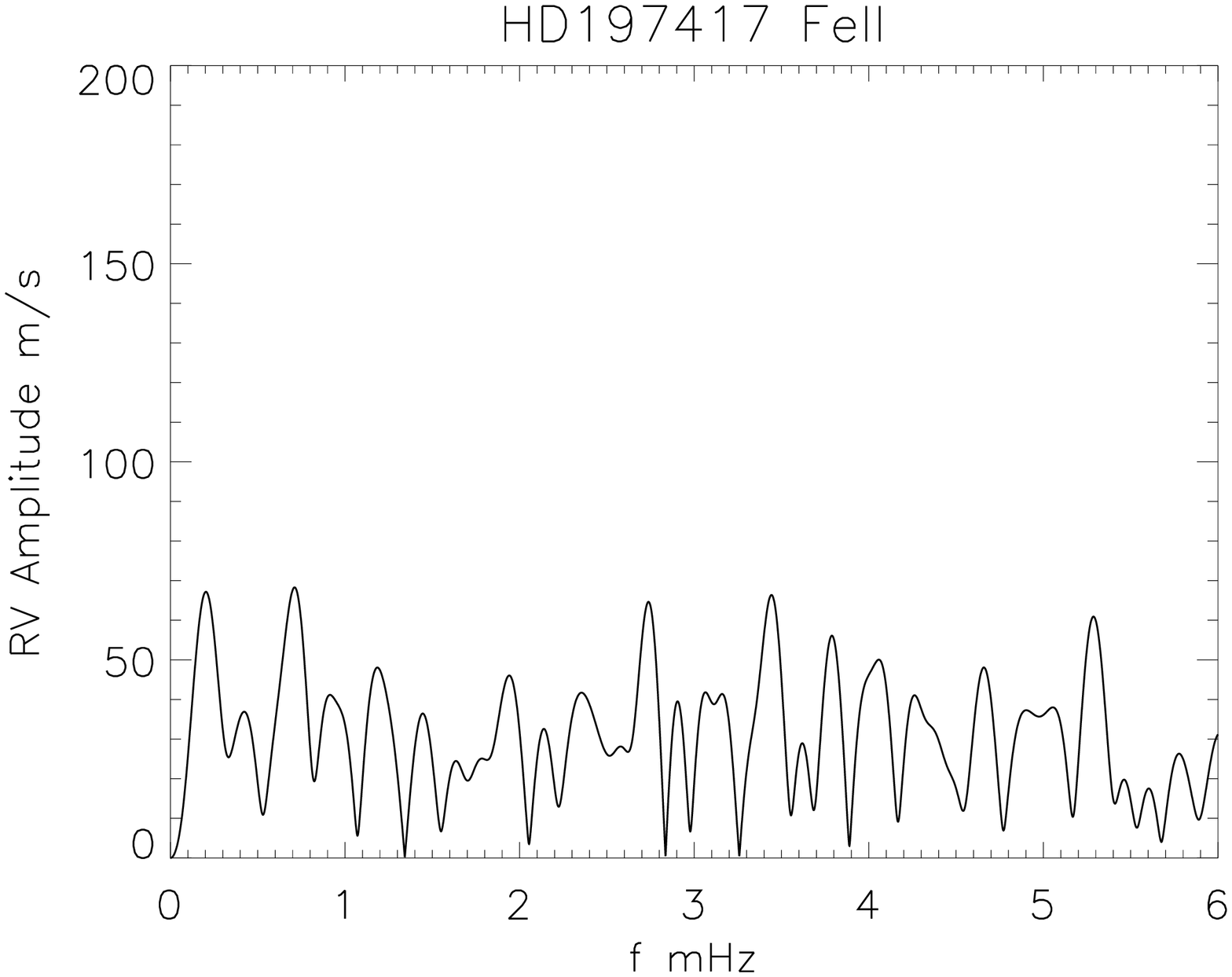}
  \includegraphics[height=5.6cm, angle=270]{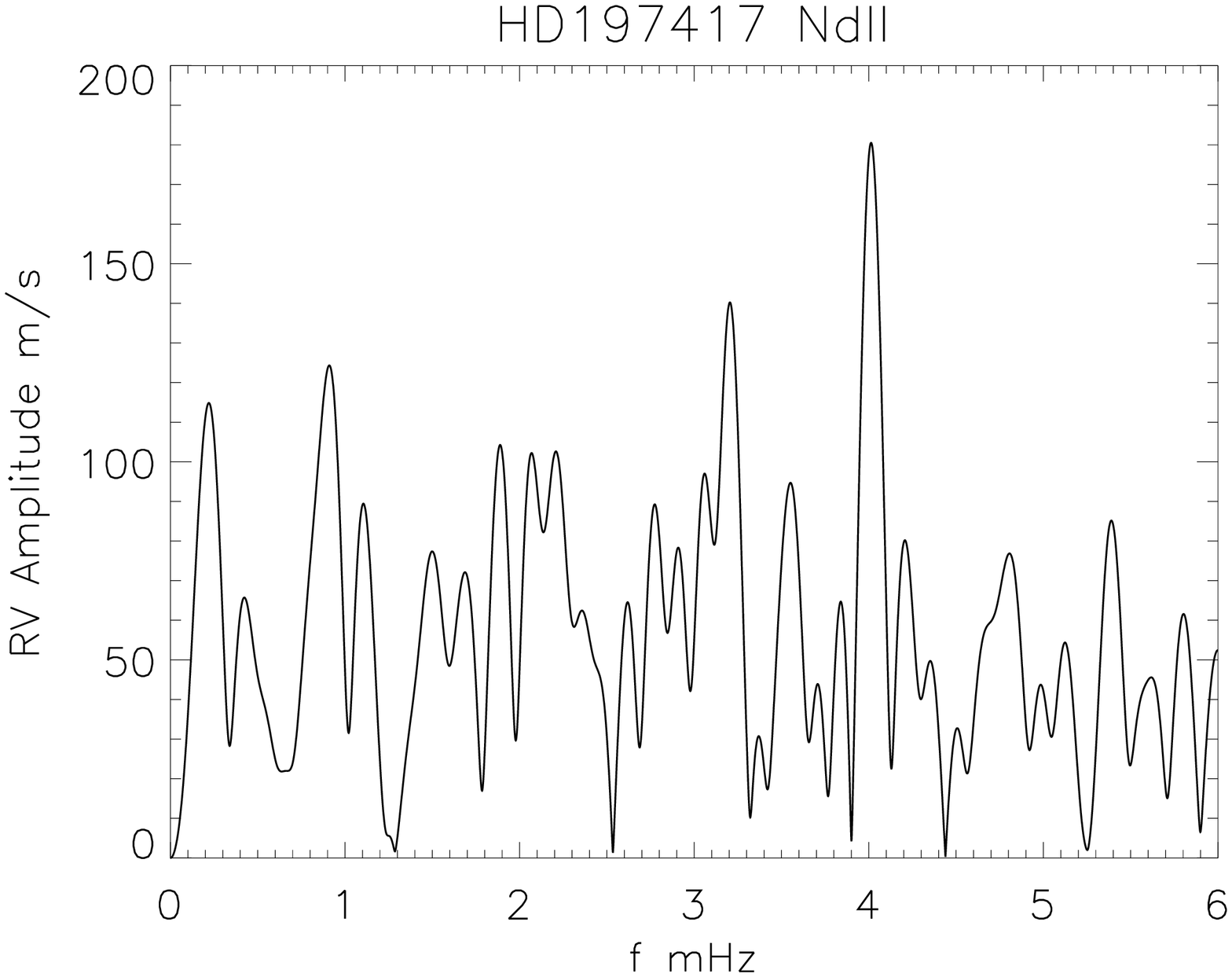}
  \includegraphics[height=5.6cm,
  angle=270]{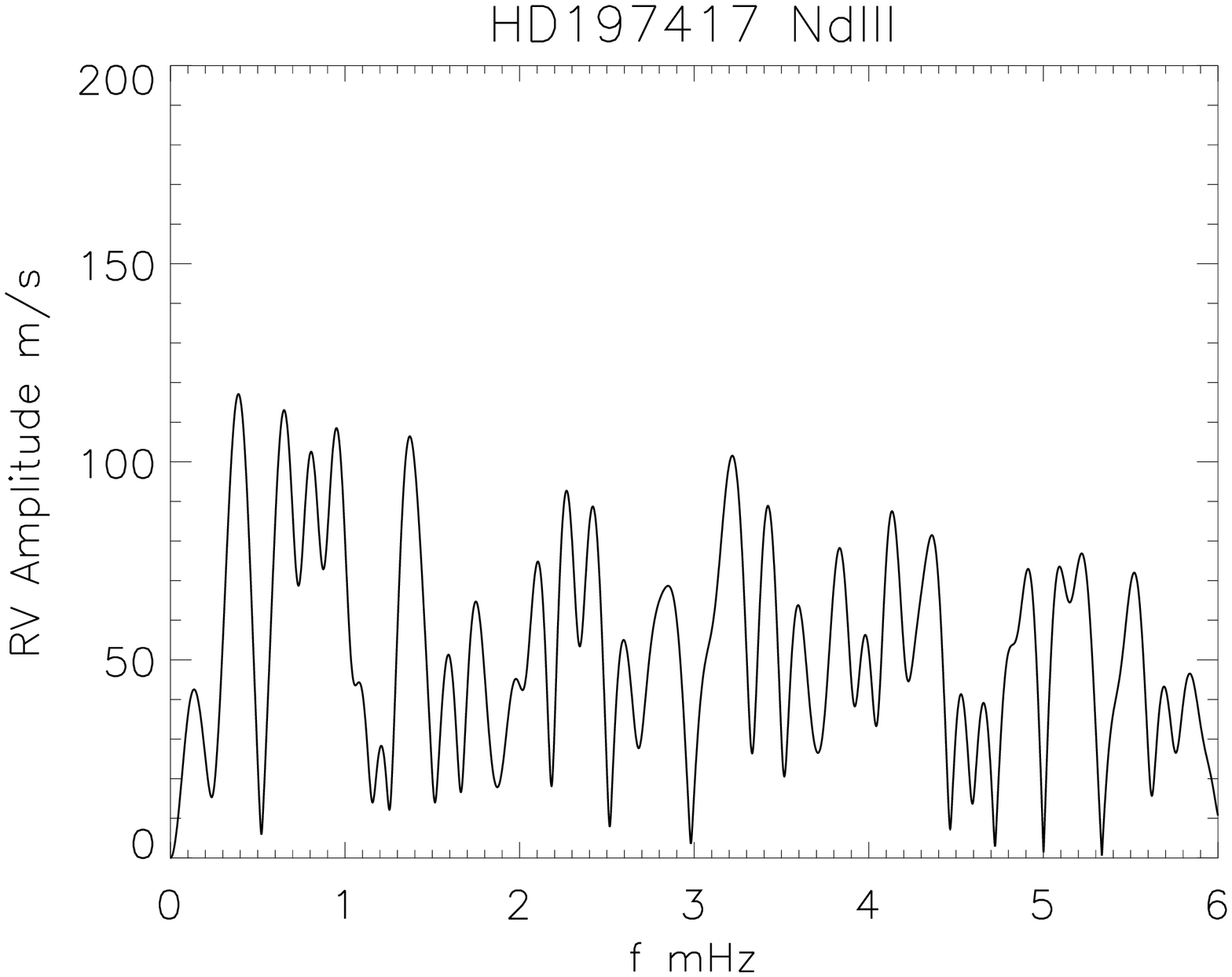}
  \includegraphics[height=5.6cm, angle=270]{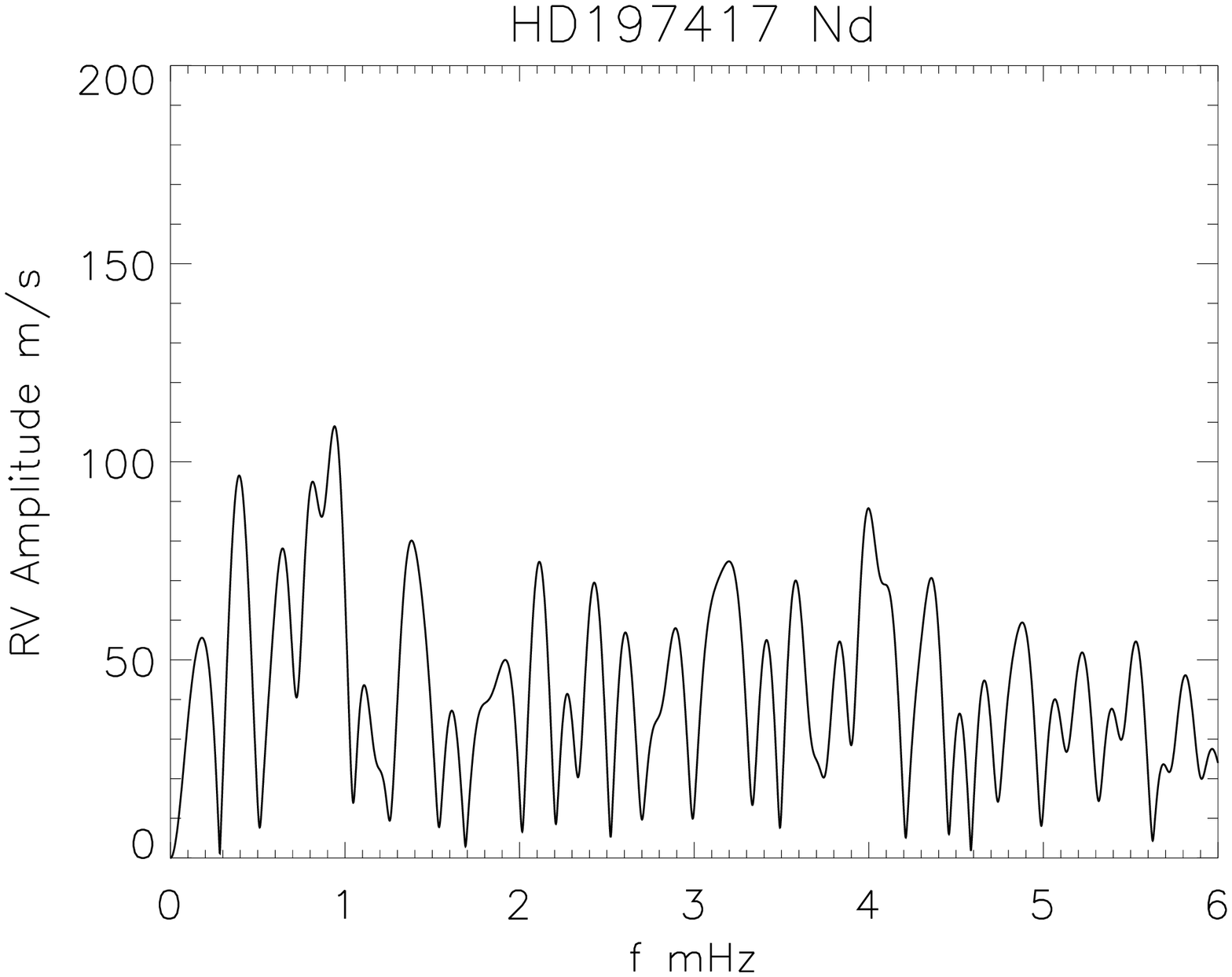}
  \includegraphics[height=5.6cm, angle=270]{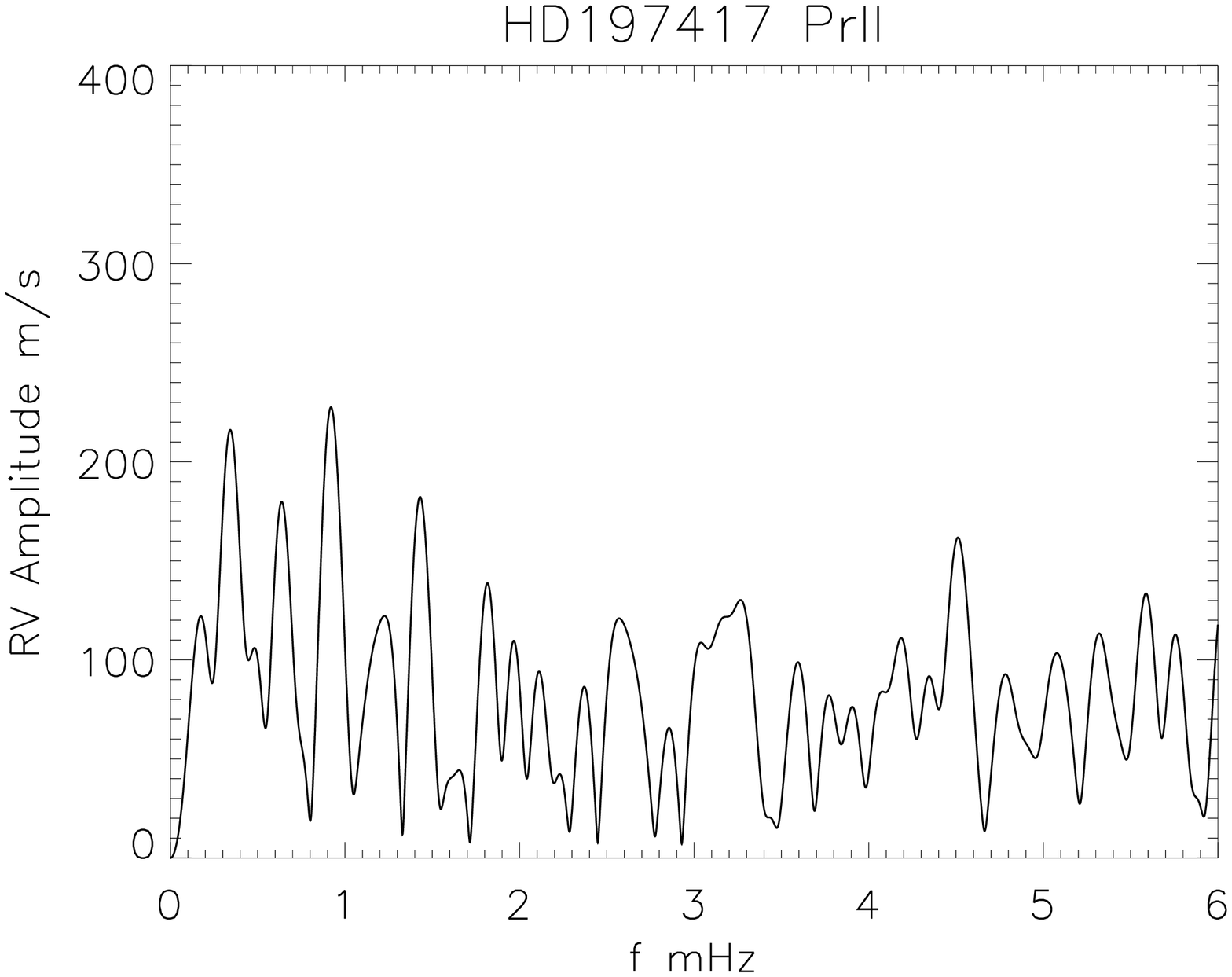}
  \includegraphics[height=5.6cm,
  angle=270]{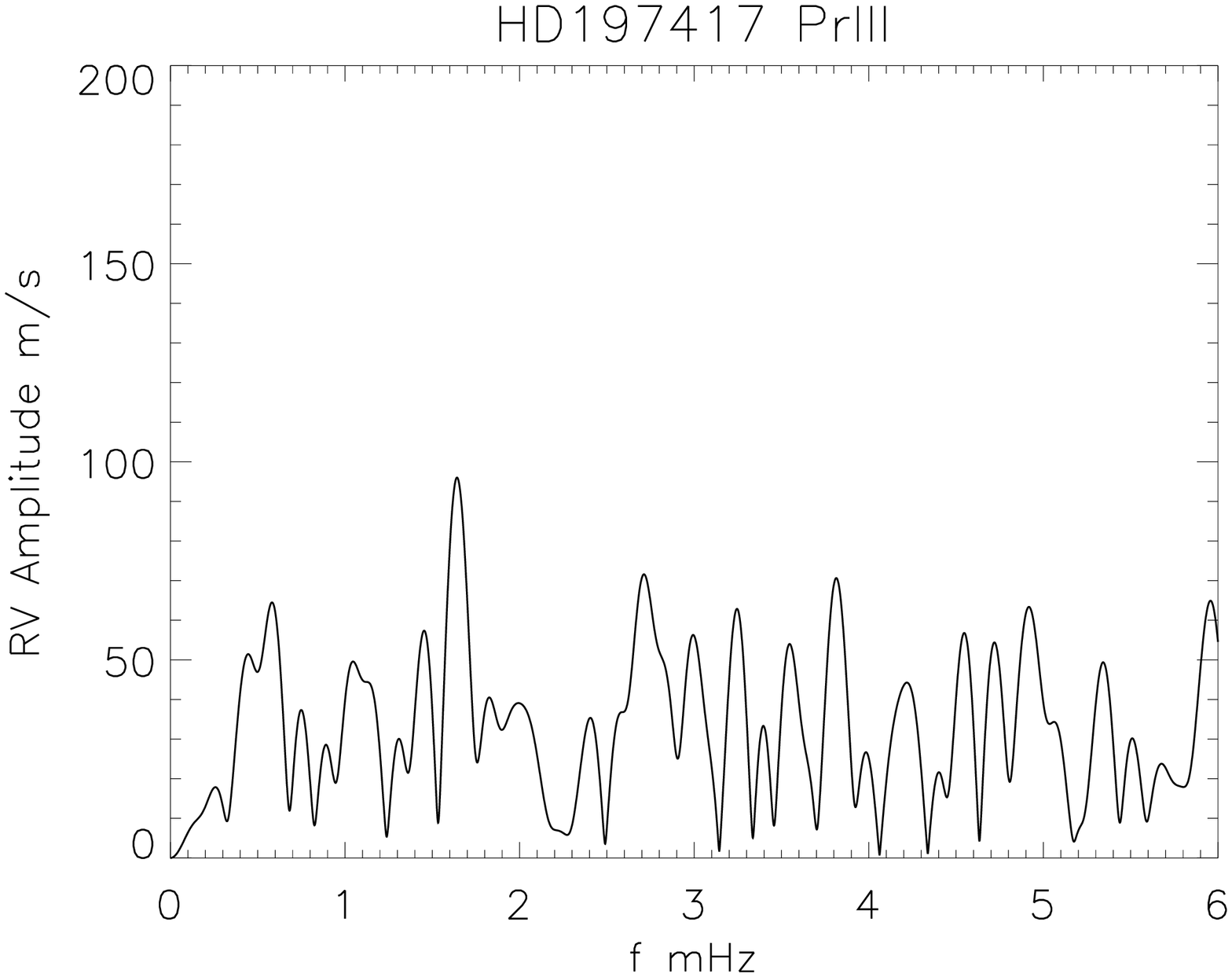}
  \includegraphics[height=5.6cm, angle=270]{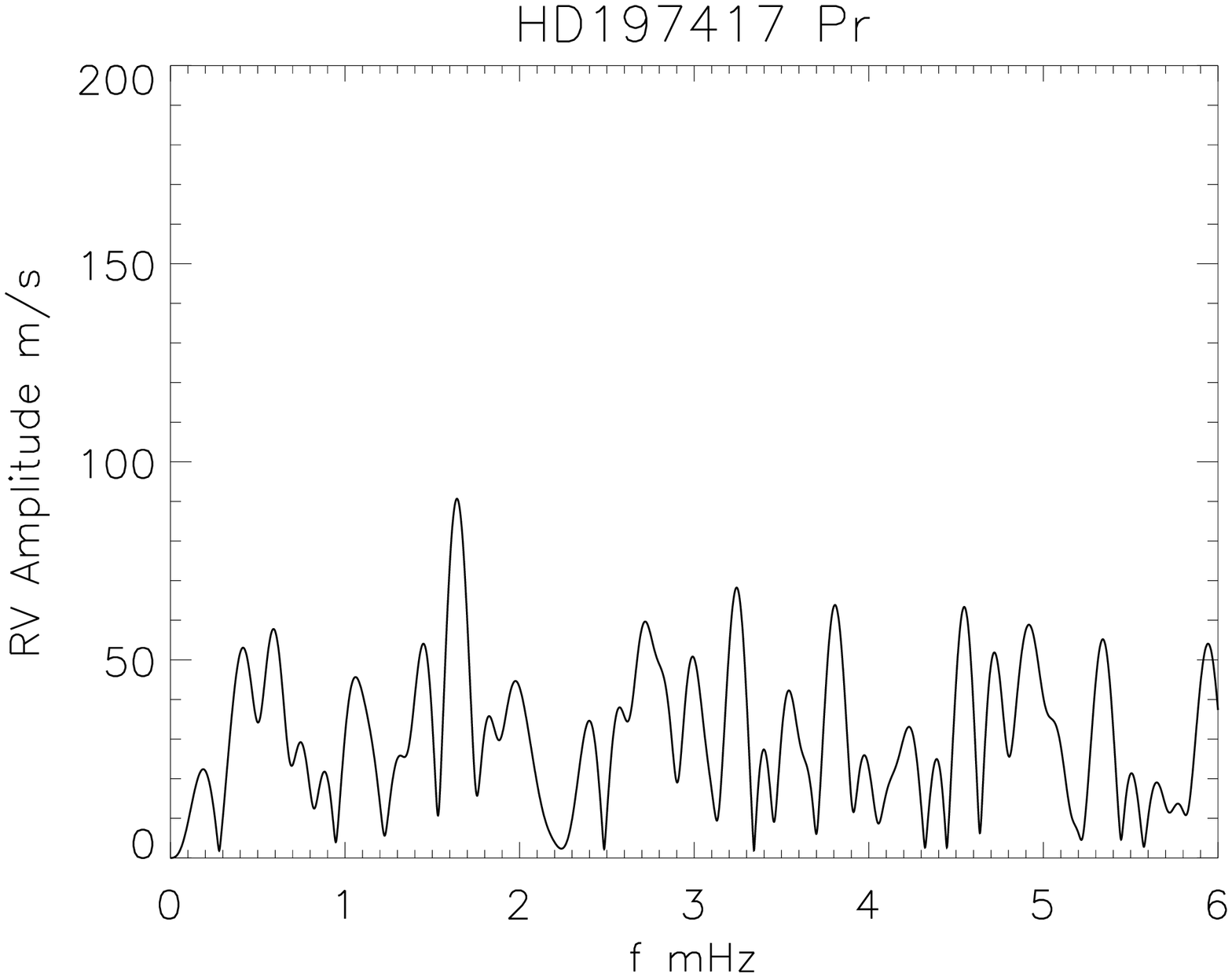}
  \includegraphics[height=5.6cm, angle=270]{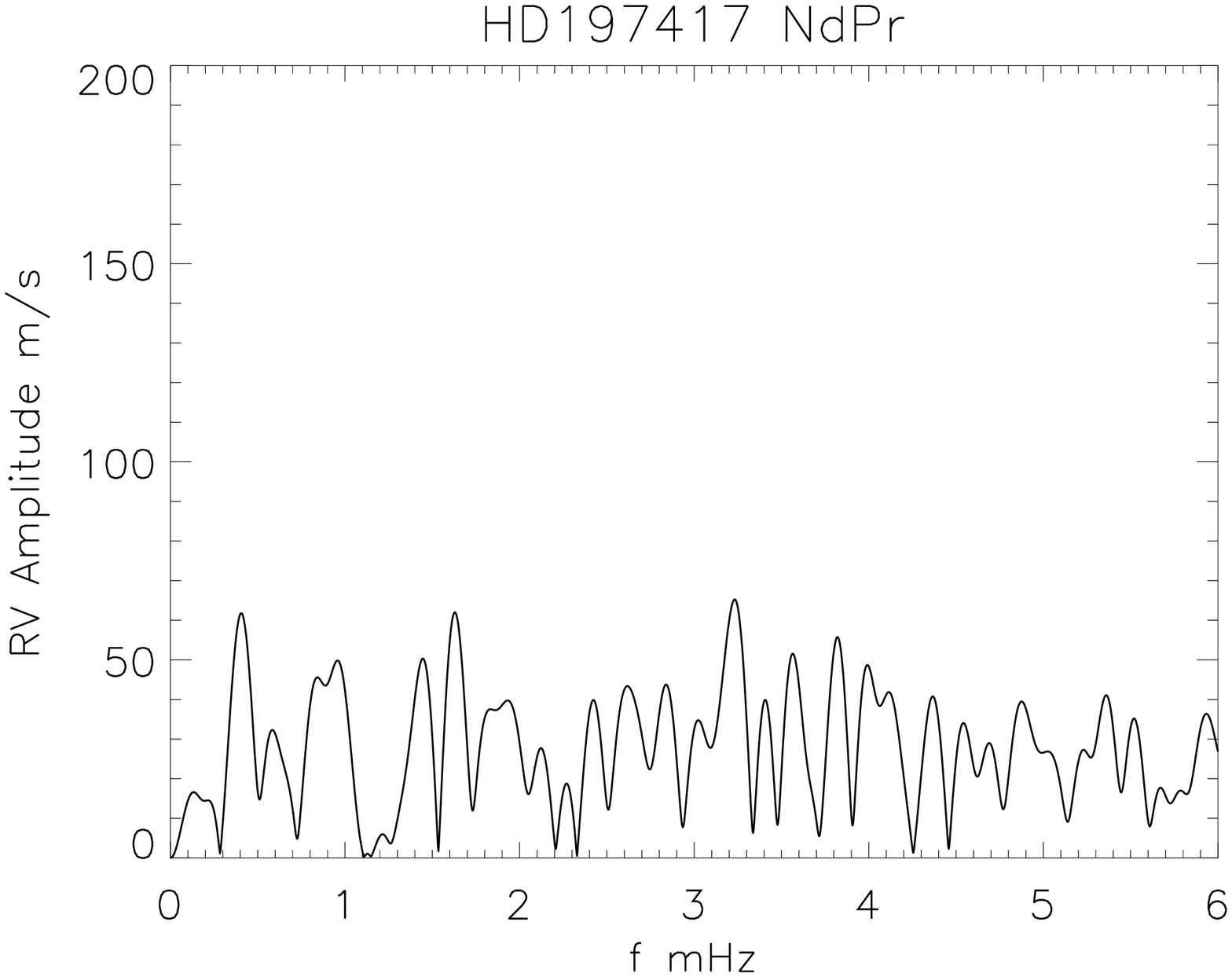}
  \includegraphics[height=5.6cm, angle=270]{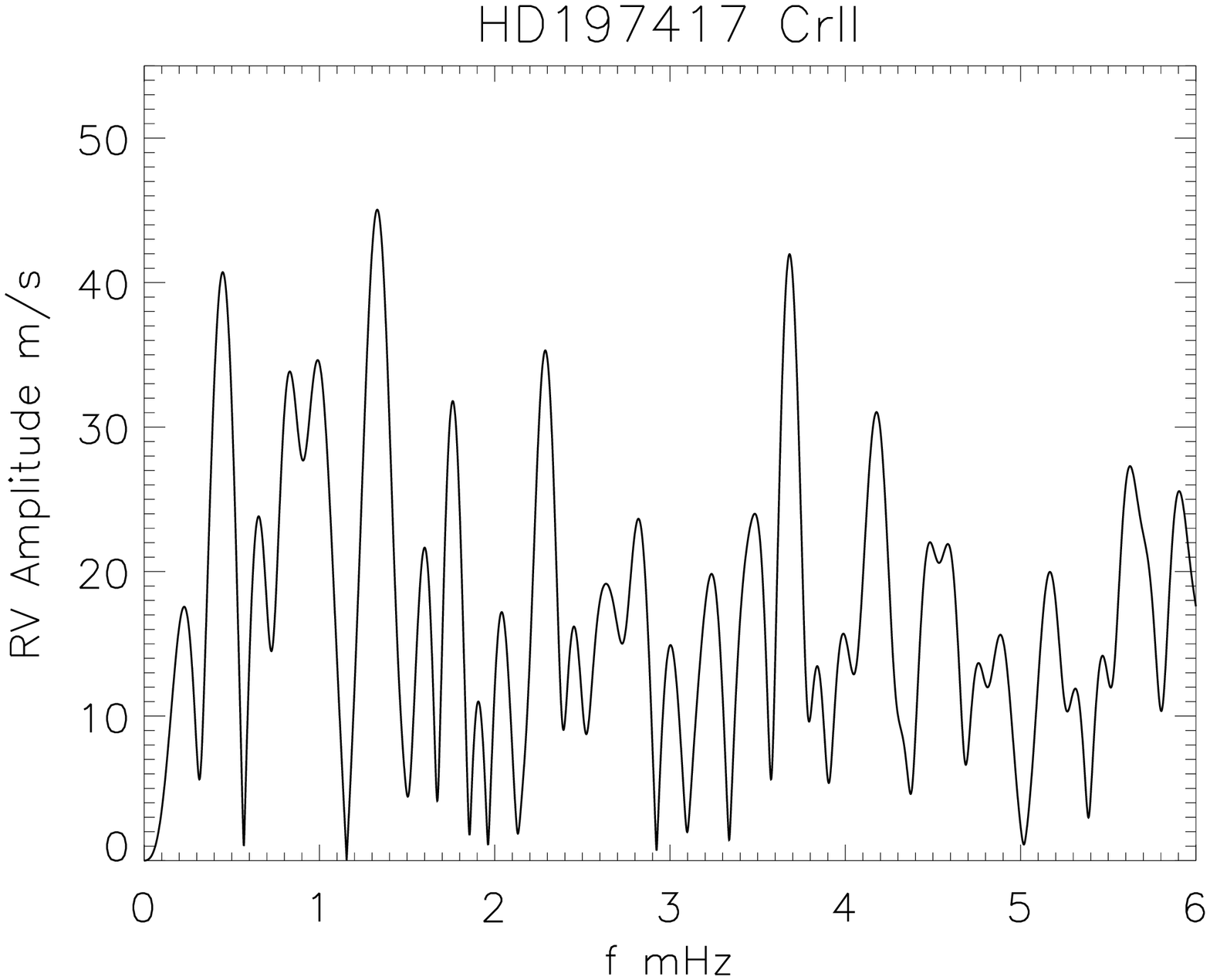}
  \includegraphics[height=5.6cm, angle=270]{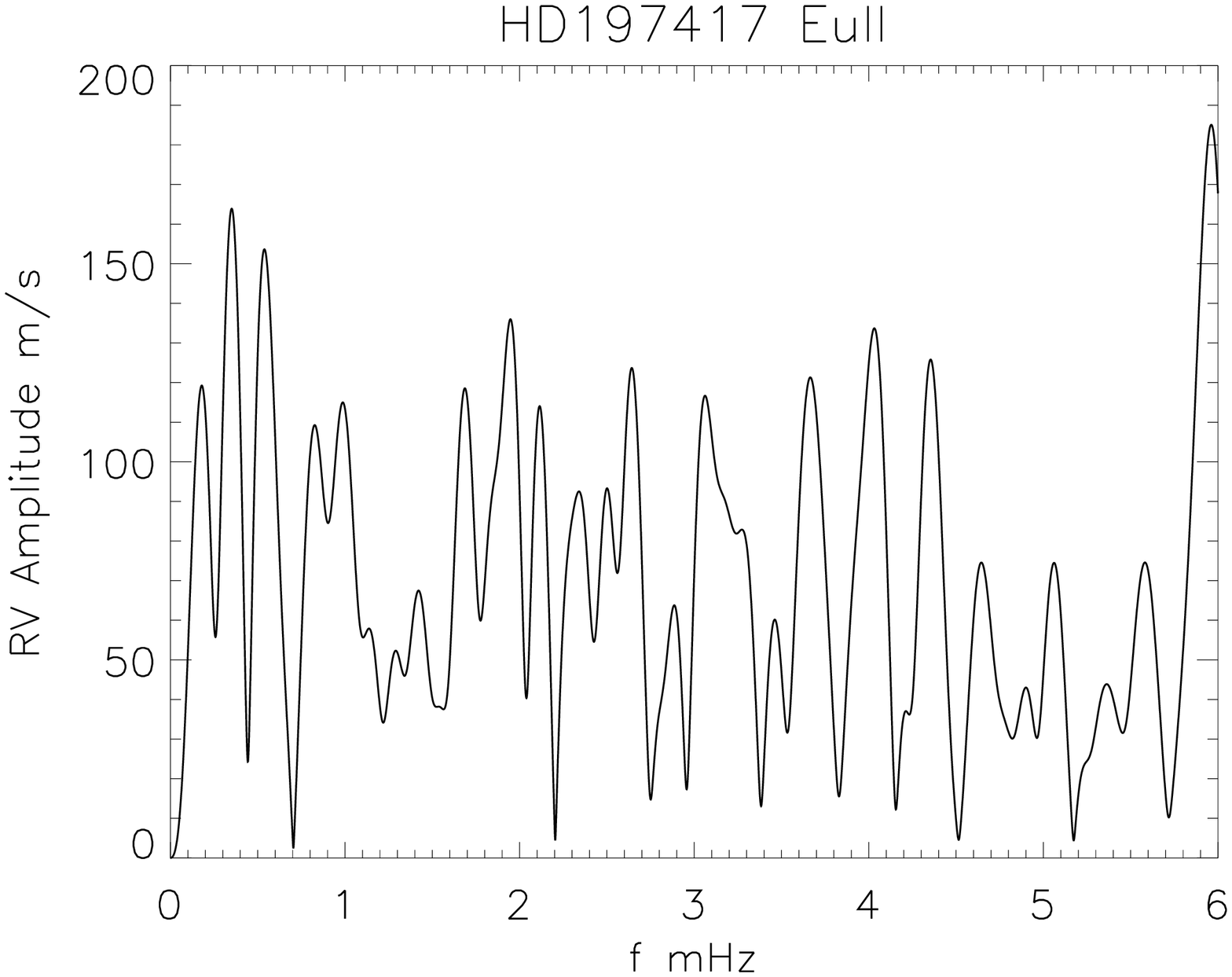}
  \includegraphics[height=5.6cm, angle=270]{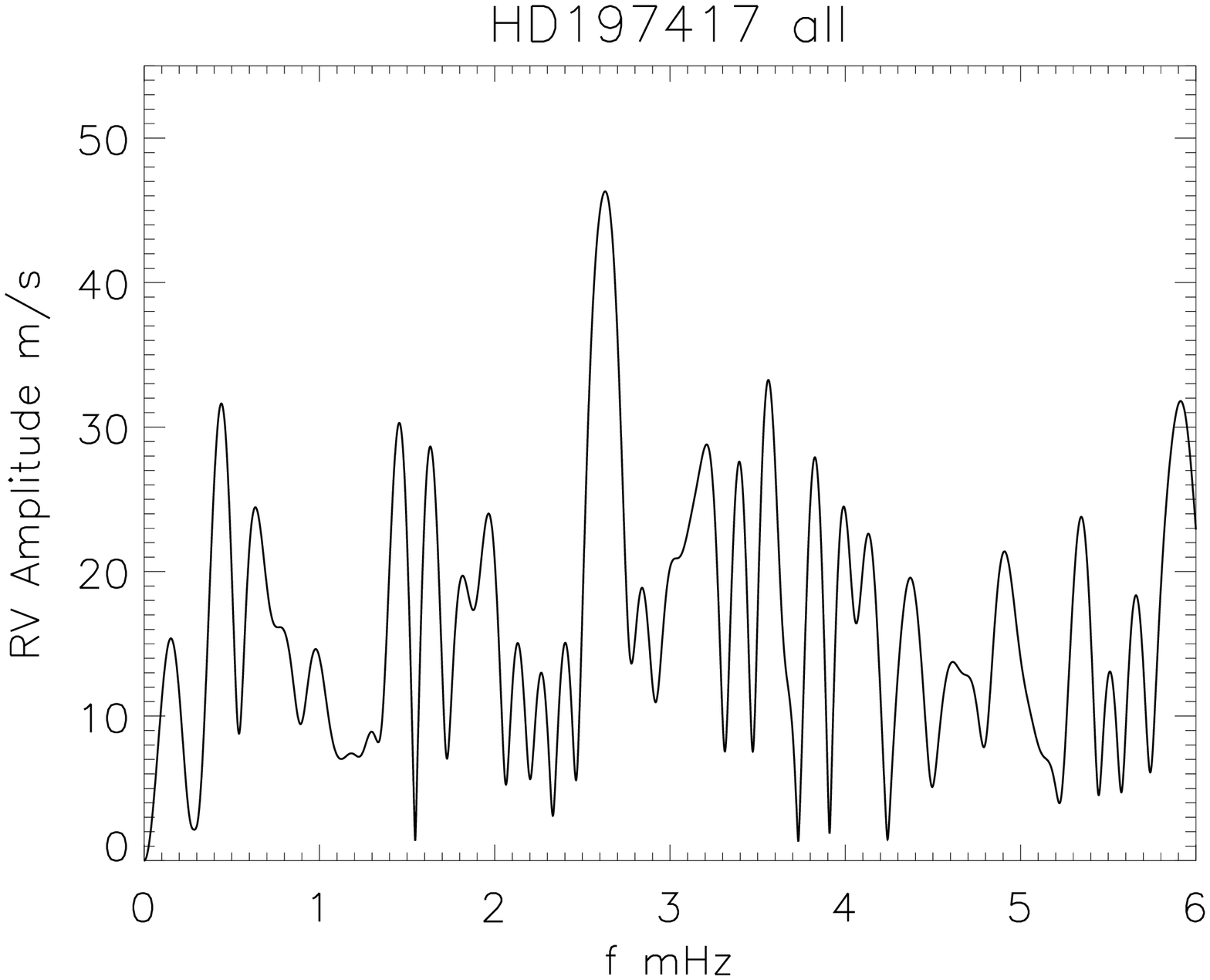}
  \includegraphics[height=5.6cm, angle=270]{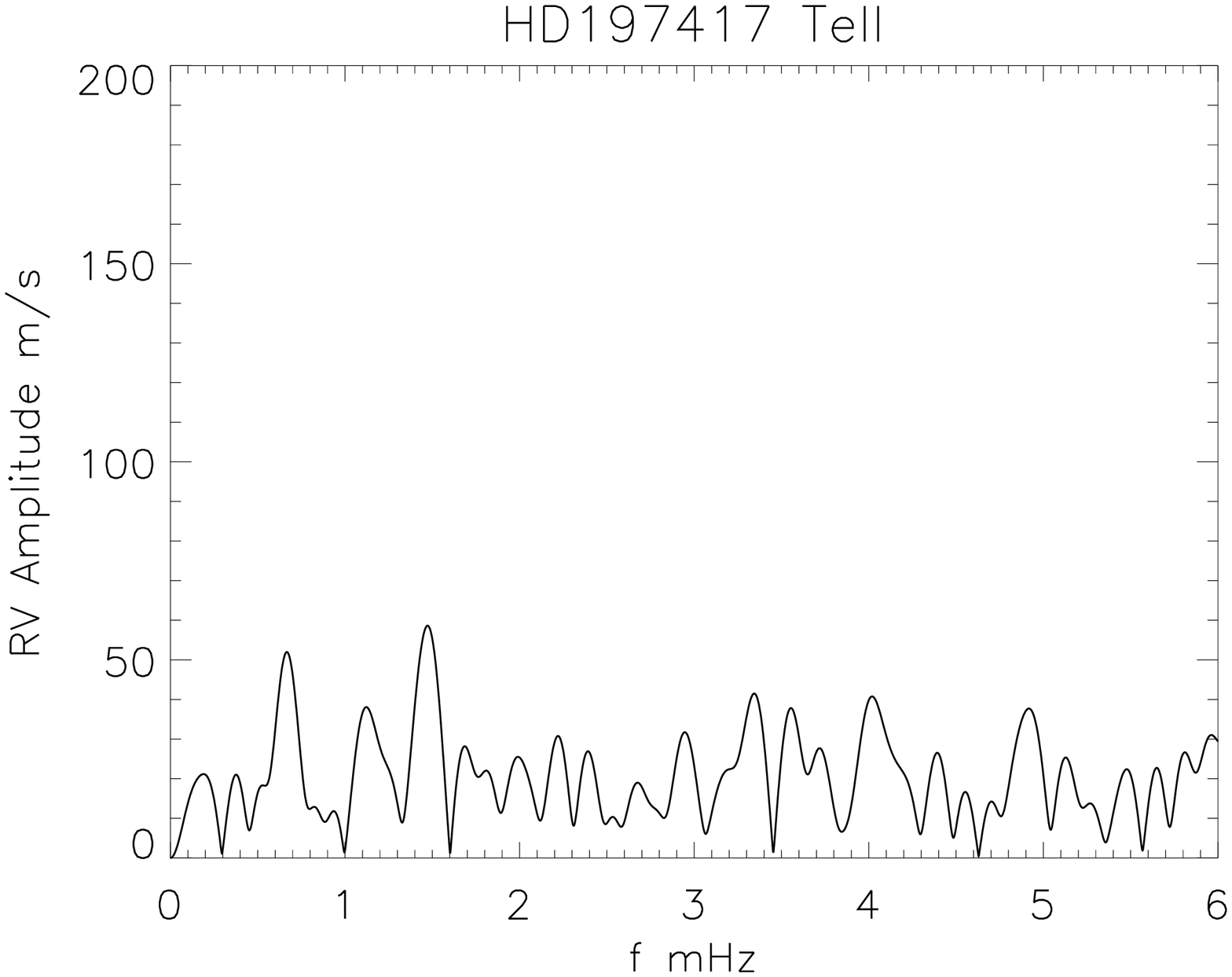}
  \hspace{5.6cm}
  \caption{\label{fig:197417cog}Same as Fig.\,\ref{fig:107107cog} but
    for HD\,197417.  }
\end{figure*}

\begin{figure*}
  \vspace{3pt}
  \includegraphics[height=5.6cm,
  angle=270]{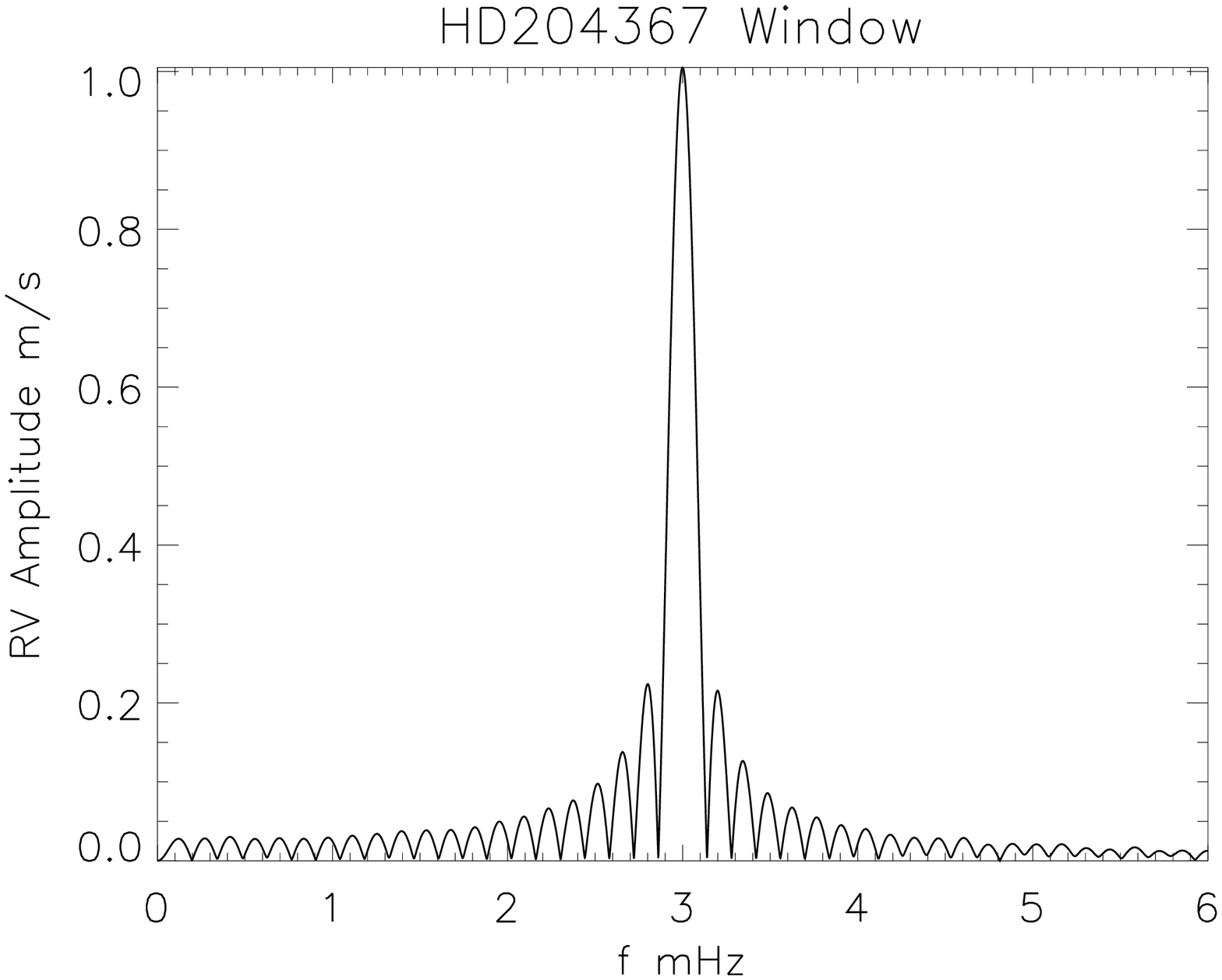}
  \includegraphics[height=5.6cm, angle=270]{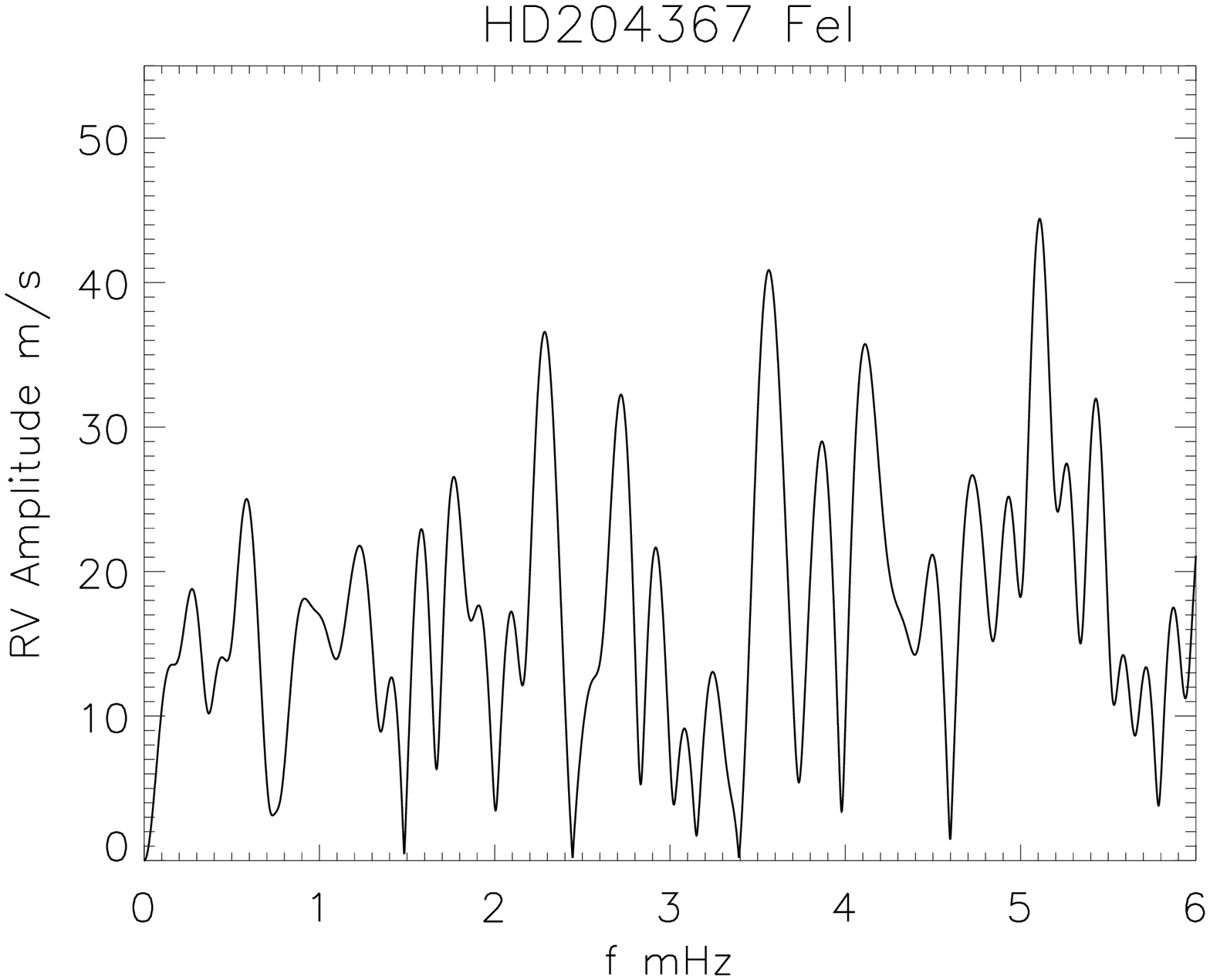}
  \includegraphics[height=5.6cm, angle=270]{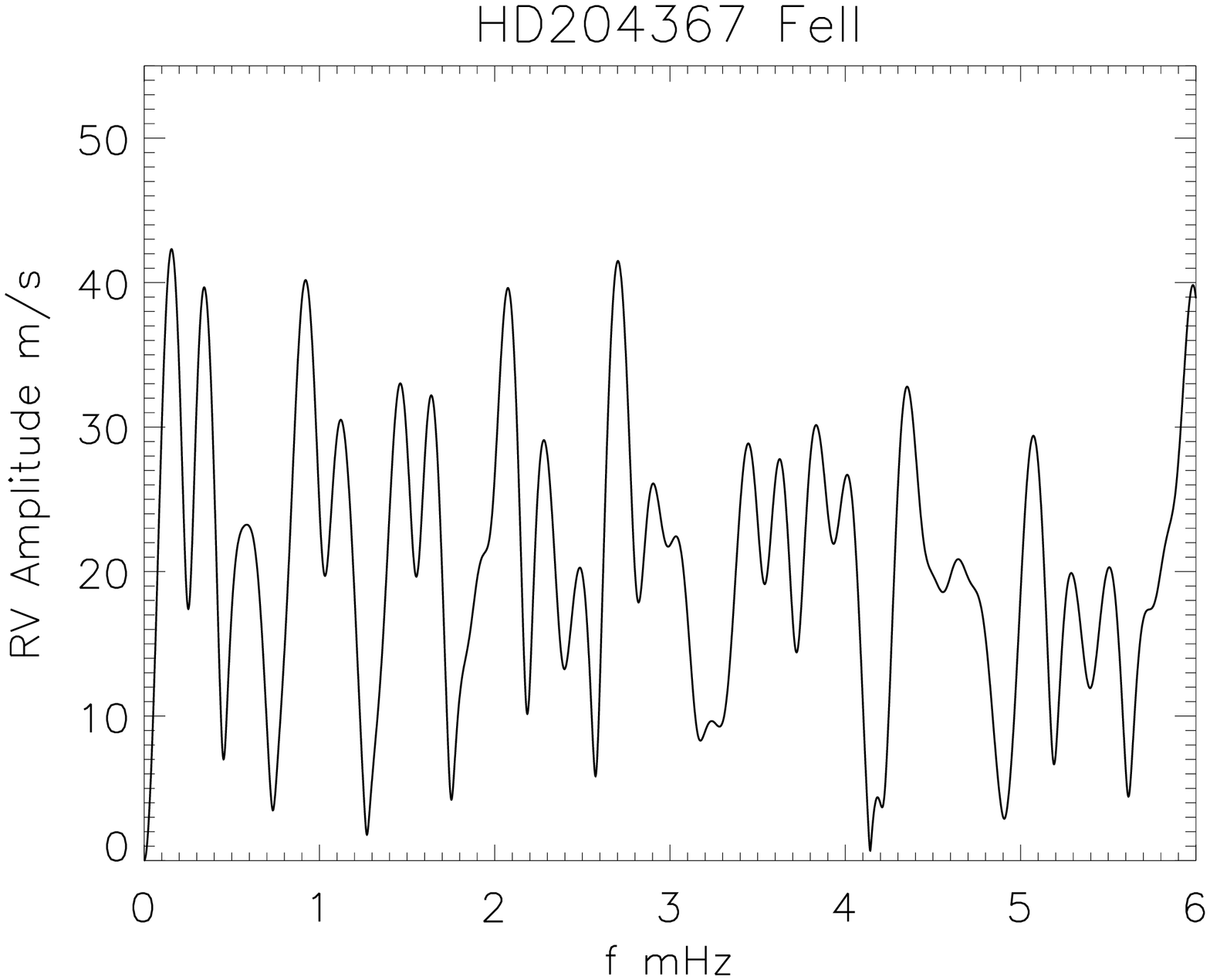}
  \includegraphics[height=5.6cm, angle=270]{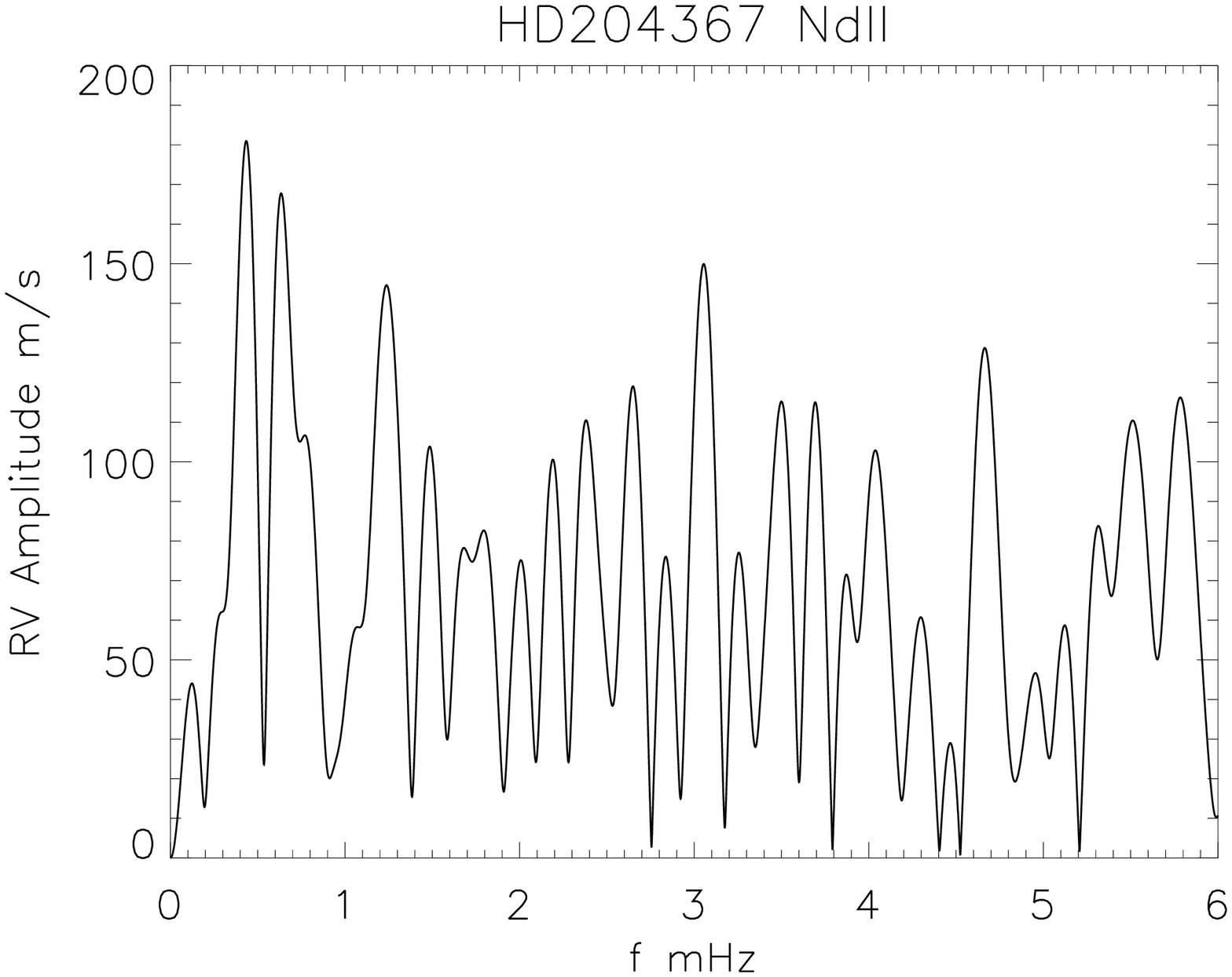}
  \includegraphics[height=5.6cm,
  angle=270]{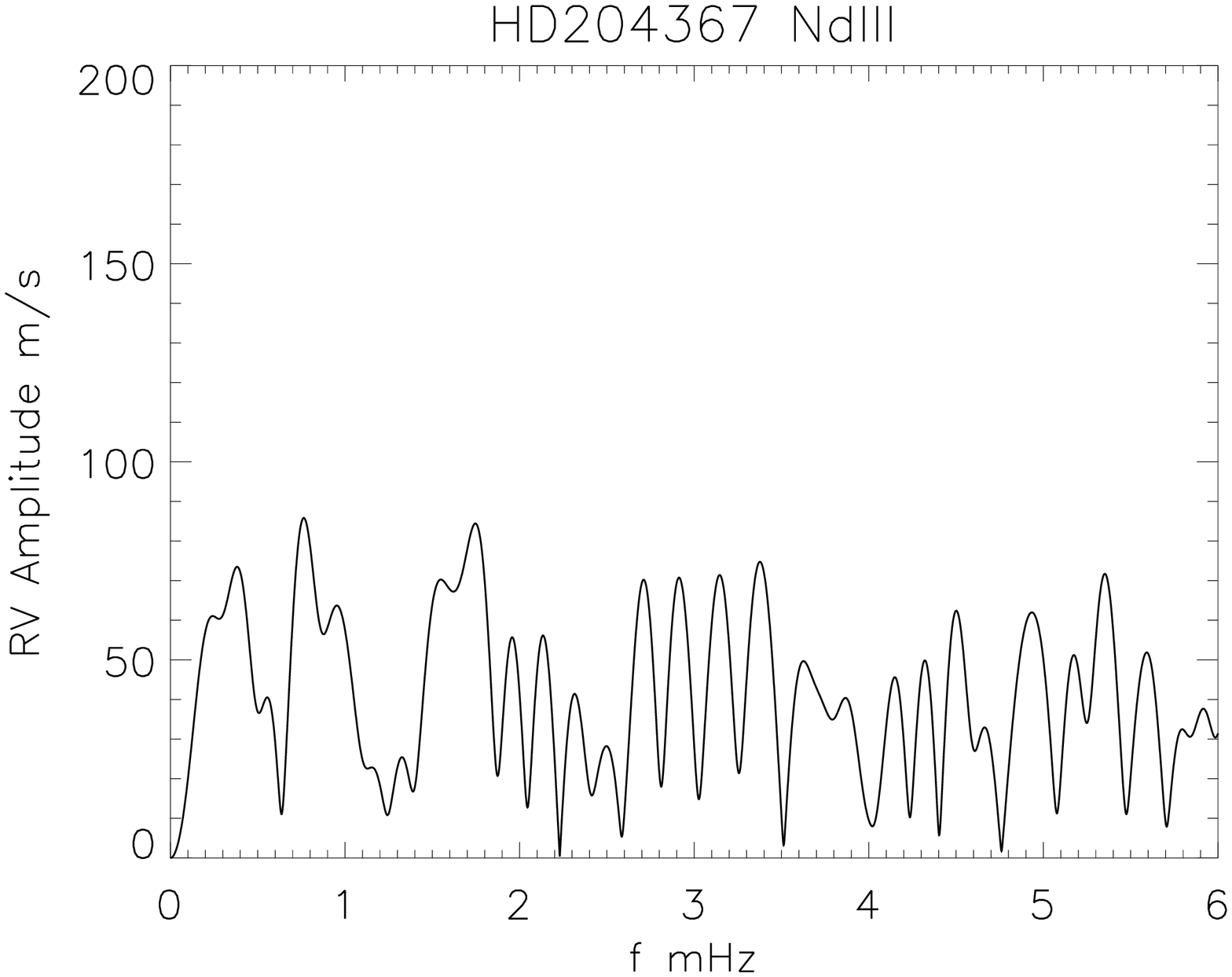}
  \includegraphics[height=5.6cm, angle=270]{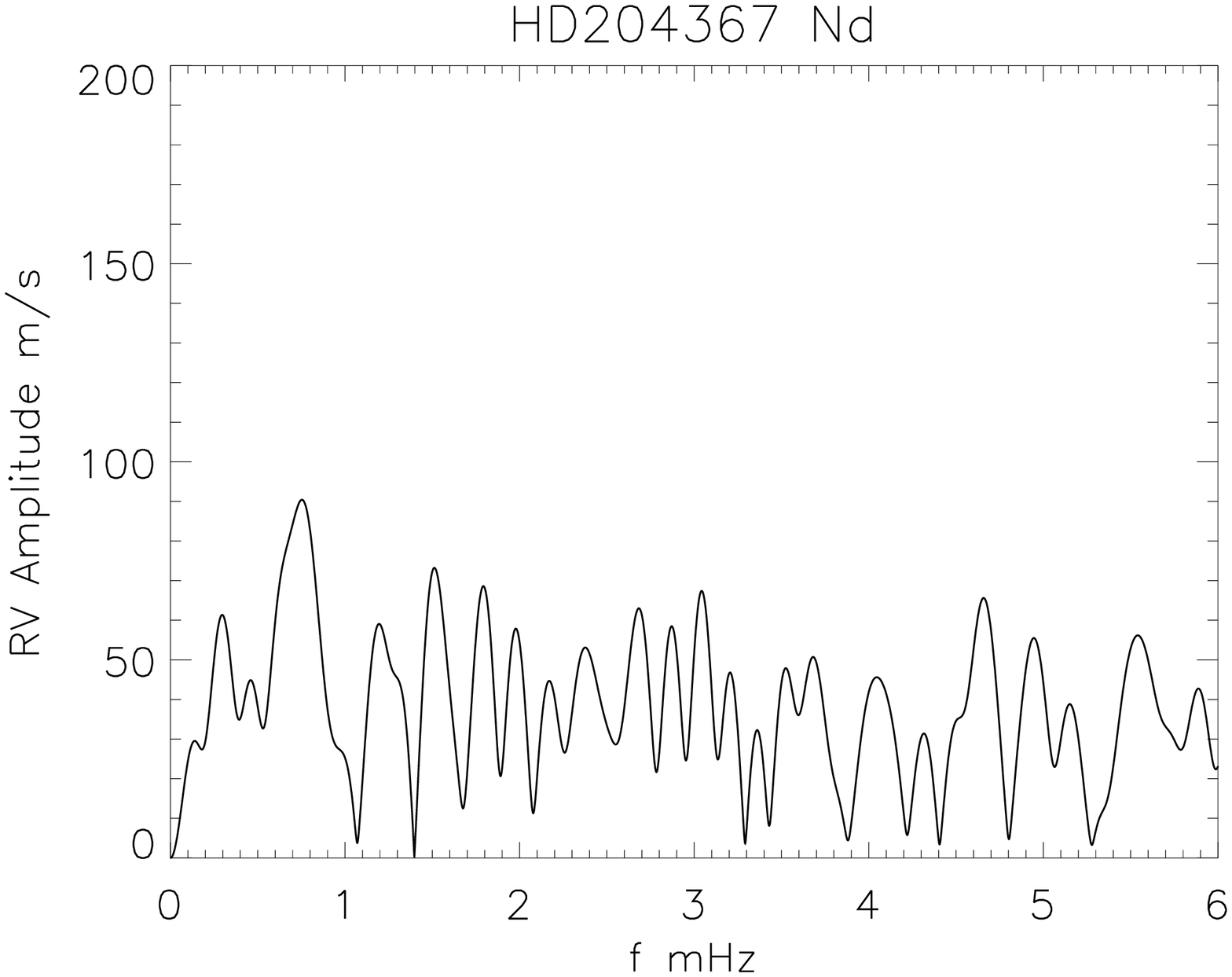}
  \includegraphics[height=5.6cm, angle=270]{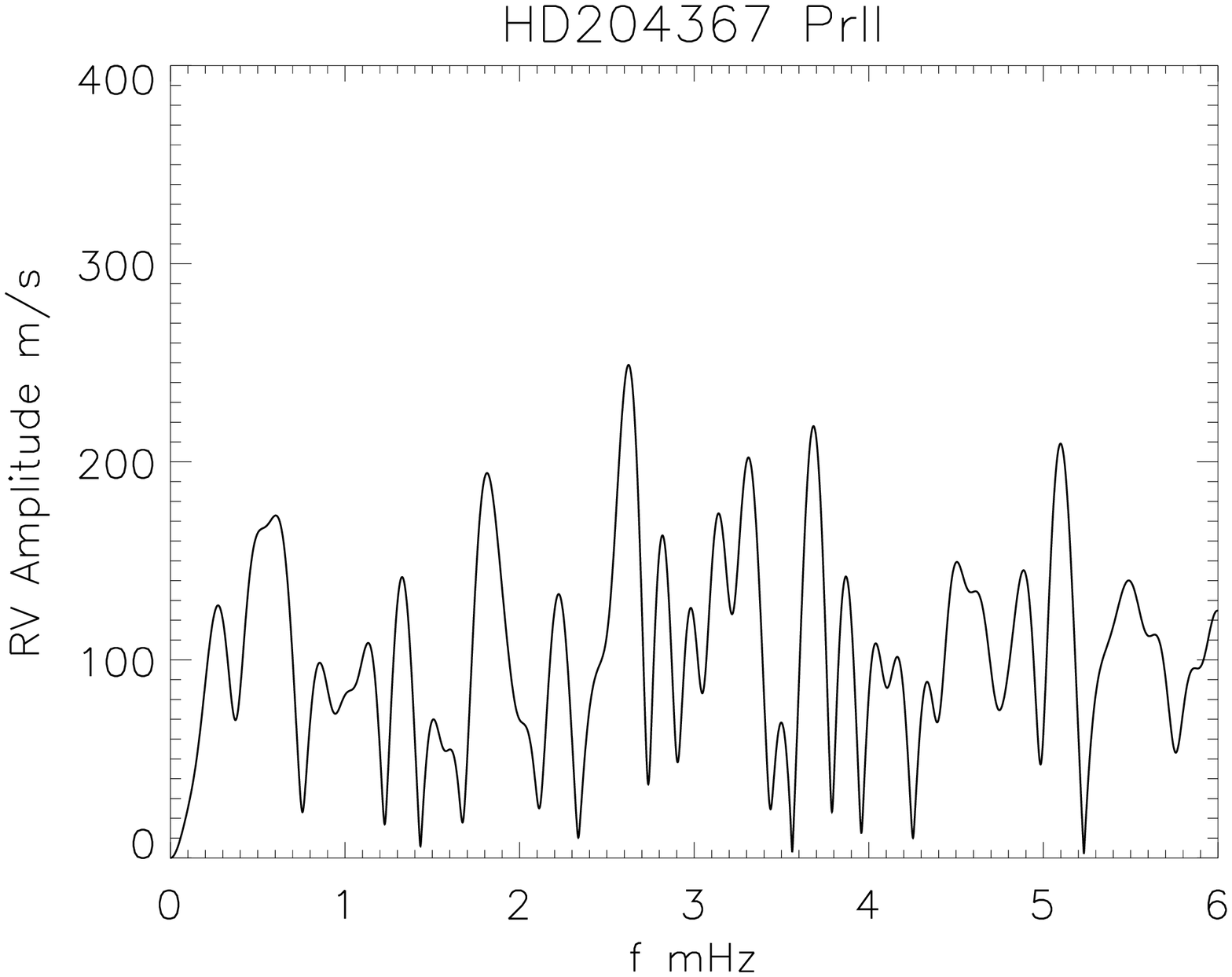}
  \includegraphics[height=5.6cm,
  angle=270]{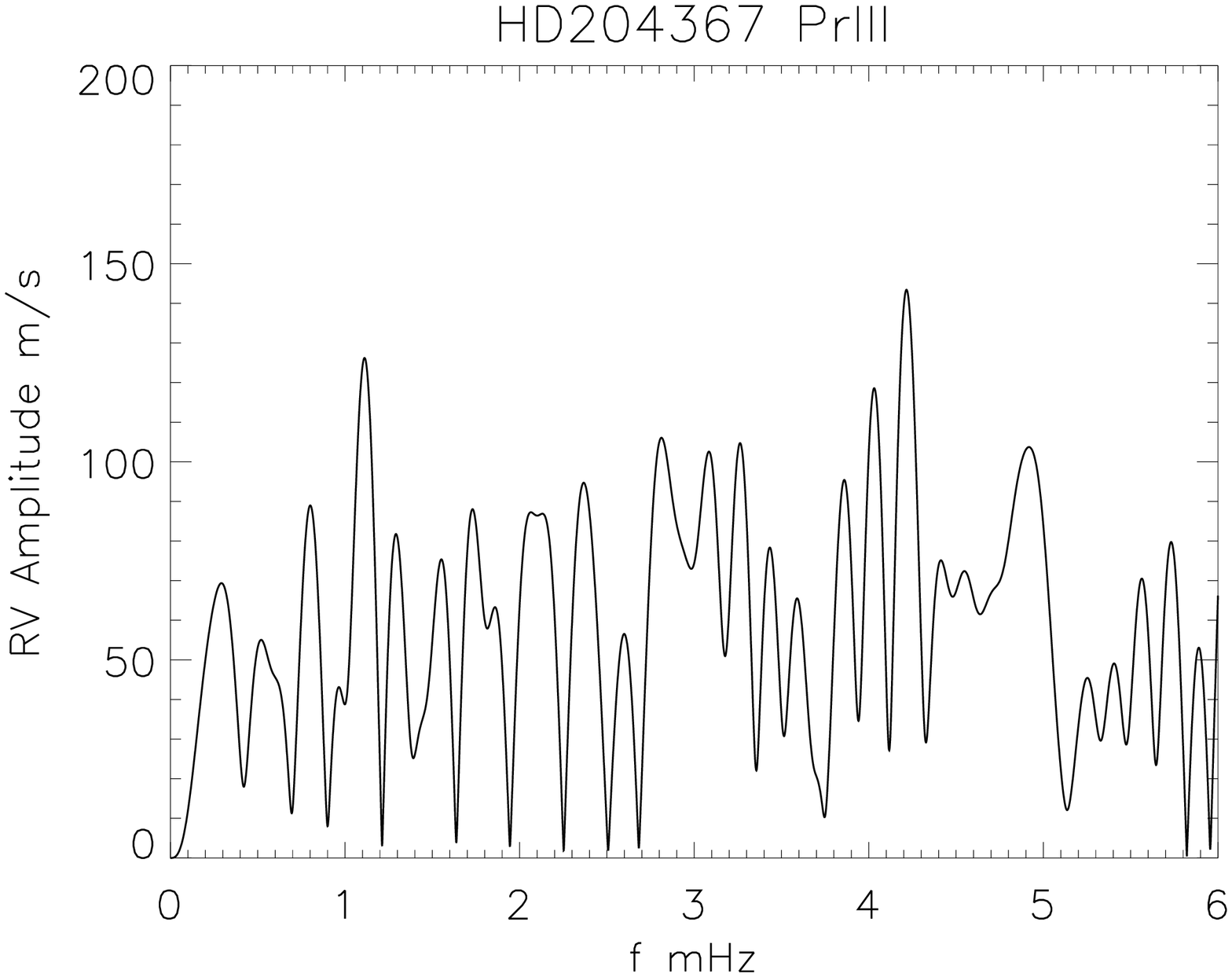}
  \includegraphics[height=5.6cm, angle=270]{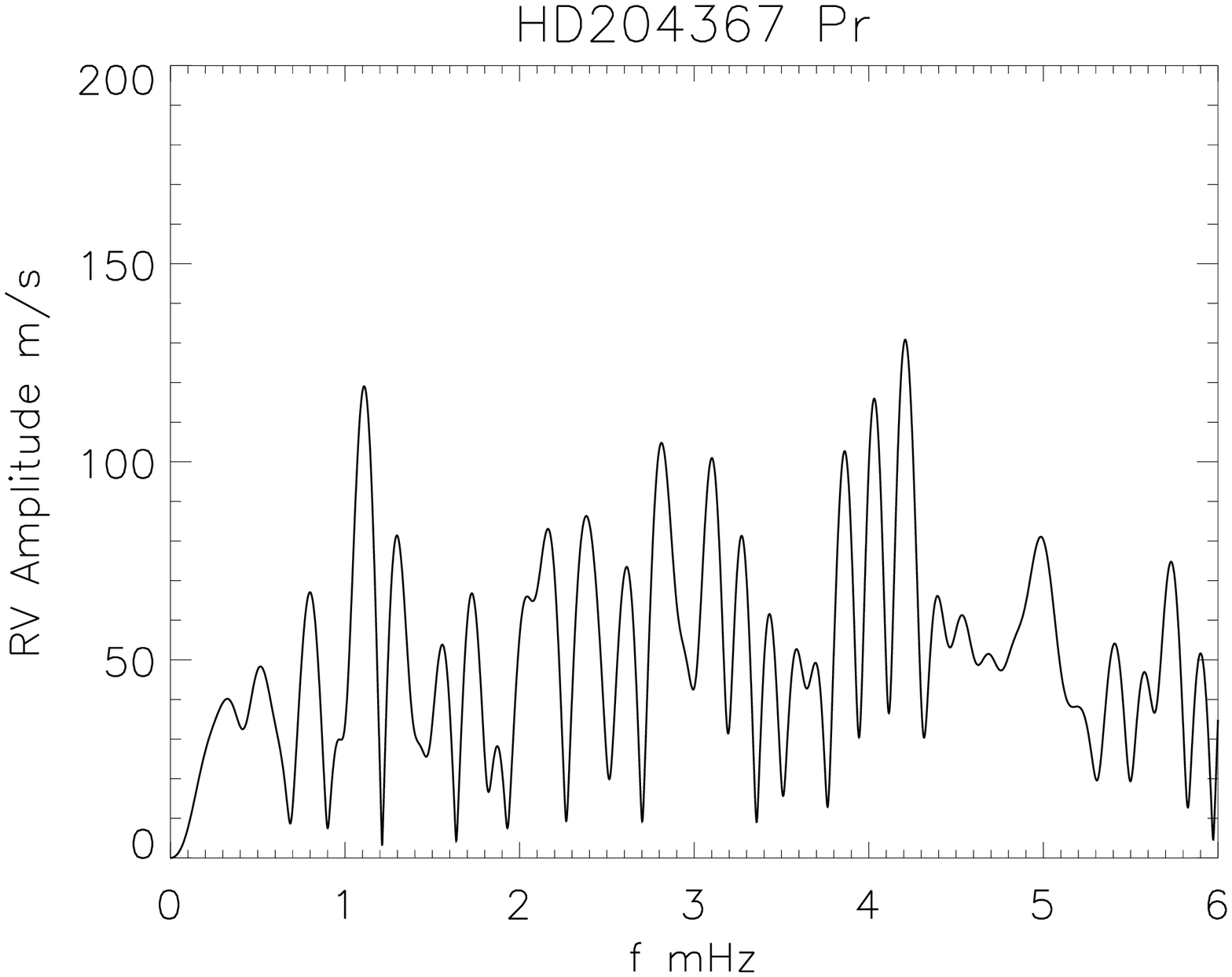}
  \includegraphics[height=5.6cm, angle=270]{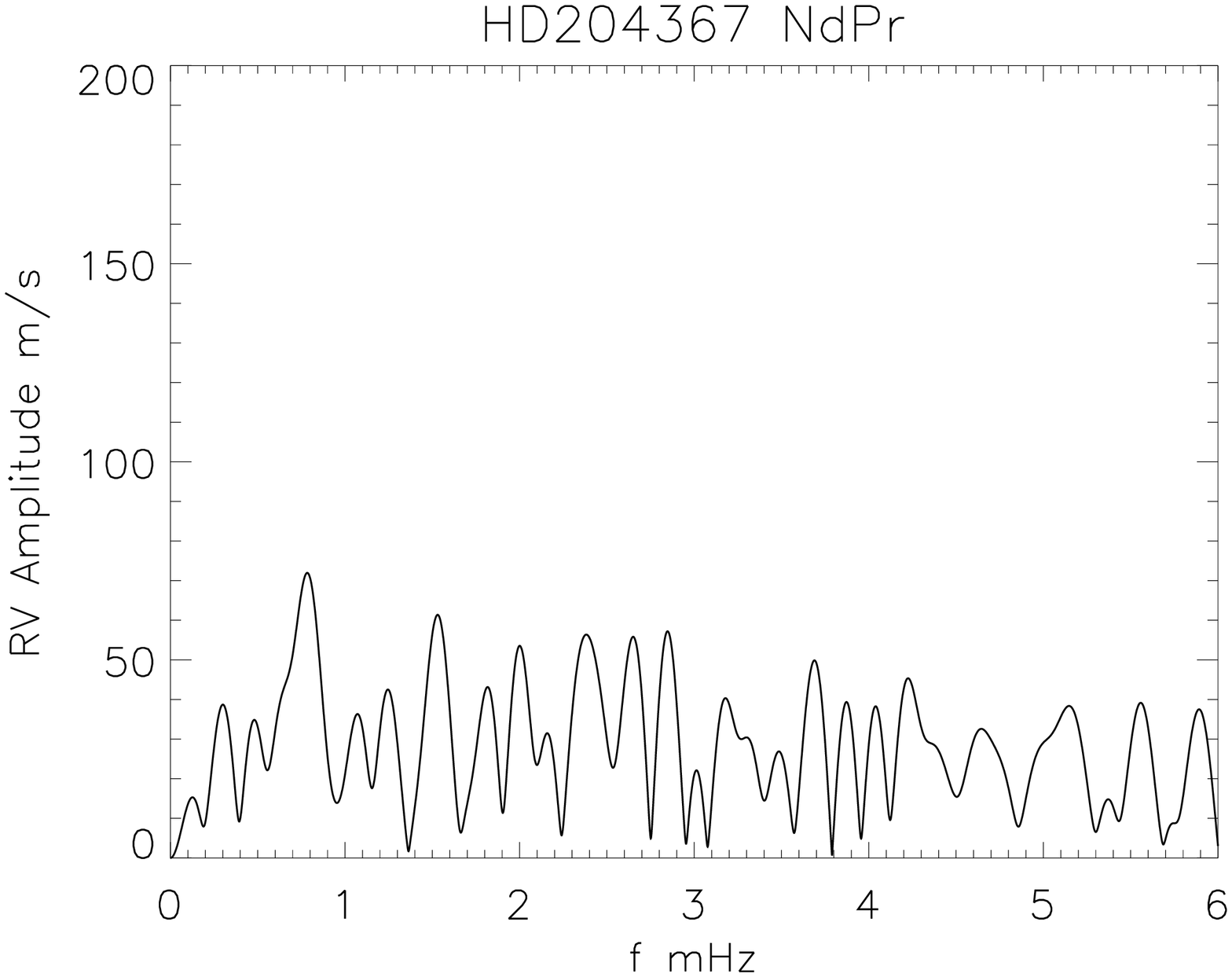}
  \includegraphics[height=5.6cm, angle=270]{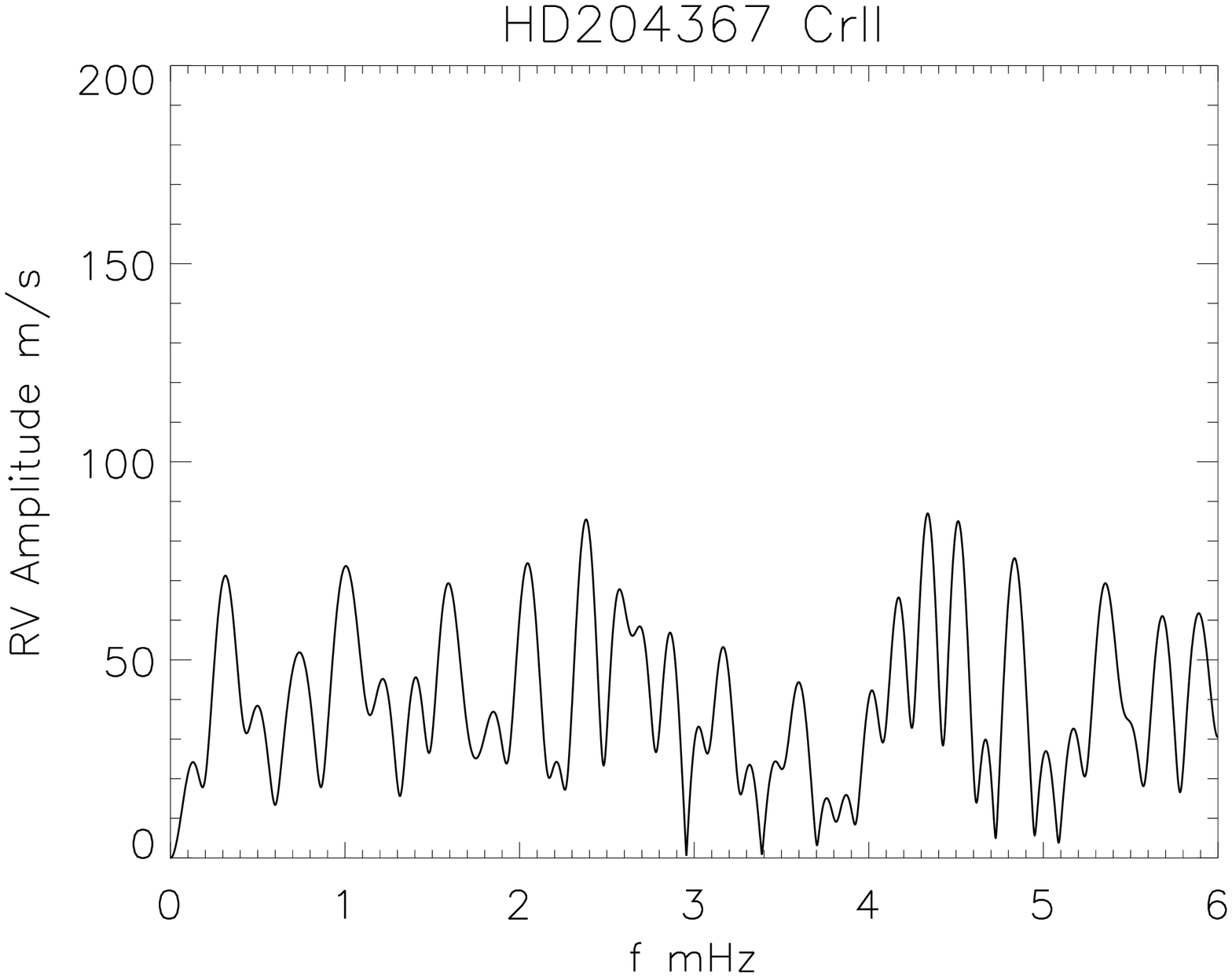}
  \includegraphics[height=5.6cm, angle=270]{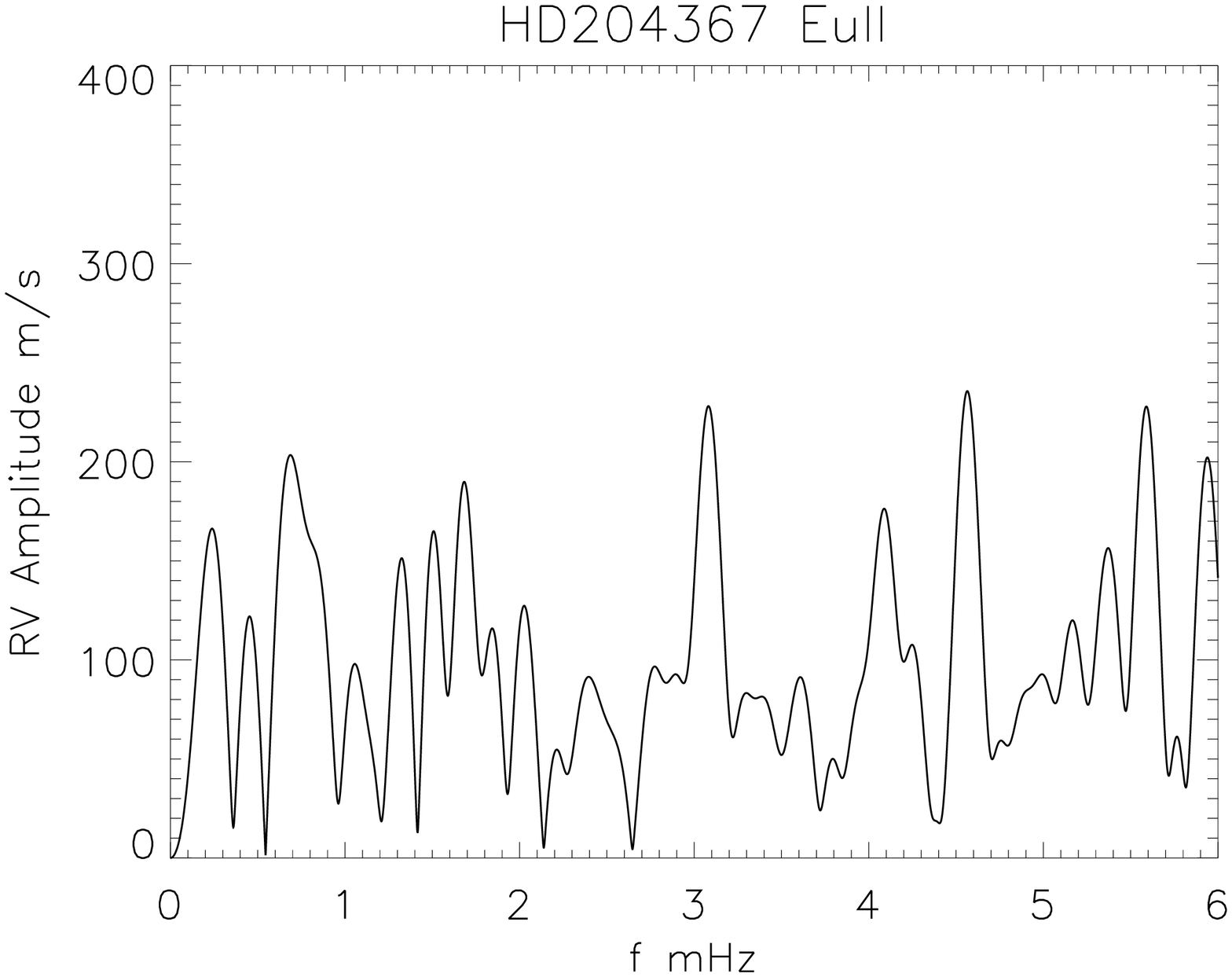}
  \includegraphics[height=5.6cm, angle=270]{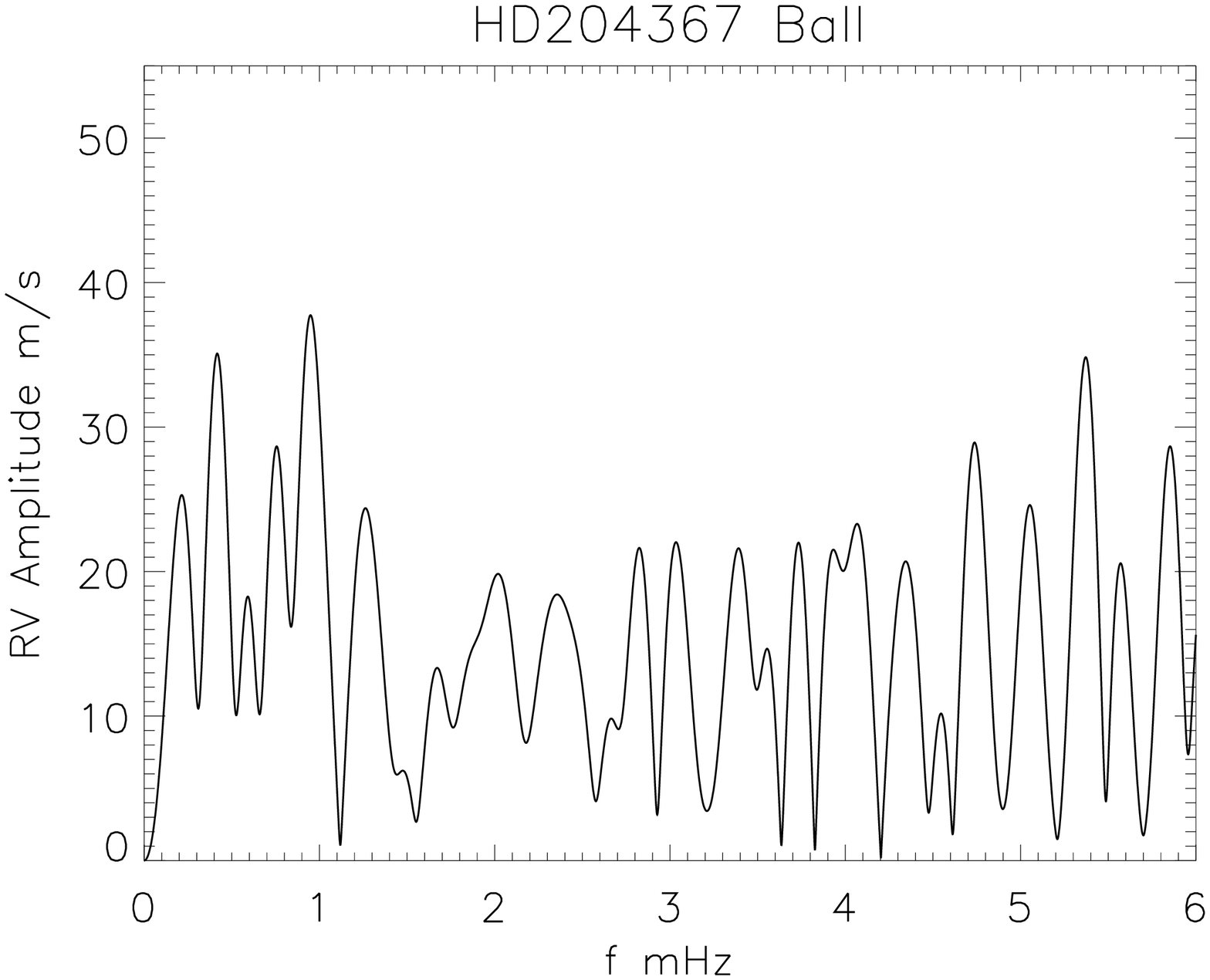}
  \includegraphics[height=5.6cm, angle=270]{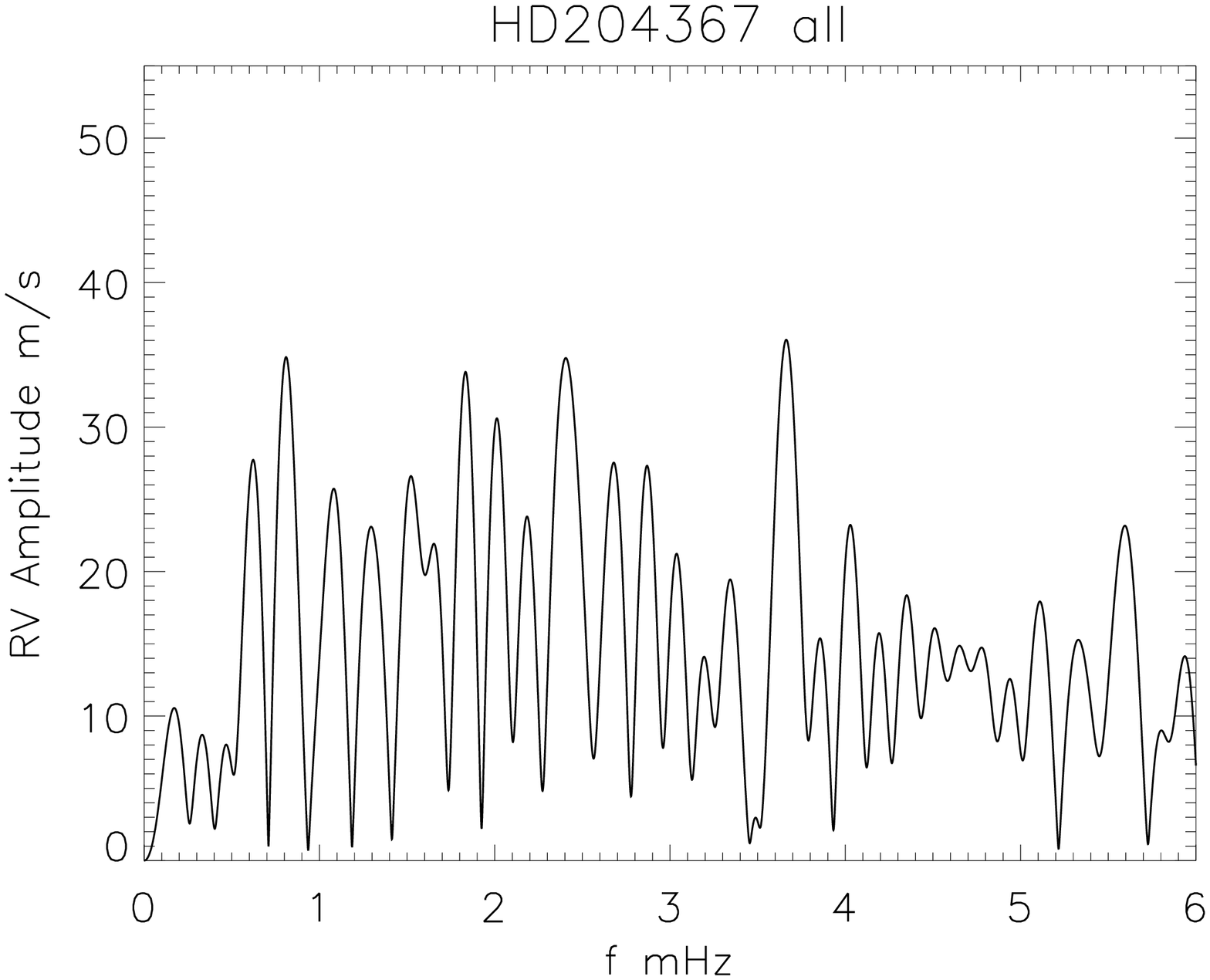}
  \includegraphics[height=5.6cm, angle=270]{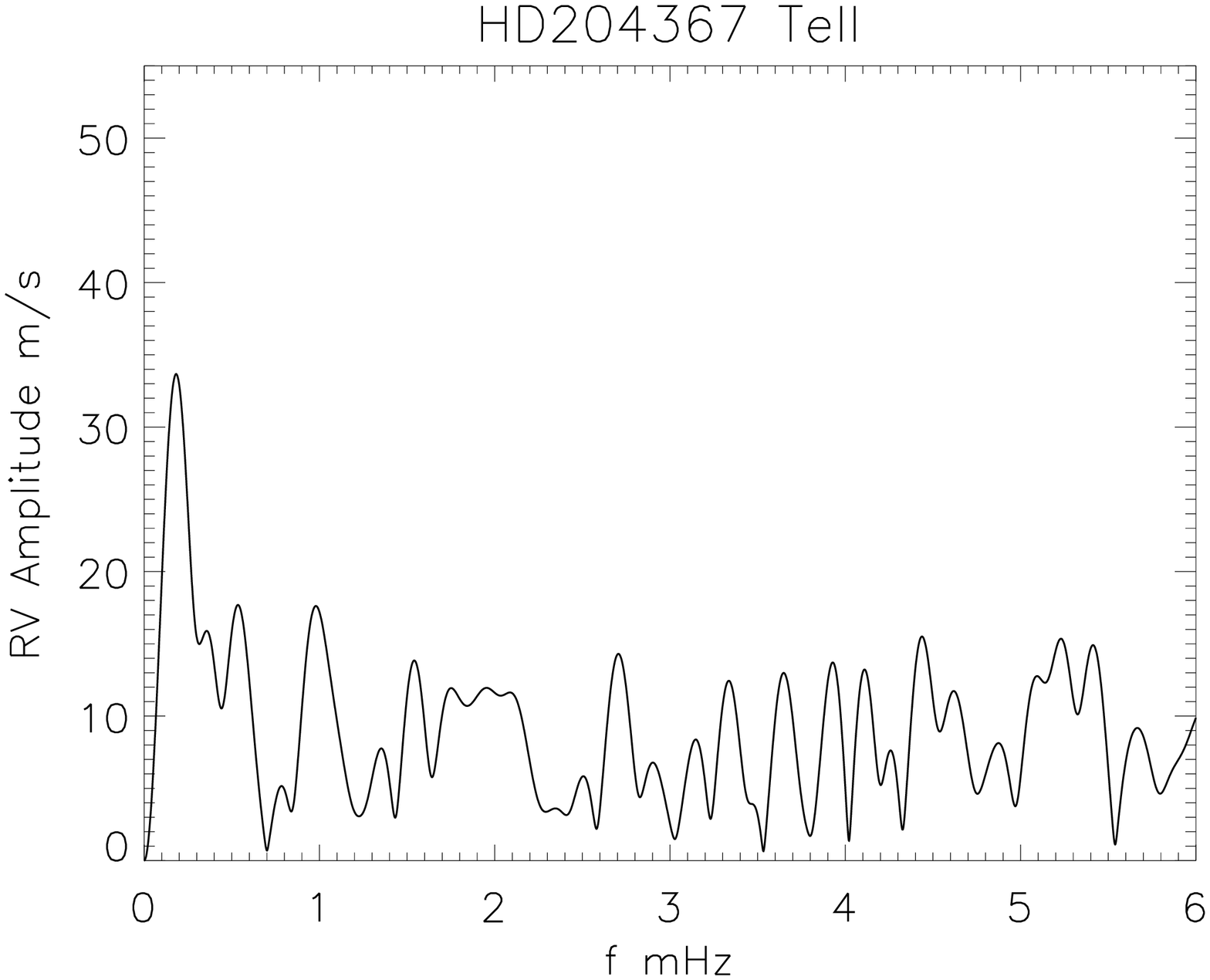}
  \caption{\label{fig:204367cog}Same as Fig.\,\ref{fig:107107cog} but
    for HD\,204367.  }
\end{figure*}

\begin{figure*}
  \vspace{3pt}
  \includegraphics[height=5.6cm,
  angle=270]{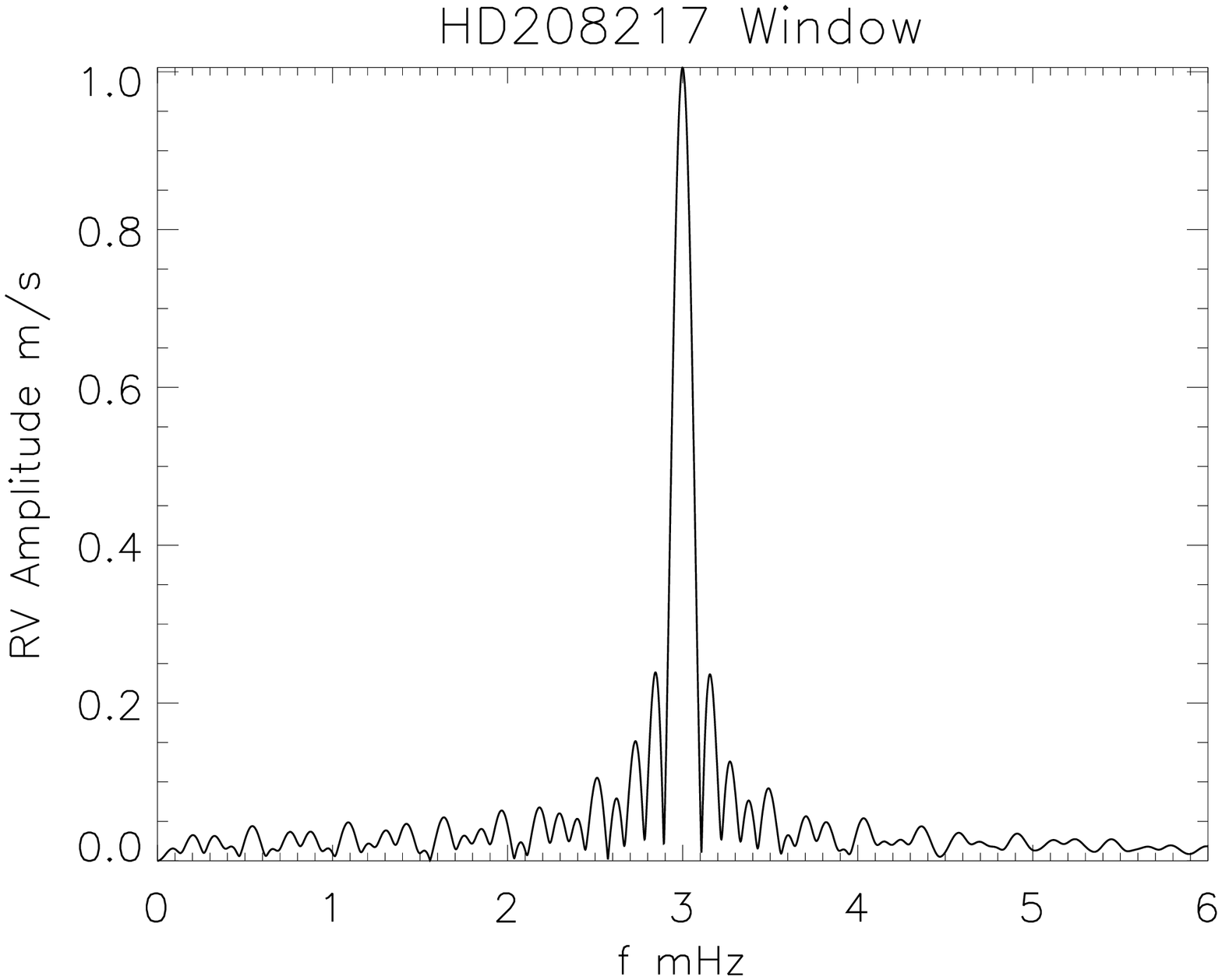}
  \includegraphics[height=5.6cm, angle=270]{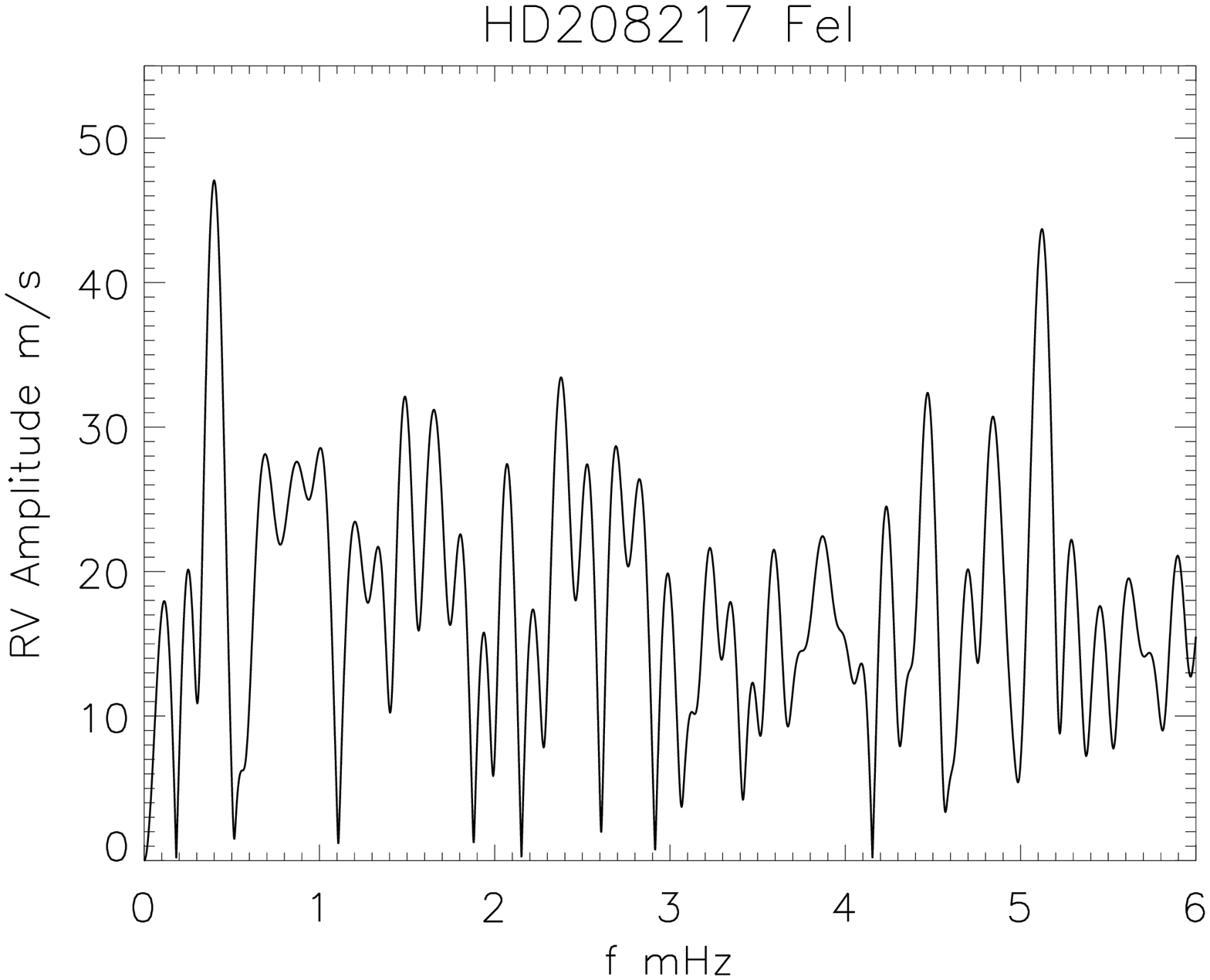}
  \includegraphics[height=5.6cm, angle=270]{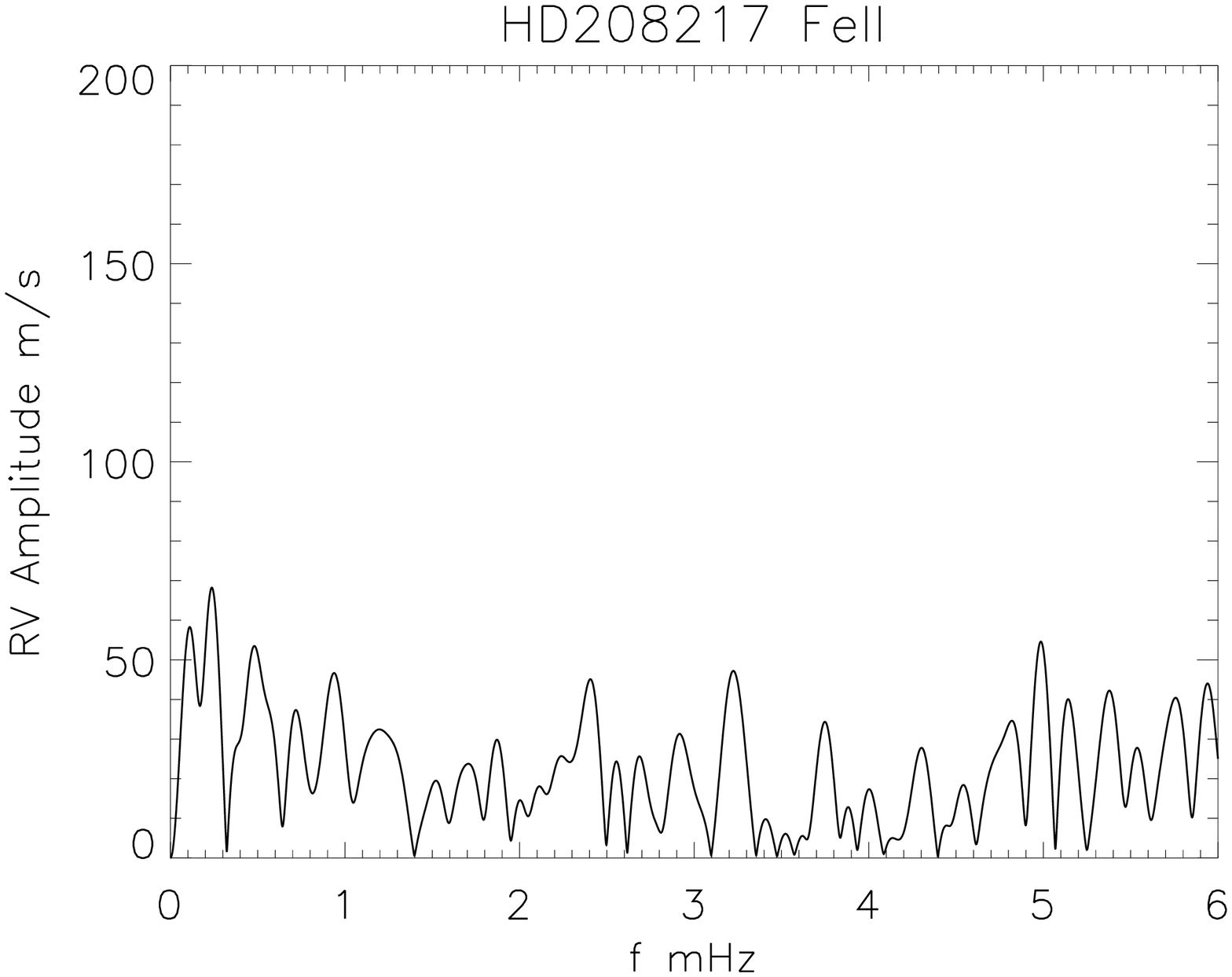}
  \includegraphics[height=5.6cm, angle=270]{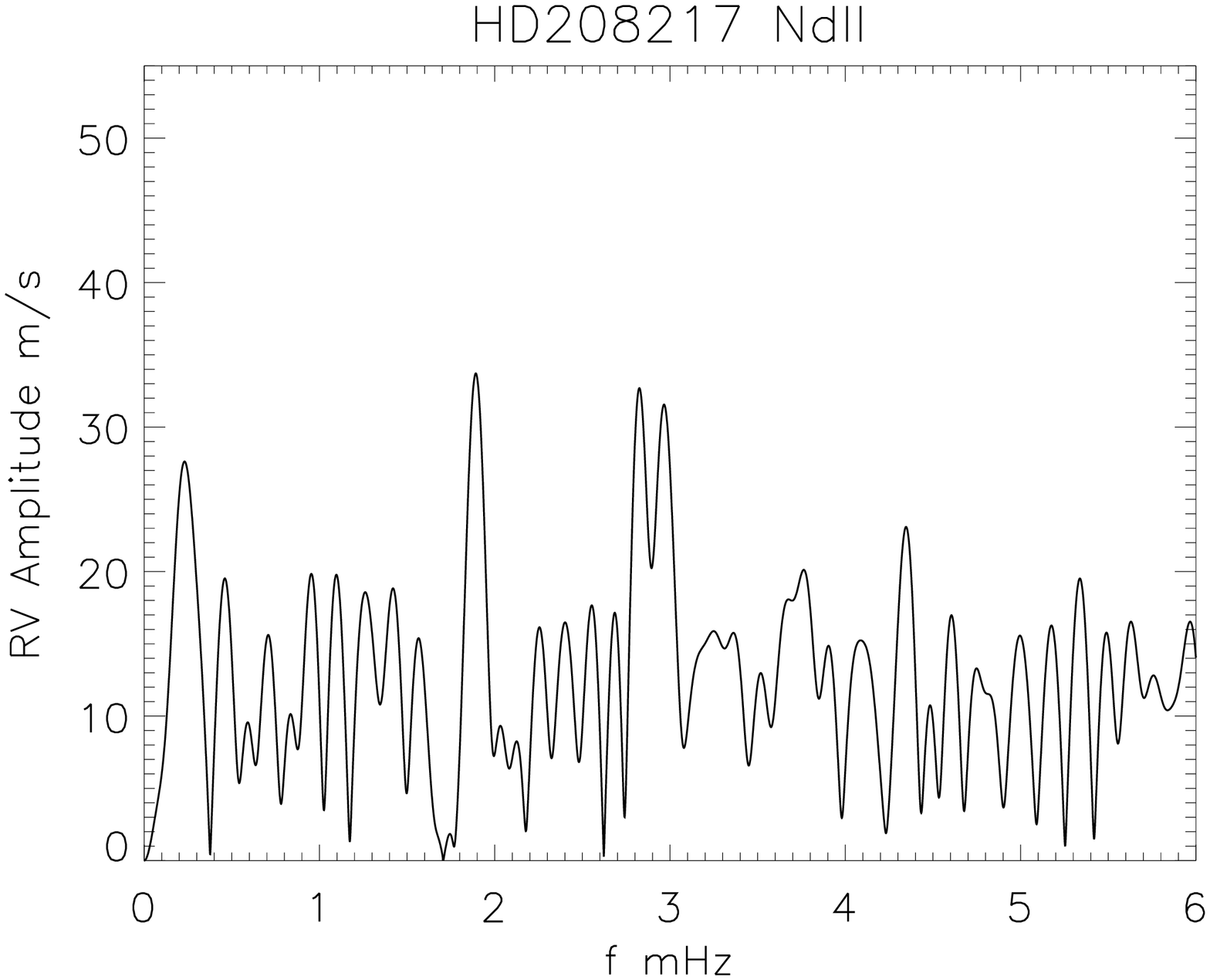}
  \includegraphics[height=5.6cm,
  angle=270]{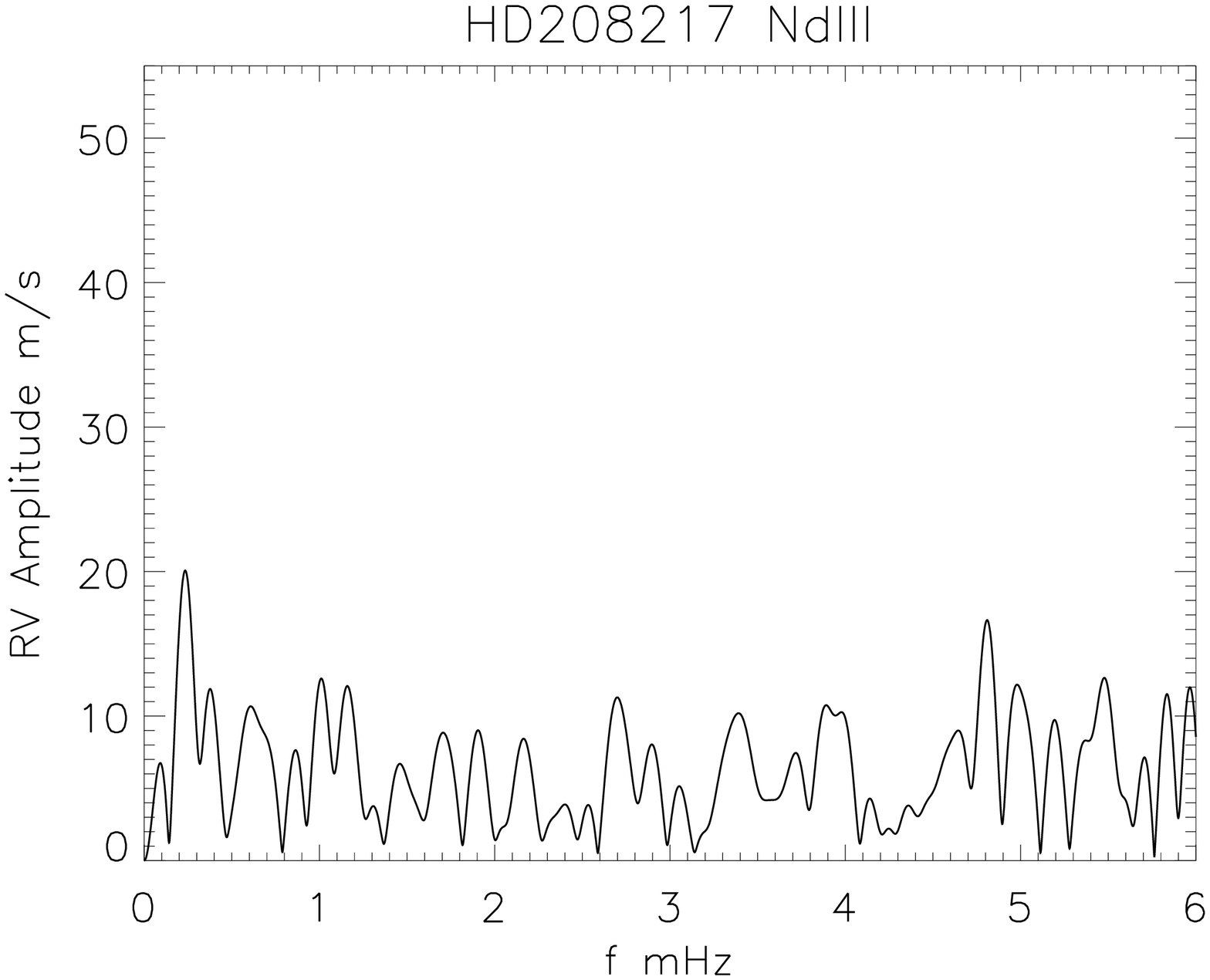}
  \includegraphics[height=5.6cm, angle=270]{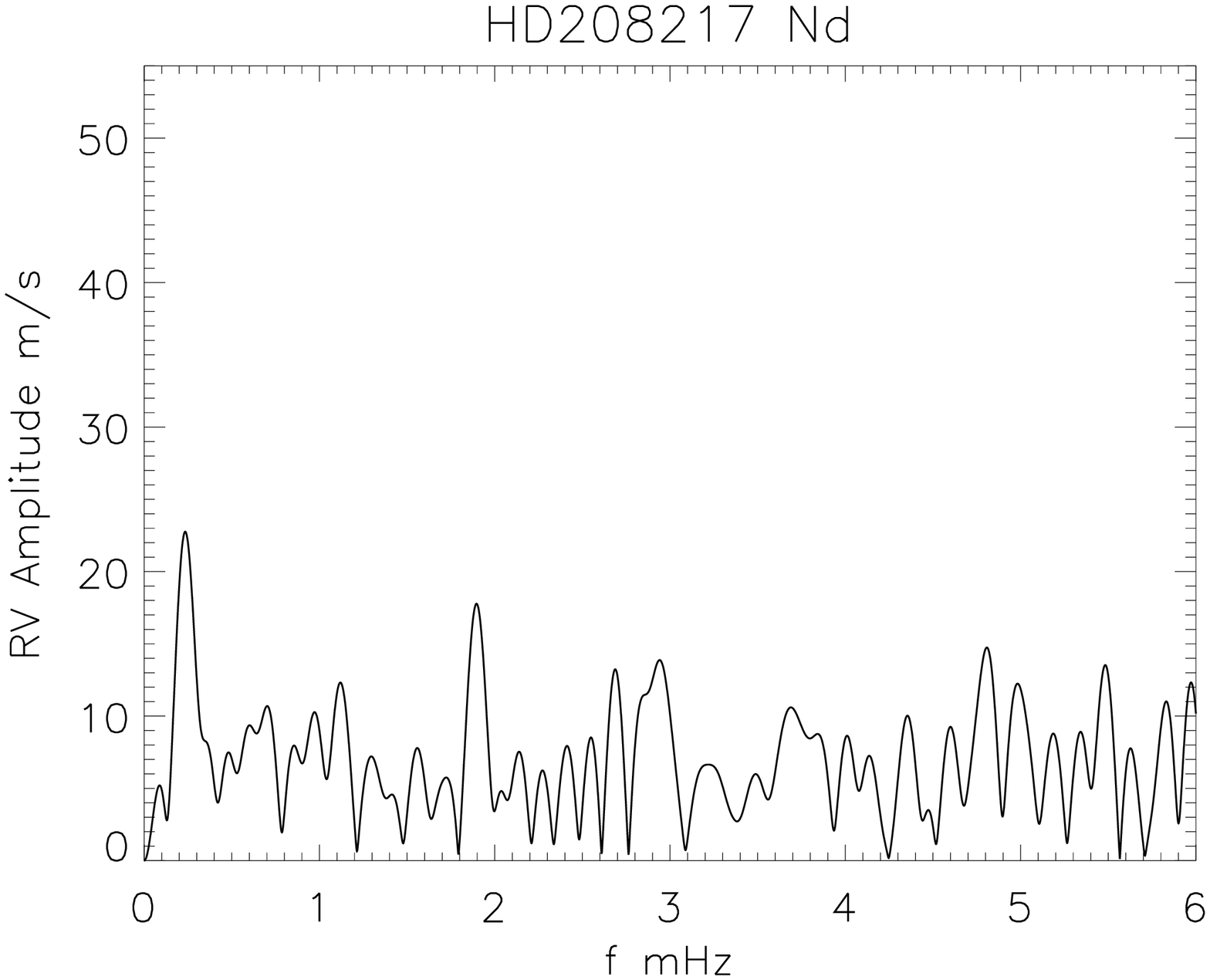}
  \includegraphics[height=5.6cm, angle=270]{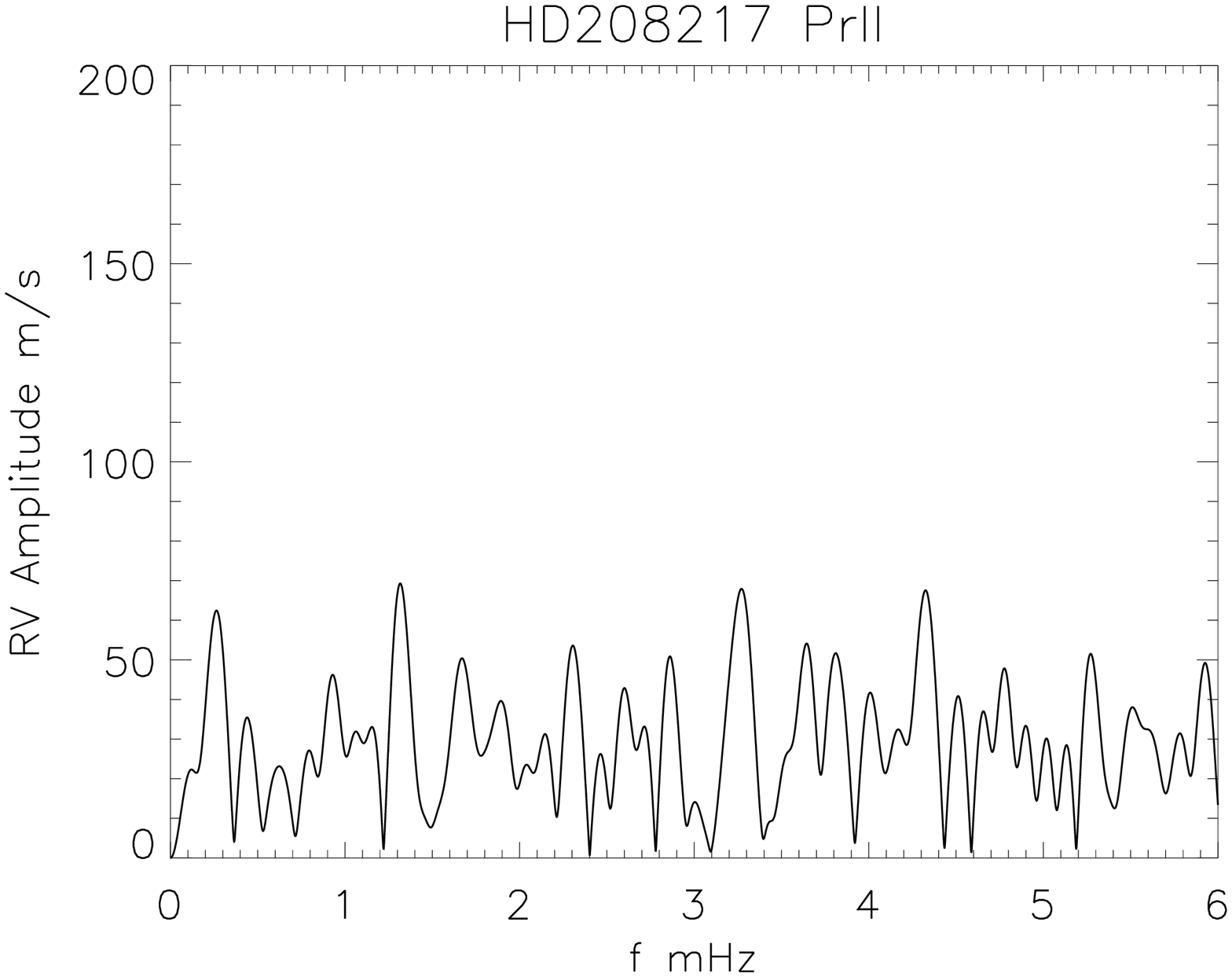}
  \includegraphics[height=5.6cm,
  angle=270]{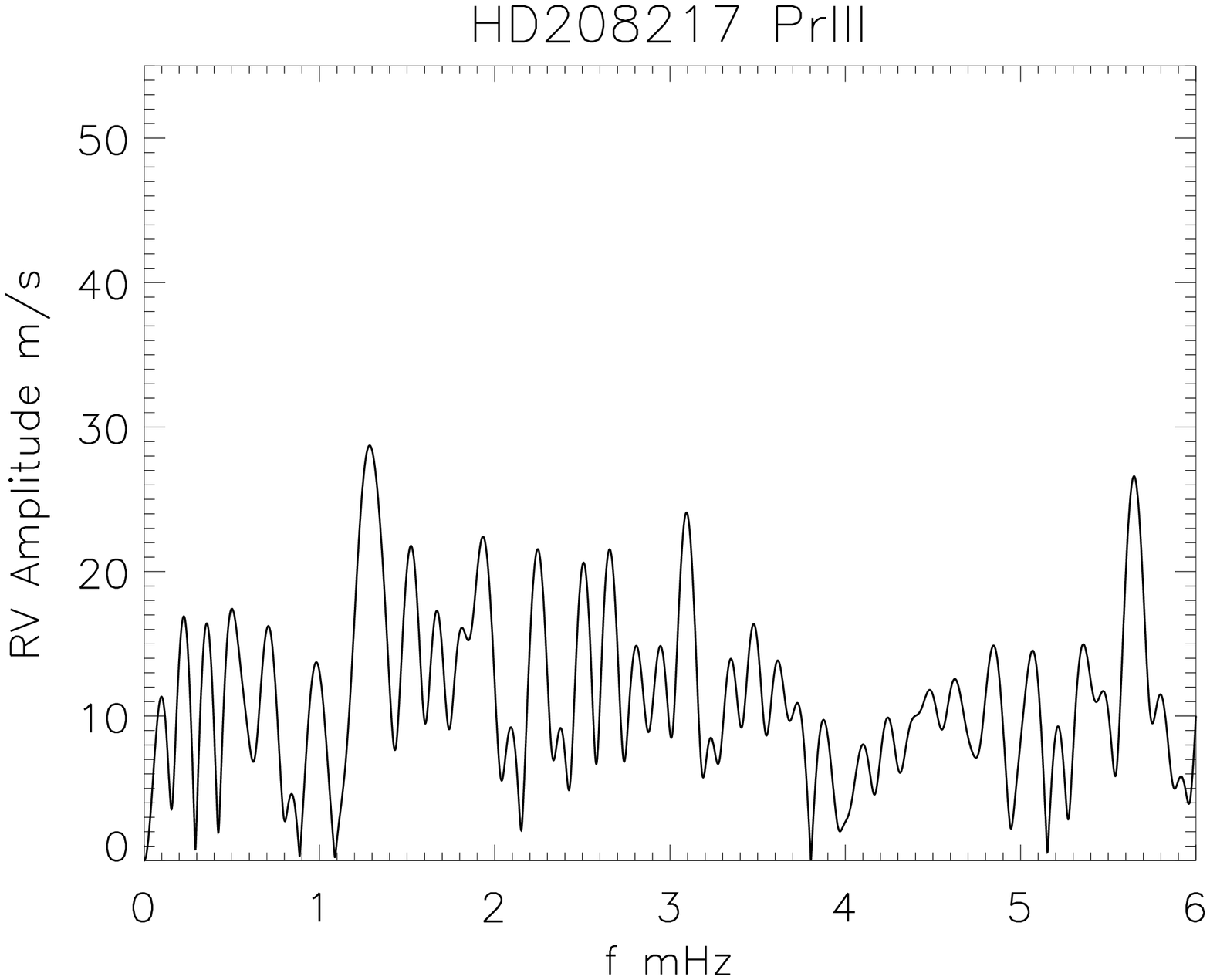}
  \includegraphics[height=5.6cm, angle=270]{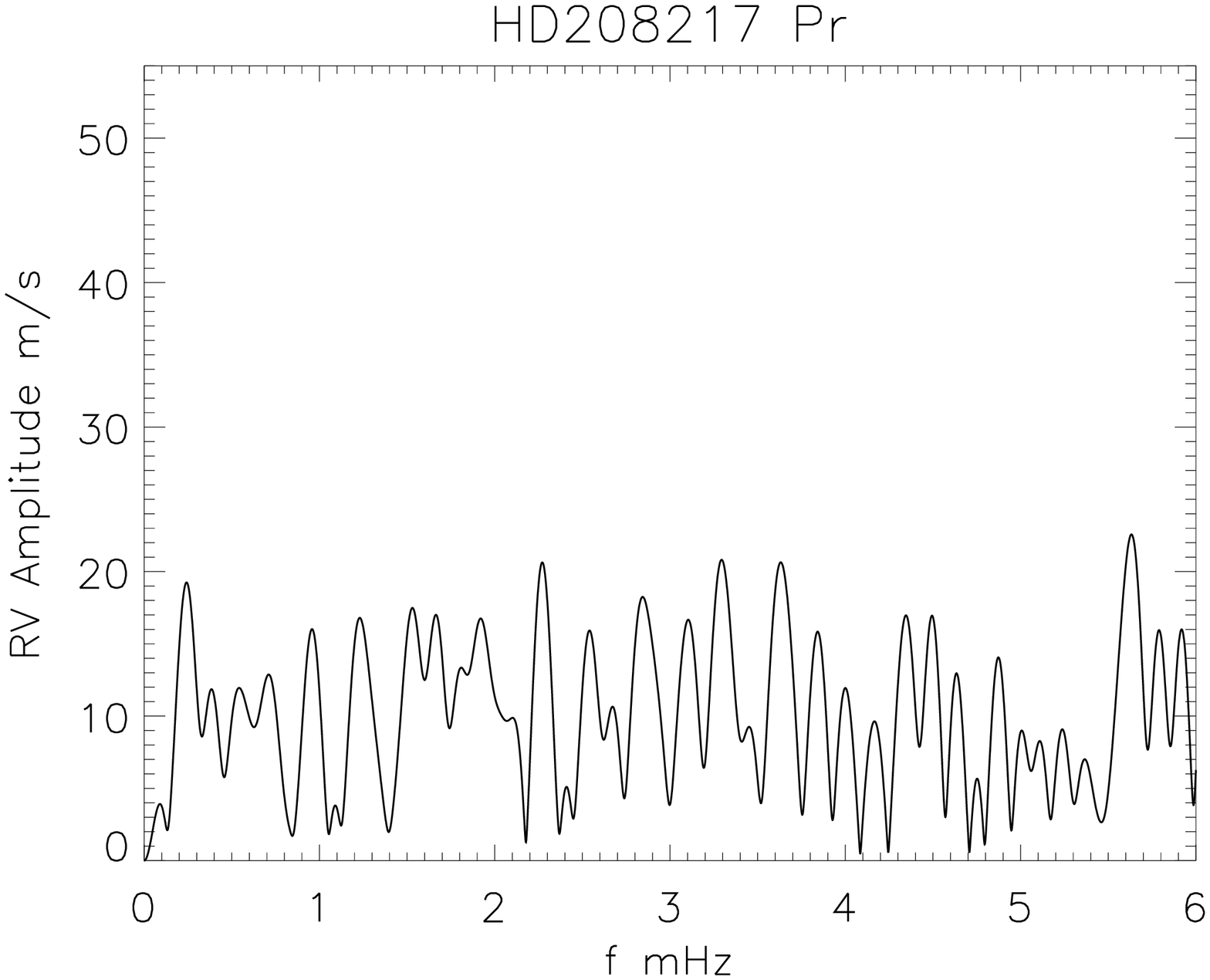}
  \includegraphics[height=5.6cm, angle=270]{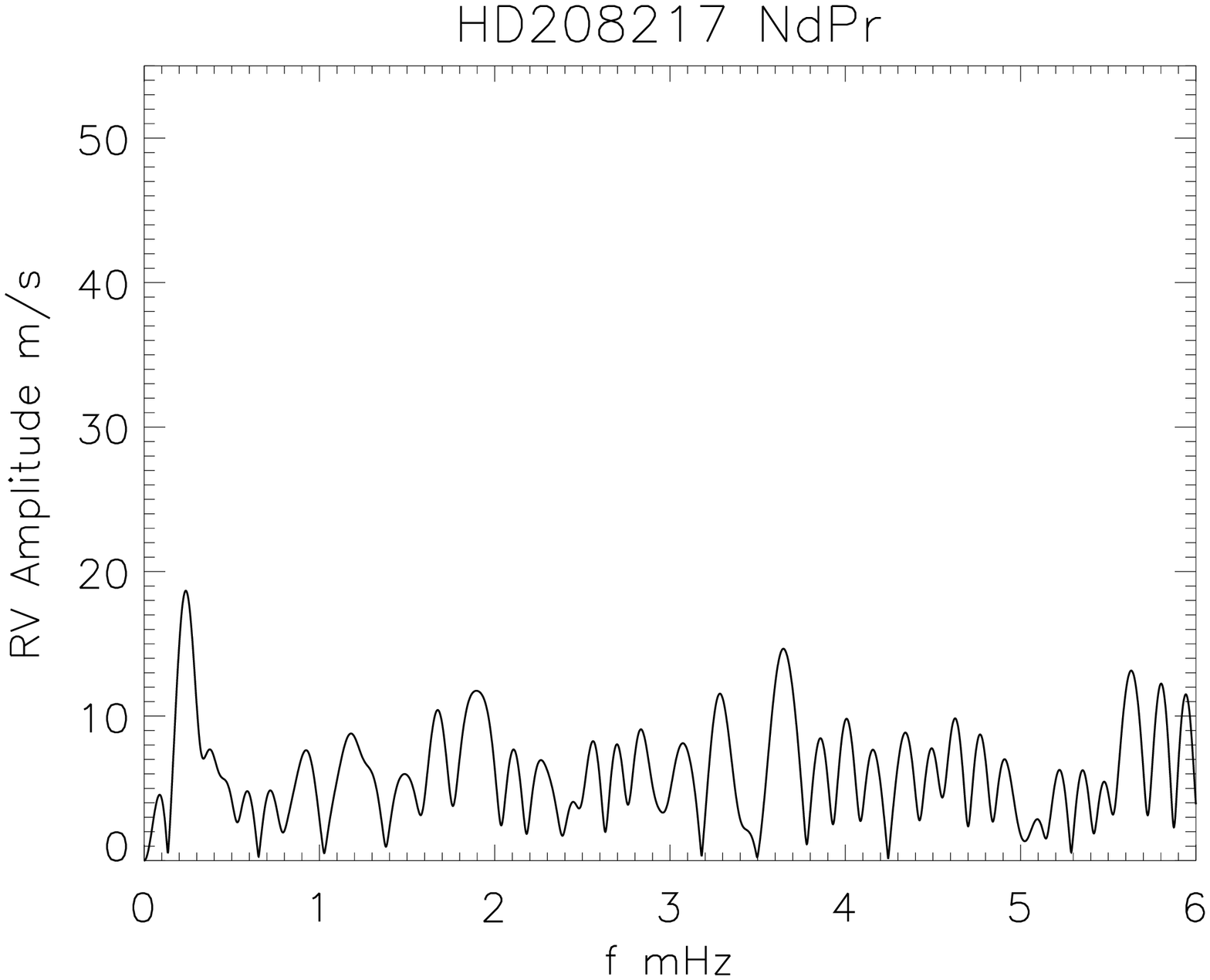}
  \includegraphics[height=5.6cm, angle=270]{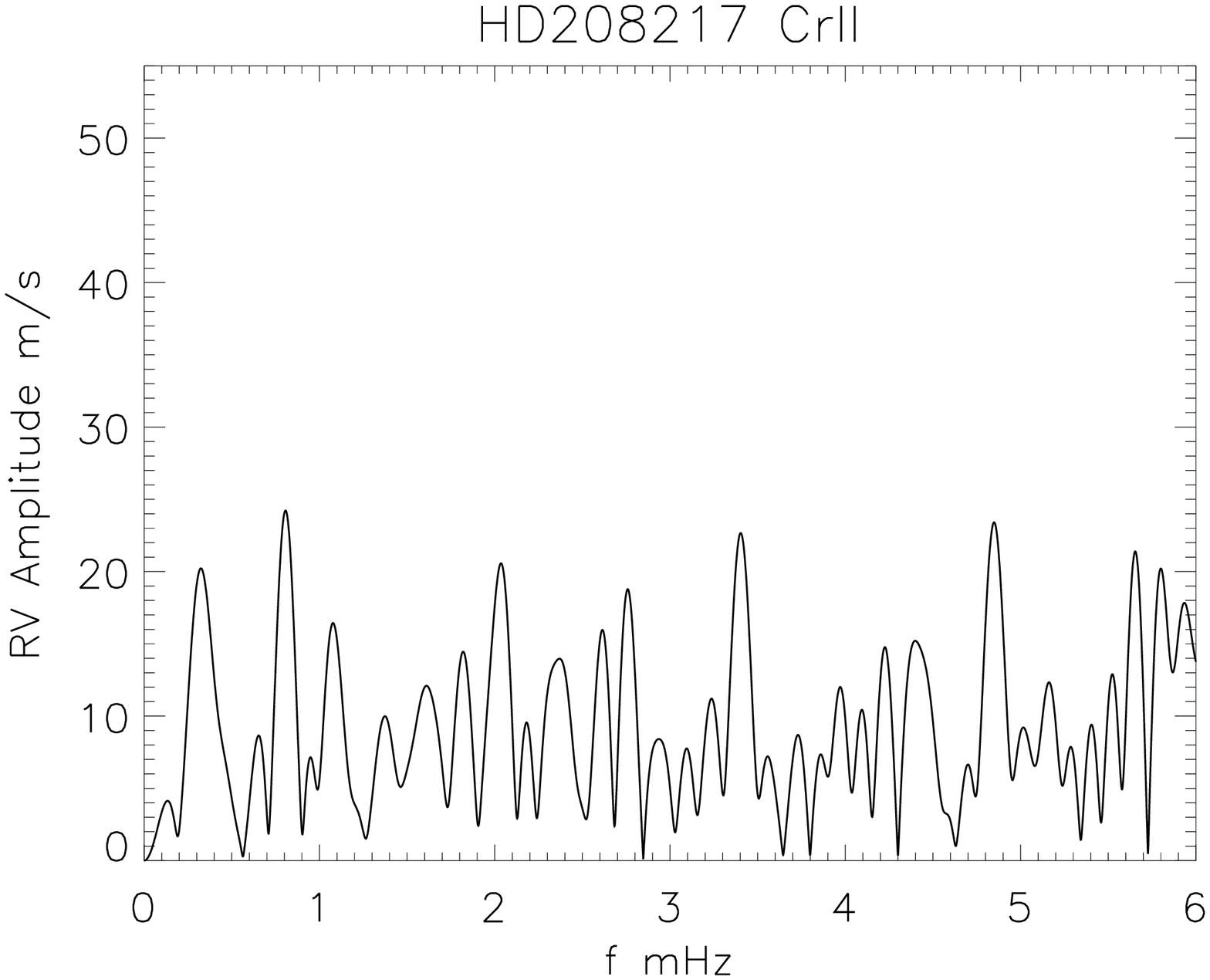}
  \includegraphics[height=5.6cm, angle=270]{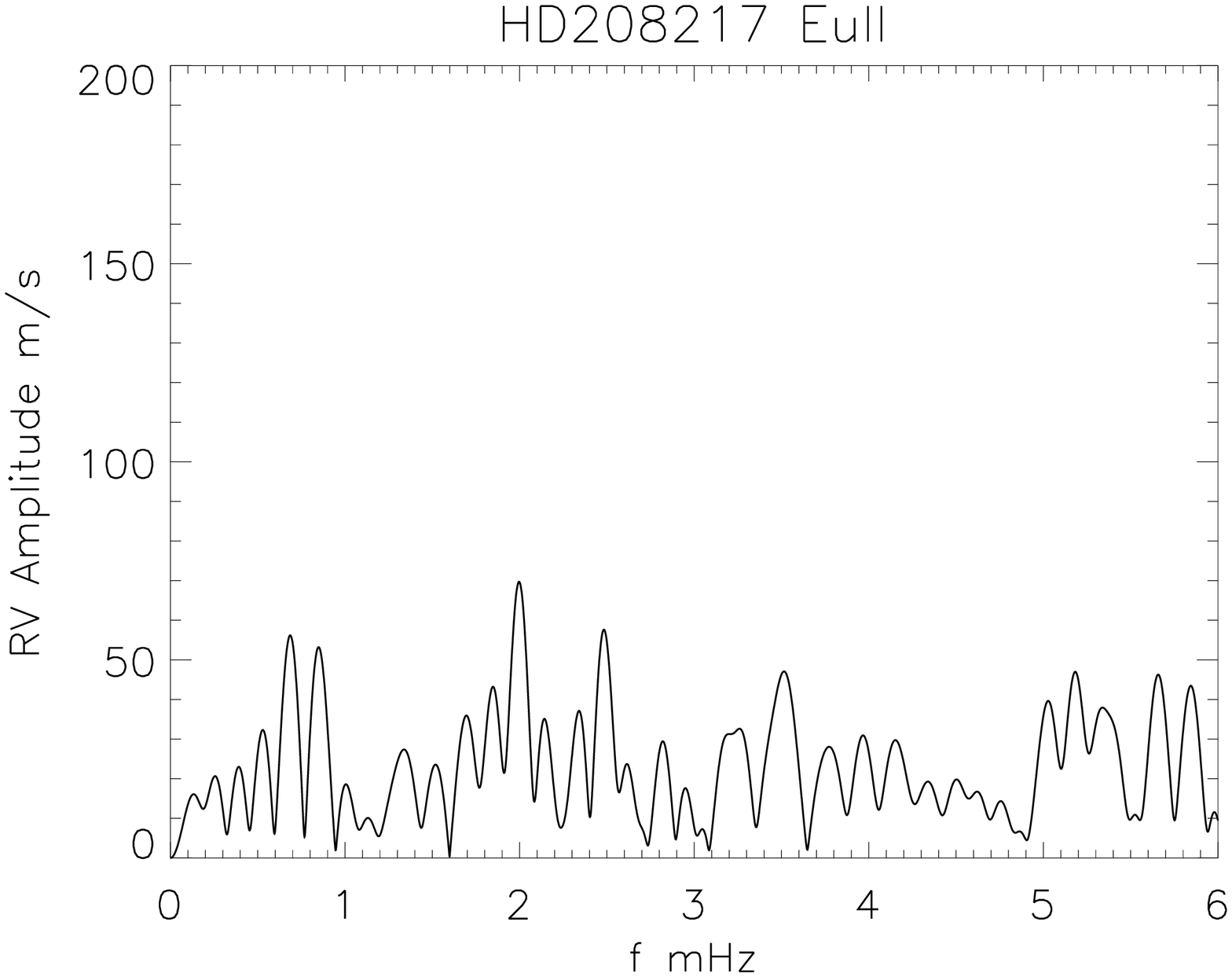}
  \includegraphics[height=5.6cm, angle=270]{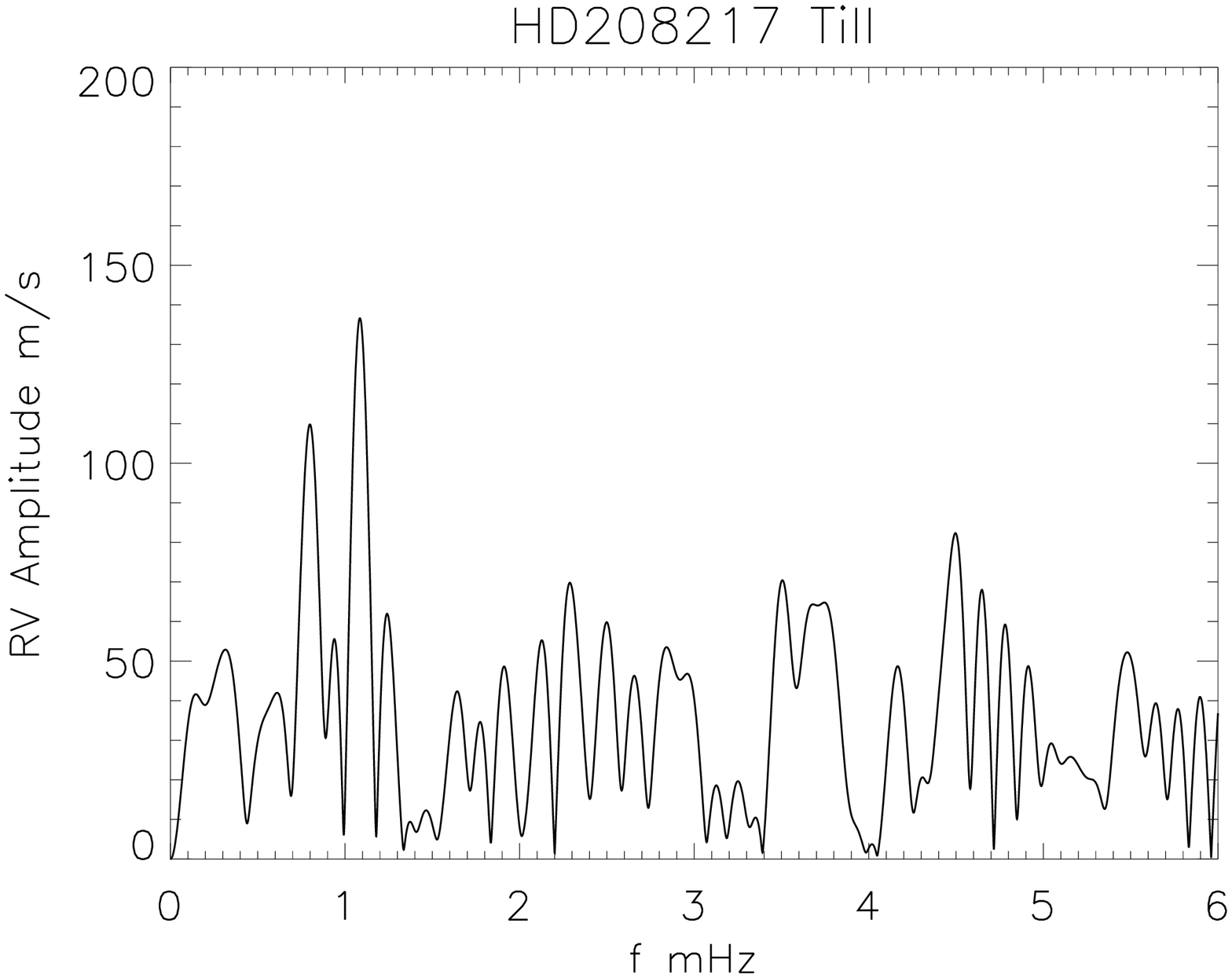}
  \includegraphics[height=5.6cm, angle=270]{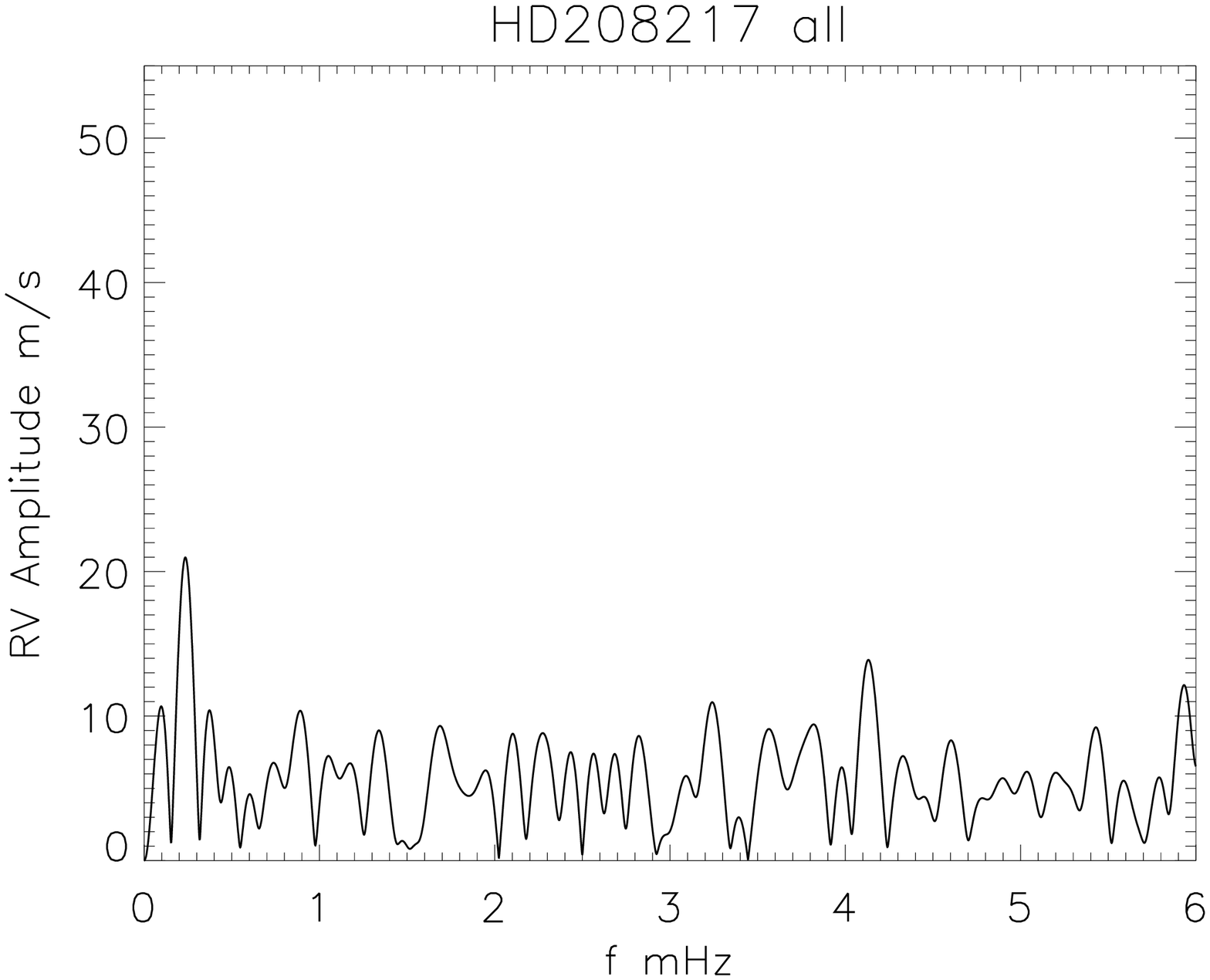}
  \includegraphics[height=5.6cm, angle=270]{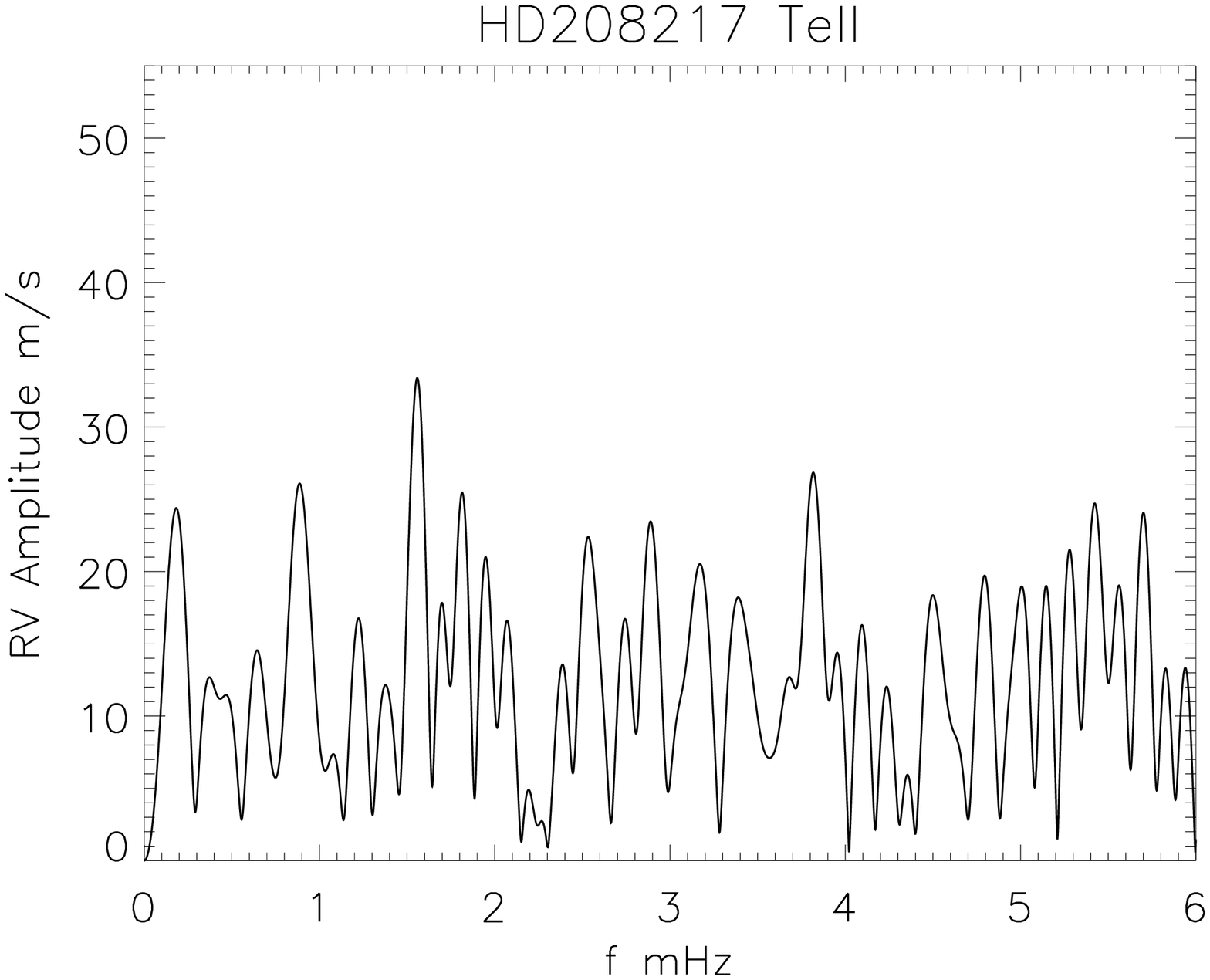}
  \caption{\label{fig:208217cog}Same as Fig.\,\ref{fig:107107cog} but
    for HD\,208217.  }
\end{figure*}

\bsp

\label{lastpage}

\end{document}